\documentclass[%
    paper=A4,               
    twoside=true,           
    openright,              
    parskip=half,           
    chapterprefix=true,     
    11pt,                   
    headings=normal,        
    bibliography=totoc,     
    listof=totoc,           
    titlepage=on,           
    captions=tablebelow,    
    chapterprefix=false,    
    appendixprefix=false,    
    draft=false,            
]{scrreprt}%

\PassOptionsToPackage{utf8}{inputenc}
\usepackage{inputenc}

\newcommand{\thesisTitle}{Black Hole Phenomenology \\ and Dark Matter Searches}
\newcommand{\thesisName}{Francesca Scarcella}
\newcommand{\thesisSubject}{Documentation}

\usepackage[english]{babel} 
\PassOptionsToPackage{
    figuresep=colon,%
    hangfigurecaption=false,%
    hangsection=true,%
    hangsubsection=true,%
    sansserif=false,%
    configurelistings=true,%
    colorize=full,%
    colortheme=bluegreen,%
}{cleanthesis}
\usepackage{cleanthesis}

\hypersetup{
    pdftitle={\thesisTitle},    
    pdfsubject={\thesisSubject},
    pdfauthor={\thesisName},    
    plainpages=false,           
    colorlinks=false,           
    pdfborder={0 0 0},          
    breaklinks=true,            
    bookmarksnumbered=true,     %
    bookmarksopen=true          %
}

\usepackage{natbib} 
\setcitestyle{square}
\setcitestyle{citesep={,}}
\setcitestyle{numbers}

\usepackage{amsmath}	
\usepackage{amssymb}	
\usepackage{subfigure}
\usepackage[capitalise]{cleveref}
\usepackage{siunitx}
\usepackage{hyperref}
\usepackage{fontawesome}
\DeclareSIUnit\parsec{pc}
\DeclareSIUnit\year{yr}
\DeclareSIUnit\solarmass{\textit{M}_\odot}

\usepackage{xspace}

\usepackage[stable]{footmisc}
\usepackage[super]{nth}

\newcommand{\fPBH}{\ensuremath{f_\mathrm{PBH}}\xspace}
\newcommand{\Msun}{\ensuremath{M_\odot}}
\newcommand{\diff}{\ensuremath{\mathrm{d}}}
\newcommand{\LCDM}{$\Lambda$CDM\xspace}

\newcommand{\rmi}{\ensuremath{\mathrm{i}}}
\newcommand{\vect}[1]{\boldsymbol{#1}}
\newcommand{\CAMB}{\texttt{CAMB}}

\newcommand{\vBH}{\ensuremath{\mathrm{v}_\mathrm{BH}}\xspace}

\newcommand{\csin}{\ensuremath{c_\mathrm{s}^\mathrm{in}}\xspace}

\begin{document}



\pagenumbering{roman}			
\pagestyle{empty}				

\pagestyle{headings}
\pagestyle{empty}

\newcommand{\HRule}{\rule{\linewidth}{1mm}}
\setlength{\parindent}{1cm}
\setlength{\parskip}{1mm}
\noindent

\noindent
\HRule
\begin{center}
\huge{\textbf{Black Hole Phenomenology\\ and Dark Matter Searches}}
 \vspace{0.2cm}
\end{center}
\HRule

\vspace{1.0cm}

\begin{center}

\large{	   
Memoria de Tesis Doctoral realizada por \\[3mm]
\textbf{\large{Francesca Scarcella}} \\[3mm]
presentada ante el Departamento de F\'isica Te\'orica \\[1mm]                  
de la Universidad Aut\'onoma de Madrid \\[1mm]
para optar al T\'itulo de Doctor en F\'isica Te\'orica \\[1mm]
}

\vspace{1.3cm}
Tesis Doctoral dirigida por  \textbf{\large{Daniele Gaggero}} \\

\end{center}

\vspace{0.7cm}

\begin{center}
{\Large {Departamento de F\'isica Te\'orica\\[1mm] Universidad Aut\'onoma de Madrid }}\\
\vspace{0.3cm}
{\Large {Instituto de F\'isica Te\'orica\\[2mm] UAM/CSIC}}\\
\end{center}

\begin{figure}[hb]
\centering
\hspace{-0.6cm}
\includegraphics[scale=.5]{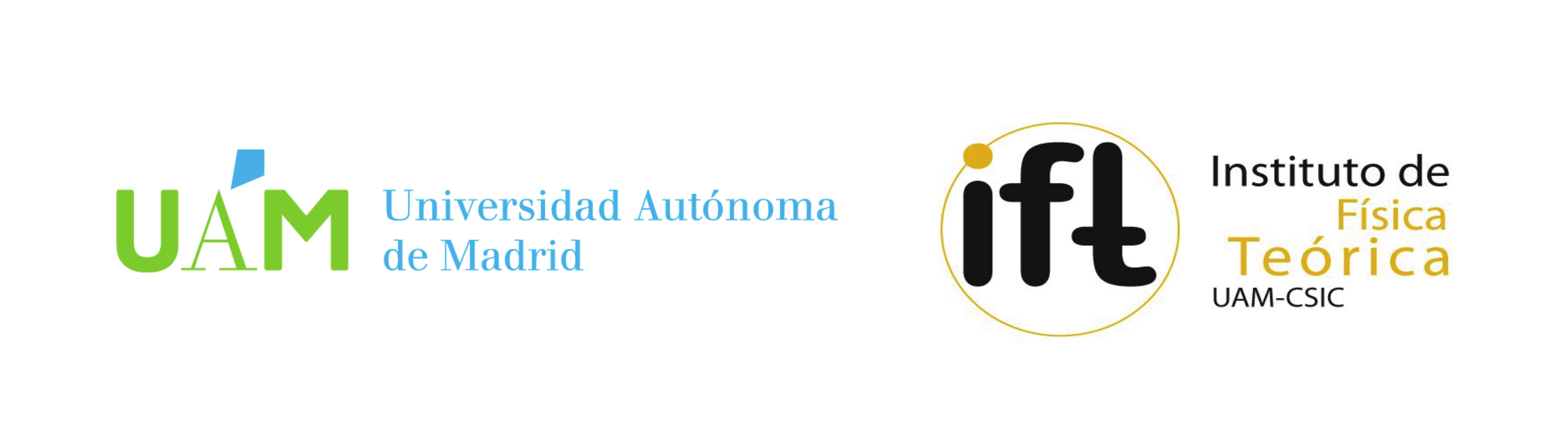}
\end{figure}

\vspace{0.2cm}

\begin{center}
{\large Madrid, Junio de 2022}
\end{center}

\newpage

\cleardoublepage

\pagestyle{plain}				
%
\pdfbookmark[0]{Abstract}{Abstract}
\addchap*{Abstract}
\label{sec:abstract}

The matter we are made of, and that we have learned to describe through the Standard Model of particle physics, makes up for only about a fifth of the total matter in the Universe.
Unravelling the nature of the remainder, the \emph{dark matter}, is undoubtedly one of the main quests of modern physics.
In this thesis we discuss some possible inter-plays between dark matter searches and the physics of black holes.

In particular, we consider black holes forming in the early Universe, long before the birth of stars and galaxies. These are known as \emph{primordial black holes}, and could originate from the gravitational collapse of large (order one) density perturbations from inflation. If black holes of this kind exist, they would make up a part, possibly even all, of the dark matter. The discovery of such objects would have far reaching consequences for the study of the dark matter, even if they were found to constitute only a subdominant component of it.  
Primordial black holes with masses ranging between a few and a hundred solar masses have recently been the subject of extensive studies and debate, following the recent detections of merging black hole binary systems through gravitational waves. In this thesis we discuss two complimentary observational channels for the observation of black holes in this mass range. 

Firstly, we consider the intense electromagnetic radiation that can be emitted by the process of gas accretion. We study the possibility of detecting isolated black holes in our Galaxy through this channel, separately examining the black hole population of astrophysical origin and an hypothetical primordial one. Multi-wavelength studies are essential for the identification of this type of sources. We consider the X-ray and radio bands, including prospects for radio detection with the future Square Kilometre Array telescope. 
Regarding the astrophysical population, our findings suggest that the detection of isolated black holes in the vicinity of the galactic centre is around the corner. We perform a complete parametric study of the uncertainty associated with this prediction. 
Turning to primordial black holes, the same observational channel can be used to constrain the abundance  of these objects in the Universe. We explore the uncertainties associated with this bound and how this dependence on the primordial black hole population model. In particular, we consider a specific well-motivated multi-modal mass distribution, which arises naturally from the thermal history of the Universe. We find that in this case constraints are significantly weakened.

In the second part of this work we turn to gravitational wave observations. At the time of writing, almost one hundred black hole merger events have been detected. While attempts have been made to disentangle the astrophysical background form a possible primordial signal in present data, any conclusion is hampered by large theoretical uncertainties on the properties of both populations. Third-generation gravitational wave detectors such as the Einstein Telescope will be able to detect mergers up to immense distances, corresponding to epochs preceding the birth of the first  stars. At such distances, the astrophysical background is expected to be absent.
We discuss the theoretical redshift dependence of the merger rates of astrophysical and primordial black holes, together with their most relevant uncertainties. Through the process of mock data generation and analysis, we assess the ability of the Einstein Telescope to identify a subdominant population of primordial black holes, disentangling it from the astrophysical one based exclusively on measurements of the distances to the events. In particular, we model and discuss the important role played by the instrumental errors on distance measurements. 
We find that the Einstein Telescope should be able to detect and constrain the abundance of primordial black holes if these constitute at least approximately one part in $  10^{5}$ of the total dark matter.

This thesis is based on publications \cite{Scarcella:2020ssk,Scarcella:2021jzp, Martinelli:2022elq, Scarcella:2022pbh}.		
%
\pdfbookmark[0]{Abstract}{Abstract}
\addchap*{Resumen}
\label{sec:abstract_esp}

La materia de la que estamos hechos, y que hemos aprendido a describir a través del Modelo Estándar de la física de partículas, constituye solo alrededor de una quinta parte de la materia total del Universo.
Desentrañar la naturaleza del resto, la \emph{materia oscura}, es sin duda una de las principales misiones de la física moderna.
En esta tesis discutimos algunas posibles interacciones entre las búsquedas de materia oscura y la física de los agujeros negros.

En particular, consideramos agujeros negros que se formaron en el Universo primitivo, mucho antes del nacimiento de las estrellas y de las galaxias. Estos se conocen como \emph{agujeros negros primordiales}, y podrían originarse a partir del colapso gravitacional de grandes perturbaciones de densidad (de orden uno) procedentes de Inflación. Si existen agujeros negros de este tipo, constituirían una parte, posiblemente incluso la totalidad, de la materia oscura. El descubrimiento de tales objetos tendría consecuencias de gran alcance para el estudio de la materia oscura, incluso si se descubriera que constituyen solo una componente menor de ella. Los agujeros negros primordiales con masas que oscilan entre unas pocas y cien masas solares han sido objeto de muchos estudios y debates en los últimos años, tras las recientes detecciones de ondas gravitacionales emitidas por la fusión de sistemas binarios de agujeros negros.  
En esta tesis discutimos dos canales complementarios para la observación de agujeros negros en este rango de masas. 

En primer lugar, consideramos la intensa radiación electromagnética que se puede emitir en el proceso de acreción de gases. Estudiamos la posibilidad de detectar agujeros negros aislados en nuestra Galaxia a través de este canal, examinando por separado la población de agujeros negros de origen astrofísica y una hipotética población primordial. 
Los estudios en múltiples longitudes de onda son fundamentales para la identificación de este tipo de fuentes. Consideramos las posibilidades de detección tanto en los rayos X como en la banda radio, incluyendo las perspectivas de detección en el radio con el futuro telescopio Square Kilometre Array. 
Con respecto a los a agujeros negros astrofísicos, nuestros hallazgos sugieren que la detección de agujeros negros aislados en las cercanías del centro galáctico está a la vuelta de la esquina. Realizamos un estudio paramétrico completo de la incertidumbre asociada a esta predicción.
En cuanto a los agujeros negros primordiales, el mismo canal de observación se puede utilizar para acotar la abundancia de estos objetos en el Universo. Exploramos las incertidumbres asociadas con este límite y su dependencia del modelo de población de agujeros negros primordiales. En particular, consideramos una  specífica distribución de masa multimodal, la cual surge naturalmente de la historia térmica del Universo. Encontramos que en este caso las cotas se relajan significativamente.

En la segunda parte de la tesis, consideramos el canal observacional de las ondas gravitacionales. Actualmente, se han detectado casi cien eventos de fusión de agujeros negros.
Si bien se han hecho intentos para diferenciar el fondo astrofísico de una posible señal primordial en los datos actuales, cualquier conclusión se ve obstaculizada por las grandes incertidumbres teóricas sobre las propiedades de ambas poblaciones. Los detectores de ondas gravitacionales de tercera generación, como el Einstein Telescope, podrán detectar fusiones hasta distancias inmensas, correspondientes a épocas anteriores al nacimiento de las primeras estrellas. A tales distancias, se espera que el fondo astrofísico esté ausente.
Discutimos la dependencia teórica en redshift de los ritmos de fusión de los agujeros negros astrofísicos y primordiales, junto con sus incertidumbres más relevantes. A través de la simulación y análisis de datos, evaluamos la capacidad del Einstein Telescope para identificar una población subdominante de agujeros negros primordiales, diferenciándola de la astrofísica, basándonos exclusivamente en mediciones de las distancias de los eventos. En particular, modelamos y discutimos el importante papel que juegan los errores instrumentales en las mediciones de distancia.
Encontramos que el Einstein Telescope debería ser capaz de detectar y medir la abundancia de agujeros negros primordiales si estos constituyen al menos alrededor de una parte en $ \sim 10^{5}$ de la materia oscura total.

Esta tesis está basada en las publicaciones \cite{Scarcella:2020ssk,Scarcella:2021jzp, Martinelli:2022elq, Scarcella:2022pbh}.

\cleardoublepage
%
%
\currentpdfbookmark{\contentsname}{toc}
\setcounter{tocdepth}{2}		
\tableofcontents				
\cleardoublepage

\pagenumbering{arabic}			
\setcounter{page}{1}			
\pagestyle{scrheadings}			

%
\chapter{The Dark Side of Matter}
\label{sec:intro}

\cleanchapterquote{How can I be substantial if I do not cast a shadow? I must have a dark side also if I am to be whole.}{C. G. Jung}{Modern Man in Search of a Soul}

\section{Black Holes}
\label{sec:intro:blackholes}

Black holes are among the most extreme yet most simple objects that can be found in the Universe. Roughly speaking, a black hole can be defined as a region of spacetime in which gravity is so strong that nothing, nor even light, can escape from within. This region is defined by the event horizon, a surface causally separating the interior from the exterior. 
By the \emph{no-hair theorem}, black holes are entirely described by only  three parameters: their mass, their spin, and their charge~\cite{Carter:1971zc,Robinson:1975bv}. 

The Schwarzschild solution of Einstein's equations in vacuum, describing a non-rotating, non-charged black hole, was found shortly after the publication of the theory of general relativity~\cite{Schwarzschild:1916uq}. The solution describing a rotating black hole was found only much later, in 1963, by Roy Kerr~\cite{PhysRevLett.11.237}. All black holes in our Universe are expected to be Kerr black holes, as we expect them to have non-zero spin and to be neutral.

In 1964 it was suggested that the highly luminous sources known as quasars were powered by an accreting supermassive black hole~\cite{1964ApJ...140..796S}.  
Soon thereafter, in the early 1970s, the X-ray source Cygnus X-1 was proposed as the first Galactic black hole candidate~\cite{1972Natur.235...37W,1972Natur.235..271B}. This source is powered by a black hole accreting gas from a companion star.

Since then, an increasing number of observations have been collected, including a large number of X-ray binaries in our Galaxy and many active galactic nuclei.\\
In the last decade, two new channels for the observation of black holes have been opened.

In 2015, the observations of gravitational waves opened a new channel for the study of black holes~\cite{Abbott:2016blz}. Almost a hundred binary black hole merger events have been observed in the few years that have followed the first detection.

\begin{figure}[tb]
	\centering
	\includegraphics[width=0.49\textwidth]{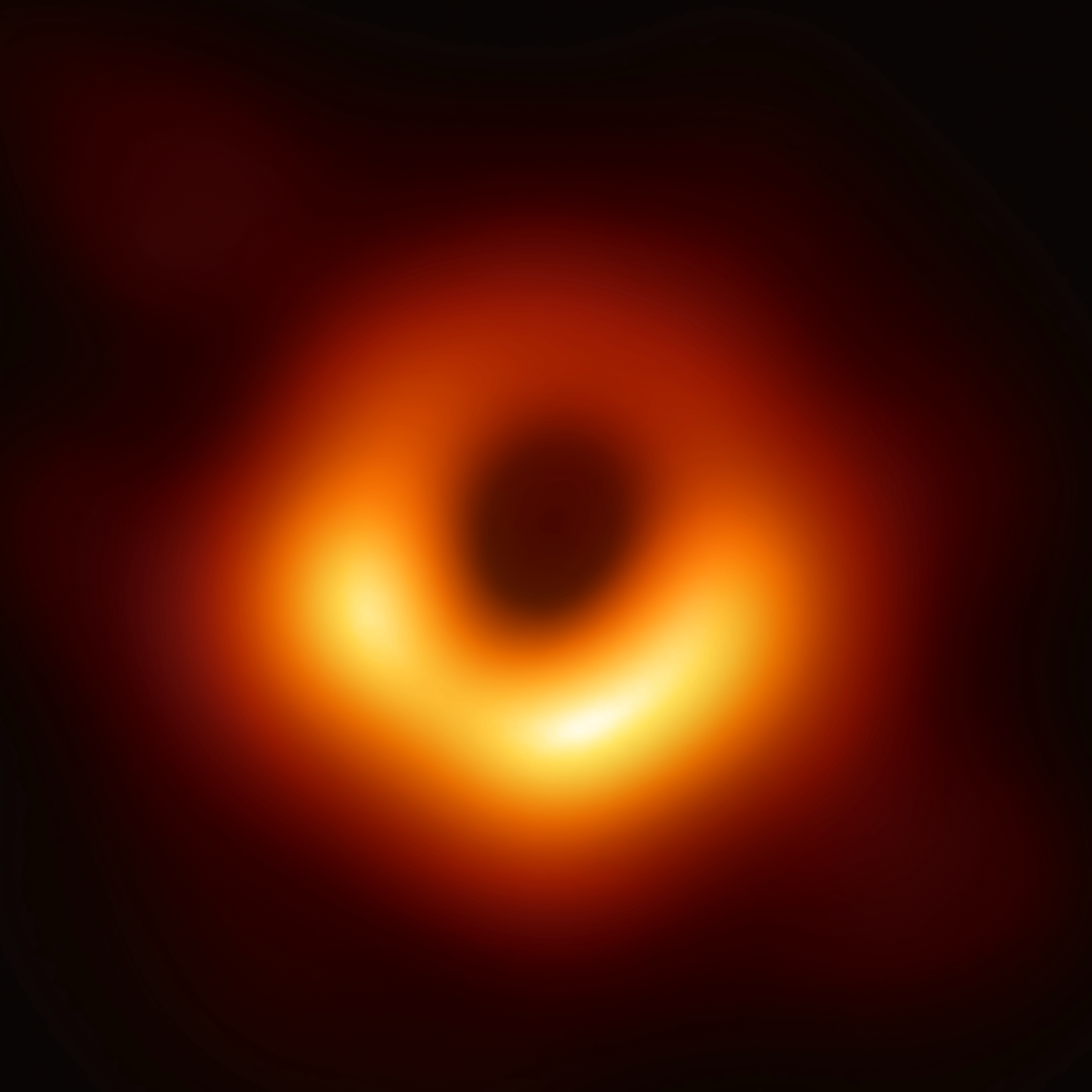}
	\hfill
	\includegraphics[width=0.49\textwidth]{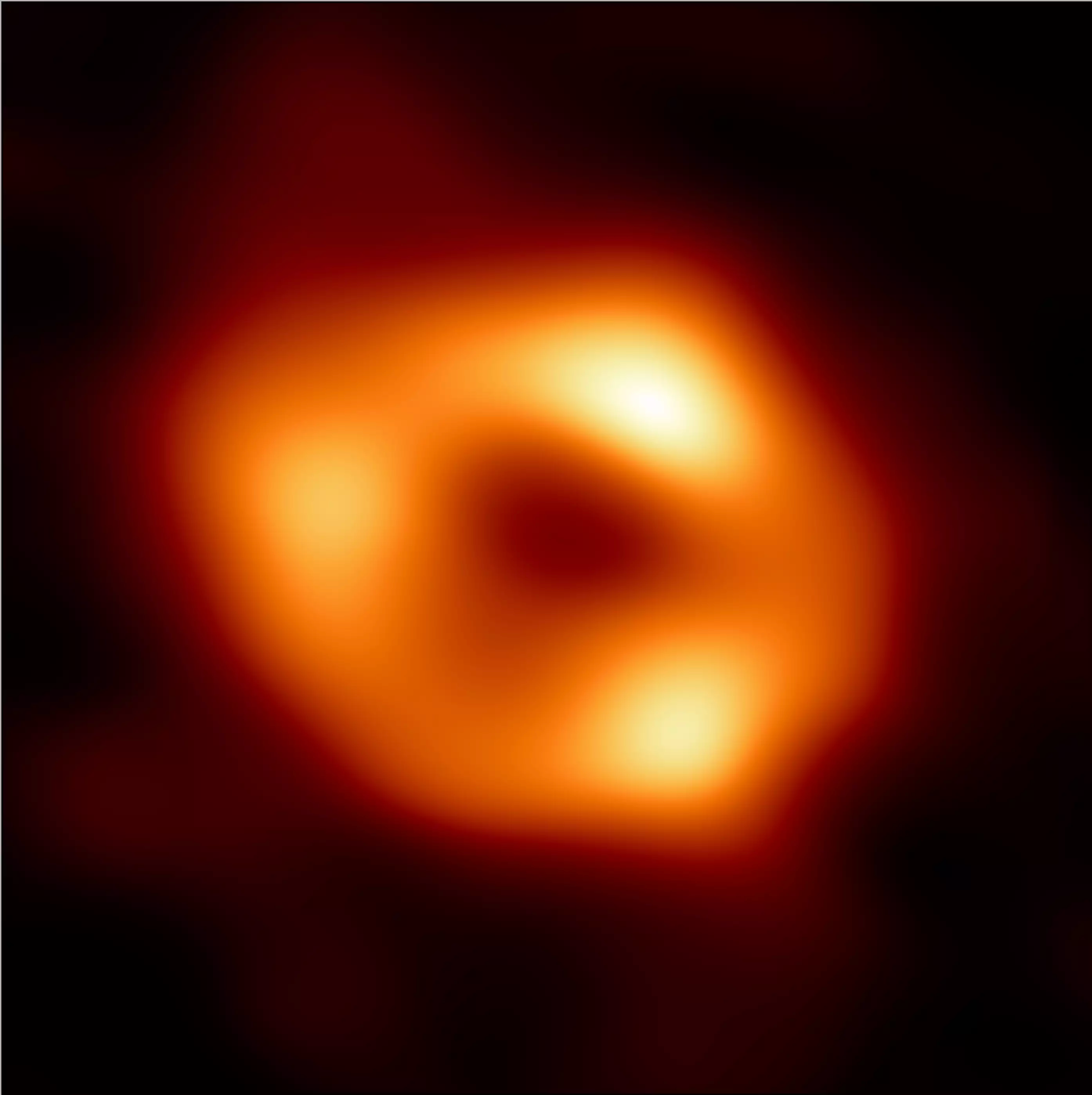}
	\caption{Direct imaging of black holes obtained by the Einstein Telescope. Left: M87. Right: Sagittarius A*.}
	\label{fig:EventHorizon}
\end{figure}

In 2019, the first direct imaging of a black hole, a supermassive black hole at the centre of the Galaxy M87, was achieved by the Event Horizon collaboration~\cite{event2019first}. During the writing of this thesis, the same collaboration also obtained an image of the supermassive black hole at the centre of our own Galaxy, Sagittarius A*~\cite{akiyama2022first}. These spectacular images are shown in \cref{fig:EventHorizon}. 

In this thesis, we discuss some of the many interplays between black holes and the mysterious substance known as the Dark Matter, which is introduced in the next section.

\section{Dark Matter}
\label{sec:intro:darkmatter}

The nature of more than one fifth of the matter composing our Universe is unknown. We know it is not made of stars, nor gas, which emit light and are visible to us. Instead, the vast majority of matter is ``dark'':  its presence is known to us only through its gravitational effects. We call it the \emph{Dark Matter}. 

The existence of the dark matter represents a central problem in modern cosmology, as well as one of the strongest indications of a limitation of our current description of fundamental physics.

Since the `30s of the last century, astronomers have dealt with the ``missing mass'' problem. The evidence in support of dark matter became increasingly extensive from the `70s on, progressively attracting the attention of a larger scientific community. 

Explanations for these observations were at first sought in the presence of a large number of dark, but ``ordinary'', objects: stars at the end of their life, cold gas, and later, neutrinos. However, it was eventually realized that none of the known forms of matter could explain the observations.
Eventually, dark matter became a field of research on its own, connecting astrophysics, early cosmology, structure formation and fundamental physics.

In this section, we briefly summarize the history and the current state of dark matter searches. Extensive reviews can be found in Refs~\cite{Bertone:2004pz, Profumo:2019ujg}, see also~\cite{Bertone:2016nfn, 2010dmp..book.....S} for historical presentations.

\subsection{Evidence}
\label{sec:intro:darkmatter:evidence}

Overwhelming evidence has been gathered for the existence of an unknown type of matter pervading our Universe, observed so far exclusively through its gravitational effects. The first hints of the existence of a large portion of invisible matter were obtained using dynamical arguments to infer the gravitational field from the motion of visible objects. The evidence for the presence of dark matter is consistent at a wide range of scales, from galactic upwards. We briefly review here the most relevant observations, in order of increasing scale.

\subsubsection{Galactic scale}

\begin{figure}[t]
	\centering
	\includegraphics[width=0.6\textwidth]{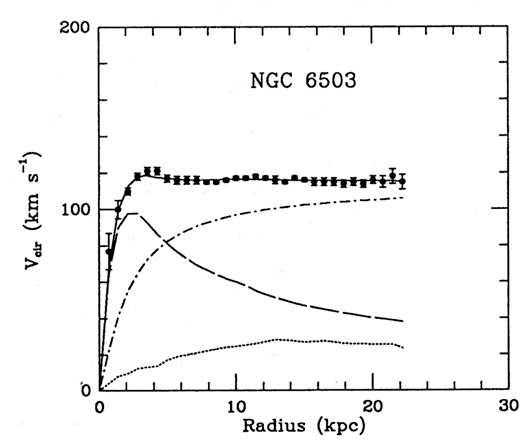}
	\caption{Rotation curve of the dwarf spiral galaxy NGC 6503. The dotted, dashed and dash-dotted lines show the contributions of gas, disk and dark matter, respectively. From Ref.~\cite{Begeman:1991iy}. }
	\label{fig:rotationcurves}
\end{figure}
%

Starting from the observations of the Andromeda galaxy performed by V. Rubin  and K.Ford~\cite{1970ApJ...159..379R}, evidence built up throughout the seventies for flat galactic rotation curves, which extended far beyond the volume occupied by the bulk of visible matter~\cite{1975ApJ...201..327R, 1978ApJ...225L.107R}, see \cref{fig:rotationcurves}.
It was soon realized that these flat curves called for the  presence of an additional matter component, whose distribution did not trace that of stars and gas~\cite{1970ApJ...160..811F,1974Natur.250..309E,  1973A&A....26..483R}.

This conclusion can be reached with a simple argument.  Based on Newtonian physics, the rotation velocity at distance $r$ from the galactic centre can be estimated as  $v \, =\, \sqrt{G\, M(r)/ r}$, where $M(r)$ is the enclosed mass and $G$ is Newton's gravitational constant.
If we assume that most of the mass is contained within a radius $R$, rotation velocities should decay as $v \propto r^{-1/2}$ beyond $R$. To obtain instead a flat curve ($v \sim \mathrm{const}$) beyond the apparent size of the Galaxy, the enclosed mass must keep growing as $M(r) \propto r $ at large distances. This implies the presence of an invisible galactic component whose density profile scales as $\rho(r) \sim r^{-2}$, a behaviour consistent with a gas of collisionless particles. \footnote{As the argument sketched above is based on the application of Newtonian dynamics, it is natural to wonder whether the application of these laws at such large scales is justified. The theory of Modified Newtonian Dynamics (MOND) was  proposed in the eighties as an alternative to the dark matter hypothesis~\cite{1983ApJ...270..365M}, suggesting that Newton's second law should be modified in the limit of small accelerations. This theory, while successful at explaining the rotation curves of galaxies, was not embedded in a consistent general relativistic model and lost attractive as further evidence for dark matter was obtained. 
Relativistic extensions of the initial MOND theory, attempting to reproduce all the data, have been elaborated in later years, see e.g.~\cite{Bekenstein:2004ne}. This partial success was obtained, however, at the cost of simplicity and with the addition of further degrees of freedom.}

It is now well established that galaxies are embedded in halos of collisionless dark matter, which can extend up to ten times the size of the hosted galaxy and account for approximately the 80\% of the total mass. In our Galaxy, the local velocity dispersion of stars allows to constrain the dark matter density in the solar vicinity. It is found to be in the range $ 0.4 - 0.7$ GeV $\mathrm{cm}^{-3}$~\cite{Benito:2020lgu}. 

\subsubsection{Galaxy clusters}

Similar dynamical arguments can also be applied at the scale of galaxy clusters. 
In his pioneering 1933 work~\cite{1933AcHPh...6..110Z}, F. Zwicky applied the virial theorem to estimate the mass of the Coma Cluster from the measurements of the velocities of the galaxies composing it.
He found a very large discrepancy between the total mass inferred this way and the sum of the estimated masses of the galaxies.
Although his estimate was rough, the conclusion was correct: most of the mass in the Coma Cluster is not contained in galaxies.
Today the total masses of galaxy clusters are estimated with much better precision through gravitational lensing, and it is consistently found that the luminous matter only accounts for a subdominant fraction of the total mass.

\begin{figure}[t]
	\centering
	\includegraphics[width=0.49\textwidth]{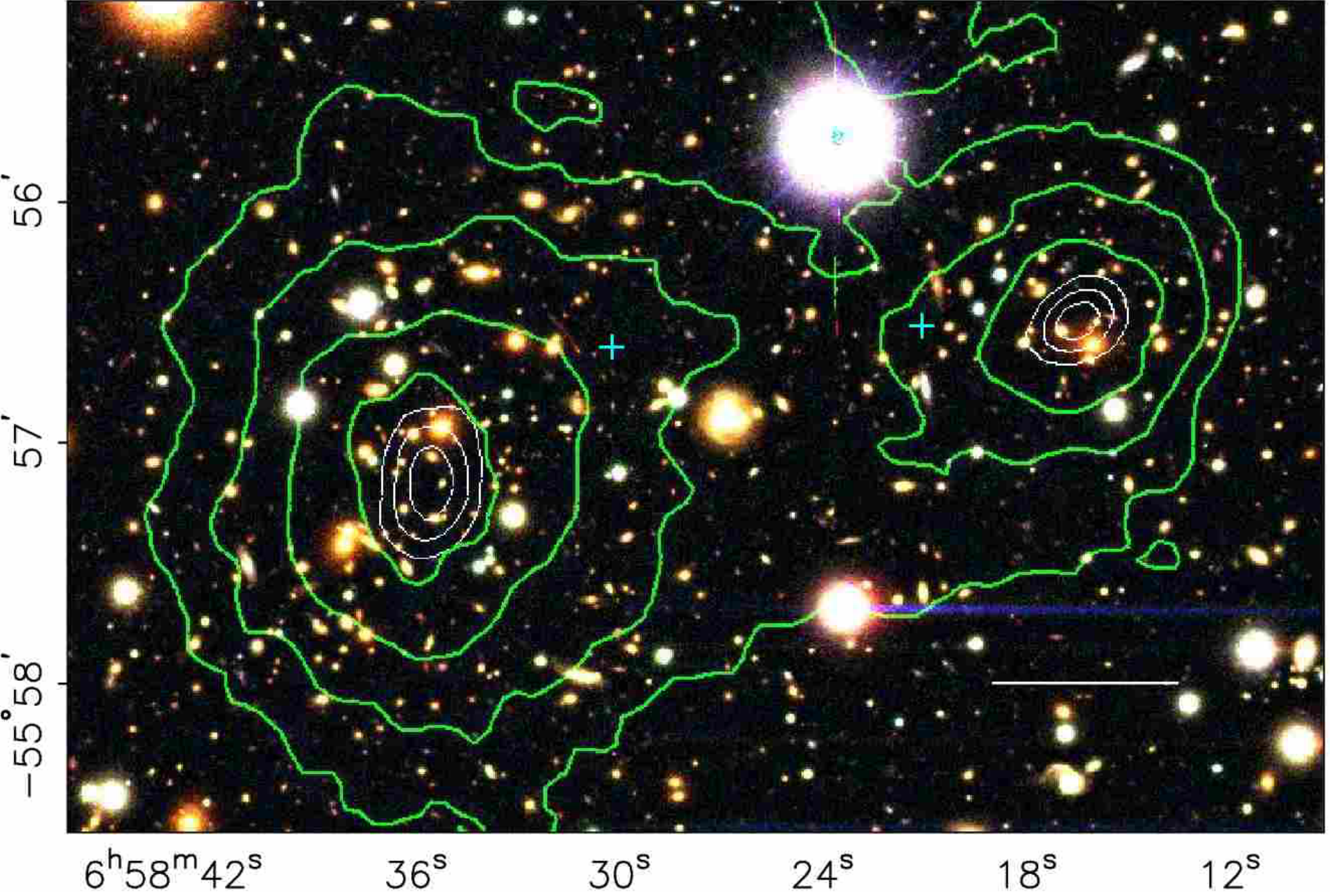}
	\includegraphics[width=0.48\textwidth]{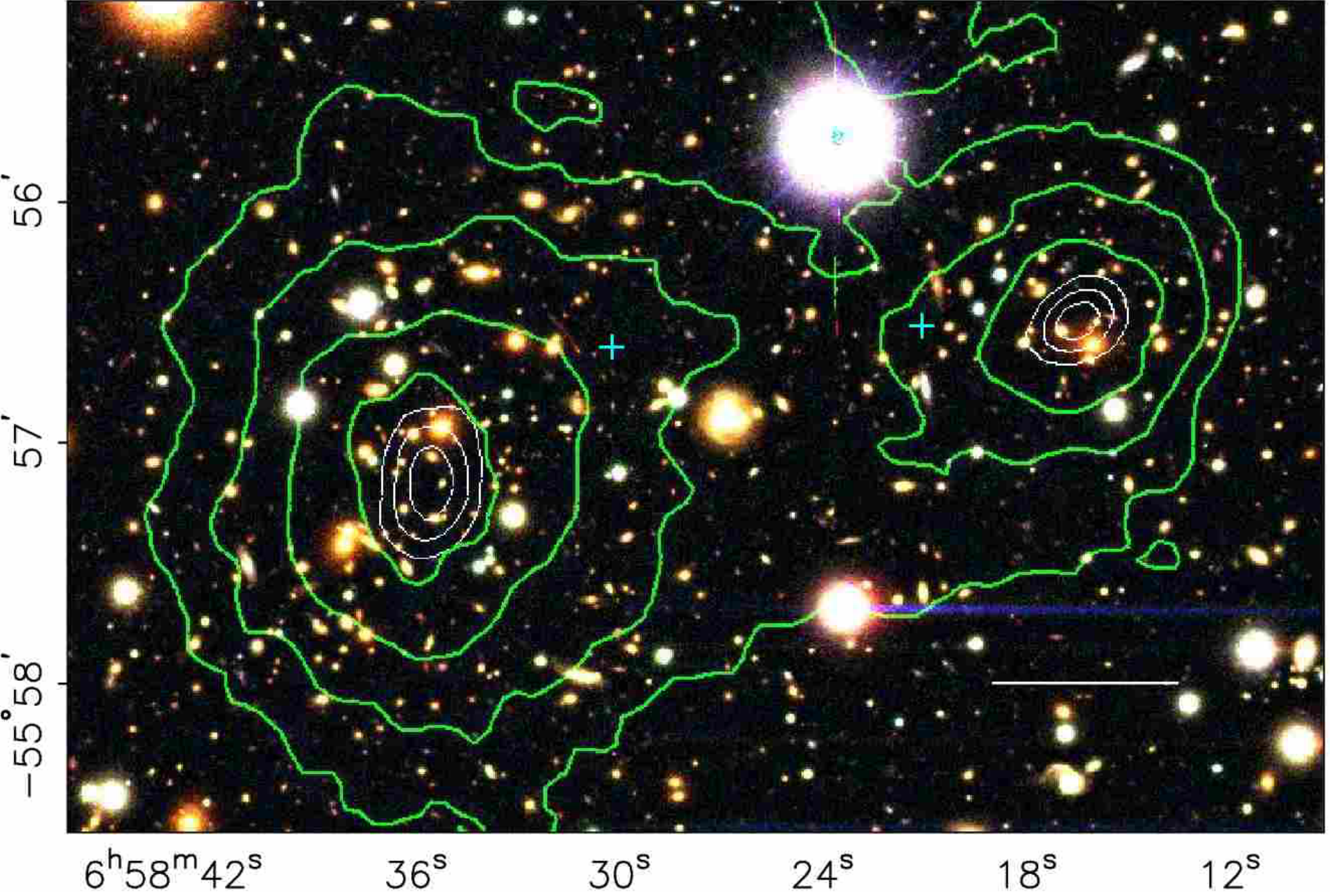}
	\caption{Observations of the Bullet Cluster reported in~\cite{Clowe:2006eq}. Left panel: optical image of the colliding clusters. Right panel: X-ray imaging of the gas component (accounting the most part of the visible matter) from Chandra. The green contours describe the gravitational field inferred form gravitational lensing.}
	\label{fig:bulletcluster}
\end{figure}

Recent and compelling evidence for the presence of dark matter comes from the observation of dynamical systems such as the Bullet Cluster~\cite{Markevitch:2001ri, Clowe:2003tk, Clowe:2006eq}, whose name is derived from its beautiful bow shock shape, see \cref{fig:bulletcluster}. This is a system of two clusters that have collided around 150 million years ago. The gas component, which accounts for the most part of luminous matter, has been observed in the X-ray wavelength, and carries the mark of the past collision. On the other hand, the mass distribution of the two clusters can be inferred by gravitational lensing. The barycentres of the two clusters are found to be displaced with respect to the gas. That is consistent with collisionless dark matter: the DM halos are able to pass through each other while the baryonic component is slowed by the interaction. Other similar colliding systems have been observed showing the same behaviour~\cite{Bradac:2008eu, Harvey:2015hha}.

\subsubsection{Large scale structure}
\begin{figure}[tb]
	\centering
	\includegraphics[width=0.6\textwidth]{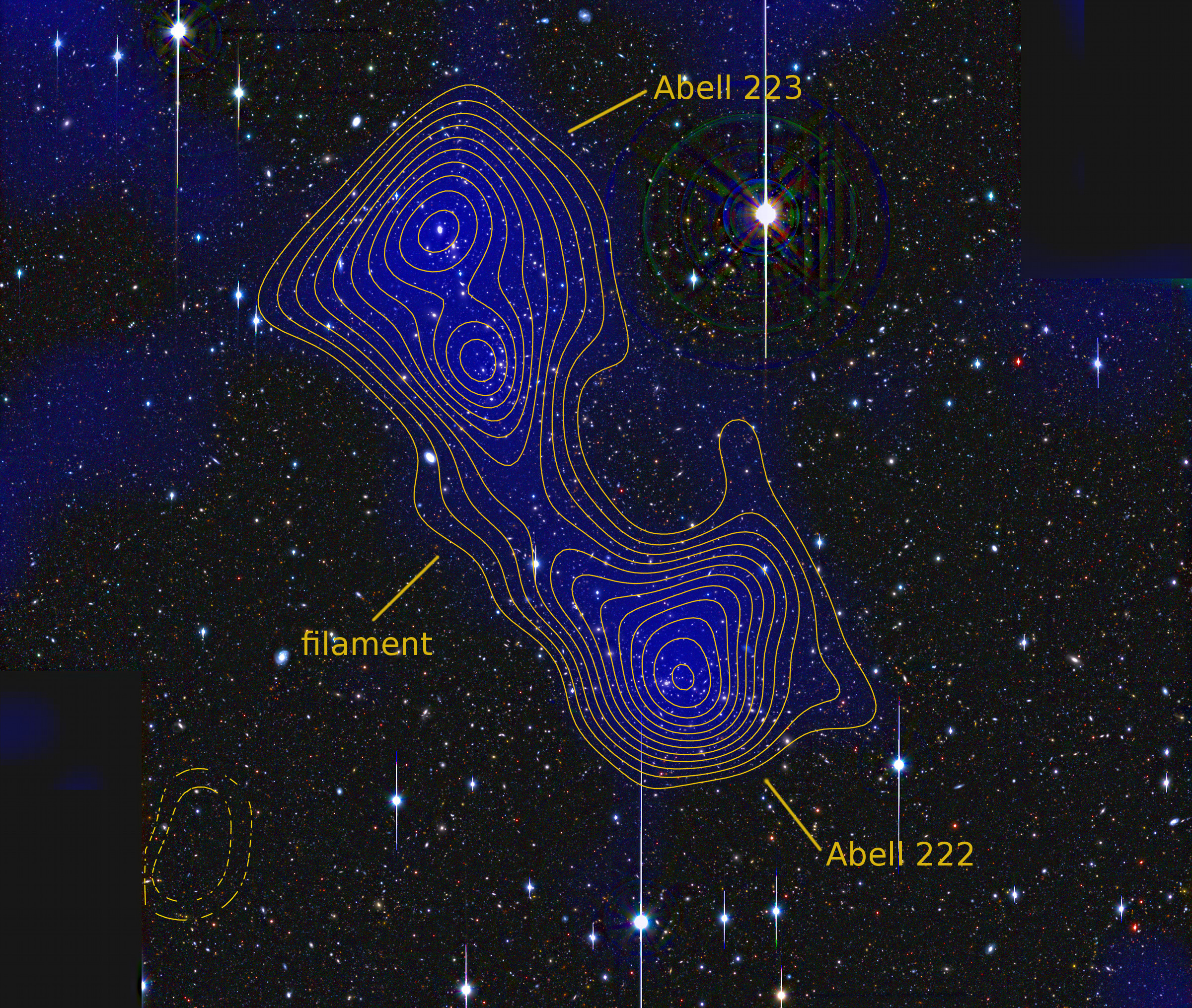}
	\caption{Image of the galaxy clusters A222 and A223 reported in~\cite{Dietrich:2012mp}. The blue colour and the contour lines indicate the reconstructed mass density profile, showing the presence of a filamentary structure connecting the two clusters.}
	\label{fig:DMfilament}
\end{figure}

At even larger scales, the observation of the properties of matter distribution provides further evidence of the presence of dark matter.  Observations of the large scale structure are extremely well reproduced by N-body simulations of the evolution of collisionless gravitational matter, such as the Millennium simulation~\cite{2012MNRAS.426.2046A}.
These simulations have allowed to establish that structures form hierarchically, with smaller ones forming first then converging to form larger ones. 
Dark matter is expected to dominate the process of structure formation. The growth of perturbations in the baryon-photon plasma is reduced by photon pressure, until the epoch of recombination -- time at which their amplitude is measured through observations of the Cosmic Microwave  Background. However, perturbations in the dark matter fluid are able to grow during this phase, as they are not coupled to radiation (or decouple from it very early, depending on the particular model). By recombination time, we expect dark matter perturbations to have grown significantly. After decoupling from the radiation, ordinary matter is rapidly drawn by gravity into the potential wells traced by the dark matter. In the absence of the dark matter component, the growth of structure would be substantially delayed in the history of the Universe, in sharp contradiction with observations.\\
Dark matter N-body simulations show that the first structure to form is a complex network of filaments pervading the Universe, sometimes called the ``cosmic web''. Halos start growing at the intersections of these filaments, then migrate along the lines of the web under mutual gravitational attraction, forming larger halos. Large scale structure observations suggest he presence of this underlying filamentary structure. A dark filament connecting galaxy clusters A22 and A223 has been detected using weak lensing~\cite{Dietrich:2004kf, Dietrich:2012mp}, see \cref{fig:DMfilament}.

\subsubsection{Cosmic microwave background}

Possibly the most striking evidence for the presence of dark matter, and the most precise measurement of its abundance, comes from the observation of the Cosmic Microwave Background (CMB). The small (one part in $10^{5}$) anisotropies in the CMB radiation trace the distribution of fluctuations in the baryon-photon plasma at the time of recombination, providing a wealth of information about the early Universe composition. The $\Lambda$ Cold Dark Matter ($\Lambda$CDM) cosmological model is in exceptionally good agreement of the CMB data and its parameters can be measured with excellent precision from its power spectrum~\cite{Planck:2018vyg}. 

\begin{figure}[tb]
	\centering
	\includegraphics[width=0.7\textwidth]{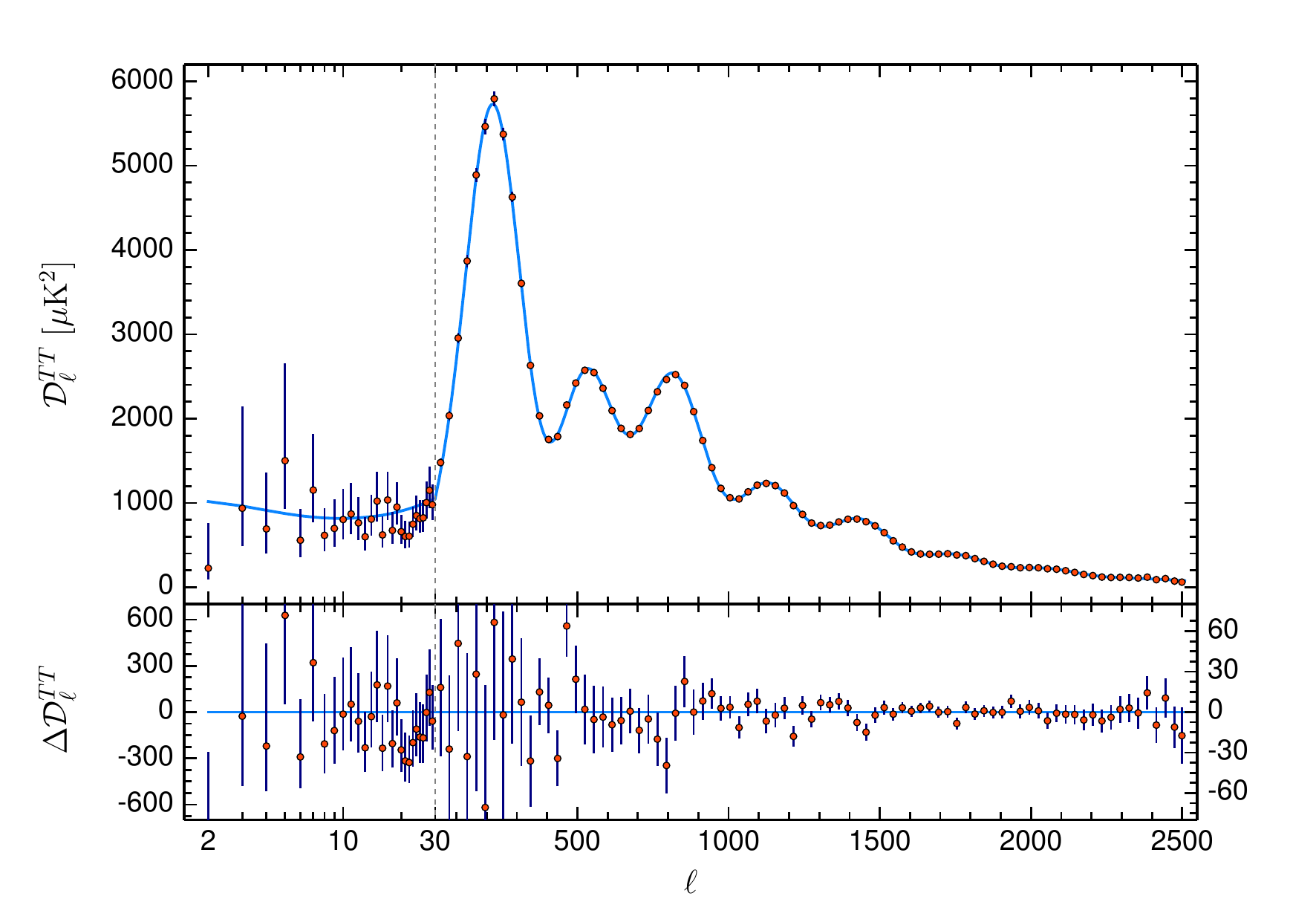}~
	\caption{The CMB angular power spectrum measured by Planck, from~\cite{Planck:2018vyg}. The best fit $\Lambda$CDM model is shown in light blue in the upper panel. The lower panel shows the residuals with respect to this model. Vertical bands indicate $\pm 1 \sigma$ uncertainties. }
	\label{fig:CMBspectrum}
\end{figure}

The presence of a  dark matter component can be inferred ``\textit{by eye}'' from the relative heights of the second and third peak in the CMB angular power spectrum, see \cref{fig:CMBspectrum}.
As we mentioned previously, the baryon-photon fluid is forced to oscillate by the interplay between its self gravity and the pressure it generates when compressed. Initially gravity due to DM contracts the fluid; pressure builds up until it overcomes gravity and the fluid density proceeds with an oscillatory dynamics until decoupling. The period of the oscillation depends on the physical size of the fluctuation. The first peak in the CMB spectrum corresponds to overdensities that had reached the state of maximum contraction at the time of decoupling; the second one, to smaller fluctuations that had contracted once and expanded back to the initial state, the maximum expansion; and so forth (the multipole index $l$ corresponds to decreasing angular size of the perturbations). 
The oscillations are damped, so that we would expect each peak to be lower than the previous one. The presence of a gravitating fluid not coupled to the photon plasma has an asymmetric effect on the peaks: the fluid is able to contract further before gravity is overcome by pressure. On the other hand, pressure then expands it back to the initial state, so that the minimum density is not affected. This implies that the odd peaks, corresponding to the maximally contracted phase, are enhanced with respect to the even ones.  Thus, the fact that the third peak is not lower than the second one is a clear indication of the presence of a dark matter component. \\
From the CMB spectrum, the dark matter abundance is measured to be
\begin{equation}
	\Omega_\mathrm{DM} \, h^2 \, = \, 0.120 \pm 0.001 \; ,
\end{equation}
where $h$ is the value of the Hubble parameter in units of \SI{100}{\kilo\meter\per\second} ($h = 67.4 \pm 0.5$ as measured by Planck), and the dark matter energy density $\rho_\mathrm{DM}$  is expressed in terms of the critical density $\rho_c  =  3H_0^2/8\pi G$ with $\Omega_\mathrm{DM} \equiv  \rho_\mathrm{DM} / \rho_c$.  The abundance of the baryonic matter can also be inferred from the CMB, giving
\begin{equation}
	\Omega_\mathrm{b} \, h^2 \, = \, 0.0224 \pm 0.0001 \; ,
\end{equation}
in agreement with the (less precise) measurement from Big Bang nucleosynthesis.
These measurements imply that the DM constitutes about the 85\% of the total matter density, a value in agreement with observations at smaller scales. 
%
%

\subsection{Properties}
\label{sec:intro:darkmatter:properties}

Observations have allowed to at least partially constrain the properties of dark matter~\cite{Taoso:2007qk,Bertone:2004pz}. In particular the observations reveal that dark matter is massive, stable over billions of years, very likely collisionless, interacting mostly gravitationally and separate from (distinct and not interacting with) baryonic matter.

\paragraph{Non-baryonic} We know that dark matter is non-baryonic, meaning it does not contribute to the cosmological baryonic energy density. The fraction of baryonic energy density is well constrained by CMB data and measurements of primordial element abundances. Furthermore, as already mentioned, dark matter potential wells are a fundamental ingredient in the formation of the structures we observe. Hence cold clouds, white dwarfs, astrophysical compact objects, etc., once believed to be possible explanations of the ``missing mass'', are excluded.
\paragraph{Cold (Warm)}  In the standard cosmological model $\Lambda$CDM, dark matter is assumed be non-relativistic by the time of structure formation, i.e. after matter-radiation equality. Relativistic dark matter would have a longer free-streaming length and wash out structures below the corresponding scale. This would imply top-down rather than hierarchical structure formation and  is in contradiction with the observed matter power spectrum~~\cite{1983ApJ...274L...1W, 1985ApJ...292..371D, Frenk:2012ph}. The possibility that standard neutrinos constituted the dark matter, very popular in the 80s, was ruled out by this observation (at the time, upper limits on their mass where less stringent and their relic abundance could account for the whole DM). Semi-relativistic dark matter (warm), such that the wash-out of structures occurs only on small enough scales, is a viable alternative within certain limits~\cite{2013MNRAS.428.1774B, Kennedy:2013uta}. 
\paragraph{Stable}  We know that dark matter is either stable or very long lived. If it can decay, its lifetime must be larger than the age of the Universe in order to account for the consistent observations of its abundance throughout different epochs. Even stronger bounds on the DM lifetime can be placed under certain assumptions, see e.g.~\cite{Audren:2014bca}.
\paragraph{Collisionless} In the standard model, dark matter is considered to behave as a collisionless fluid, meaning that its self-interaction is neglected. However, models of self-interacting dark matter, with small cross sections, can be compatible with observations (see Ref.~\cite{Tulin:2017ara} for a review).
Lower limits on DM self interaction can be obtained, for example, by the study of colliding galaxy clusters~\cite{1985ApJ...299..633L}.

\paragraph{Neutral} We know that dark matter interacts little or not at all with photons. That is necessary to explain why we do not detect it electromagnetically, and crucial to early Universe physics and structure formation~\cite{Boehm:2001hm}. Direct interaction in the form of an integer electric charge is hence excluded. Nevertheless, ``milli-charged'' dark matter scenarios have been proposed. Mass dependent upper limits on the electromagnetic charge of DM particles can be placed from early Universe observations as well as from direct detection experiments~\cite{McDermott:2010pa, Sanchez-Salcedo:2010gfa}.
More generally, a small interaction with ordinary matter can be present, provided that the decoupling from the baryon-photon fluid happens early enough.
Limits on the interaction with ordinary matter, in the case of particle dark matter, are further set by both direct and indirect detection experiments. 

\paragraph{Mass range}
A general lower limit on the dark matter mass can be placed by the requirement that its De Broglie wavelength $\lambda = h/ p$, where $h$ is the Planck constant and $p$ is the particle momentum, should be smaller than the size of dwarf spheroidal galaxies (the smallest scales on which it is observed in bound systems). Considering the typical scale of these objects, $\lambda \sim 1$~kpc and a typical velocity dispersion of $\sim100$~km/s, one obtains a lower bound of $m \gtrsim 10^{-22}$~eV.  Particles in this mass range would exhibit collective quantum properties, being described by Bose-Einstein/Fermi-Dirac statistics if they are bosons/fermions. In fact, for fermions a more stringent lower limit is obtained from the Pauli exclusion principle. In this case, the Gunn-Tremaine bound sets a lower limit of $m \gtrsim10 $~eV, 23 orders of magnitudes stronger than the bosonic case.\\
On the other hand, so-called dynamical constraints, related to the manifestation of granularity in the DM distribution, set an upper limit of $\sim 10^3 \, \Msun$~\cite{1985ApJ...299..633L, Profumo:2019ujg}. 
In conclusion, not much can be said about the mass of the dark matter components, which can span about eighty orders of magnitude.

\subsection{Candidates}
\label{sec:intro:darkmatter:candidates}


There are, of course, much more questions than answers, as the properties of dark matter are largely unconstrained. If DM is in the form of an elementary particle, this is not present in the Standard Model (SM) of particle physics. As we have mentioned, the three known neutrinos, because of their extremely light mass, behave as ``hot'' DM; furthermore, again because of their lightness, their relic abundance makes them a strongly subdominant component of the DM, $ \Omega_\nu h^2 \lesssim 10^{-2}$.
The evidence for DM is then one of the strongest motivations to search for ``New Physics''.\\  A plethora of DM candidates can and have been proposed which are able to explain DM observations. Without entering in any detail, we briefly summarize here some of the most successful classes of models.
\paragraph{WIMPs}
One of the most popular dark matter particle scenario is that of WIMPs, which stands for Weakly Interactive Massive Particle. The term WIMP is used more specifically to designate a stable particle that interacts through the weak force only, and has a mass around the electro-weak scale (about 200 GeV). For such particles, the relic abundance from thermal decoupling corresponds to that of the dark matter, a fact known as the ``WIMP miracle''. WIMPs appear naturally in supersymmetric extensions of the SM and the Higgs hierarchy problem provided strong motivation for the existence of new physics at the electro-weak scale. While the WIMP scenario still maintains part of its appeal, strong limits on the phase space of this class of models have been placed by the null results of searches in particle accelerators and direct detection facilities. See Ref.~\cite{Arcadi:2017kky} for a review of WIMP models and the present status of experimental searches.
\paragraph{Sterile neutrinos}
Another attractive possibility is that of sterile neutrinos. The modelling of neutrinos in the SM is incomplete, as their mass term is not included. This can be solved by adding a Yukawa coupling term with the Higgs boson, but the possibility exists of adding a Majorana mass term to the SM Lagrangian as well. In this case, an extra family of neutrinos appears; the so-called Seesaw mechanism explains the exceptional lightness of the observed neutrinos by setting a heavy mass scale for this additional family. These massive states are related to the light ones through a mixing angle, which decreases with the mass ratio. Such heavy sterile neutrinos and their generalizations can provide in general good dark matter candidates. For reviews see e.g.~\cite{Abazajian:2017tcc, Boyarsky:2018tvu, Lesgourgues:2006nd}.
\paragraph{Axions}
Amongst the most poplar dark matter candidates we find axions, pseudo-Goldstone bosons arising from the spontaneous breaking of the Peccei-Quinn symmetry (the Peccei-Quinn theory introduces in the SM a new complex scalar field to explain dynamically the absence of CP violation in the strong sector). Observations constrain the axions to be very light $m_a \lesssim \SI{1}{eV}$ and exclude thermal production. Light axions and axion-like particles (ALPs) are stable and are a popular dark matter candidate. See~\cite{Irastorza:2018dyq} for a recent review. 
\paragraph{Primordial black holes}
An alternative possibility is that dark matter is composed of compact objects. Massive Compact Halo Objects (MACHOs) of astrophysical origin, once considered a possible candidate, are now excluded because of their baryonic nature. However \emph{primordial black holes}, formed in the first instants of the history of the Universe, do constitute a viable candidate. These objects can be formed by the collapse of large overdensities in the primordial plasma, their mass being set by the size of the perturbation. As PBHs are central to the works presented in \cref{sec:accretion} and \ref{sec:GWPBH}, we discuss them in detail in the next section.

\section{Primordial black holes}
\label{sec:intro:pbhs}

The idea that a population of black holes may have formed in the primordial Universe dates back to the pioneering works of late 1960s and early 1970s~\cite{Zeldovich:1967lct,Hawking:1971ei,Chapline:1975ojl, Carr:1974nx,1975ApJ...201....1C}. 
A first wave of interest in the topic ensued in the 1990s, following the results of microlensing observations by the MACHO collaboration~\cite{MACHO:1996qam}. An excess of events was reported, compatible with around half of the Galactic halo being in the form of compact objects with mass $\sim 0.5  \, \Msun$ and non-astrophysical origin~\cite{Fields:1999ar}. However, subsequent microlensing results~\cite{MACHO:2000qbb,EROS-2:2006ryy} have further constrained the allowed fraction of halo in the from of compact objects (see \cref{sec:intro:pbhs:constraints}).\\
With the dawn of the era of gravitational wave observations, a new, powerful channel for the study of these objects has been opened. The first observations of binary black hole mergers have originated a second, ongoing wave of interest in primordial black holes~\cite{LIGOScientific:2016aoc,Bird:2016dcv, Clesse:2016vqa}.

A large variety of mechanisms can lead to the formation of primordial black holes. The most straightforward of these is the gravitational collapse of over-dense regions in the primordial plasma. An alternative class of scenarios involves first-order phase transitions and the collapse of topological defects~\cite{PhysRevD.26.2681, 1982PThPh..68.1979K,HAWKING1989237, Jenkins:2020ctp}, and a variety of other mechanisms have been proposes in the context of beyond the standard model theories (see~\cite{Carr:2020xqk} for a review).

Primordial black holes can form over a very wide range of masses. Provided that they are massive enough to be stable (see \cref{sec:intro:pbhs:constraints}), they constitute a good dark matter candidate: on large scale they behave as a collisionless, cold dark matter fluid. As they are formed deep in the radiation era, their role in the thermal history of the Universe is that of a non-baryonic fluid. Part of the appeal of the idea of PBH forming the dark matter is that it does not require the introduction of any new physics, but only the existence of large small-scale density fluctuations in the primordial plasma.

The present abundance of primordial black holes is usually expressed in therms of the fraction of dark matter in the form of PBHs
\begin{equation}
	\fPBH \,  \equiv \, \dfrac{\Omega_\mathrm{PBH}}{\Omega_\mathrm{DM}} \, .
\end{equation}
An upper limit on PBH abundance is given by the requirement that it does not exceed the observed dark matter density, that is $\fPBH \leq 1$. A variety of observations further constrain the abundance in different mass ranges (see \cref{sec:intro:pbhs:constraints}).

It is important to stress that PBHs are a relevant topic of study even if they do not constitute the whole of the dark matter. Even if they were to represent only a subdominant portion of it, the discovery of a population of PBHs would have profound implications for fundamental physics.  \\
In fact, the formation of PBHs holds precious hints about inflationary and early-Universe physics~\cite{Dolgov1993,Jedamzik:1996mr,Garcia-Bellido:1996mdl,Garcia-Bellido:2017mdw,Ballesteros:2017fsr,Pattison:2017mbe}; their subsequent evolution could impact structure formation~\cite{Inman:2019wvr} and solve long-standing puzzles related to the early formation of supermassive black holes~\cite{Volonteri:2021sfo}. \\
More closely related do dark matter searches, their detection could strongly constrain models of particle dark matter which invoke new physics at the weak scale, notably including the WIMP scenario~\cite{Lacki:2010zf,Bertone:2019vsk, Adamek:2019gns}. On the other hand, PBH evaporation in the early Universe could provide a mechanism for dark matter production~\cite{Fujita:2014hha, Hooper:2019gtx,Gondolo:2020uqv,Cheek:2021cfe}. 

In this section we briefly discuss the formation of PBHs from the collapse large density fluctuations, and review present constraints on their abundance. More complete discussions can be found in~\cite{Green:2020jor,Carr:2020xqk}, see also~\cite{Byrnes:2021jka}.


\subsection{Formation from density fluctuations}
\label{sec:intro:pbhs:formation}

The primary mechanism for the formation of PBH is the gravitational collapse of high density regions in the radiation-dominated Universe.
Provided that the density contrast is large enough, the collapse occurs when the size of such regions becomes comparable to the Hubble radius. The mass of the resulting black hole is then related to the total mass within a Hubble patch at the time of formation. As the Universe expands, small perturbations enter the horizon first, so that light PBHs are the first to (possibly) form. \\
 For perturbations that enter the horizon at matter-radiation equality, the corresponding PBH mass is around $ M_\mathrm{PBH} \sim 10^{16} M_\odot$. As BHs of this mass are not observed, we restrict our discussion to PBH forming before equality, in the radiation-dominated era. We notice, however, that light PBHs can be also formed in non-standard early matter-dominated epochs~\cite{Khlopov:1980mg, Harada:2016mhb, PhysRevD.96.063507}. 

\subsubsection{Density threshold for PBH formation}
The density threshold for PBH formation during the radiation dominated era was first estimated by B.~Carr in~\cite{1975ApJ...201....1C}. He found that the density contrast $\delta \equiv \delta\rho/\rho$ at horizon crossing must be larger than $\delta_\mathrm{c} = c_\mathrm{s}^2$, where $c_\mathrm{s}$ is the sound speed of the fluid. In the radiation era $c_\mathrm{s} = 1/ \sqrt{3}$ , hence the critical density contrast is of order one, $\delta_\mathrm{c} = 1/3$.
This first estimate has been since refined both analytically~\cite{Harada:2013epa} and through numerical methods~\cite{Musco:2018rwt,Yoo:2020lmg}, see also~\cite{Escriva:2021aeh} for a review. These and other studies found that the threshold value depends only mildly on the density profile of the fluctuation. More precisely, the threshold is lowest for broad profiles, $\delta_\mathrm{c} \simeq 0.41$ and highest for peaked profiles $\delta_\mathrm{c} \simeq 0.67$ (the lower value agrees with analytical estimates).

\begin{figure}[t]
	\centering
	\includegraphics[width=.7\linewidth]{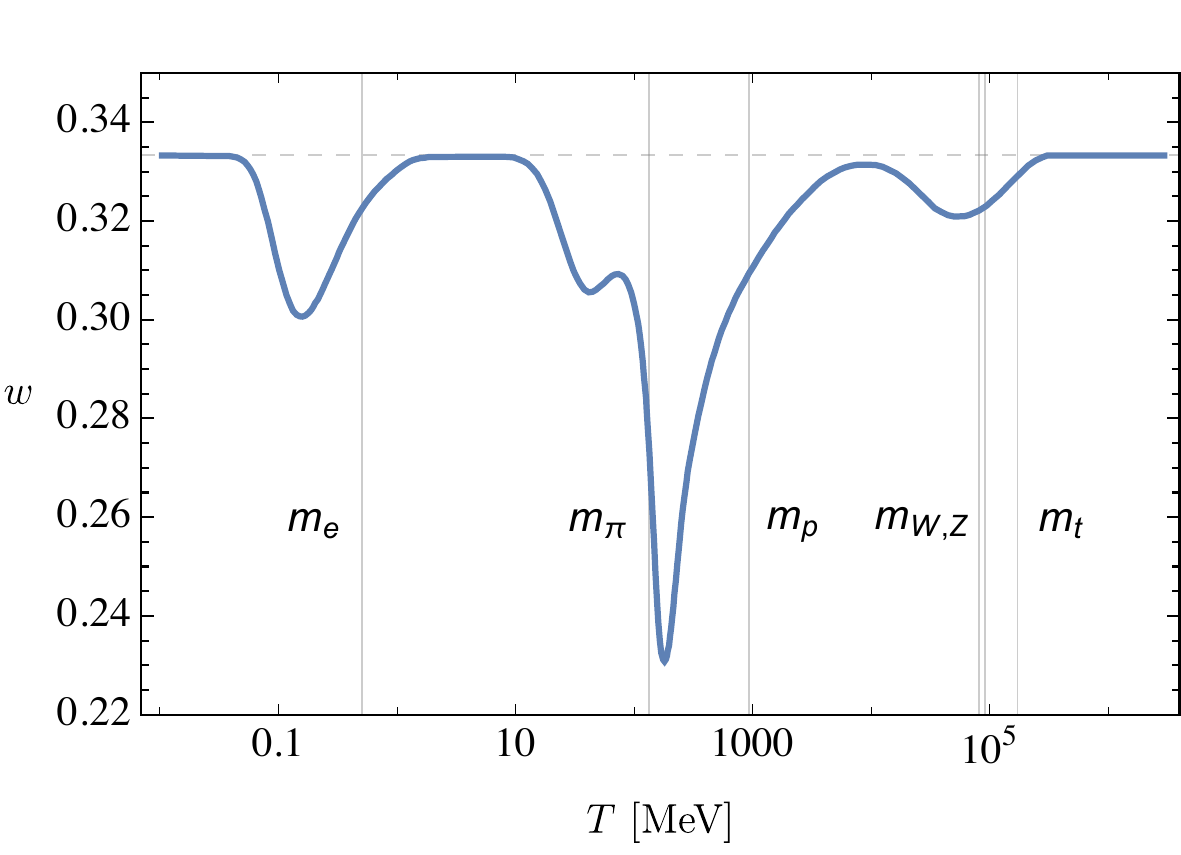}
	\caption{The equation of state parameter $w$ as a function of temperature in the radiation era. The grey vertical lines correspond to the masses of the electron, pion, proton/neutron, W/Z bosons and top quark, respectively. The grey dashed horizontal lines corresponds to $w$ = 1/3. From Ref.~\cite{Carr:2019kxo}}
	\label{fig:eqstate}
\end{figure}

There is a significant dependence of the density threshold $\delta_\mathrm{c} $ with the equation of state $w = P/\rho$, where $P$ and $\rho$ are the pressure and energy density of the primordial plasma. In particular for $w \rightarrow 0$, the relation $\delta_\mathrm{c} = c_\mathrm{s}^2$ is no longer a good approximation~\cite{1975ApJ...201....1C,Harada:2013epa}.  This has interesting consequences, since whenever a drop in the number of relativistic degrees of freedom occurs, the effective equation of state parameter $w$ temporarily decreases. These pressure drops result in a significant enhancement of the probability of PBH formation for some particular masses. The most significant of these drops occur during the QCD transition ($t \sim 10^{-6}$ s, $E \sim 150$ MeV), with a large dip corresponding to a horizon mass of $\sim 1 M_\odot$ and a smaller one corresponding to $\sim 30 M_\odot$ ~\cite{Jedamzik:1996mr,Byrnes:2018clq}.
Other transitions can have a similar albeit milder effect: the early electro-weak transition ($t \sim 10^{-12}$ s, $ E \sim 100$ GeV) and the later electron-positron annihilation ($t \sim 6$ s, $E \sim 500$ keV)~\cite{Carr:2019kxo}, see \cref{fig:eqstate}.

Despite the details mentioned above, the requirement of an $\mathcal{O}(1)$ density fluctuation for PBH formation during radiation domination remains true. This may not seem a very stringent condition, but we have learned from CMB observations that the early Universe presents an exceptional homogeneity, with variations of the order of one part in $10^{5}$ (on the scales probed). We are therefore led to expect PBH formation to be a very rare event. 

\subsubsection{Primordial black hole mass}

The mass of the PBH is related to the mass contained in the Hubble patch at the time of horizon crossing, $M_H$. Since the physical horizon is $r_H = c/H$, where $H$ indicates the Hubble parameter $H= \dot{a}/a$ and $a$ is the scale factor, we have in first approximation
\begin{equation}
	\label{eq:MPBHapprox}
	M_\mathrm{PBH} \, \simeq \, M_H  \, \simeq \, \rho \left( \dfrac{c}{H} \right)^3 \; .
\end{equation}
From here, we can estimate how the PBH mass varies with formation time. In the radiation era, the density decreases as $\rho \propto a^{-4}$, while from the Friedman equation we have $ H^2 \propto \rho$. Hence the volume of the Hubble patch increases as $H^{-3} \propto \rho^{-3/2} \propto a^6$, and we find
\begin{equation}
	M_\mathrm{PBH} \, \propto \, a^2  \, \propto t\; .
\end{equation}
Inserting numerical factors, the mass of a PBH formed at scale factor $a_\mathrm{f}$ is approximately
\begin{equation}
	M_\mathrm{PBH} (a_\mathrm{f}) \, \approx \, \left( \dfrac{a_\mathrm{f}}{a_\mathrm{eq}} \right)^2 \, 10^{16} M_\odot\; .
\end{equation}
For the comoving wavenumber at formation $k = (a H)|_f$, we have
\begin{equation}
	k ( M_\mathrm{PBH}) \simeq \, 10^{-7 }\, \mathrm{Mpc}^{-1} \left( \dfrac{M_\mathrm{PBH}}{M_\odot} \right)^{-1/2} \; .
\end{equation}
The formation of PBH of mass $M_\mathrm{PBH} \sim 100 ~Msun$ corresponds to comoving scales $1/k \sim \mathcal{O}$(pc). Notice that these are much smaller than the scales probed by CMB observations ($\gtrsim 1$ Mpc).

The relation between mass and formation time is actually somewhat more complicated than \cref{eq:MPBHapprox}; it can be parametrized as~\cite{Niemeyer:1997mt, Niemeyer:1999ak}
\begin{equation}
	M_\mathrm{PBH}  \, = \kappa M_H \left( \delta - \delta_\mathrm{c} \right)^\gamma \, 
	\label{eq:PBHmass}
\end{equation}
where $M_H $ is the horizon mass at the time of formation, $\delta > \delta_\mathrm{c}$ is amplitude of the fluctuation and $\kappa , \gamma $ are parameters of order 1, which depend on the shape of the fluctuation and on the background equation of state~\cite{Musco:2004ak, Musco:2008hv}.

\subsubsection{Primordial black hole abundance}
Estimating the abundance of PBH amounts to computing the probability of the density contrast  exceeding the threshold for collapse, starting from a given power spectrum for the perturbations.\\
We can make a first rough estimate of the probability of  PBH formation as follows.
Given a probability distribution $P(\delta)|_R$ for the density fluctuations at comoving scale $R = 1/k $, the probability of forming a PBH can be obtained through the Press-Schechter formalism~\cite{1974ApJ...187..425P}
\begin{equation}
	\beta|_R \, = \dfrac{\rho_\mathrm{PBH}}{\rho_\mathrm{tot}} \bigg\rvert_R \, =  \, \int_{\delta_c}^\infty \, \diff \delta \, P(\delta)|_R \; . 
\end{equation}
If we further assume that the p.d.f. is Gaussian, $ P(\delta)|_R \propto  \exp{ ( \delta^2/ 2 \sigma_R^2 )}$, we obtain
\begin{equation}
	\label{eq:betaofk}
	\beta|_R \, = \, \frac{1}{2} \, \mathrm{erfc} \left( \dfrac{\delta_\mathrm{c}}{\sqrt{2} \sigma_R} \right) \, \approx \, \dfrac{\sigma_R}{\sqrt{2\pi } \delta_\mathrm{c}} \exp\left( - \dfrac{\delta_\mathrm{c}^2}{2 \sigma_R^2} \right)\; .
\end{equation}
In the last step we have assumed $\delta_\mathrm{c} \gg \sigma_R$, as we expect the threshold to be much higher than the average density contrast. We see that the abundance scales exponentially with both $\sigma_R$ and the threshold for collapse, $\delta_\mathrm{c}$. \\

From here, one can make a first estimate of the variance $\sigma_R^2$ that corresponds to $\fPBH\sim1$ at a given mass.
In first approximation, the fraction $\beta|_R$ represents the PBH abundance for a mass of $M\sim M_H(k) $, where  $R = 1/k $ and $M_H(k)$ is the horizon mass at $k = aH$. 
This abundance is obtained at the time of formation, and it is significantly smaller than the present day one. In fact, during radiation era, the background density scales as $a^{-4}$, while the PBH density only decreases as $a^{-3}$, meaning that even a very small fraction at formation can become important. We have\footnote{neglecting accretion effects and hierarchical mergers that can affect the PBH mass}
\begin{equation}
	\beta (M )\, = \,  \dfrac{a_\mathrm{f}(M)}{a_\mathrm{eq}} \dfrac{\rho_\mathrm{PBH}(M)}{\rho_\mathrm{tot}} \bigg\rvert_\mathrm{today}\ \, \approx \, 0.85 \, \dfrac{a_\mathrm{f}(M)}{a_\mathrm{eq}} \, \fPBH(M) \; ,
\end{equation}
where we have indicated with $a_\mathrm{f}$ the scale factor corresponding to an horizon mass $M_H \sim M$.

Consider for example PBHs of one solar mass. Their time of formation is $a_\mathrm{f} \approx 10^{-8} \, a_\mathrm{eq}$, so producing enough PBH to account for all the DM requires $ \beta (1 \, M_\odot) \approx 10^{-8}$. Assuming $\delta_\mathrm{c} = 0.45 $,  this corresponds to $\sigma_R^2\sim 10^{-2}$.  It is important to notice that, because of the exponential dependence of $\beta$ on $\sigma_R $, perturbations of the \emph{same order of magnitude} are required to generate even a strongly subdominant fraction of PBHs in DM (for $\sigma_R^2 \sim 10^{-3}$ we obtain $\fPBH \sim 10^{-40}$). 

As a first approximation, one can identify the variance $\sigma_R^2$ in \cref{eq:betaofk} with the curvature power spectrum $\mathcal {P}_\mathcal{R} (k)$.  This is the Fourier transform of the two-point correlation function for the curvature $\mathcal{R}$
\begin{equation}
	\langle \mathcal{R} ( \vec{k}) \mathcal{R} ( \vec{k}^\prime)  \rangle \, = \, \dfrac{2 \pi^2}{k^3}  \delta\left( \vec{k} + \vec{k}^\prime \right)  {P}_\mathcal{R} (k) \; ,
\end{equation}	
where $\delta$ indicates the three-dimensional Dirac delta function.The primordial curvature power spectrum inferred by CMB observations (on scales $k \sim 10^{-4} - \SI{1}{\per\mega\parsec}$) can be parametrized in terms of two parameters, the amplitude $A_s $, and the scalar spectral index $n_\mathrm{s}$:
\begin{equation}
	P_\mathcal{R}(k) \, =  A_s \left( \dfrac{k}{k_0} \right)^{n_\mathrm{s}-1} \;, 
\end{equation}
where $k_0$ is an arbitrary reference scale. The latest Planck analysis has found $A_s \simeq 2 \times 10^{-9}$, while the scalar spectral index has been measured to be  $n_\mathrm{s}\simeq0.96$, corresponding to a near scale-invariance (exact scale-invariance has been ruled out with high significance)~\cite{Planck:2018jri}.

Simply extrapolating the behaviour observed on CMB scales down to the ones relevant for PBH formation results in a flat $\fPBH = 0$. Obtaining a significant fraction of PBH requires instead $\sigma_R \sim 10^{-2}-10^{-3}$. Therefore, a boost of around 7 orders of magnitude is required between CMB scales and PBH formation scales.\\

In the details, relating the PBH abundance to the power spectrum is quite more complex than in the simple estimate presented above.
First, the density fluctuation $\delta_R$ in a given region of size $R = 1/k$ is given by the smoothed density contrast
\begin{equation}
	\delta_R \, = \,  \int \diff^3 x^\prime \, W_R( \vec{x}-\vec{x}^\prime) \, \delta(x^\prime) \; ,
\end{equation}
where $W_R(r)$ is a window function on the given scale $R$. The variance of the smoothed distribution is then given by
\begin{equation}
	\sigma_R^2 = \langle \delta^2|_R \rangle \, = \,  \dfrac{1}{2 \pi} \int \dfrac{\diff k }{k}\, \tilde{W}_R^2( k)   P_\delta (k)\; ,
\end{equation}
where $\tilde{W}_R( k)$ is the window function in momentum space
and $P_\delta (k)$ is power spectrum of the density fluctuations. The value of this variance can vary by almost an order of magnitude depending on the choice of the window function, and a correct choice is not established~\cite{Ando:2018qdb}.\\
The density perturbation $P_\delta (k)$  must then be related to the curvature power spectrum $P_\mathcal{R}(k)$. At linear order, we have
\begin{equation}
	P_\delta (k) \,  = \,  \dfrac{16}{81}  \left( \dfrac{k}{a \,H} \right)^{4}\, P_\mathcal{R}(k)\; .
\end{equation}
Since we are interested in the power spectra at the time of horizon crossing, $ k \sim a H$, the two are related by an order-one factor.
The evolution of the power spectrum between the end of inflation and the time of formation must also be taken into account, that is $P_\mathcal{R}(k, t) \propto T ^2(k,t) P_\mathcal{R}(k)$, where $T(k)$ is the corresponding transfer function.
Furthermore, when obtaining the initial fraction $\beta(M)$, one should consider that there is not a one to one correspondence between the horizon size and the PBH mass,  \cref{eq:PBHmass}; this induces additional complications in the abundance calculation.\\
Most importantly, the density fluctuations responsible for the formation of PBHs belong to the high amplitude tail of their probability distribution. The computation of the abundance is very sensitive to small changes in this tail, therefore, dropping the simplifying assumption of a Gaussian spectrum can have a significant impact on the abundance prediction. \\
Many studies have been dedicated to the impact of non-Gaussianities on PBH formation~\cite{Bullock:1996at, Ivanov:1997ia,Young:2015kda, Franciolini:2018vbk, Atal:2018neu, Meerburg:2019qqi, Young:2019gfc, Inomata:2021tpx}, particularly in the context of inflation.
Even assuming $P_\mathcal{R}(k)$ to be Gaussian, inevitable non-Gaussianities arise from taking into account its non-linear relation to the density power spectrum $P_\delta (k)$. Such non-linear effects have been shown to result in lower PBH abundances~\cite{DeLuca:2019qsy, Young:2019yug, Kawasaki:2019mbl}. \\
In summary, primordial non-Gaussianities are an active topic of research, and their impact on PBH formation is still currently under debate.

\subsubsection{PBHs from inflationary scenarios}

The idea of an early phase of accelerated expansion of the Universe provides a simple explanation of the exceptional homogeneity of the Universe, its observed flatness and the correlation of its features on large scales (the so-called ``horizon problem''). Generically speaking, this accelerated expansion is achieved through the introduction of a field $\phi$, the inflaton, which dominates the energy density and evolves within a potential $V(\phi)$, before decaying into radiation and matter at the end of the inflationary period.\\
While the existence of this inflationary phase is not demonstrated by observations, it is commonly included in the standard cosmological picture. Beyond its ability to solve the aforementioned problems, its popularity is also due to the fact that it provides a mechanism to generate the initial inhomogeneities in the primordial plasma from quantum fluctuations of the inflaton field.
Within this picture, the initial curvature power spectrum $ P_\mathcal{R}(k) $ is a prediction of inflation, which depends on the particular model.\\

The simplest inflationary scenario is known as slow-roll inflation. The inflaton is described as a single field with a slowly varying potential. In this case the power spectrum can be obtained as
\begin{equation}
	\label{eq:PRinflation}
	P_\mathcal{R}(k) \,  \simeq \,  \dfrac{H^2}{M_\mathrm{Pl}^2 \, \epsilon} \,
\end{equation}
where $\epsilon$ is the slow-roll parameter 
\begin{equation}
	\epsilon \,  \equiv \, - \dfrac{\dot{H}}{H^2} \,\;
\end{equation}
This parameter measures the ratio of kinetic to potential energy in the inflaton field, $ \epsilon \approx K/ V $. Inflation requires $ K/V < 1$, and ends when $\epsilon$ becomes of order one. In the slow-roll scenario, one further has  $\epsilon \sim $ const. Since $H$ is also slowly varying during inflation, we can see from \cref{eq:PRinflation} that slow-roll inflation leads to a near-to scale invariant power spectrum, that is $n_\mathrm{s} \sim 1$.
In this case, given the measurements on Planck scales, no PBHs are expected to be produced. 

While a near scale invariant  power spectrum is in agreement with observations, many different inflationary models can be conceived, able to predict an enhancement of the power spectrum at small scales.

Perhaps the simplest case is that of ultra slow-roll inflation~\cite{Kinney:2005vj,PhysRevD.50.7173}, where the presence of a feature in the potential causes the inflaton field to quickly loose kinetic energy. 
In that case we no longer have $\epsilon \sim $ const; in fact $\epsilon \sim $ can decrease exponentially. Naively, one can see from \cref{eq:PRinflation} that this leads to an exponential growth of the power spectrum on the corresponding scales (the computation of the power spectrum in ultra-slow roll is actually more complicated). %
The ultra slow-roll phase cannot last too long, since it can easily cause PBH overproduction. 
It has been argued that taking into account quantum diffusion of the inflaton field~\cite{Ezquiaga:2019ftu, Pattison:2017mbe}, results in the tails of the curvature power spectrum being exponential rather than Gaussian. This can boost the production of PBHs by several orders of magnitude.
\begin{figure}[h]
	\centering
	\includegraphics[width=.7\linewidth]{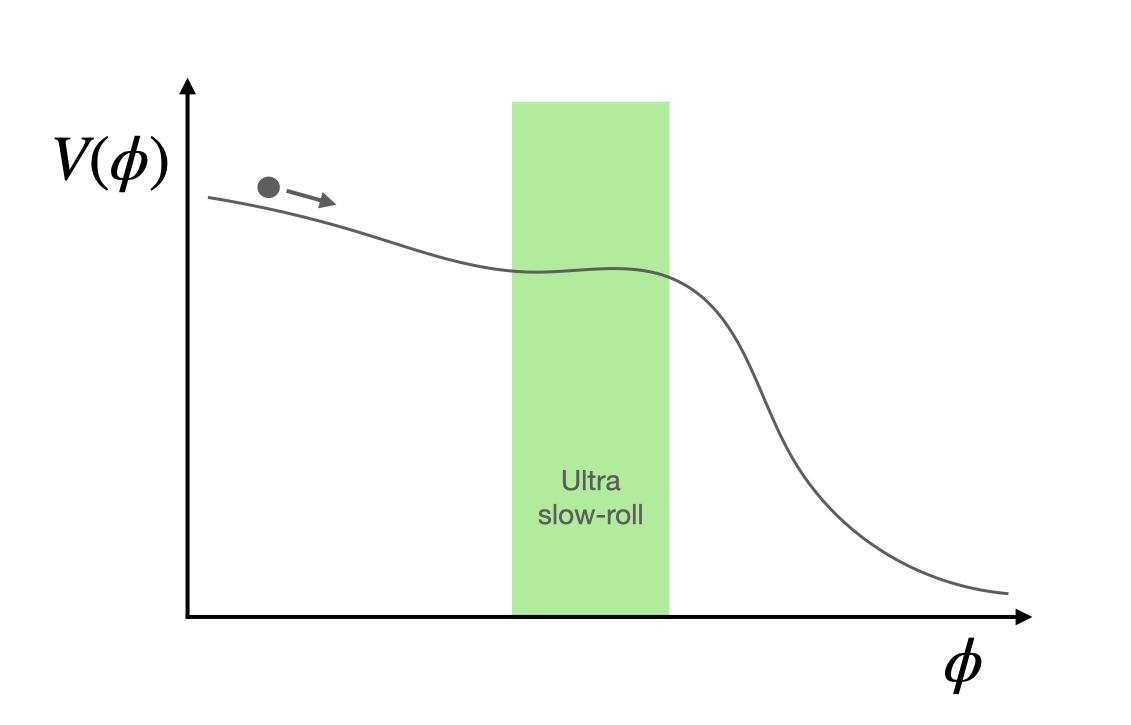}
	\caption{An illustrative potential for the inflaton, presenting a feature that can induce an ultra slow-roll phase.}
	\label{Fig:ultraslow-roll}
\end{figure}
Beyond slow-roll inflation, a large variety of models, often involving more than one inflaton field, are able to produce significant abundances of PBHs (e.g.~\cite{Garcia-Bellido:1996mdl,Braglia:2020eai}, see~\cite{Green:2020jor} for a review).

\subsection{Properties}
\label{sec:intro:pbhs:properties}

PBHs can form in a wide range of masses, and their mass distribution is strongly model-dependent. From the discussion of the previous section, it is clear that the often considered mono-chromatic mass function, i.e. a population of PBHs all having the same mass, is not realistic. Even in the case of near-critical collapse where all PBHs form at the same time, a significant tail at low masses is expected~\cite{Niemeyer:1997mt}.
If the primordial power spectrum has a broad peak, PBHs form at a range of different times. In this case the PBH mass function can be obtained using a binned approach, computing the abundance for each bin of the power spectrum and summing the resulting mass functions. In this case the mass function can be approximately fitted by a lognormal distribution~\cite{Kannike:2017bxn, Green:2016xgy}, see also~\cite{Gow:2020cou}.\\
As we mentioned, an enhancement of the probability of PBH formation is expected in correspondence to the pressure drops in the thermal history of the Universe. Taking this into account results in a strong modulation of the mass function, even when it is computed starting from an approximately flat power spectrum~\cite{Jedamzik:1996mr,Byrnes:2018clq, Carr:2019kxo}. \\
We notice that the initial mass distribution can be significantly modified in time, notably through mergers (which may become relevant in the case of PBH clusters) and matter accretion. The latter has a significant impact for BHs of large masses $M \gtrsim 1 -10 ~Msun$~\cite{DeLuca:2020fpg}.

The density perturbations from which PBHs form are close to spherically symmetric. Therefore the torques on the perturbation are small and PBHs are expected to from with small initial spin~\cite{Mirbabayi:2019uph,DeLuca:2019buf}. However, this may be significantly by PBH evolution, in particular by accretion~\cite{DeLuca:2020bjf}.

A Poisson distribution of PBH is often considered to be the initial configuration. It has been shown that this is the case for PBH arising from Gaussian curvature perturbations with a narrow peak~\cite{Ali-Haimoud:2018dau, Desjacques:2018wuu}. A significant initial clustering, on the other hand, is expected to be present for broad mass distributions~\cite{Ballesteros:2018swv} or local non-Gaussianities~\cite{Tada:2015noa}.

\subsection{Observations}
\label{sec:intro:pbhs:constraints}
\begin{figure}[h]
	\centering
	\includegraphics[width=.8\linewidth]{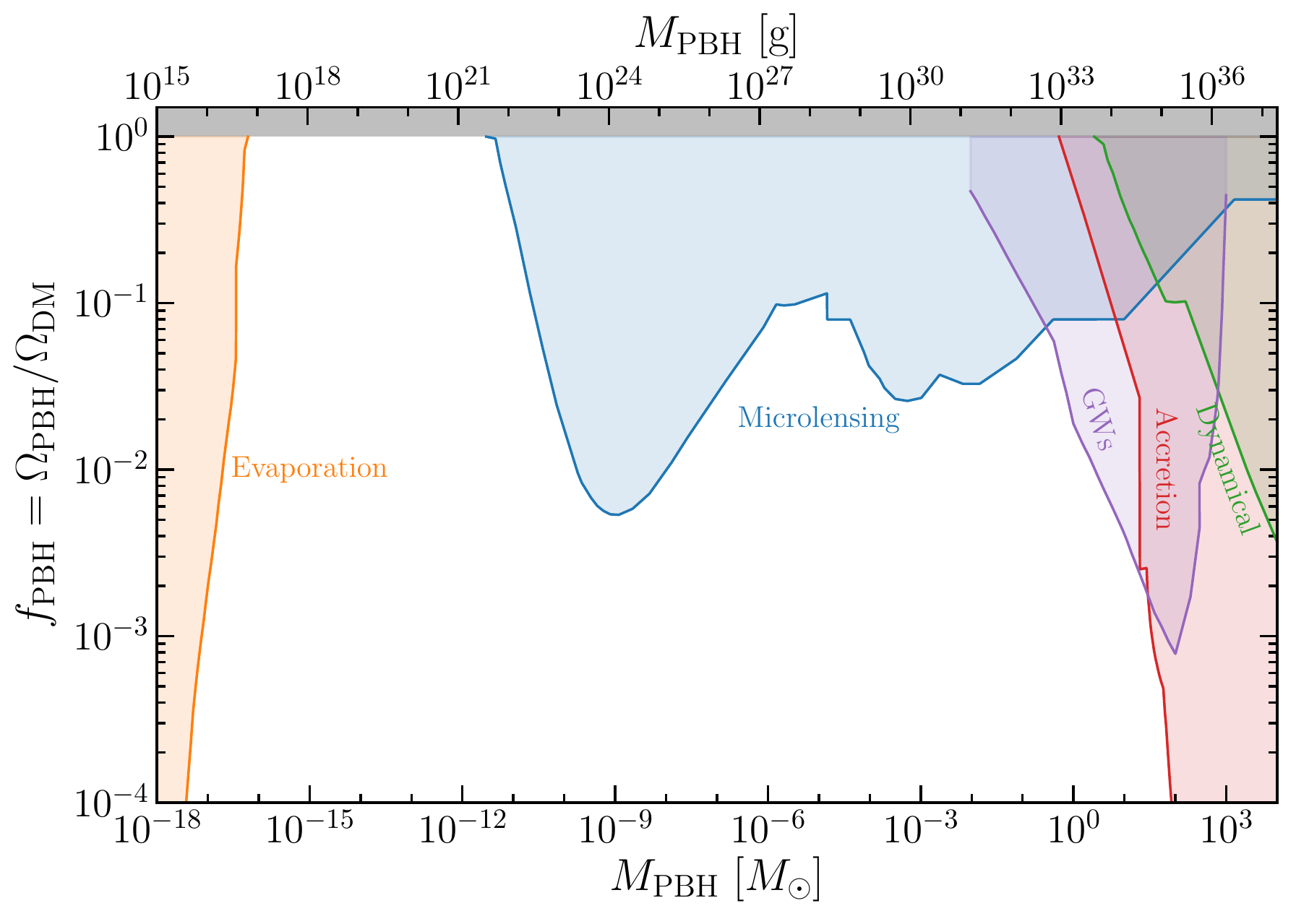}
	\caption{The most relevant experimental bounds on the fraction of dark matter in the form of PBHs \fPBH. The bounds are obtained assuming that the mass function is monochromatic. Credit:~\cite{Green:2020jor} and  \href{https://github.com/bradkav/PBHbounds }{github.com/bradkav/PBHbounds }}
	\label{fig:PBHbounds}
\end{figure}

A variety of searches has allowed to constrain the fraction of DM in the form of PBHs, over different mass ranges. We emphasize that the robustness of many of these bounds is currently the subject of close scrutiny, in particular with respect to assumptions such as PBH clustering or non-trivial mass functions. Without any pretence to a comprehensive account, in this section we discuss the most relevant and widely accepted constraints on \fPBH. \Cref{fig:PBHbounds} shows these bounds, obtained under the assumption of a monochromatic mass function. See~\cite{Carr:2017jsz, Bellomo:2017zsr} for a discussion on how  constraints are to be modified when considering a broad mass function, e.g for a lognormal distribution. Another word of caution regards the redshift dependence that constraints, as these are obtained through observations at different epochs of the Universe history. This point can be relevant for mass functions that evolve through history, e.g. through mergers or accretion~\cite{DeLuca:2020fpg}. 
See for example~\cite{Green:2020jor, Carr:2020gox} for a more complete review of PBH constraints.

\paragraph{Evaporation}

 Hawking predicted that BHs radiate energy with a blackbody spectrum at temperature $T \propto 1/M$~\cite{Hawking:1975vcx,1974Natur.248...30H}. This temperature is completely negligible for astrophysical-size BHs, but becomes relevant for very small masses. \\
 Considering that a blackbody emits at rate per unit area $ \propto T^4$ and the area of a BH is $ A \propto r_s^2 \propto M^2$, the radiation emitted is emitted at a rate $ \propto M^{-2}$. The BH losses mass at he same rate, $ \diff M/\diff t \propto - M^{-2}$, hence the evaporation time scales as  $t_\mathrm{evap} \propto M^3$.  Plugging in numerical factors we have 
 \begin{equation}
 	t_\mathrm{evap} \simeq \SI{13}{\giga\year} \, \left(  \dfrac{M}{  10^{15} \mathrm{g}   }\right)^3
 \end{equation}	
 PBHs lighter than $M \approx 10^{15}$ g would have evaporated by today, which provides a lower limit on the mass of PBH as DM candidates. \\
Slightly larger masses would be evaporating today. Their Hawking radiation would be observable in the extra-Galatic gamma-ray background, as well as in the $e^\pm$ and antiproton fluxes. The non observation of these fluxes allow to set bounds on the abundance of light PBHs~\cite{1991ApJ...371..447M, Carr:2009jm, Laha:2019ssq, Boudaud:2018hqb}. The expected flux of radiation today is negligible for masses  $M \gtrsim 10^{17}$ g.  The bound is shown on the left of \cref{fig:PBHbounds}. \\
Hawking radiation can also be used to constrain the abundance of PBHs of lower mass (which are not a dark matter candidate). In the early Universe, PBH radiation emitted during the re-ionization epoch would affect the 21 cm line~\cite{Clark:2018ghm} and energy injection during recombination or BBN would affect the CMB spectrum~\cite{Poulin:2016anj}. These observations allow to constrain PBH abundance down to $M \sim 10^{10}$ g -- PBHs lighter than this would have evaporated before big bang nucleosynthesis  and are extremely hard to constrain.  

The process of Hawking evaporation eventually leads into the quantum regime as the BH mass reaches the Planck scale, $M_\mathrm{P}=\sqrt{c \hbar/G} \sim 10^{-5} $ g. It is then impossible to say whether BHs evaporate completely or rather leave behind Planck-mass ``relics''. These remnants, if generated before the BBN, could themselves constitute a viable dark matter candidate~\cite{1987Natur.329..308M, Lehmann:2021ijf}.

\paragraph{Microlensing}

\begin{figure}[h]
	\centering
	\includegraphics[width=.6\linewidth]{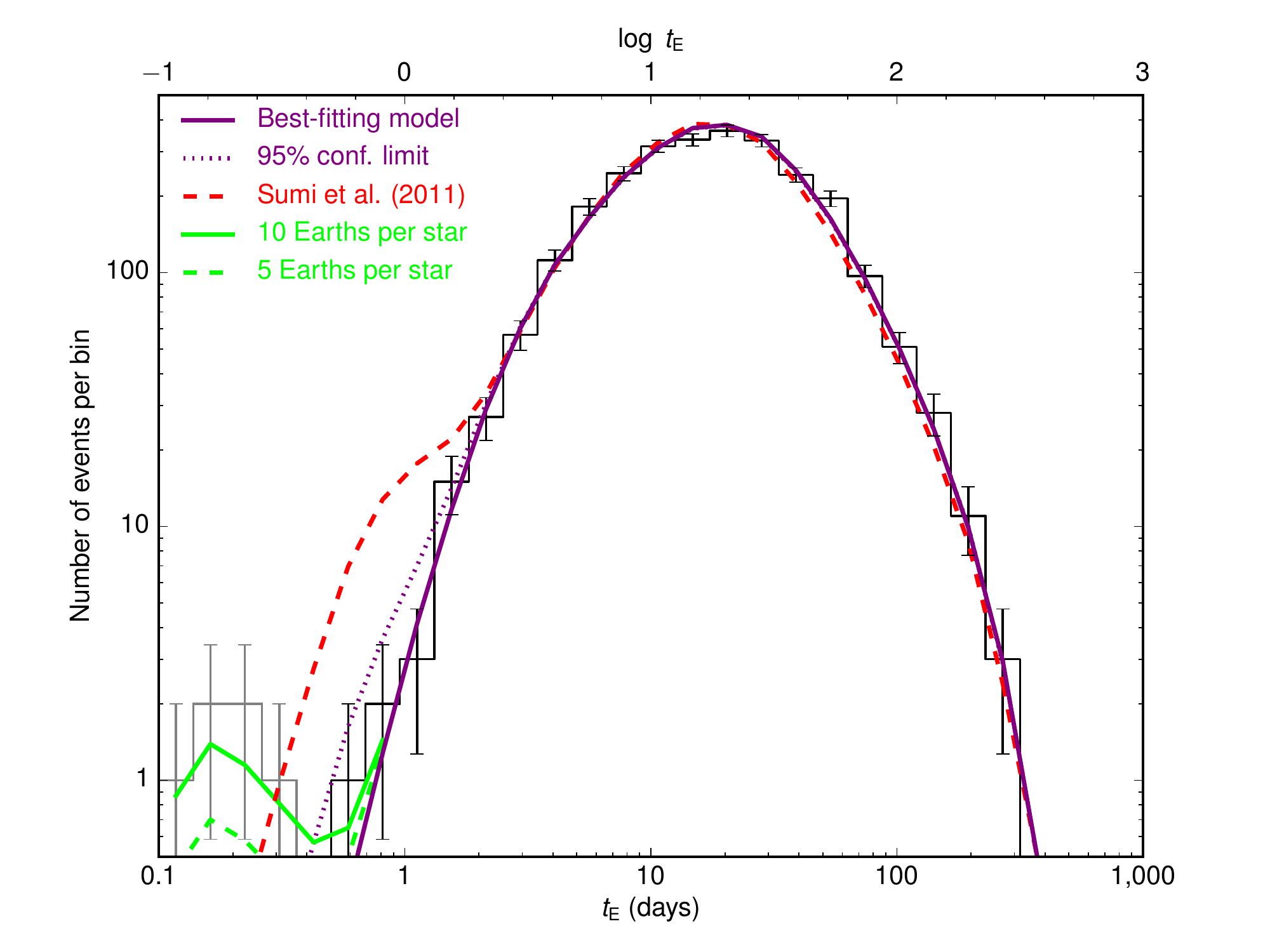}
	\caption{Distribution of micro-lensing events observed by OGLE. Figure from~\cite{2017Natur.548..183M}}
	\label{fig:OGLE}
\end{figure}

A luminous object in a nearby galaxy or cluster can be magnified by gravitational microlensing if a compact object crosses the line of sight~\cite{1986ApJ...304....1P} (this is a particular case of strong lensing, in which the instrumental resolution is not sufficient to separately resolve multiple images). Microlensing surveys observe a background distribution of stars and compare the image at different time, looking for magnification caused by compact objects in our own Galaxy. The range of mass probed by the survey depends on the frequency and on the overall duration of the survey. \\
For small masses of the compact objects the event duration is small and a detection requires high observation frequency (few hours for $M \sim 10^{-6}$) (earth mass). Higher masses may instead require long observation times for the changes to become significant: the time-scale is of the order of months for $M \sim 10 \, M_\odot$. This is the mass range probed by surveys such as MACHO~\cite{Macho:2000nvd} and EROS~\cite{EROS-2:2006ryy}, that observed the Large Magellanic Cloud for a few years with approximately daily cadence, and the more recent OGLE survey of the Galactic centre. OGLE was optimized for observing short events, in the planetary mass range. They interestingly detected an excess of ultra-short events (hours), corresponding to earth-mass objects~\cite{2017Natur.548..183M}. While these events are compatible with the lens being a free-floating planet rather than a compact object, the excess can be interpreted as a hint of the presence of light PBHs~\cite{ Niikura:2019kqi}, see \cref{fig:OGLE}.
A lower mass range has been probed by the Subaru/HSC survey of the Andromeda Galaxy, using a $\sim 2$ minutes sampling rate~\cite{Niikura:2017zjd}. This has allowed to set a quite stringent constraint in the range $10^{-11} - 10^{-6} \, M_\odot$. \\
For yet lower masses of the compact objects, this observation channel is limited by the finite source effect: magnification is drastically reduced when the apparent size of the lens becomes comparable to that of the background object. \\
It has been pointed out that uncertainties in the DM distribution can affect lensing constraints~\cite{2015A&A...575A.107H,Green:2017qoa}; these might also be altered if the compact objects are clustered~\cite{Calcino:2018mwh, Petac:2022rio,Gorton:2022fyb}.

 \paragraph{Accretion}

Gas being accreted by BHs is heated and emits high energy ionizing radiation. 
At present day, accretion on BHs in our Galaxy can emit detectable radiation. This is expected to be particularly relevant in regions of high density gas such as the Galactic centre. Observations in the  X-ray and radio bands have been used to place bounds on the PBH abundance~\cite{Gaggero:2016dpq,Inoue:2017csr, Manshanden:2018tze}. We discuss present day accretion in detail in \cref{sec:accretion}.\\ 
However, PBHs are expected to start accreting gas at early times, soon after matter-radiation equality. The emission of ionized radiation would alter the opacity of the gas in the long period between recombination and reionization. CMB temperature and polarization fluctuations are extremely sensitive to this phenomenon and can be used to constrain it~\cite{1981MNRAS.194..639C, Ricotti:2007au, Ali-Haimoud:2016mbv, Poulin:2017bwe, Serpico:2020ehh}. 
Accretion effects on CMB observation provide the most stringent bound for masses $M \gtrsim 100 \, ~\Msun$.

\paragraph{Gravitational waves}

Mergers of PBH binaries in the range $1-100 \, \Msun$ could be detected by present day observatories.
The rate of mergers form PBHs formed in the early Universe was first computed by Nakamura et al~\cite{Nakamura:1997sm, Ioka:1998nz}. Comparison with the merger rate inferred by observation places constraint on the PBH abundance of the order  $\fPBH \lesssim 10^{-3}$~\cite{Sasaki:2016jop, Ali-Haimoud:2017rtz, Kavanagh:2018ggo, Wong:2020yig, Raidal:2018bbj, Vaskonen:2019jpv, Hall:2020daa}. 
There are significant uncertainties in the calculation of the PBH merger rate, particularly in relation to the effects of structure formation on the evolution of binaries. We will discuss gravitational wave searches for PBHs in detail in \cref{sec:GWPBH}.

\paragraph{Constraints on the power spectrum}

\begin{figure}[h]
	\centering
	\includegraphics[width=.8\linewidth]{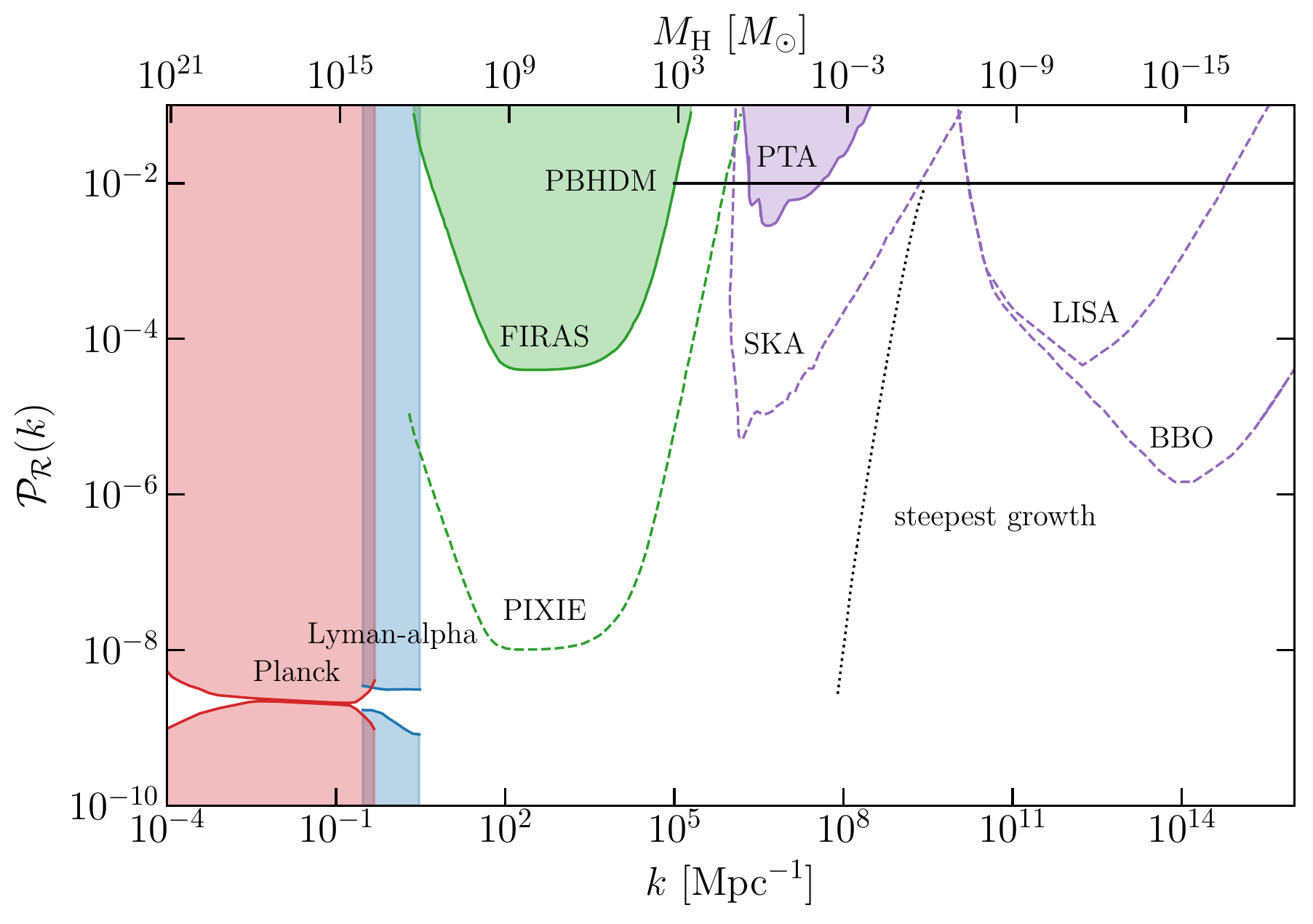}
	\caption{Present and prospect constraints on the primordial curvature power spectrum. Figure from~\cite{Green:2020jor} }
	\label{fig:PSbounds}
\end{figure}

Indirect constraints on PBH abundance can be obtained as bounds on the initial power spectrum, shown in \cref{fig:PSbounds}. \\
Large scalar fluctuations can generate tensor perturbations at second order, which in turn can result  in a detectable primordial stochastic gravitational wave background~\cite{Ananda:2006af,Baumann:2007zm}, see~\cite{Domenech:2021ztg} for a review. 
Pulsar timing arrays (PTA, see~\cref{sec:GWPBH:detection}) can probe the initial power spectrum on scales corresponding to the formation of PBHs in the mass range around $1~\Msun$. SKA is expected to improve significantly the constraints in a similar mass range, while LISA will be able to constrain a smaller mas range (see~\cite{Gow:2020bzo} and references therein).\\
Furthermore, the dissipation of large perturbations can also cause distortions on the CMB spectrum. The curvature power spectrum is constrained by the COBE/FIRAS searches for $\mu$-distortions to be $P_\mathcal{R}(k) \lesssim 10^{-4}$,~\cite{Fixsen:1996nj,1994ApJ...420..439M}, which in turn implies a bound on PBHs larger than $M_\mathrm{PBH}\gtrsim 10^4$\Msun~\cite{Kohri:2014lza,Nakama:2017xvq,Gow:2020bzo}\\
Finally, we mention that conversely, PBHs searches can be used to probe the primordial power spectrum on small scales, presently not accessible by other experiments.

\section{Dark matter structures around black holes}
\label{sec:intro:spikes}

Cold dark matter is expected, under quite general hypothesis, to forms large overdensities around BHs (``spikes''). 
These dark matter overdensities around BHs have been extensively studied in the context of searches for annihilating/decaying DM, as the high densities of the spikes are interesting because they result in boosted signals.

More recently, it has been shown that the dense environments of these structures can leave an imprint on the gravitational waves that are emitted when two BHs inspiral and merge~\cite{Eda:2013gg}\cite{Yue:2017iwc}\cite{Kavanagh:2020cfn}. If the merging BHs are similar in mass, the spikes will be quickly disrupted. However, if the mass ratio is large, the spike around the more massive BH will cause the light one to loose energy through gravitational friction. This will cause it to inspiral faster, resulting in a \textit{dephasing} of the GW signal with respect to the in-vacuum scenario. For an extreme mass-ratio inspiral (EMRI), the loss of energy through gravitational friction is negligible with respect to the emission in GW. The largest effects on the GW signal are obtained instead for intermediate mass-ratio inspirals (IMRI), with mass ratios $q \sim 10^3 - 10^5$. If one of the two bodies is a solar-mass compact object (BH or neutron star), such mass ratios are achieved in an inspiral with an intermediate mass black hole (IMBH).\\
The inspiral of an IMBH with a solar mass BH will emit low frequency GWs, falling in the sensitivity band of the planned Laser Interferometer Space Antenna (LISA)~\cite{Babak:2017tow}. Preliminary studies indicate that LISA will have the potential to reveal the presence of a DM spike around the IMBH through its effect on the GW signal. This opens an exciting perspective: the possibility of using future gravitational wave observatories to constrain the properties of dark matter~\cite{2021arXiv210804154C}\cite{Hannuksela:2019vip}.\\
For this to program to be put into practice, it is necessary to predict the modified gravitational waveforms with high precision. Consequently, it is essential to perform an accurate modelling of the DM spike around IMBHs. In general, we expect the properties of the spike to depend, on the one hand, on the characteristics of the dark matter and, on the other hand, on the formation and evolution history of the system. 

To obtain the slope of the dark matter spike, the pioneering work of Gondolo and Silk~\cite{Gondolo:1999ef} considered the adiabatic growth of a seed black hole at the centre an isotropic and spherically symmetric halo. Assuming an initial power-law profile and a final state dominated by the central BH potential, they computed the steepening of the inner part of the halo. They found that an initial profile with slope $\gamma $, $0 < \gamma < 2$, is steepened to create a spike with slope $\gamma_\mathrm{ sp} = (9-2 \gamma) / (4 - \gamma)$. For instance, an NFW initial inner profile ($ \gamma =1$) is transformed into a spike with slope $\gamma_\mathrm{ sp} = 7/3$. The final profile is only mildly dependent on the initial slope: $2.25 < \gamma_\mathrm{ sp} < 2.5$ for all values of $\gamma $ between 0 and 2.\\
Ullio et al.~\cite{Ullio:2001fb} later showed that the formation of such a steep DM spike relies strongly on the conditions that most of the BH growth takes place adiabatically and close to the halo centre. 
They showed that instantaneous growth would produce a less pronounced spike, with~$\gamma_\mathrm{ sp} \sim 4/3$. Off-centre formation, in a region of uniform background density, would also result in a shallower spike,~$\gamma_\mathrm{ sp} \sim 3/2$~~\cite{Zhao:2005zr}. 

It appears  then that the shape of the DM overdensity  has an important dependency on the way the IMBH is formed. 
Among the proposed formation channels, one possibility is to obtain IMBHs from population III stars, which can have masses up to $10^3 M_\odot$. In this case, the BHs would typically be formed off-centre of the host halo. Another possible channel is the \emph{direct collapse} of most of the gas in a large halo. The BHs formed this way are expected to be more massive and, as they form at the halo centre, carry steeper spikes~\cite{volonteri2021origins, Wanders:2014xia}. 
Beyond formation, the subsequent evolution of the system until observation must be taken into account. Mergers occurring during the hierarchical process of structure formation can disrupt the spikes~\cite{Merritt:2002vj, Wanders:2014xia}. Nevertheless, a significant fraction of these is expected to survive until present time~\cite{Wanders:2014xia}.

%
\chapter{Gas accretion: black holes in shiny dresses}
\label{sec:accretion}

The process of matter accretion onto BHs is extremely efficient in converting gravitational energy into heat and radiation. The gas being accreted is heated up to millions of Kelvin, producing substantial amounts of non-thermal radiation (mainly from bremsstrahlung and inverse Compton scattering), which are able to escape the gravitational pull of the black hole and can be detected, mainly in the X-ray and radio band.\\
This highly luminous radiation provides a mean to probe, through electromagnetic observations, these otherwise completely dark objects \cite{1973A&A....24..337S}. 
Until recently, all detections of BHs were achieved through the accretion process. These include distant accreting supermassive black holes, known as  Active Galactic Nuclei~\cite{1984ARA&A..22..471R}, and a significant number ($\sim 70$) of X-ray binaries~\cite{Tauris:2003pf, Corral-Santana:2015fud}, which are Milky Way BHs accreting gas from a companion star.

The observed X-ray binaries represent only a very small fraction of the black holes present in our Galaxy. Around total $10^8$ astrophysical BHs are estimated to be present in the Milky Way. While a large fraction of stars are found in binary systems, BHs receive what is known as a \emph{natal kick} at their formation: a large impulse that is likely to separate them from their companion. Hence, most astrophysical BHs are expected to be isolated.\\
On top of this, if PBH are a significant fraction of the DM, we expect an further large population of isolated BHs to be present the Galactic halo (the estimated mass of the DM halo is around $10^{12 } ~\Msun$ ).\\
Recent tentative detections of isolated black holes have been obtained through microlensing (see e.g. \cite{Wyrzykowski:2019jyg,Karolinski:2020jey} and \cite{Calcino2018} for a discussion of the constraints that can be derived from these observations). 

This chapter is dedicated to discussing the possibility of detecting \emph{isolated} BHs through the accretion process. Our Galaxy is rich in dense clouds of molecular hydrogen and, under certain conditions, a BH accreting gas from such clouds can be luminous enough to be detected. We consider the sensitivities of present and future surveys, and also discuss how multi-wavelength studies can be used to unambiguously identify sources as isolated accreting BHs.

We start by describing a state-of-the-art model for the accretion process~\cite{Park:2012cr, Sugimura:2020rdw}, which is able to account for the role of radiative feedback on the interstellar medium.
We then apply this phenomenological model to searches for astrophysical black holes in section \cref{sec:accretion:ABH}, based on Ref.~\cite{Scarcella:2020ssk}.
This is an interesting problem \emph{per se} but also relevant as the background for PBH searches.
We turn to the search of primordial black holes in \cref{sec:accretion:PBH}, based on Ref.~\cite{Scarcella:2021jzp} and Ref.~\cite{Scarcella:2022pbh}.\\

\section{ Modelling the accretion process}
\label{sec:accretion:models}

We start by discussing how to model the accretion rate $\dot{M}$ for a BH accreting from a gas cloud, since, as we will see in \cref{sec:accretion:emission}, the expected luminosity of the system scales with $\dot{M}$. We first introduce the textbook Bondi-Hoyle-Lyttleton (BHL) model, then show how this must be modified to account for the fact that the radiation emitted by the accreting gas creates a ionized region around the BH, following \cite{Park:2012cr, Sugimura:2020rdw}.

\subsection{The Bondi-Hoyle-Lyttleton model}
\label{sec:accretion:models:BHL}

According to the BHL accretion model \cite{Hoyle1939,Bondi1944}, the rate of accretion onto an isolated compact object of mass $M$ moving at a constant speed $\mathrm{ v}_\mathrm{BH}$ is given by:
\begin{equation}
	\label{eq:bondi}
	\dot{M}_\mathrm{BHL} = 4 \pi \frac {(GM)^2 \rho} {(\mathrm{ v}_\mathrm{BH}^2 + c_\mathrm{s}^2)^{3/2}} \,\,\, ,
\end{equation}
where $\rho$ and  $c_\mathrm{s}$ are, respectively, the density and the sound speed that characterize the ambient medium and $G$ is the gravitational constant. 
The accretion rate scales linearly with the density of the interstellar medium, and increases for lower velocities, $\dot{M }\sim \mathrm{ v}_\mathrm{BH}^{-3}$. The peak value is achieved for $\mathrm{ v}_\mathrm{BH} = 0$.\\
This formula is usually re-scaled by introducing a suppression factor $\lambda$. Ref.~\cite{Fender:2013} for instance, found that values larger than $\lambda=0.01$ are excluded, under realistic assumptions, by observations of the local region, where a significant population of isolated BHs should be present. Refs~\cite{Tsuna:2019kny} and \cite{Tsuna:2018oqt} considered values between $\lambda=0.1$ and $\lambda=10^{-3}$. The $\lambda$ parameter may effectively capture the outflow of material that is expelled from the Bondi sphere, as suggested by several authors \citep{Blandford:1999}. Its introduction is also supported by the non observation of a large population isolated neutron stars in the local region \citep{Perna:2003} and the studies of nearby AGNs \citep{Pellegrini:2005} as well as the supermassive BH at the Milky Way centre, Sagittarius A* \citep{2003ApJ...591..891B}. In conclusion, it seems that the BHL accretion formula may overestimate the accretion rate by orders of magnitude, although the physical mechanism behind this deviation is still disputed. One such mechanism is introduced in PR13, as we discuss below.

\subsection{ The Park-Ricotti model}
\label{sec:accretion:models:PR}

The hydrodynamical simulations performed in PR13 \cite{Park:2012cr, Sugimura:2020rdw} capture 
the ionizing effect of the high energy radiation emitted during the accretion process. \\

\begin{figure}[ht]
	\centering
	\includegraphics[width=.49\linewidth]{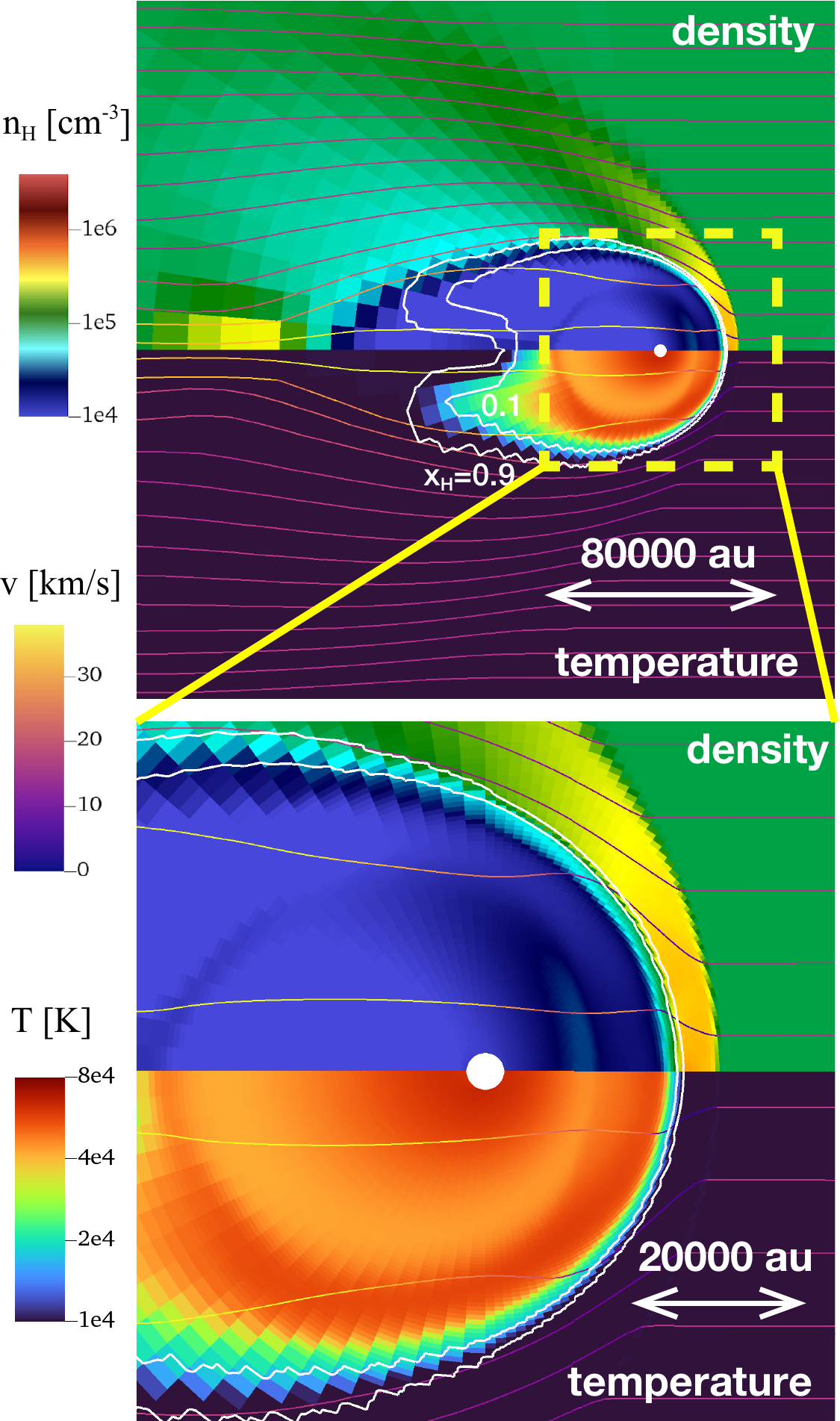}
	\caption{ Figure from \cite{Sugimura:2020rdw} showing the simulation of gas flow around a BH for $\mathrm{ v}_\mathrm{BH}=2 {c}_\mathrm{s}$. The gas is moving from the right side to the left, with the BH located at the centre of the sink region (white circle). The bottom panel is a zoom-in view of the top panel where the entire ionized bubble is displayed. In each panel, the density (upper) and the temperature (lower) are shown, together with the velocity streamlines. One can notice the formation of an under-dense, hot ionized region around the BH, preceded by a bow shock. }
	\label{fig:PRsimulation}
\end{figure}

\clearpage

A cometary-shaped ionized region is generated around the BH, as it moves through the interstellar medium. Furthermore, for certain values of the BH velocity, the formation of a bow-shock preceding the ionization front is observed, see \cref{fig:PRsimulation}.
Taking into account this \emph{radiative feedback} drastically modifies the accretion rate with respect to the prediction of the BHL model. Specifically, it reduces, by many orders of magnitude, the accretion rate for low BH speeds.

\begin{figure}[t]
	\centering
	\includegraphics[width=.49\linewidth]{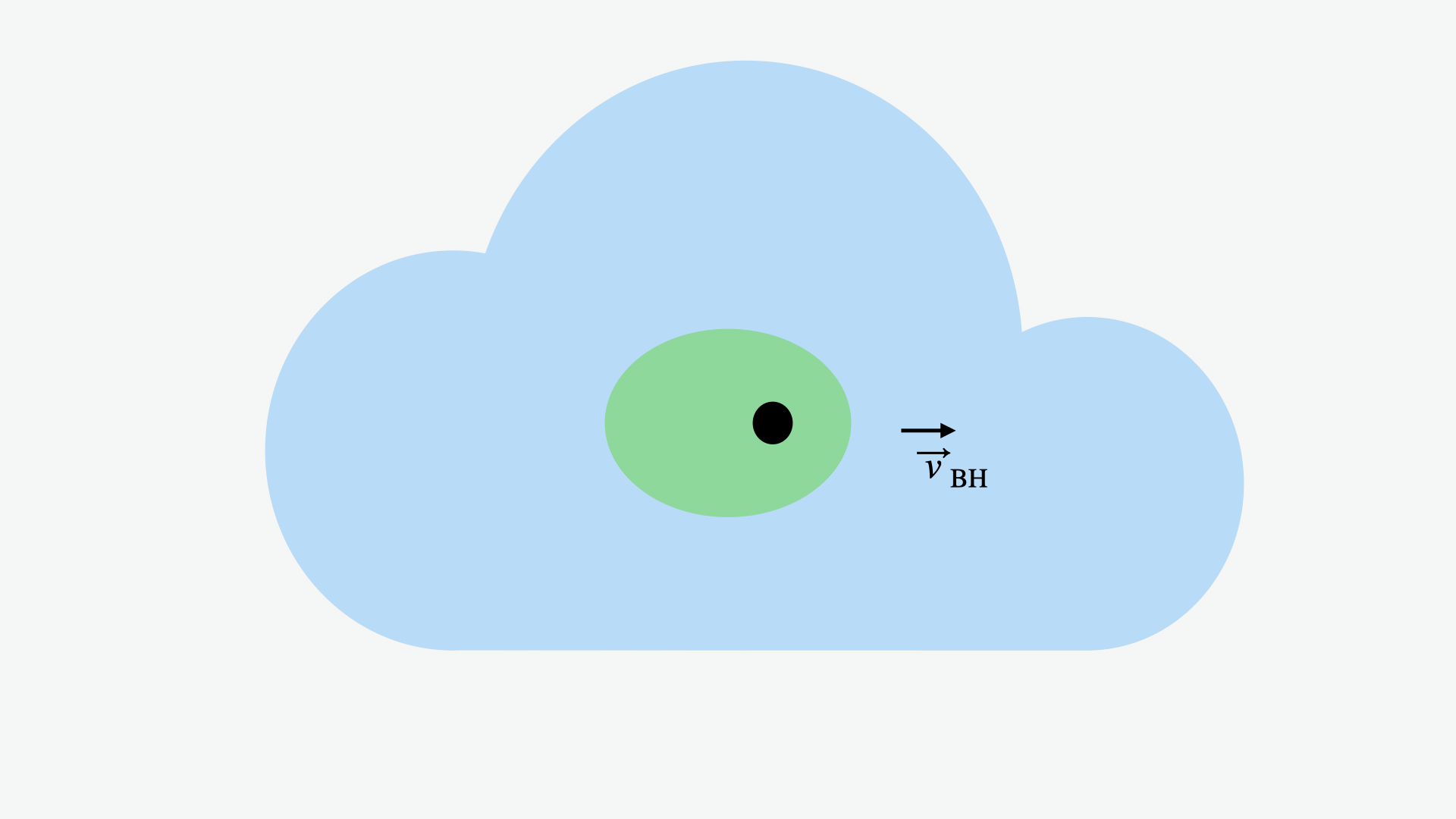}
	\includegraphics[width=.49\linewidth]{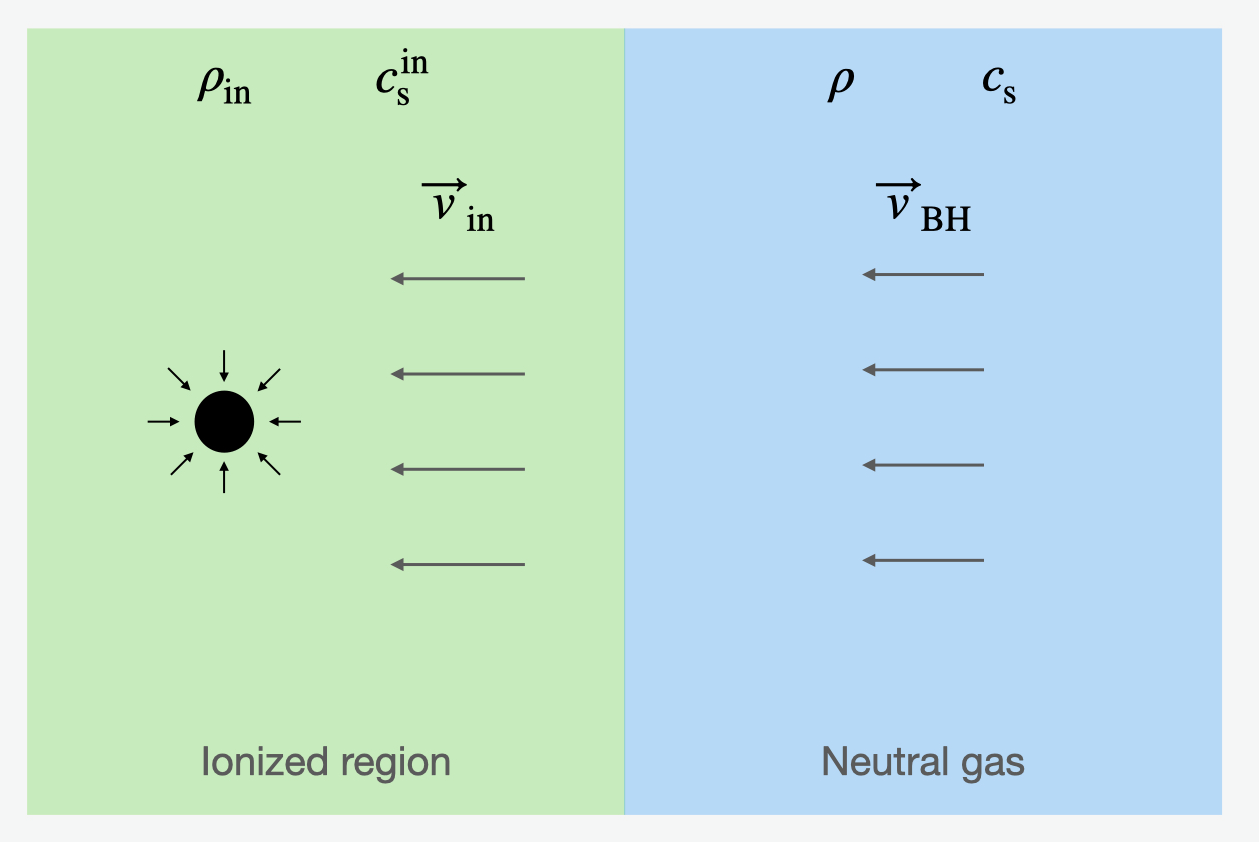}
	\caption{Schematic representation of the formation of an ionized bubble around a BH as it moves through a dense cloud (left) and the one dimensional model for the ionization front (right). }
	\label{fig:ionizationfront}
\end{figure}

In the same work~\cite{Park:2012cr}, the authors combined the BHL formula \ref{eq:bondi} with the modelling of the ionization front to obtain an analytical formula in agreement with the results of the simulations. We will refer to this as the PR13 model. \\
Within the ionized region surrounding the BH, BHL accretion is assumed to hold. However, within this region the density and the speed of the gas relative to the BH are different from their asymptotic values (before the ionized front), see \cref{fig:ionizationfront}. The PR13 accretion rate can be written from \cref{eq:bondi} as:
\begin{equation}
	\label{eq:ricotti}
	\dot{M}_\mathrm{PR13} = 4 \pi \frac {(GM)^2 {\rho}_\mathrm{in}} {(\mathrm{ v}_\mathrm {in}^2 + (c_\mathrm{s}^\mathrm{in})^2)^{3/2}} \,\,\, ,
\end{equation}
where $c_\mathrm{s}^\mathrm{in}$ and ${\rho}_\mathrm{in}$ are the sound speed and density of the ionized medium, and  $\mathrm{ v}_\mathrm {in}$ is its velocity relative to the BH. The sound speed $c_\mathrm{s}^\mathrm{in}$ is on the order of few tens of km/s (the temperature of the HII gas is expected to be around $4 - 5 \times 10^4$ K \cite{Sugimura:2020rdw}), significantly higher than the external one (molecular clouds have temperatures around 10~K, corresponding to $ c_\mathrm{s}\sim 1 $~km/s). 

The problem is then to relate ${\rho}_\mathrm{in}$ and $\mathrm{ v}_\mathrm {in}$ to the corresponding quantities referred to the external neutral medium: the cloud density $\rho$ and the velocity of the BH relative to it, $\mathrm{ v}_\mathrm{BH}$, taking into account the different temperature of the two regions.

This can be done starting from Euler's equations for mass and momentum conservation applied to the ionization front, assuming a one-dimensional steady flux. The resulting ``jump conditions'' are 
\begin{align}
	\label{eq:masscons}
	&{\rho}_\mathrm{in}\mathrm{ v}_\mathrm {in}  = {\rho}\,{\mathrm{ v}_\mathrm{BH}}  \,\,\, ,      \\[5pt]
	\label{eq:momcons}
	&{\rho}_\mathrm{in}\mathrm{ v}_\mathrm {in}^2+P_\mathrm{in}  =  {\rho}\mathrm{ v}_\mathrm{BH}^2+P_\mathrm {out} \,\,\, ,
\end{align}
where $P$ indicates pressure. The second of these equations can be written, assuming ${c}_\mathrm{s}^2 =  P/\rho $, as
\begin{equation}
	{\rho}_\mathrm{in}(\mathrm{ v}_\mathrm {in}^2+(c_\mathrm{s}^\mathrm{in})^2)  =  {\rho}(\mathrm{ v}_\mathrm{BH}^2+{c}_\mathrm{s}^2) \,\,\, .
\end{equation}
This system of equations  has the following solutions:
\begin{equation}
	\begin{aligned}
		&\rho_\mathrm{in } = \rho_\mathrm{in }^{\pm} \equiv \rho \, \frac{\mathrm{ v}_\mathrm{BH}^2+c_\mathrm{s}^2 \pm \sqrt{\Delta}}{2 \, (c_\mathrm{s}^\mathrm{in})^2 } \; , \qquad \Delta \equiv (\mathrm{ v}_\mathrm{BH}^2+c_\mathrm{s}^2)^2 - 4 \,\mathrm{ v}_\mathrm{BH}^2 \, (c_\mathrm{s}^\mathrm{in})^2 \label{eq:deltarho}\\
		&\mathrm{ v}_\mathrm {in} = \frac{{\rho}}{{\rho}_\mathrm{in}}\mathrm{ v}_\mathrm{BH}  \; ,
	\end{aligned}
\end{equation}
for $\mathrm{ v}_\mathrm{BH} \leq \mathrm{ v}_\mathrm{D}$  or $\mathrm{ v}_\mathrm{BH} \geq \mathrm{ v}_\mathrm{R}$, with $\mathrm{ v}_\mathrm{D}$ and $\mathrm{ v}_\mathrm{R}$ the roots of $\Delta$. Since typically $c_\mathrm{s}^\mathrm{in}\sim\mathcal{O}$(10)~km/s while $c_\mathrm{s}\sim\mathcal{O}$(1)~km/s, we have:
\begin{equation}
	\label{eq: velocities}
	\begin{aligned}
		&\mathrm{ v}_\mathrm{R}
		\approx 2 c_\mathrm{s}^\mathrm{in} \, , \\
		&\mathrm{ v}_\mathrm{D}
		\approx \frac{c_\mathrm{s}^2}{2 c_\mathrm{s}^\mathrm{in}} \; \ll 1 \mathrm{km/s} .
	\end{aligned}
\end{equation}
We can distinguish three regimes:

\paragraph{Low velocity regime}

At very low velocities, $ \mathrm{ v}_\mathrm{BH} \leq \mathrm{ v}_\mathrm{D} \approx  1 \mathrm{km/s} $, the ionization front is best described by by $\rho_\mathrm{in}^+$ \citep{Park:2012cr}. Due to the higher pressure, the density in the ionized region is significantly lower than the external one and, correspondingly, the velocity increases by the same factor (\cref{eq:masscons}). The combination of these two factors results in a significantly suppressed accretion rate (\cref{eq:ricotti}). In this regime, the accretion rate increases -- as for the BHL model -- towards lower velocities. However, the peak it reaches at $\mathrm{ v}_\mathrm{BH}=0$ is significantly lower than in the BHL case.

\paragraph{Intermediate velocity regime}

As the BH velocity increases, so does the  --~much higher~-- velocity of the gas within the ionized bubble . When $ \mathrm{v}_\mathrm{BH} \sim \mathrm{v}_\mathrm{D} $, $ \mathrm{v}_\mathrm{in } $ reaches the speed of sound.  At this point, a bow shock front forms in front of the ionized bubble. 
\begin{figure}[t]
	\centering
	\includegraphics[width=.49\linewidth]{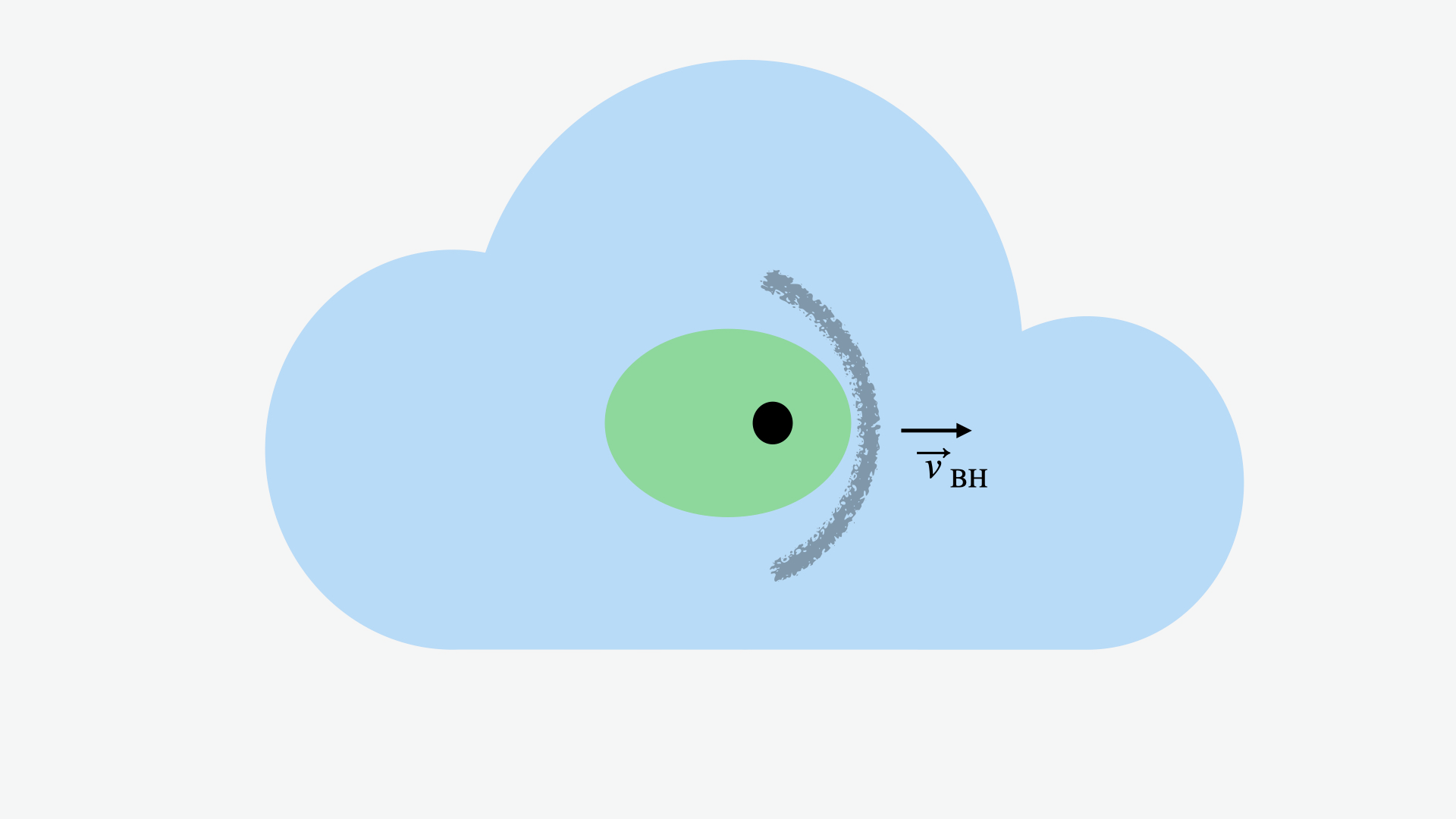}
	\includegraphics[width=.49\linewidth]{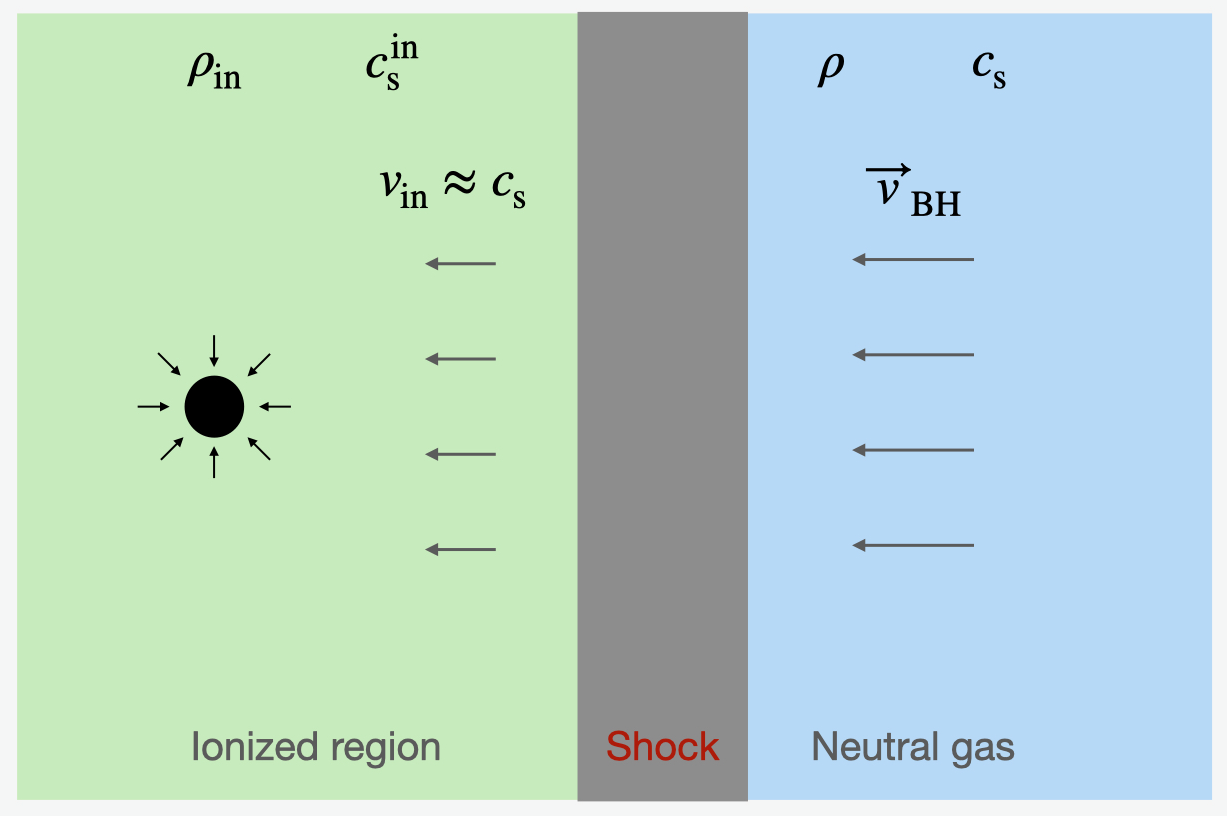}
	\caption{Appearance of a shock in the intermediate velocity regime. }
	\label{fig:ionizationfront2}
\end{figure}
The gas experiencing the shock forms a dense shell between the shock and the ionization front. 

The velocity of the gas beyond the shock is reduced 
and the gas accumulates between the two fronts, while its flux is in part deviated from the direction of the BH displacement (see \cref{fig:PRsimulation,fig:ionizationfront2}). \\
In this regime, which occurs for $\mathrm{ v}_D < \mathrm{ v}_\mathrm{BH} < \mathrm{ v}_R \approx 2 c_\mathrm{s}^\mathrm{in}  $ , the gas velocity inside the ionized region remains approximately fixed at $\mathrm{ v}_\mathrm {in} \approx c_\mathrm{s}^\mathrm{in} $. This relation, promoted to an equality, can be used together with Eq. \ref{eq:momcons} to compute the density $\rho_\mathrm{in}$. This way, for  $\mathrm{ v}_D \leq \mathrm{ v}_\mathrm{BH} \leq \mathrm{ v}_R$, we obtain:
\begin{equation}
	\begin{aligned}
		&\rho_\mathrm{in } = \rho_\mathrm{in }^0 \equiv \rho \frac{\mathrm{ v}_\mathrm{BH}^2+c_\mathrm{s}^2 }{2 \, (c_\mathrm{s}^\mathrm{in})^2 } \;, \\
		&\mathrm{ v}_\mathrm {in} = {c}_\mathrm{s,in} \, .
	\end{aligned}
\end{equation}
Notice how in this regime, while the velocity $\mathrm{ v}_\mathrm {in}$ remains approximately constant, the density $\rho_\mathrm{in}$ increases with the BH velocity. Comparing with \cref{eq:ricotti}, one can see that in this phase the accretion rate \emph{increases} with the BH speed, contrary to the usual behaviour. In fact in this regime one has
\begin{align}
	\dot{M}_\mathrm{PR13} = \pi \frac{(GM)^2 \rho (\mathrm{ v}_\mathrm{BH}^2 + c_\mathrm{s}^2)}{\sqrt{2} \, (c_\mathrm{s}^\mathrm{in})^5} \, ,
	\label{eq:intermediatePR13}
\end{align}
which increases quadratically with the BH velocity, in sharp contrast to the behaviour of the BHL rate, which decreases with velocity.

\paragraph{High velocity regime}

The accretion rate reaches its peak at $\mathrm{ v}_\mathrm{BH}  = \mathrm{ v}_R \approx 2 c_\mathrm{s}^\mathrm{in} $. At higher velocities, the shock front disappears and the flux can be again considered one-dimensional. One has in this case a weak ionization front, given by the solution $\rho_\mathrm{in}^-$ in \cref{eq:deltarho}. At high velocities, the pressure term in \cref{eq:momcons} becomes negligible, and one recovers the values of velocity and density of the external medium. Correspondingly, one finds agreement with the BHL description.

\noindent
In summary the PR13 model is given by \cref{eq:ricotti} and
\begin{equation}
	\rho_\mathrm{in} \quad 
	= \quad
	\begin{cases}
		\quad
		\rho_\mathrm{in}^- \; , \qquad 
		& \, \mathrm{ v}_\mathrm{BH} \geq \mathrm{ v}_\mathrm{R} \;  \\[5pt]
		\quad  
		\rho_\mathrm{in}^0 \; , \qquad 
		& \, \mathrm{ v}_\mathrm{D } < \mathrm{ v}_\mathrm{BH} < \mathrm{ v}_\mathrm{R} \;  \\[5pt]
		\quad
		\rho_\mathrm{in}^+ \; , \qquad
		& \, \mathrm{ v}_\mathrm{BH} \leq \mathrm{ v}_\mathrm{D } \; 
	\end{cases}\\[20pt]
\end{equation}
\begin{equation}
	\mathrm{ v}_\mathrm {in} \quad 
	= \quad
	\begin{cases}
		\quad
		\dfrac{\rho}{{\rho}_\mathrm{in}}\mathrm{ v}_\mathrm{BH} \; , \qquad
		& \, \mathrm{ v}_\mathrm{BH} \geq \mathrm{ v}_\mathrm{R} \;  \\[5pt]
		\quad  
		c_\mathrm{s}^\mathrm{in} \; , \qquad
		& \, \mathrm{ v}_\mathrm{D } < \mathrm{ v}_\mathrm{BH} < \mathrm{ v}_\mathrm{R} \;  \\[5pt]
		\quad
		\dfrac{\rho}{{\rho}_\mathrm{in}}\mathrm{ v}_\mathrm{BH} \;, \qquad
		& \, \mathrm{ v}_\mathrm{BH} \leq \mathrm{ v}_\mathrm{D } \; 
	\end{cases}\\[20pt]
\end{equation}
where $\rho_\mathrm{in}^\pm$ are defined in \cref{eq:deltarho}.

\begin{figure}[tb]
	\centering
	\begin{subfigure}
		\centering
		\includegraphics[width=.49\linewidth]{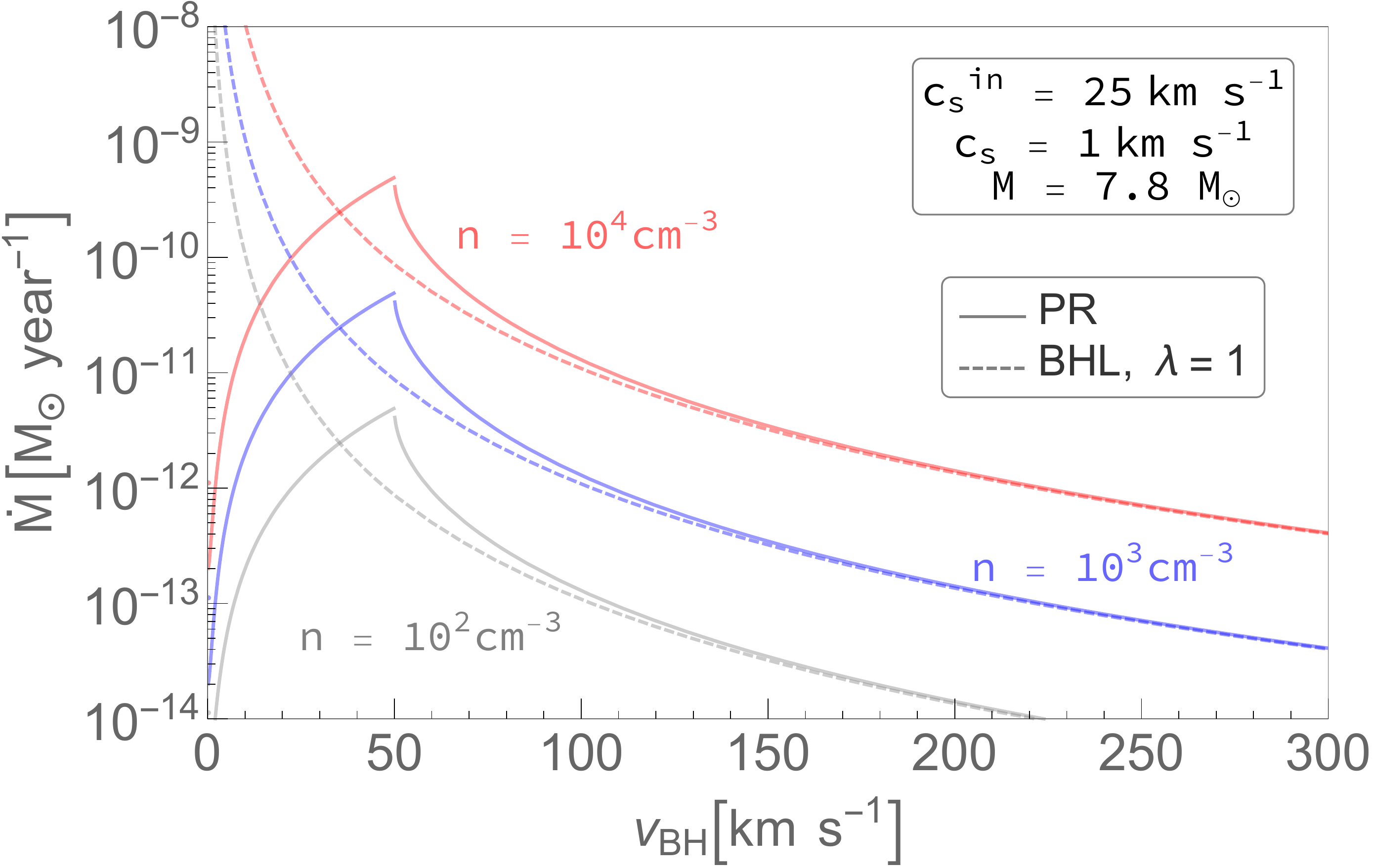}
	\end{subfigure}
	\begin{subfigure}
		\centering
		\includegraphics[width=.49\linewidth]{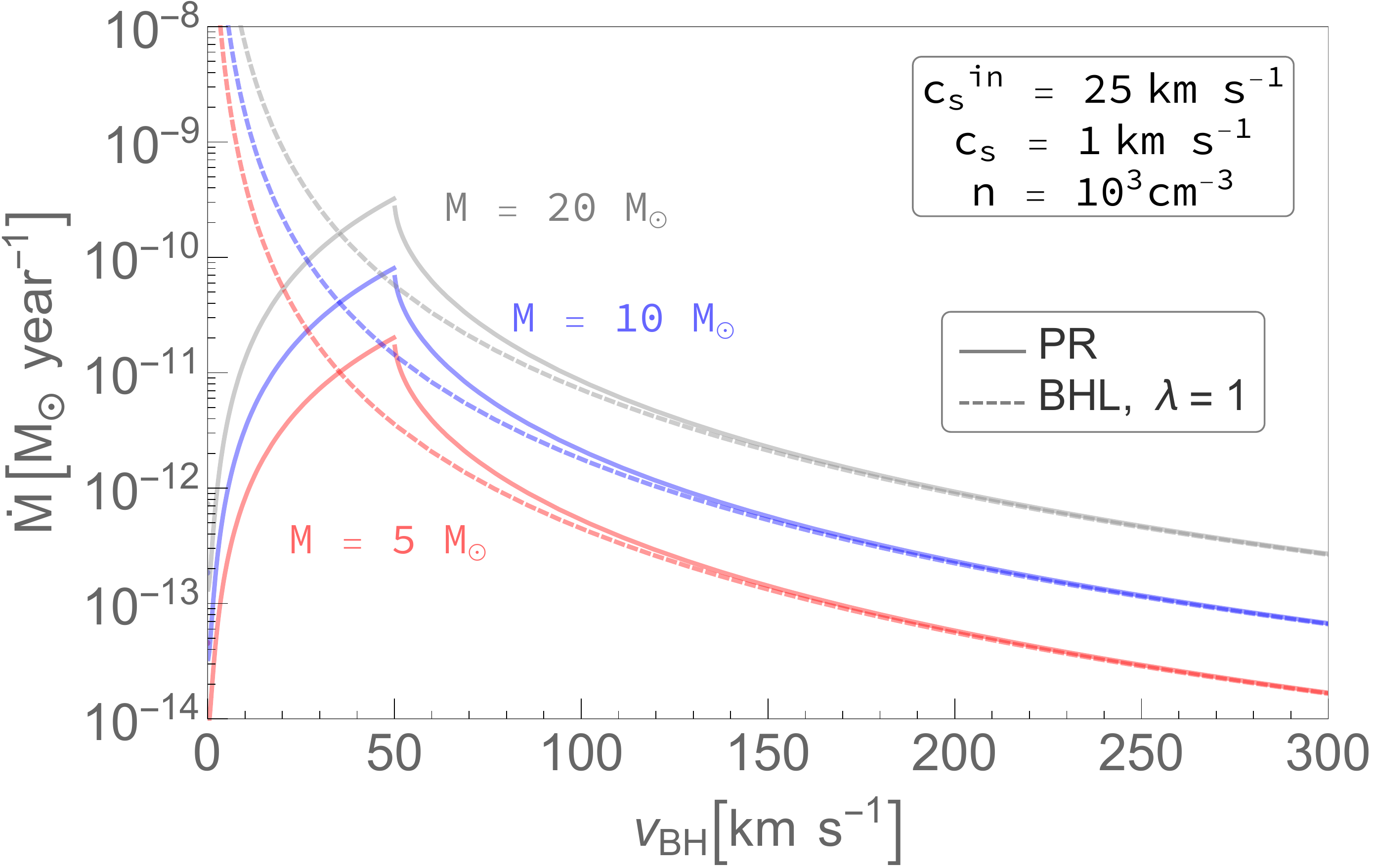}
	\end{subfigure}
	\caption{Accretion rate obtained according to the PR13 and BHL models, as a function of the BH speed. For the BHL rate, we set the suppression factor $\lambda =1$, to allow for a more direct comparison. The two models agree at high velocity, but predictions differ by many orders of magnitude in the low velocity range. The left and right panels show the scaling of the accretion rate with the cloud density and the BH mass, respectively.}
	\label{fig:accretion1}
\end{figure}

\begin{figure}[tb]

		\centering
		\includegraphics[width=.6 \linewidth]{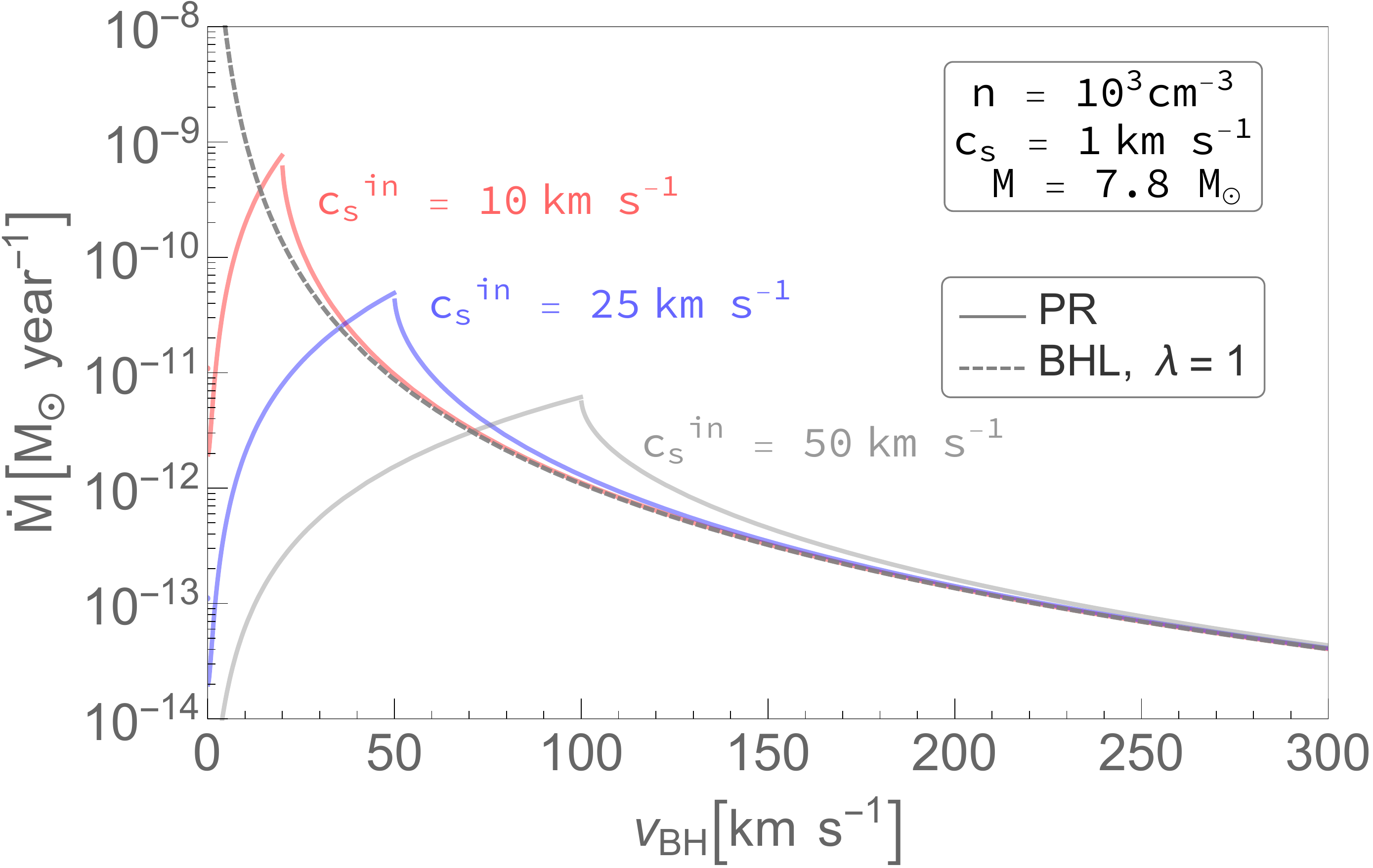}
	\caption{Accretion rate obtained according to the PR13 and BHL models, considering different values for the sound speed within the ionized region \csin.  In the PR 13 model, the peak of the accretion rate is reached when $\vBH \approx 2 \csin$.}
	\label{fig:accretion2}
\end{figure}

The complete velocity dependence of the PR13 rate is shown in \cref{fig:accretion1,fig:accretion2}, for varied gas densities, BH masses, and sound speeds of the ionized region. For comparison, the BHL rate with $\lambda = 1$ is also shown. We can observe in this figure how the  BHL rate decreases monotonically with velocity, whereas the PR13 rate peaks at $ \mathrm{ v}_\mathrm{BH} = \mathrm{ v}_\mathrm{R} = 2 \, c_\mathrm{s}^\mathrm{in}$ and is suppressed at lower velocity by the presence of the bow shock.
In the low velocity regime, $v < \mathrm{ v}_\mathrm{D} $, the rate increases again. However, this applies to velocities $\lesssim 0.1$ km/s (see Eq. \ref{eq: velocities}), which are of little relevance for this work and not shown in figure. 

The different velocity dependence of the PR13 rate compared to the BHL rate has important consequences when studying the emission properties of a BH population characterized by a given velocity distribution. According to the BHL prescription, the low-velocity tail of the population is the easiest to detect. Following PR13, the highest emissions are instead associated to BHs with intermediate velocities. \\
Furthermore, BHL can predict very high accretion rates if the speeds are low enough, which can easily lead to overshooting experimental bounds, while the highest rates predicted by PR13 are orders of magnitude smaller. \\

\section{Emission mechanism}
\label{sec:accretion:emission}

Using \cref{eq:intermediatePR13} to express the peak of the PR13 rate in terms of the Eddington accretion rate $\dot{M}_\mathrm{Edd}$ 
\begin{align}
	\frac{\left. \dot{M}_\mathrm{PR13} \right\rvert_{\mathrm{ v}_\mathrm{BH}=\mathrm{ v}_\mathrm{R}}}{\dot{M}_\mathrm{Edd}} \approx 10^{-4}
	\left( \frac{M_\mathrm{BH}}{10 \, M_\odot} \right) \left( \frac{\rho/m_\mathrm{p}}{10^3  \, \mathrm{cm}^{-3}} \right) \left( \frac{c_{s,\mathrm{in}}}{25 \, \mathrm{km/s}} \right)^{-3}  \, ,
	\label{eq:PR13_rate}
\end{align}
one can see that the accretion rate will always be highly sub-Eddington in the range of BH masses $M_\mathrm{BH}$, gas densities $\rho$ and ionized sound speeds $c_{s,\mathrm{in}}$ we consider in this work. 

In order to translate the predicted accretion rates of a population of isolated BHs into a prediction of detectable sources we need estimates of the associated bolometric luminosity of a given source, and the Spectral Energy Distribution (SED). Our focus is on the X-ray and radio luminosity, $L_\mathrm{X}$ and $L_\mathrm{R}$, of the accreting BH. We can estimate both by considering the SED of the BH at varying accretion rates. We are particularly interested in emission from highly sub-Eddington accreting BHs (as predicted by Eq.~\ref{eq:PR13_rate}). Therefore in this section we describe our simple framework of the properties of sub-Eddington flows onto isolated BHs based on our prior understanding of known weak accretors in nature: Galactic BH-XRBs, and low-luminosity AGN. 

Radiatively Inefficient Accretion Flow (RIAF) models have spearheaded many studies regarding the emission mechanisms associated with such sub-Eddington accretion flows, with a focus both on Galactic BH-XRBs at low accretion rates, and the supermassive BH in the Galactic centre, Sgr~A* \citep{Narayan:1994,yqn03}. \\
The regularly cited RIAF models which first considered the emission processes in inefficiently radiating accreting BHs are classed as advection-dominated accretion flows (ADAFs; \cite{Narayan:1994,Esin1997}). In the ADAF model, Bremsstrahlung radiation dominates the observed spectrum at the lowest accretion rates, and thus the spectrum has some curvature and its peak energy depends on the thermal gas temperature. As the accretion rate and gas density increase, Inverse Compton (IC) scattering begins to dominate (likely via SSC, i.e., Synchrotron photons become scattered), though if densities remain low enough, the spectrum will still show come curvature. At higher accretion rates, the IC scattering process is more efficient, resulting in a power-law-like spectrum in the X-ray band.

The key thermodynamic property of RIAFs is the inefficient cooling of ions due to the Coulomb decoupling of electrons and ions in the accreting, low-density plasma. Due to this decoupling, such flows are likely well described by a two-temperature electron-ion plasma, with only electrons radiating via Bremsstrahlung, Synchrotron, and inverse Compton scattering \cite{Esin1997}. The models, as well as a plethora of observational evidence, show that such flows display radiative inefficiency, with $\eta=0.1\dot{M}/\dot{M}_\mathrm{crit}$, where $\eta$ is defined by:
\begin{equation}
	L = \eta \dot{M} ,
\end{equation}
and $\dot{M}_\mathrm{crit}$ is the accretion rate below which we have a RIAF. We assume $\dot{M}_\mathrm{crit}$ is the accretion rate corresponding to $1\%$ of the Eddington luminosity with $\eta_\mathrm{Edd }= 0.1$ \citep{Fender:2013}. The bolometric luminosity $L$ in such inefficient states is thus given by
\begin{equation}
	L = 0.1 \frac{\dot{M}^2}{\dot{M}_\mathrm{crit}}, \label{eq:ineffacclum}
\end{equation}
thus exhibiting a quadratic scaling $L\propto \dot{M}^2$ \citep{Esin1997}, as opposed to the linear behaviour observed at higher efficiency. 

\paragraph{X-ray band}

A significant fraction of this bolometric luminosity is typically assumed to fall in the X-ray band. Following \cite{Fender:2013}, we assume a $L_\mathrm{X}/L$ fraction of $\simeq 30\%$, allowing for a direct calculation of the X-ray luminosity from Eq. \ref{eq:ineffacclum}. 

\paragraph{Radio band}

Radiatively inefficient Galactic BH-XRBs also almost ubiquitously launch outflows. Such outflows are typically in the form of steady, self-absorbed relativistic jets \citep{Fender:2001}. Ballistic, transient jets are also observed during spectral state transitions at higher accretion rates. Such jets are typically identified via their radio emission, and are shown to be present during quiescence, at luminosities below $10^{-8}~L_\mathrm{Edd}$ \citep{Plotkin2015}. These steady jets exhibit a flat-to-inverted spectrum from radio through to IR frequencies---a consequence of self-absorbed synchrotron emission through the optically-thick regions of the jet (see, e.g., \cite{Markoff2005}). \\
In addition, multiple studies have now established that the luminosity of such radio-emitting jets scales in a consistent and predictable way with the X-ray emitting plasma in the inner regions of the flow. This scaling, $L_\mathrm{R} \propto L_\mathrm{x}^{0.7}$ \citep{Corbel2000,Corbel2003,Gallo2003,Corbel2008,MillerJones2011,Gallo2014}, is known as the radio-X-ray correlation, and applies to Galactic BH-XRBs (though other compact objects have been tracked in this phase space, showing similar but distinct trends of their own). Therefore, by using this analogy between isolated BHs and their low-accretion rate counterparts in binaries, we can infer that isolated BHs will similarly launch these steady jets, and thus their radio and X-ray luminosities will scale according to this radiatively inefficient track we observe in Galactic BH-XRBs. \\
However, in order to invoke a mass scaling and capture low-luminosity accretion onto a mass-variable population of isolated BHs, one must refer to the established connection between Galactic BH-XRBs and AGN---the Fundamental Plane of Black Hole Activity (FP; \cite{Merloni:2003,F04,Plotkin:2012}). The FP is, in a sense, an extension of the radio-X-ray correlation discovered in Galactic BH-XRBs to their supermassive counterparts. The FP is an empirical, parametrised relation between the X-ray luminosities, radio luminosities, and masses of hard state Galactic BH-XRBs with steady jets, and AGN of types which display similar X-ray emission characteristics. These include low-luminosity AGN (LLAGN), low-ionization nuclear emission-line regions (LINERS), Faranoff-Riley type I (FRI) and BL Lacs. It is understood that the fundamental connection between these types of AGN and Galactic BH-XRBs is simply their sub-Eddington accretion rates, and the presence of radio-emitting jets. All sources in the FP display something resembling a power law spectrum in the X-ray band, and a flat-to-inverted radio spectrum.\\

To determine the radio emission, we invoke this scaling relation, thus assuming that an isolated population of low-luminosity accreting BHs will display similar spectral properties as the aforementioned systems. In particular, this thus relies on the assumption that the BHs will launch jets. We remark that this assumption (and the presence of an accretion disk on small scales) is not incompatible with a spherical accretion pattern on large scales.

We adopt the scaling relation determined in one of the most recent such FP studies \citep{Plotkin:2012}: 
\begin{equation}
		\log L_\mathrm{ X}= (1.45 \pm 0.04) \log L_\mathrm{R}- (0.88 \pm 0.06) \log  \left( \frac{M_\mathrm{BH}}{M_\odot} \right) - 6.07 \pm 1.10,
	\label{eq:FP}
\end{equation}
where $L_\mathrm{X}$ is the X-ray luminosity in the 2-10 keV band and $L_R$ is the radio luminosity at 5 GHz, both expressed in $\mathrm{erg} \, \mathrm{cm}^{-2} \mathrm{s}^{-1}$. \\
Thus we have a simple prescription for both the X-ray luminosity, $L_\mathrm{X}$, and radio luminosity, $L_\mathrm{R}$, of an isolated accreting BH. \\

\section{ Accretion signals  from  Milky Way black holes}
\label{sec:accretion:MilkyWay}

In this section we describe the procedure we follow to estimate the expected number of luminous sources in two regions of interest: the vicinity of the solar system and the Central Molecular Zone, located at the Galactic centre.

\subsection{Molecular clouds}
\label{sec:accretion:MilkyWay:clouds}

A fundamental ingredient in the study of accretion signals is an accurate modelling of the interstellar medium in the Galaxy. This is usually described as composed by three different components: the ionized component, the neutral gas, and the molecular clouds. In the PR13 picture the probability of obtaining large enough fluxes from a BH found in the first two of these components turns out to be negligible. We will therefore consider only densities associated to molecular clouds.
We identify two regions of interest: the local region, for its vicinity, and the Galactic centre, for its high gas concentration as well as the high expected number density of black holes in this region

\paragraph{Local region} We assume two types of molecular clouds to be present, each type with uniform density. These are: a) warm clouds, of number density $10^2$~cm$^{-3}$, sound speed 0.9 km/s, filling factor $5 \times 10^{-2}$, and b) cold clouds, of number density $10^3$~cm$^{-3}$, sound speed 0.6 km/s and filling factor $5 \times 10^{-3}$ \citep{Fender:2013}. This combination results in an average density $\bar{n}= 10 \, \mathrm{cm}^{-3}$. 

\paragraph{Central Molecular Zone} The Central Molecular Zone (CMZ), is a cloud complex with an approximately cylindrical shape that extends up to $\approx 160$ pc away from the galactic centre and for a total height of $\approx 30$ pc. This large reservoir of molecular gas, located close to the gravitational centre of the Galaxy, is a particularly promising target for BH searches.

The CMZ has been the object of extensive multi-wavelength observational campaigns over the years, aimed at characterizing its structure and physical properties, including star formation rate (see for instance \cite{2019MNRAS.484.5734K} and references therein for a recent analysis). 

The molecular clouds in this region are known to be denser than the Galactic average.
We assume an average number density of molecular hydrogen of $\bar{n}= 150\, \mathrm {cm}^{-3}$ \citep{Ferriere2007} and consider two types of clouds: warm, less dense clouds, and cold, denser ones. 

The warm clouds are of secondary importance given the large distance from Earth (see figure \ref{XFluxCMZ}). We set their density to $n= 10^{2.5} \, \mathrm{ cm }^{-3}$ and consider a filling factor of 14\% \citep{Ferriere2007}. 
Regarding the cold clouds, we choose to adopt a more refined modelling based on a power law density distribution between $n_\mathrm{min}$ and  $n_\mathrm{max}$ :
\begin{equation}
	P(n) \propto n^{- \beta}
\end{equation}
Based on \cite{Ferriere2007}, we set $n_\mathrm{min}=10^{3.5} \, \mathrm{ cm }^{-3}$ and  $n_\mathrm{max}=10^5 \, \mathrm{ cm }^{-3}$, while we treat $\beta$ as a free parameter. The filling factor for the cold clouds is then obtained by requiring  $\bar{n}= 150 \, \mathrm{cm}^{-3}$. For a reference value of  $\beta =2.4$ \citep{HeRG:2019,HeRG:2020} we obtain a filling factor around $1 \%$ and an average cloud density of around $8 \times 10^3 \, \mathrm{cm}^{-3} $.\\
A further free variable is the temperature of the clouds, which enters the accretion rate through the sound speed $c_\mathrm{s}$. This variable has little impact on the predictions and we set it to 1 km/s throughout.

\subsection{Estimating the number of visible sources}
\label{sec:accretion:MilkyWay:estimate}

We now turn to assessing the number of isolated astrophysical black holes that are potentially detectable by the current (and forthcoming) generation of X-ray (with particular focus on the hard X-ray band) and radio experiments. In this section we summarize the main points of our methodology.
Our main observable is the number of sources associated to a radiation flux above detection threshold, i.e. that satisfy $ \phi > \phi^*$, where $ \phi$ is the flux at Earth defined as:
$
\phi= L/ (4 \pi r^2)
$
where r is the distance to the source from Earth, and $ \phi^*$ is the experimental threshold value. 

Previous studies (\cite{Fender:2013, Tsuna:2018oqt, Tsuna:2019kny}) have obtained their estimates through Monte Carlo simulations of a BH population. We adopt instead a semi-analytical approach, aiming to perform a comprehensive parametric study associated to the physical model of accretion physics described above and assessing its uncertainties.

Here we describe the general prescription for computing the number of visible sources, which we will apply with some variation to the two regions of interest for this work: the local region (section \ref{sec:accretion:ABH:local}) and the central molecular zone (CMZ, section \ref{sec:accretion:ABH:CMZ}).

We obtain the probability for a BH to emit above threshold by integrating the joint PDF of the relevant randomly distributed variables over the volume defined by the condition $ \phi > \phi^*$ . 

Regarding the BH population, the random variables entering the flux expression are: speed, mass and distance from Earth. We neglect any possible correlation between these variables. In addition, we also treat as a random variable the density of the interstellar medium at each BH location. We make the simplifying assumption that the gas density and BH position are not correlated. 
Therefore, we obtain the expected number of luminous sources $N_\mathrm{sources}$ by computing the following integral:
\begin{equation}
	\label{eq:Nsources}
	\begin{split}
		&N^\mathrm{sources}  ( \phi^*, \{ p_i\}, \{ q_i\}) \,\,=\\
		&N^\mathrm{ tot}  \int_{\phi (\mathrm{ v}_\mathrm{BH}, \, M,\,  d ,\,  \{ p_i\}) > \phi^*} P(\mathrm{ v}_\mathrm{BH})P(M)P(r)P(n)  \, \diff \mathrm{v}_\mathrm{BH} \,\diff M \,\diff r \,\diff n \, .
		\end{split}
\end{equation}
Here $P(\mathrm{ v}_\mathrm{BH})$, $P(M)$, $P(r)$ and $P(n)$ are the normalized p.d.fs describing respectively the BH speed, mass, distance from Earth and the interstellar medium density at its location. With $\{p_i\}$ we indicate the free parameters entering the expression for the flux, and which we consider to be fixed for all BHs. These are: the sound speed of the neutral medium $c_\mathrm{s}$, the sound speed in  the ionized region \csin, the fraction of bolometric luminosity $L_\mathrm{X}/L$ and, in the case of BHL accretion, the suppression factor $\lambda$ (regarding the first of these, we remark that variations of the sound speed of the neutral medium $c_\mathrm{s}$ have a negligible impact on the flux magnitude). Finally, the parameters that define the PDFs for the the random variables, which we indicate with $\{q_i\}$, should also be regarded as free parameters of the model. \\

\section{ The astrophysical background\footnote{Adapted from \cite{Scarcella:2020ssk}}}
\label{sec:accretion:ABH}

In this section, we apply the analysis detailed above to the astrophysical black hole (ABHs) population of the Milky Way. The search for isolated astrophysical black holes can be considered a relevant problem \textit{ per se}, and is also crucial as a background in the context of the quest for primordial black holes, which will be discussed in the following Section.

\subsection{Characterizing the astrophysical black hole population}
\label{sec:accretion:ABH:bhpopulation}


\paragraph{Mass} 
We assume BH masses to follow a normal distribution of mean $\mu_\mathrm{mass} = 7.8 \, M_\odot$ and $\sigma_\mathrm{mass} = 1.2 \, M_\odot$. This distribution was obtained from the study of X-ray binaries by \cite{_zel_2010} and confirmed in an independent analysis by \cite{Farr_2011}. It has also been employed by previous studies on this topic such as the works of \cite{Fender:2013} and \cite{Tsuna:2018oqt}. 

\paragraph{Speed} 
The BH velocity is given by the combination of two components: the velocity of the progenitor star, $\mathrm{ v}_\mathrm{star}$, and the kick the BH receives at birth due to the supernova explosion, $\mathrm{ v}_\mathrm{kick}$. 
As for the former, we are concerned with the velocities relative to the molecular gas, so we ignore the rotational component along the Galactic disk and consider exclusively the velocity dispersion $\sigma_{D}$. This is well measured for both nearby stars and stars in the Galactic centre~\cite{2018}.\\
On the other hand, little is known about the magnitude of the natal BH kicks. A study of pulsars proper motions by \cite{Hobbs_2005} concluded that neutron star kicks obey a Maxwell Boltzmann distribution with $\sigma = 265$ km/s. Re-scaled with the average masses \citep{Fender:2013}, this gives $\sigma_\mathrm{kick} \approx 50 $ km/s, corresponding to a mean of $\mu_\mathrm{kick} \approx 75$ km/s. We consider here as reference values $\mu_\mathrm{kick} =50$ km/s (\textit{"low kick"}) and $\mu_\mathrm{kick} =100$ km/s (\textit{"high kick"}). \\
The BH speed, resulting from the combination of these two independent components, is then distributed following a Maxwell Boltzmann of mean:
$
\mu_\mathrm{BH}=\sqrt{\mu_\mathrm{star}^2+\mu_\mathrm{kick}^2},
$.
Hence, the uncertainties in both the kick and the progenitor star's velocity dispersion are enclosed in only one parameter, $\mu_\mathrm{BH}$. 

\paragraph{Distance from Earth} 
For what concerns the CMZ, we can safely approximate $d \sim 8.3 $ kpc for all sources. In the local region a more accurate description is required and we  the obtain distance distribution assuming a uniform distribution in space.

\subsection{Searching for black holes in the local region}
\label{sec:accretion:ABH:local}

We start by applying our model to the study of the innermost 250 pc around Earth, following the work of Fender and Maccarone \cite{Fender:2013}. In this study, the authors applied the BHL model to describe accretion and estimated the number of visible sources through Monte Carlo simulations. They considered a few combinations of the values of the mean BH speed $\mu_\mathrm{BH}$ and of the suppression factor $\lambda$, obtaining a bound on the combination of these parameters. Here we reproduce the same setup to obtain a prediction for the number of detectable X-ray sources, verifying  the result for the BHL model and comparing it to the predictions of the PR13 model. 

\subsubsection{Setup}
\label{sec:accretion:ABH:local:setup}

First, we estimate the number $N^\mathrm{ local}$ of BHs present in the spherical region around Earth ($\text{r}=0$) between $70\, \mathrm {pc} < \text{r} <250\, \mathrm {pc}$ (we exclude the innermost 70 pc to account for the local void). As a rough estimate, we assume the fraction of BHs contained in this region to be equivalent to that of the progenitor stars. The total stellar mass in the Milky way is $\mathrm M^*_\mathrm{MW } = 6 \times 10^{10} M_\odot $ \citep{2015ApJ...806...96L}, while the mass density of stars in the local region is around $7 \times 10^{-2} \mathrm M_\odot / \mathrm {pc}^3 $  \citep{McKee_2015}.
Assuming a total of $10^8$ BHs in the galaxy we obtain:
\begin{equation}
	\mathrm{ N}^\mathrm{local}= 10^8 \frac{ M^*_\mathrm{local}}{ M^*_\mathrm{MW}} \approx 7 \times 10^3 \; .
\end{equation}
While an accurate estimate of the number of BHs in the local region is beyond the scope of this work, we should notice that we are not taking into account the larger scale height of the BH population distribution with respect to the stellar one (a consequence of the natal kicks), which would result in a reduced local density. Our estimate is thus likely overestimating the true number of BHs by a factor $\mathcal{O}(1)$.\\
We assume a uniform spatial distribution, a normal mass distribution and a MB velocity distribution of average $\mu_\mathrm{BH}$, as discussed in the previous section. \\
The integral in \cref{eq:Nsources} becomes:
\begin{equation}
	N^\mathrm{ sources}  = \sum_\mathrm{warm, cold} N^\mathrm{ clouds} \int_{\phi > \phi^*} P(\mathrm{ v}_\mathrm{BH})P(M)P(r) \,\diff \mathrm{ v}_\mathrm{BH}\, \diff M\, \diff r  \; .
\end{equation}
Here $N_\mathrm{clouds} $ indicates the number of BHs we expect to find in each type of cloud: 
\begin{equation}
	N^\mathrm{ clouds}=f_\mathrm{clouds} \, N^\mathrm{ local},
\end{equation}
where $f_\mathrm{clouds} $ is the associated filling factor, defined as the fraction of volume occupied by the clouds. We obtain $N_\mathrm{clouds} \approx 30$ in the cold clouds and $N_\mathrm{clouds} \approx 300$ in the warm ones.\\
Finally, the bolometric fraction of luminosity in the X-ray band is set to $0.3$.

\subsubsection{Results}
\label{sec:accretion:ABH:local:results}

\begin{figure}
	\centering
	\includegraphics[width=.7\linewidth]{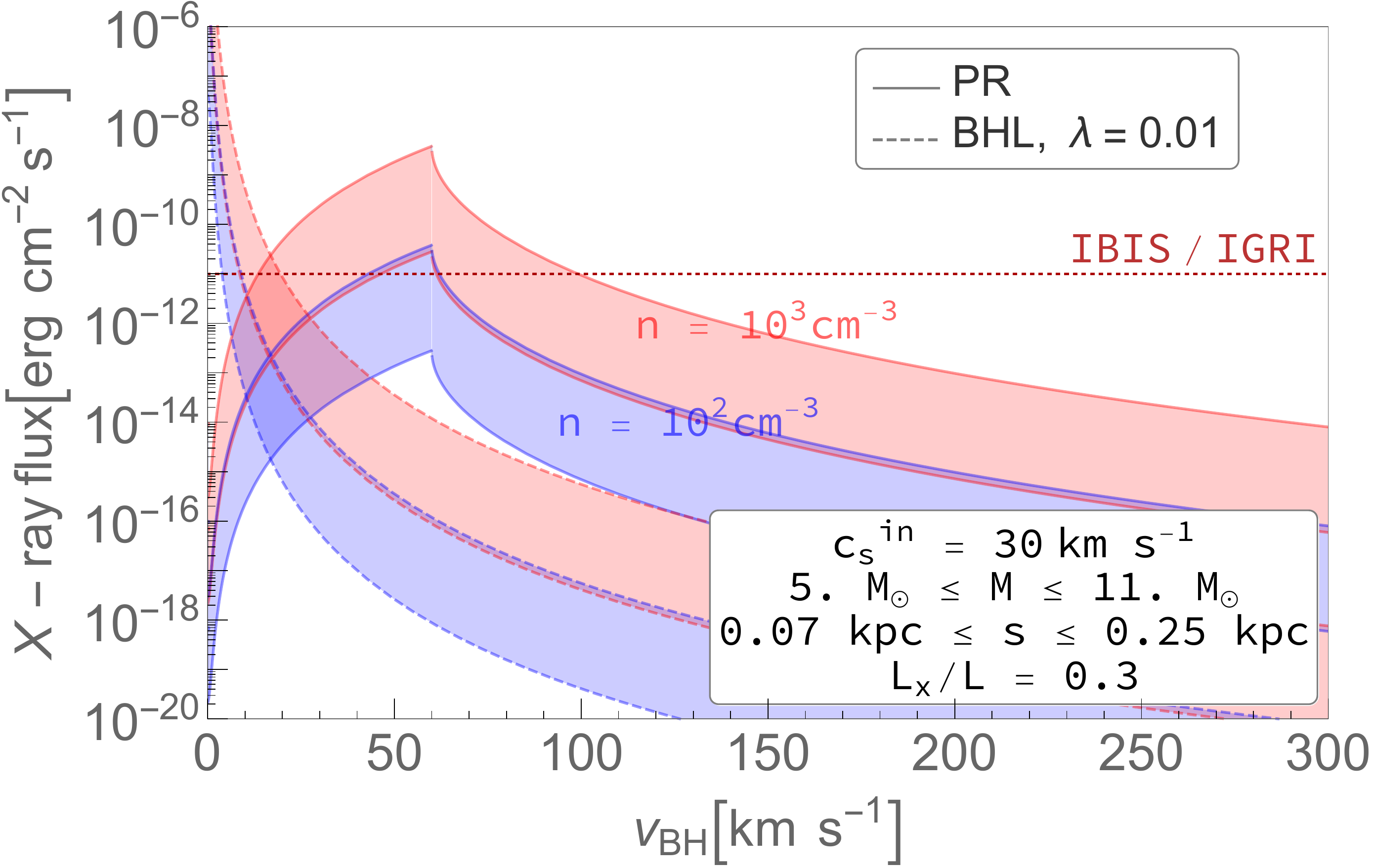}
	\caption{ \textbf{X-ray flux from BHs in the local region.} We show the X-ray flux associated to a BH accreting from a molecular cloud in the local region, as a function of the relative velocity with respect to the gas cloud. The coloured bands describe the two types of clouds considered and their width corresponds to a range of BH masses and distances. We show the fluxes for the Park-Ricotti and the suppressed ($\lambda=0.01$) Bondi-Hoyle-Littleton models. As a reference, we show the detection threshold associated to the IBIS/IGRI survey \citep{Bird_2009}.}
	\label{XFluxLocal}
\end{figure}

\begin{figure}
	\centering
	\includegraphics[width=.7\linewidth]{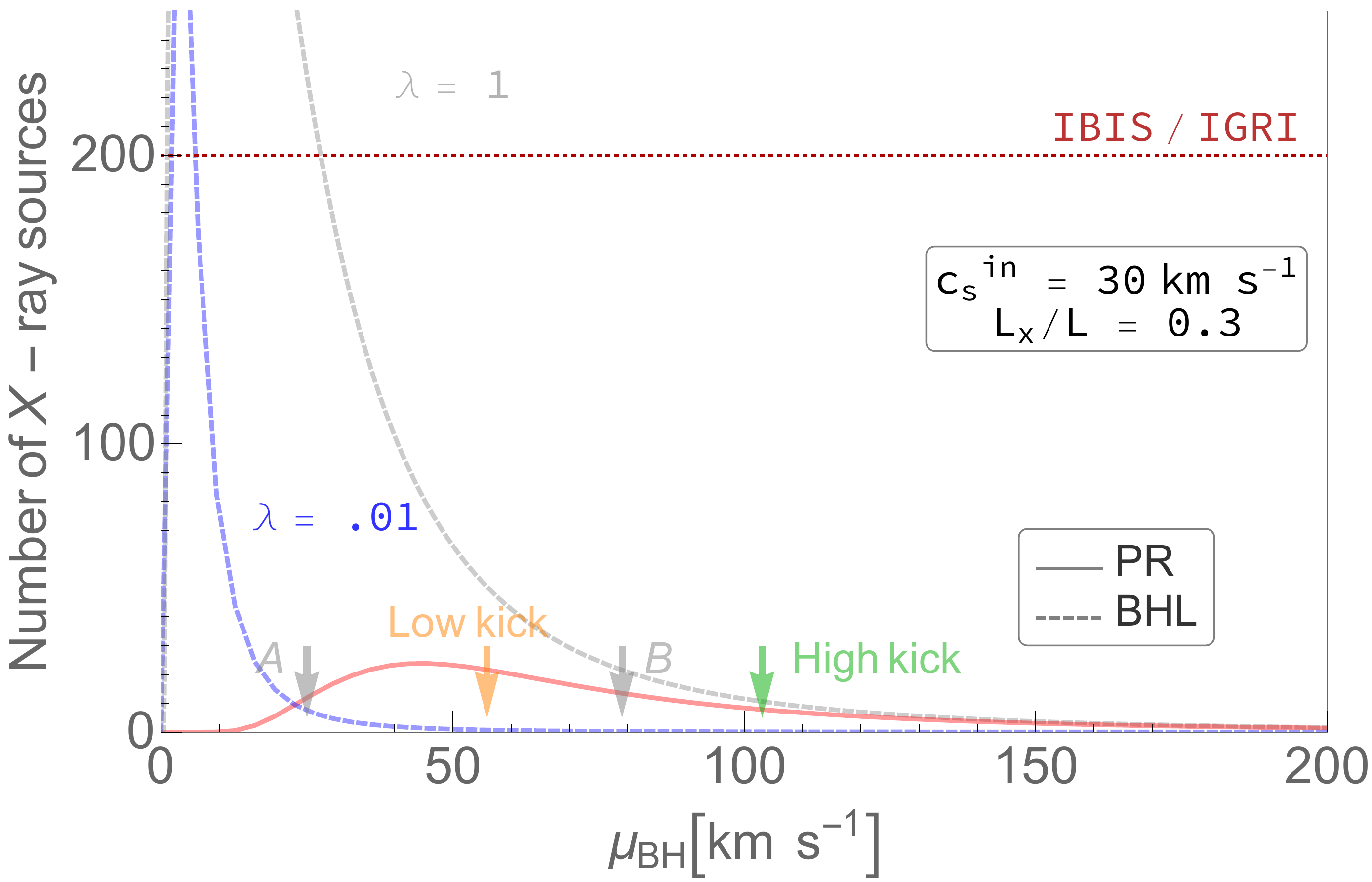}
	\caption{ \textbf{Total number of X-ray sources in the local region (full sky, d < 250 pc).} We show the expected number of BH observed in the X-ray as a function of the average BH speed according to a) the PR13 model (red, solid) b) the BHL model with $\lambda$=1 (grey, dashed) c) the BHL model with $\lambda$=0.01 (blue, dashed).  
	}
	\label{fig:NLocal}  
\end{figure}

Let us first consider the X-ray flux associated to a generic isolated BH located in the local region and accreting gas from a molecular cloud as a function of the BH speed. In figure \ref{XFluxLocal} we show this flux for both the PR13 accretion scenario and the conventional modelling based on the BHL formalism, the latter suppressed by a factor $\lambda =0.01$. The two coloured bands correspond to the two densities considered for molecular clouds. The width of the band is obtained considering distances between 70 and 250 pc and masses between 5 and 11 solar masses.  We show for reference the detection threshold associated to the 4th INTEGRAL IBIS/ISGRI soft gamma-ray
survey catalogue (17–100 keV; \cite{Bird_2009}) 
, which is approximately $10^{-11}$ erg/cm$^2$/s. 
We want to emphasize once again the distinct behaviour associated to the two accretion scenarios.
In the Bondi picture, the slower sources are very luminous. This is particularly relevant in this context, since velocities in the local region are expected to be on the low end of the range shown in this figure: the unsuppressed BHL scenario predicts a huge number of sources (see figure \ref{fig:NLocal}) and can easily overshoot the bound. As a consequence of the introduction of a suppression factor, only very slow sources are expected to be visible. This is no longer true when the PR13 scenario is considered, since it naturally predicts a flux suppression for the slower sources. Instead, the population of BHs emerging above threshold and showing up in the X-ray sky will present relatively high speeds (around $50$ km/s). Furthermore, we can notice how, based on the Bondi picture, we expect to detect BHs in both types of clouds. According to the PR13 model, however, only BHs located in the denser clouds can be detected.

Following the procedure described in the previous section, we can now integrate the distributions that characterize the BH population to obtain the number of sources visible in the X-ray sky with the IBIS/IGRI survey. 
In figure \ref{fig:NLocal} we show the dependence of number of sources on the average speed $\mu_\mathrm{BH}$. This prediction is obtained for three accretion scenarios: \textit{ a)} the PR13 model; \textit{ b)} the BHL model with suppression factor $\lambda$= 0.01 c) the BHL model with perfect efficiency $\lambda$= 1. \\
We indicate with orange and green arrows the reference values given by the \textit{ "low kick"} the \textit{ "high kick"} scenarios, assuming a velocity dispersion of the progenitor stars of $15$ km/s (corresponding to an average speed of $\approx 25$ km/s). For comparison, the average speeds corresponding to the \textit{ "no kick"} and \textit{"high kick"} ( $\mu_\mathrm{ kick}$ = 75 km/s) scenarios considered in \cite{Fender:2013} are also shown and indicated with A and B, respectively.

\noindent
Our results can be summarized as follows:
\begin{itemize}
	\item { As far as the BHL scenario is concerned, we notice that the number of sources can easily overshoot the observational constraints in absence of a significant natal kick. We also point out that our findings are in broad agreement with the ones reported in \cite{Fender:2013}: i.e., a few tens of visible sources in scenario A (no kick, $\lambda = 0.01$ ).   
	}
	\item In the PR13 scenario, the introduction of a suppression factor is not necessary. While being compatible with the experimental bound, our model nevertheless predicts a significant number of bright sources. {This prediction corresponds to a population of  $\sim10\mbox{--}30$ accreting isolated black holes in the existing catalogues, taking as reference the \textit{ "no kick"} and \textit{"high kick"} scenarios }. We will discuss in more detail the consequences of this result in the next Sections, and compare this prediction with the one associated to the Central Molecular Zone.
\end{itemize}

\subsection{Searching for black holes in the central molecular zone}
\label{sec:accretion:ABH:CMZ}

In this Section we turn our attention to the inner part of the Galactic bulge, in particular to the aforementioned cloud complex known as the Central Molecular Zone (CMZ).

\subsubsection{Setup}
\label{sec:accretion:ABH:CMZ:setup}

To model the CMZ, we employ a simplified version of the model by \cite{Ferriere2007}. We describe it as  a cylindrical region of half height 15 pc and radius 160 pc. \\
The number of BHs contained in this region is very uncertain, but we can make a naive order-of-magnitude estimate as follows.
We assume the BHs to be generated following the distribution of stars in the Galaxy. Here we are interested in the bulge component, which accounts for around $15 \%$ of the total stellar mass and can be modelled  with a spherical exponential with scale radius $R_\mathrm{bulge} = 120 $ pc \citep{Sofue_2013}.  Integrating the bulge spherical exponential distribution over the CMZ volume gives us the fraction of ABHs born in this region: around $2 \%$ of the bulge component. Assuming a total of $10^8$ BHs present in the whole Galaxy, this corresponds to  $3 \times 10^5$ BHs. \\
We must however take into account that, due to the large initial natal kicks, the initial spatial distribution is modified \citep{Tsuna:2018oqt}. ~
To quantify this, we performed a simulation of the evolution in the Galactic potential (modelled following~\cite{Irrgang_2013}, model II)  of 1000 BHs. These are initially distributed uniformly in the CMZ region with an average speed of 130 km/s and given an average natal kick of 75 km/s. The simulation shows that only $\approx 25 \%$ of the BHs remain in the region: we observe in particular a spreading in the direction perpendicular to the Galactic plane. On the other hand, we do expect some BHs born outside of the CMZ to enter the region under the effect of the gravitational potential. However, these objects cross the CMZ close to the periastron of their orbit, with very large velocities and hence suppressed accretion rates. We can therefore neglect the latter effect. However the former is significant and we take it into account. We thus update our naive estimate to $ N ^\mathrm{CMZ} \approx 7.5 \times 10^4$. \\

Regarding the speed distribution, the average speed $\mu_\mathrm{BH}$ is treated as a free parameter. Nevertheless, we can make some estimates of reasonable values. The most recent observations suggest a velocity dispersion of the central bulge stars of average  $\mu_\mathrm{bulge} \approx 130$ km/s (\cite{Sanders_2019}, \cite{2018}). This gives, for reference, $\mu_v$= 140 km/s in the \textit{ "low kick" } scenario and $\mu_v$= 160 km/s in the \textit{ "high kick" } scenario.\\
Following \cref{eq:Nsources}, we compute the number of sources as 
\begin{equation}
	\begin{split}
		N^\mathrm{ sources}  =  N^\mathrm{ clouds}_\mathrm{warm} \int_{\phi > \phi^*} P(\mathrm{ v}_\mathrm{BH})P(M) \, \diff \mathrm{ v}_\mathrm{BH}\, \diff M \; + \\
		N^\mathrm{ clouds}_\mathrm{cold} \int_{\phi > \phi^*} P(\mathrm{ v}_\mathrm{BH})P(M)P(n) \, \diff \mathrm{ v}_\mathrm{BH}\, \diff M\, \diff n
	\end{split}
\end{equation}
The expected number of black holes in the clouds is obtained as:
\begin{equation}
	N^\mathrm{ clouds}=f_\mathrm{clouds} \, N^\mathrm{ CMZ},
\end{equation}
where $f_\mathrm{clouds} $ is the associated filling factor.

\subsubsection{Results}
\label{sec:accretion:ABH:CMZ:results}
\begin{figure}[tb]
	\centering
	\includegraphics[width=.7\linewidth]{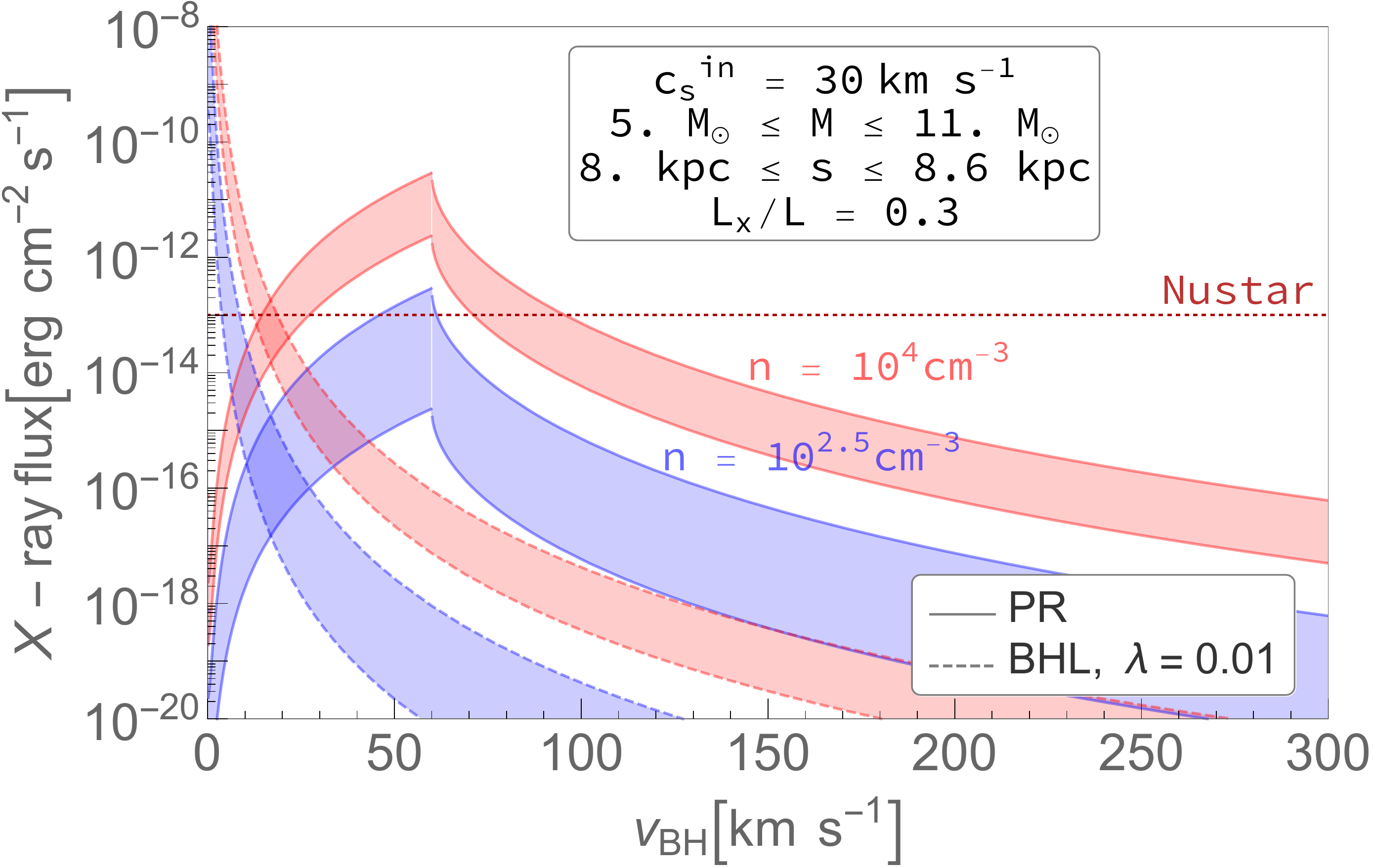}
	\caption{ \textbf{X-ray flux from BHs in the CMZ.} We show the X-ray flux associated to a BH orbiting in the Central Molecular Zone molecular cloud complex, for different values of the relative velocity with respect to the gas cloud, and the gas density in the cloud. We consider the Park-Ricotti and the Bondi-Hoyle-Littleton formalism. As a reference, we show the detection threshold associated to the NuSTAR Galactic Center survey \citep{Hong:2016}.}
	\label{XFluxCMZ}
\end{figure}
We start by considering the X-ray flux associated to a generic isolated BH located in the Central Molecular Zone and accreting gas from a molecular cloud as a function of the speed and of the molecular cloud density. We show this observable in figure \ref{XFluxCMZ}. As a reference, we show the detection threshold associated to the NuSTAR Galactic Center survey in the 3-40 keV band \citep{Hong:2016}. The same trends highlighted in Sec. \ref{sec:accretion:ABH:local} can be noticed in this plot. In particular, we point out that, if the cloud density is high enough, there is a wide range of BH speed associated to emission above threshold within the PR13 formalism. On the other hand, no sources are expected to be detected in the lower density clouds.

\begin{figure}[t]
		\centering
		\includegraphics[width=.7\linewidth]{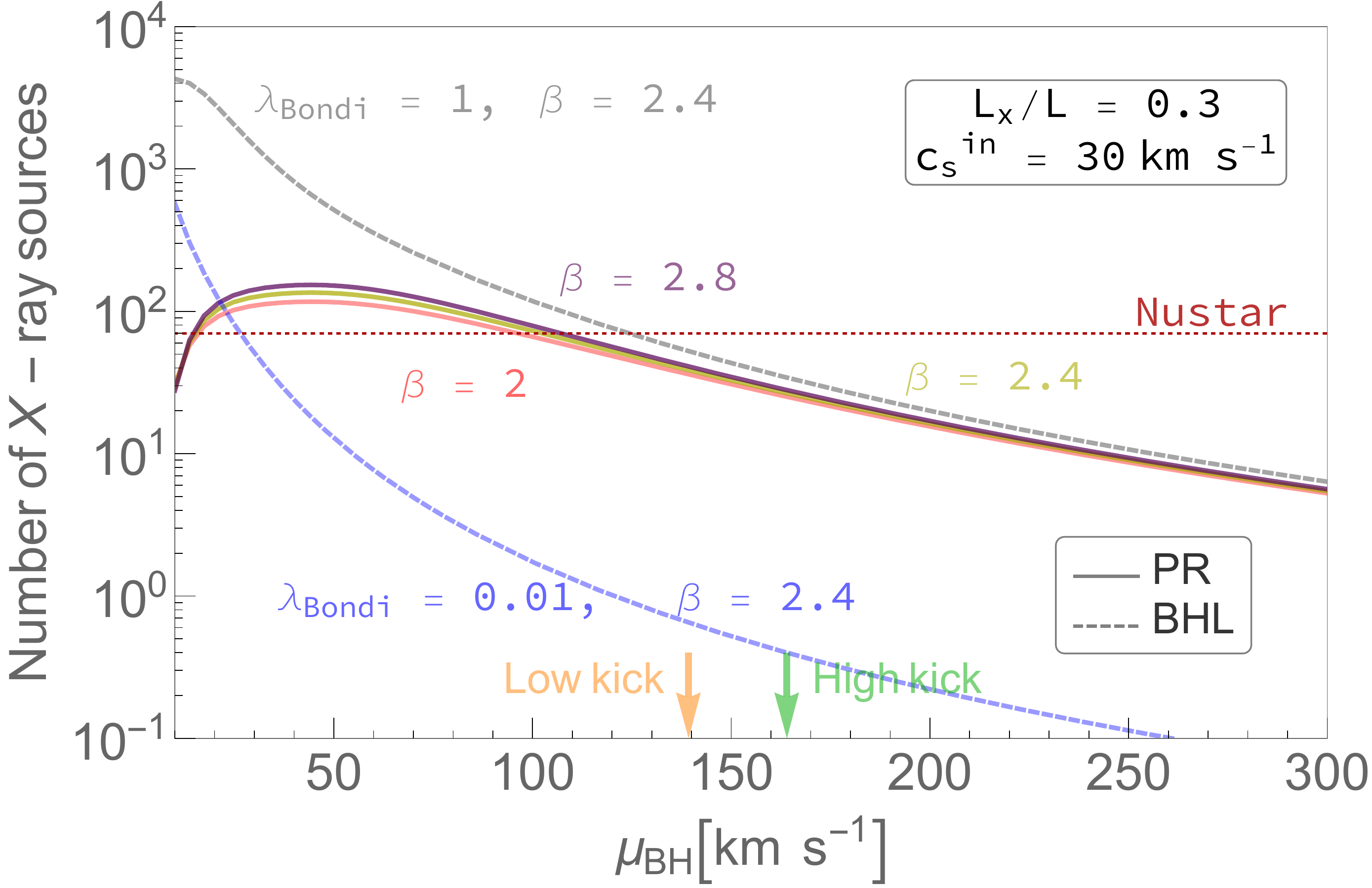}
	\caption{ \textbf{Parametric study of the number of X-ray sources in the CMZ (a)}. Number of X-ray sources as a function of the average speed, compared to the number of sources detected by NuSTAR in the CMZ region \citep{Hong:2016}.}
	\label{fig:NCMZa}
\end{figure}

\begin{figure}
	\centering
	\includegraphics[width=.7\linewidth]{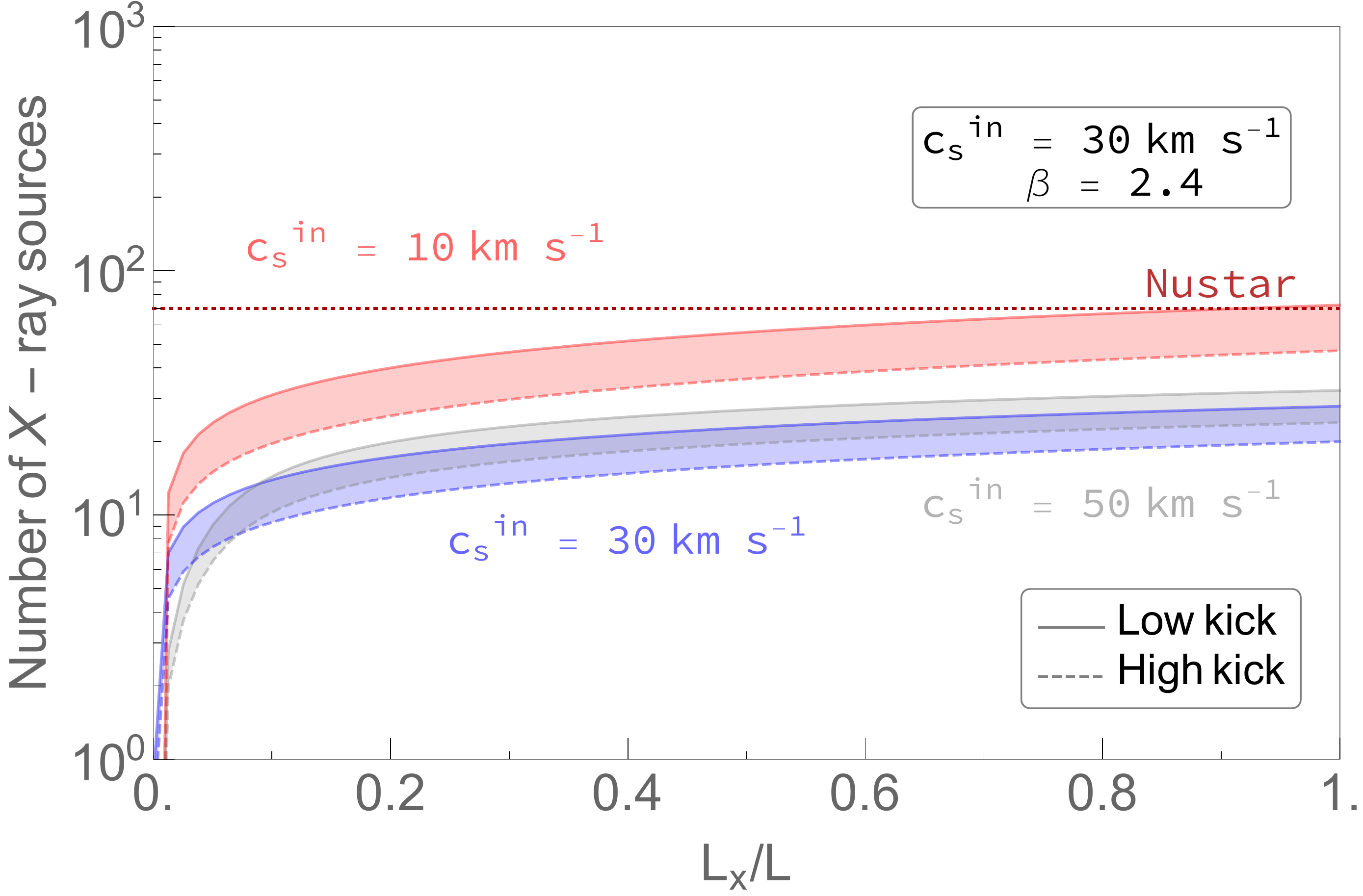}
	\caption{ \textbf{Parametric study of the number of X-ray sources in the CMZ (b)}. Number of X-ray sources as a function of the bolometric luminosity that is radiated in the X-ray band of interest, compared to the number of sources detected by NuSTAR in the CMZ region \citep{Hong:2016}.}
	\label{fig:NCMZb}
\end{figure}

\begin{figure}
	\centering
	\includegraphics[width=.7\linewidth]{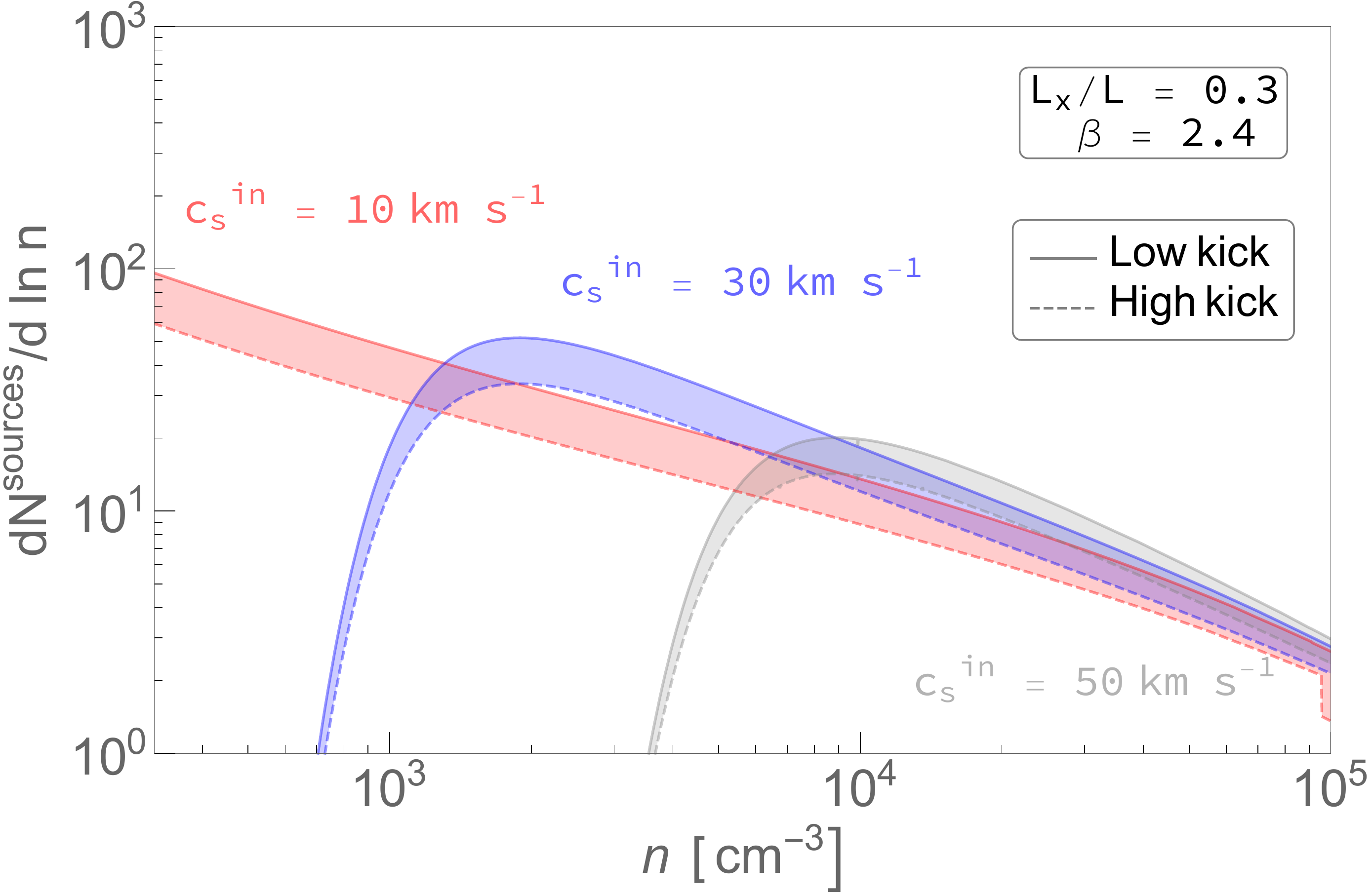}
	\caption{ \textbf {Distribution of bright X-ray sources as a function of clump density.} We show the log-differential number of sources detected by NuSTAR as a function of the clump density. A power law distribution for the clump density is assumed.}
	\label{fig:dNdlnn}
\end{figure}

We apply the procedure described in the previous sections, and compute the number of X-ray sources associated to accreting BHs in this region.
The results are shown in figure \cref{fig:NCMZa,fig:NCMZb} and \ref{fig:dNdlnn}, illustrating the impact of the different free parameters on this key observable.
In particular, we show in \cref{fig:NCMZa} the number of X-ray sources in the CMZ region as a function of the average speed of the BH population adopting both the BHL and the PR13 accretion models. We show the result for various choices of the $\beta$ parameter defining the density distribution, and for two choices of the $\lambda$ parameter in the BHL case. We can notice how, in this setting, the PR13 scenario predicts as many sources as the $\lambda =1$ BHL, due to the fact that high speeds are prevalent among the BH population of the CMZ. However, we have seen that the unsuppressed BHL scenario is excluded by previous studies on nearby compact objects and complementary studies focused on AGN populations as mentioned in Sec. \ref{sec:accretion:models}. 
The comparison should therefore be made between the predictions of the PR13 model and the suppressed BHL model, also shown in figure for $\lambda =0.01$. Then, the PR13 model predicts a significantly greater number than those obtained assuming BHL accretion. (See also the work of \cite{Tsuna:2018oqt}, in which $\mathcal{O}$(1) sources were predicted in the Galactic centre using BHL accretion).  \\

In \cref{fig:NCMZb} we focus on the dependence of the same observable with respect to the fraction of the bolometric luminosity that is radiated in the X-ray band of interest. In figure \ref{fig:dNdlnn} we consider the differential contribution of the clumps of different density, again for different reference values of the ionized sound speed in the BH vicinity $c_\mathrm{s}^\mathrm{ in}$.

We can conclude from the results visualized in \cref{fig:NCMZa,fig:NCMZb,fig:dNdlnn} that, despite the relevant uncertainties associated to the modelling of such a complex setup (and despite the existence of a threshold effect due to the peak in the accretion rate), the prediction of few tens of sources is solid with respect to these uncertainties. In particular, we notice from panel (a) that a number of bright X-ray sources comprised between $10$ and $20$ is expected for our reference value of $c_\mathrm{s}^\mathrm{ in}$ and for $\mu_\mathrm{BH} = 150$ km/s. As far as the speed distribution is concerned, we remark that the two reference values we have chosen to bracket the uncertainty both lead to similar predictions, while large deviations from this range would require to assume unrealistically high speeds. The same line of thought applies to the slope associated to the clump density distribution.
From \cref{fig:NCMZa} we also notice a lower number of sources may only arise if we were to assume either a very high temperature in the ionized region, or else a very low fraction of bolometric luminosity radiated in the X-ray band (we remark, however, that the prediction scales linearly with the expected number of BHs in the region, $N^\mathrm{ CMZ}$, which does carry a large uncertainty).

In summary, the PR13 model (together with our rather conservative assumptions regarding the X-ray emission) leads to the  \textit{ prediction of a significant number of X-ray sources in the CMZ region} associated to isolated BHs accreting from molecular clouds. The total number of sources is of the same order of magnitude as the one predicted in the local region and discussed in the previous section. However, while the previous section was dealing with a full-sky analysis, in this case we are focusing on a specific region of interest (ROI) with small angular extent, which is ideal for multi-wavelength observational campaigns.

The NuSTAR survey has detected $70$ hard X-ray sources sources in this region of interest. The nature of most of the sources in the catalogue in not precisely known, although a significant population of cataclysmic variables and X-ray binaries is expected. Our finding seems to suggest the need of a careful data analysis, in order to identify the possible presence of a relevant population of isolated accreting black holes in the already existing data. Following this line of thought, and in the prospect of a discovery, a multi-wavelength analysis is compelling, so we will turn our attention to the radio domain in the next section.

\subsection{Multi-wavelength prospects for the Square Kilometre Array}
\label{sec:accretion:ABH:SKA}

\begin{figure}
	\centering
	\includegraphics[width=.7\linewidth]{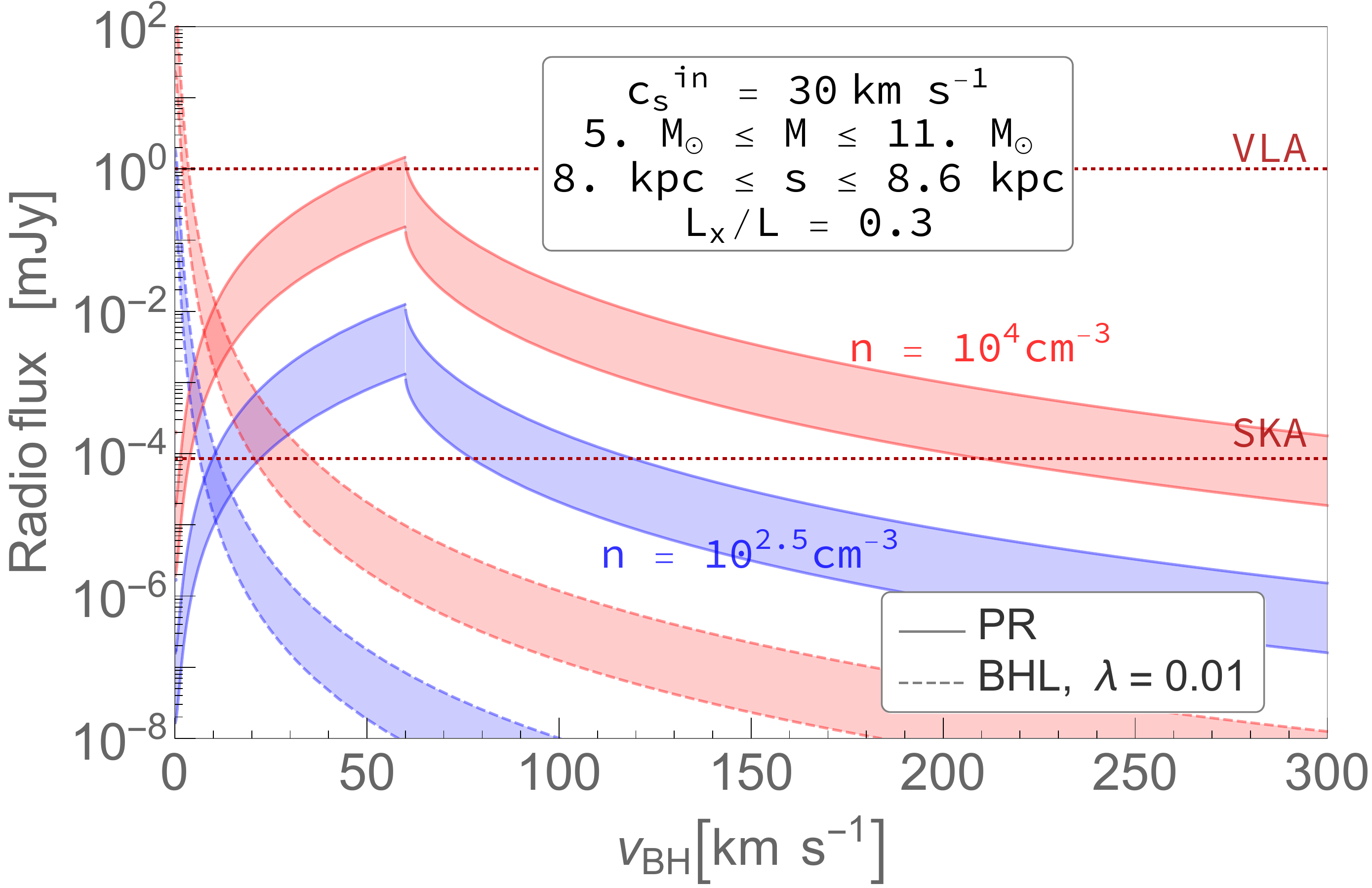}
	\caption{ \textbf{Radio flux from BHs in the CMZ}: We show the radio flux for a generic BH in the CMZ. The present VLA sensitivity and the prospective sensitivity for SKA are shown. }
	\label{fig:radioflux}
\end{figure}
\begin{figure}
	\centering
	\includegraphics[width=.7\linewidth]{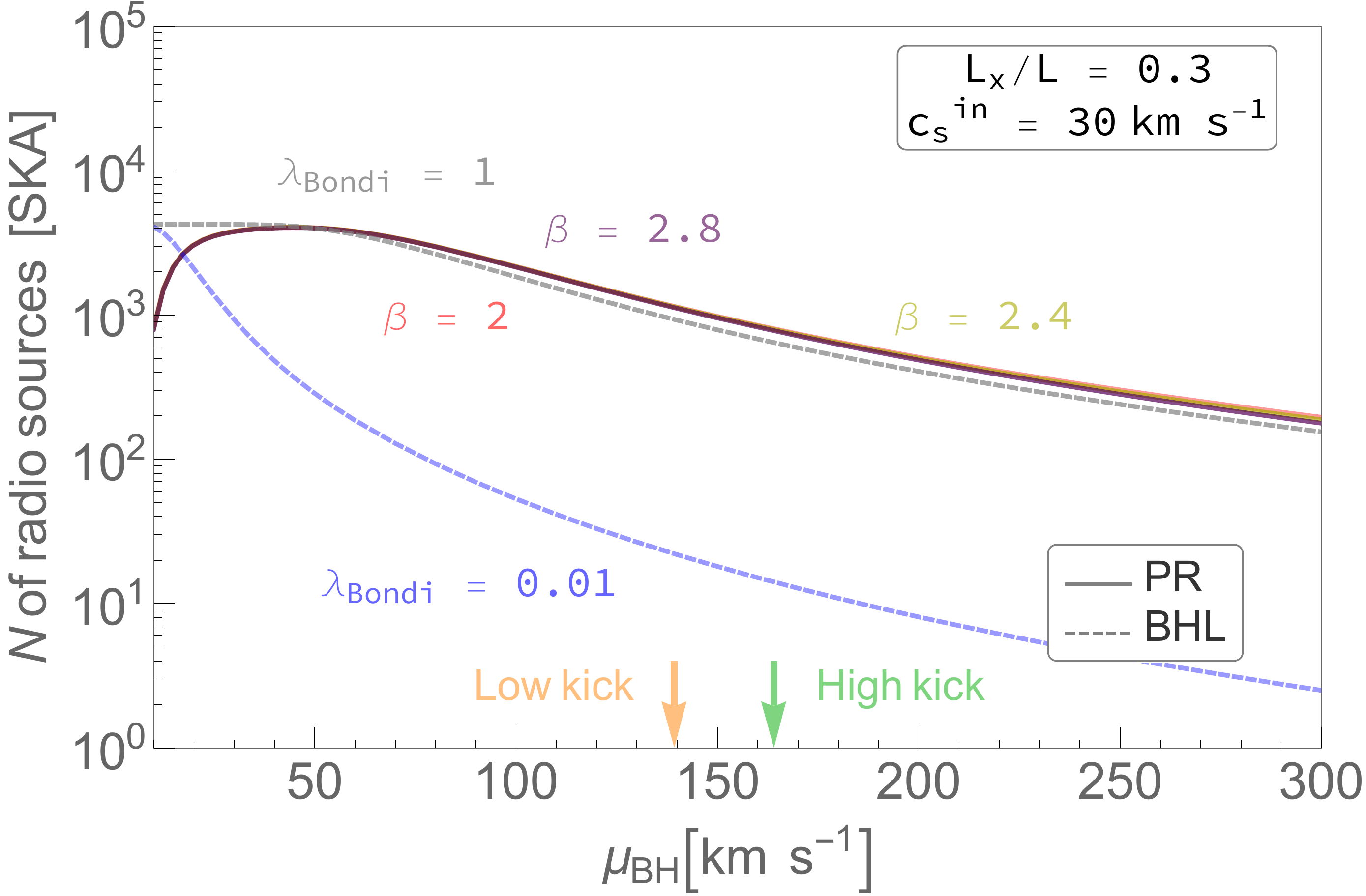}
	\caption{ \textbf{Radio prediction (I)}: We show the number of radio sources expected from the CMZ as a function of the number density of gas in molecular clouds, for different choices of the relevant parameters, obtained without taking into account the information from the X-ray band. The combination of this prediction with the X-ray bound is presented in figure \ref{fig:SKA2}. 
	}
	\label{fig:SKA1}
\end{figure}
In the previous section we have shown that, in a wide portion of the parameter space associated to our problem, a large number of bright X-ray sources are expected in the Galactic Center region.
However, in order to pinpoint a source as an accreting black hole, a careful multi-wavelength study has to be performed. The GHz radio band is particularly interesting in this context. As in previous works (\cite{Gaggero:2016dpq,Manshanden:2018tze}), the empirical scaling between X-ray and radio flux (\cite{Plotkin:2012}) discussed in detail in Sec. \ref{sec:accretion:emission} (Eq. \ref{eq:FP}) is employed to obtain the radio flux. \\
A remarkable increase in the sensitivity is expected in the radio domain over the coming decade, thanks to the development of the Square Kilometre Array (SKA) project. This experiment has a huge potential towards shedding light on key problems of fundamental physics, cosmology and astrophysics \citep{Bull:2018lat}. Here we focus on the discovery potential of the population of astrophysical black holes in the CMZ at $1.4$ GHz band, and provide estimates of the number of potentially detectable sources by the SKA1-MID facility.
Assuming gain $G = 15$ K/Jy, receiver temperature $T_\mathrm{rx} =  25$ K, sky temperature towards the Galactic Center $T_\mathrm{sky} = 70$ K, and bandwidth $\delta \nu = 770$ MHz (as in \cite{2016ApJ...827..143C}), we obtain an instrumental detection sensitivity of $2.7 \mu$Jy for a one-hour exposure. 
In the following, we assume an optimistic $1000$ h exposure time and consistently adopt a potential detection threshold of $85$ nJy. 

\begin{figure}
   \centering
	\includegraphics[width=.7\linewidth]{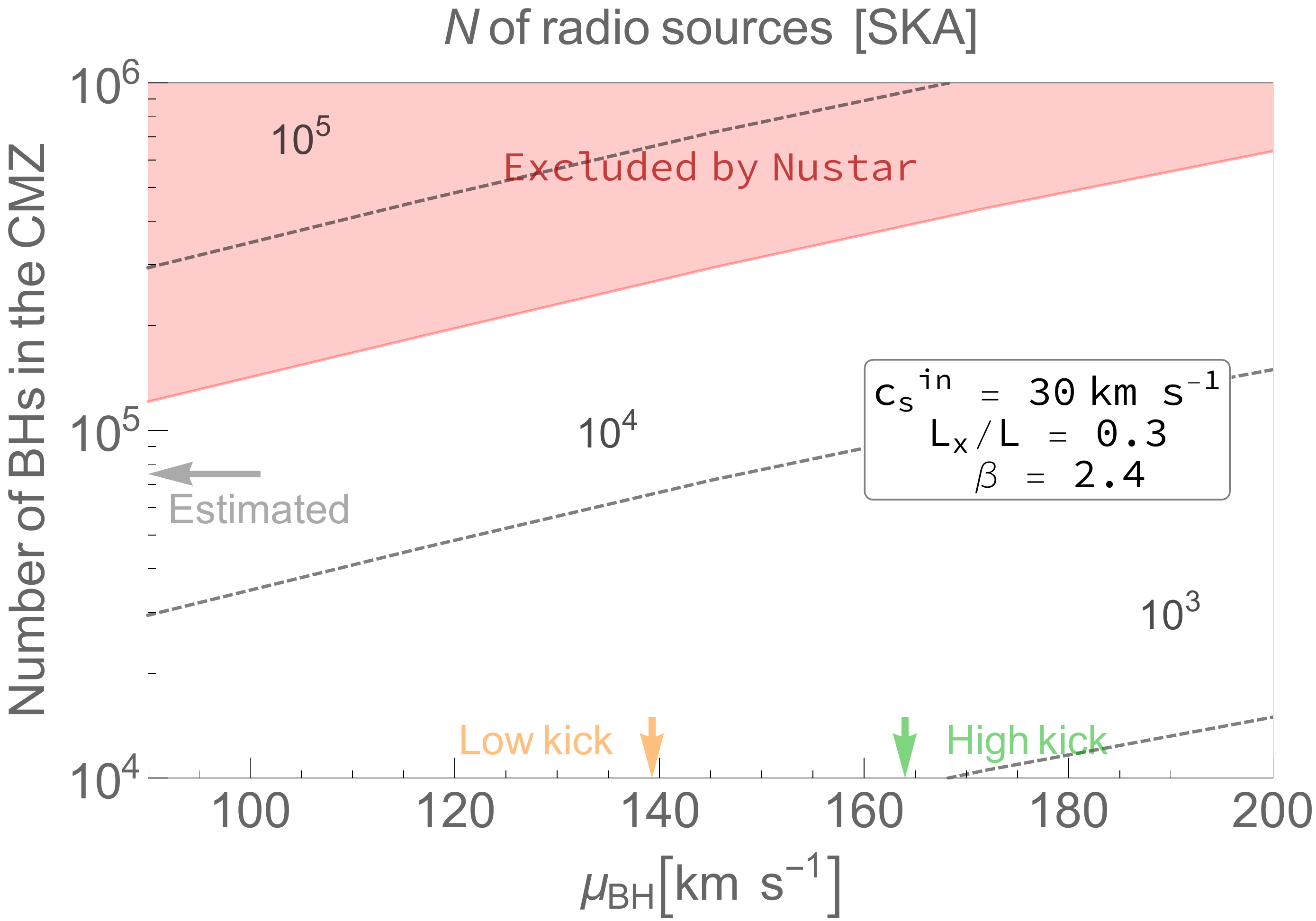}
	\caption{ \textbf{Radio prediction (II)}: We show the contour lines for the number of sources potentially detectable by the Square Kilometre Array as a function of cloud density and total number of BHs in the CMZ clouds. The red region corresponds to the parameter space region excluded by NuSTAR observations. The grey arrow indicates the estimate of the number of BHs in the CMZ region used as reference in the previous section. 
		}
	\label{fig:SKA2}
\end{figure}

In figure \ref{fig:radioflux} we show the radio flux associated to a generic BH in the CMZ cloud, and compare it with the prospective SKA sensitivity and with the threshold of the existing VLA catalogue \citep{2008ApJS..174..481L}. Focusing on the PR13 scenario, we can conclude from this plot that, while we do would not expect to detect any isolated BH given the VLA sensitivity, SKA would be able to unveil a huge population of isolated BH in the Galactic centre. In fact, most of the BHs accreting from the cold clouds would emit above the SKA threshold.\\
We show our prediction for the number of radio sources detectable with SKA in \cref{fig:SKA1,fig:SKA2}, obtained with the same setup described in section \ref{sec:accretion:ABH:CMZ}.
In particular, in figure \ref{fig:SKA1} we show the number of radio sources detectable by SKA as a function of the BH speed, for various choices of the parameters discussed in the previous section. Our model predicts thousands of visible sources for the two reference values we have chosen to bracket the uncertainty on the speed distribution.

We recall that $N^\mathrm{ CMZ}$, the total number of BHs in the CMZ encloses a significant part of the uncertainties of our model. Let us now widen our perspective and promote it to a free parameter. In figure \ref{fig:SKA2} we show the number of bright radio sources in the ($\mu_\mathrm{BH}$, $N^\mathrm{ CMZ}$) space.
In this parameter space, we may use the comparison with NuSTAR observations discussed in the previous section to obtain a bound: in particular, we can identify an upper limit on the number of black holes present in the CMZ, by requiring not to overshoot the number of sources observed by that experiment.\\

We find that, in the portion of parameter space compatible with the NuSTAR bound, $\mathcal{O} (10^4)$ radio point sources could be detected by SKA. Such order of magnitude estimate is solid with respect to the uncertainties of the remaining parameters of the model.\\
This prediction relies on the assumption that the FP scaling between radio and X-ray emission holds, which in turn requires assuming the emission of a jet. While this is not disproven, we are neither able to assert with certainty that every accreting isolated BH will emit a jet. If no jet is emitted, we would expect a lower radio flux. 
It is interesting to assess the prospects in case of a lower radio emission with respect to the FP prediction. The SKA sensitivity we consider here corresponds to a flux about $ 10^{8}$ times lower than the one detectable with the NuSTAR sensitivity. The FP predicts a luminosity ratio $L_\mathrm{R}/L_\mathrm{X}$ of around $10^{-6}$ at peak luminosity. Therefore, 1\% of the radio luminosity predicted by the FP scaling would be sufficient for all NuSTAR isolated BH sources to have a radio counterpart detectable by SKA. The most luminous sources would still be visible if the radio flux is 0.01\% of that predicted by the FP scaling.

In conclusion, we have shown that a scenario based on the PR13 accretion model and our best estimate of the number of BHs in the CMZ naturally implies a \textit{ large number of detectable X-ray and radio sources}, for any reasonable choice of the other free parameters involved in the problem. This relevant result is highly suggestive that a clear identification of a population of isolated, accreting BHs in the Galactic Center region -- by means of the current or forthcoming generation of radio and X-ray experiments -- may be around the corner. 

\subsection{Discussion}
\label{sec:accretion:ABH:Comments}

\paragraph{Number of black holes in the clouds}
We have concluded that our predictions for the CMZ are stable in a very wide region of the parameter space of the model. However, we should note that we made a number of assumptions regarding the distribution of black holes in the CMZ. 
Our estimate of the total number of BHs in the CMZ is obtained with a simple model for the initial distribution and subsequent orbit evolution. 
Furthermore, we have assumed that both the BHs and the clouds are uniformly distributed in the CMZ. \\
These simplifications eventually affect the number of BHs in the clouds, and therefore the final predictions on the number of sources, which scale linearly with this quantity.\\

\paragraph{The accretion scenario}
We have shown in Section \ref{sec:accretion:ABH:local} that, based on local observations, the PR13 accretion model does not require the introduction of a suppression factor $\lambda$. However, since the PR13 model relies on the BHL model to describe accretion within the ionized region (see Sec. \ref{sec:accretion:models}), one may wonder whether the BHL suppression factor should be introduced at that stage. With respect to this, we should take into account that this suppression factor is usually associated to a very strong outflow of matter from within the Bondi radius, associated to a fraction of $1-\lambda$ of the total matter initially accreted. Such a large outflow is not observed in the simulations of \cite{Park:2012cr}, where most matter crossing the Bondi radius within the ionized region ultimately reaches the BH. We conclude from this that introducing a suppression factor in the expression would not be physically justified.\\
We remark that the simulations considered here represent the state of the art in the field. However, some physical ingredients are still not captured: we mention in particular the role of magnetic fields. Future works will assess the potential impact of a full magneto-hydrodynamical treatment in this context.\\

\paragraph{The emission mechanism and spectrum}
In Sec.~\ref{sec:accretion:emission} we also assume that $\sim30\%$ of the total bolometric luminosity of all our isolated BHs falls within the observable X-ray band, that is $L_\mathrm{X}=0.3~L_\mathrm{bol}$. This assumption was based on previous estimates, for example those presented by \cite{Fender:2013}. However, it is worth assessing the accuracy and precision of this assumption. We assumed a flat radio spectrum ($L_{\nu}\propto\nu^0$), and then an additional inverted power law in the X-ray band. The assumption that $30\%$ of the observed bolometric luminosity falls in the X-ray band depends on the spectral index of this additional power law, $\alpha$, the high-energy cut off of the power law, $E_\mathrm{cut}$, the absorbing hydrogen column density along the line of sight, $N_\mathrm{H}$, and the observing energy band (which we conservatively adopted as $3\mbox{--}40$~keV for the NuSTAR X-ray telescope). Inefficiently radiating Galactic BH-XRBs are dominated by power law emission with a typical range for the spectral index given by $0.6 \le\alpha\le 1.0$ (or a photon index of $1.6 \le\Gamma\le 2$ in common X-ray astronomy parlance). However, the power law spectra of Galactic BH-XRBs can become softer in intermediate spectral states, or very quiescent states, so we can extend this out to an upper bound of $\alpha\le 1.4$ or $\Gamma\le 2.4$. Adopting a rough estimate of the jet break frequency of $\sim10^{14}$~Hz (infrared frequencies; \cite{Russell2013}), a typical high-energy cut off in the X-ray of $100$~keV, and a high hydrogen column density of $10^{23}~\mathrm{ cm^{-2}}$, we calculate that $L_\mathrm{X}\sim(0.24\mbox{--}0.35)~L_\mathrm{bol}$. The corresponding unabsorbed range of fractional luminosities is then $0.27\mbox{--}0.46~L_\mathrm{bol}$. These predictions, however, depend strongly on the cutoff energy. For example, just reducing the cutoff energy from $100$~keV to $80$~keV can increase the fractional luminosity to $\sim60\%\mbox{--}70\%$. Therefore, our assumption that $L_\mathrm{X}\sim0.3~L_\mathrm{bol}$ is rather conservative, even if we assume strong absorption along the line of sight. It is also based upon a relatively hard power law spectrum in the X-ray band. Nevertheless, we have shown in Sec. \ref{sec:accretion:ABH:CMZ}, that our predictions are solid with respect to variations of this parameter, even for values as large as $70 \%$.

For the radio flux calculations through the Fundamental Plane of Black Hole Activity (FP; \cite{Merloni:2003,F04,Plotkin:2012}) we had to assume the presence of a jet emitting in the GHz radio band, following \cite{Fender:2013, Gaggero:2016dpq, Manshanden:2018tze}. This crucial assumption is motivated by results of \cite{Fender:2001}, which established that all highly sub-Eddington accreting BHs in X-ray binary systems are accompanied by such radio emission. 
Further emission could come from other types of outflow that are less collimated or relativistic. Such alternative emission could constitute a lower limit on the expected radio emission, which would be of particular interest in the absence of a jet. Ref.~\cite{Tsuna:2019kny} assumed a spherically symmetric outflow based on losing a fraction (1-$\lambda$) of the accreting matter before reaching the BH, where $\lambda$ is the suppression factor. Assuming a power-law radial dependence of the accretion rate, they used the escape velocity to conservatively estimate the power associated to such an outflow. It should be noted that this radial power-law assumption (with slope larger than $1$) implies that most matter would be lost close to the BH, which requires a powerful outflow. The shock from the collision of this outflow with the ISM could 
accelerate non-thermal electrons through diffusive shock acceleration and subsequently produce radio-synchrotron emission. The precise emission profile will be highly dependent on the fraction of energy going into the electrons and the magnetic field.\\
However, as highlighted above, such a relevant outflow is not observed in the simulations presented in \cite{Park:2012cr} and hence this scenario is hard to unify with the PR13 accretion model. 
Instead, it would be interesting to further investigate what type of emission could be associated to the shock on the boundary of the cometary region itself, which would likely require additional detailed simulations.\\

\section{ Primordial black holes: abundance constraints and observational prospects\footnote{Based on \cite{Scarcella:2021jzp,Scarcella:2022pbh}}}
\label{sec:accretion:PBH}

Multiple connections between primordial black hole searches and  the study of e.m. radiation in different bands has been considered in the literature (see e.g. \cite{Kashlinsky:2019kac} and references therein). In particular, in this thesis we are focusing our attention to the X-ray and radio band, and to the Galactic environment. As argued in the previous section, the Central Molecular Zone (CMZ) appears to be the most promising region for accretion based searches, both because of the presence of a large amount of dense molecular clouds, and because it is a preferential target for extensive multi-wavelength observational campaigns. Furthermore, we expect a large population of these objects to be inhabiting the central part of the Galaxy. 

The potential of this observational channel in constraining the abundance of PBHs through observations of the inner part of the Galaxy, with particular focus on the CMZ region, was first pointed out in \cite{Gaggero:2016dpq}, within the framework of the Bondi-Hoyle-Littleton formalism. 
In \cite{Manshanden:2018tze} the impact of the Park-Ricotti formalism was assessed in detail. The main result of that work is a strong upper bound on the abundance of PBHs (in the context of a non-clustered distribution of PBHs with monocromatic, log-normal and power-law mass functions). The upper bound was set at the level of $10^{-3}$ of the dark matter in the form of PBHs, which would correspond to roughly $10^8$ objects in the Milky Way for a $10$ M$_{\odot}$ reference mass. 

In the work presented in this section, we aim to apply the semi-analytical treatment described in \cref{sec:accretion:MilkyWay} to assess the uncertainties associated with the bounds of \fPBH obtained through the accretion channel. We start by discussing the expected number of sources detectable by the NuSTAR survey as a function of the relevant parameters and the corresponding constraints on PBH abundance, considering a monochromatic mass function. The NuSTAR survey has detected $\sim 70$ hard X-ray sources sources in this region interest. The nature of most of the sources in the catalogue in not precisely known; we obtain the bound conservatively by assuming that all the sources are accreting isolated PBH, hence neglecting the astrophysical background. \\

We will in particular discuss the appearance of a threshold effect due to the particular behaviour of the PR accretion rate, which makes PBHs of $M \lesssim 1 ~\Msun$ difficult to detect. 

Motivated by this, we will consider a specific multimodal mass function~\cite{Jedamzik:1996mr,Byrnes:2018clq, Carr:2019kxo}, which arises naturally when carefully considering the thermal history of the early Universe (see \cref{sec:intro:pbhs:formation}) and which presents its major peak at one solar mass. We will discuss the constraining power of accretion-based searches on a population of PBH described by this mass function and the related uncertainties.

\noindent
The results presented in this section are based on \cite{Scarcella:2021jzp,Scarcella:2022pbh}.

\subsection{The primordial black hole population and the central molecular zone}
\label{sec:accretion:PBH:population}

\paragraph{Cloud distribution}

\begin{figure}
	\centering
	\includegraphics[width=.6\linewidth]{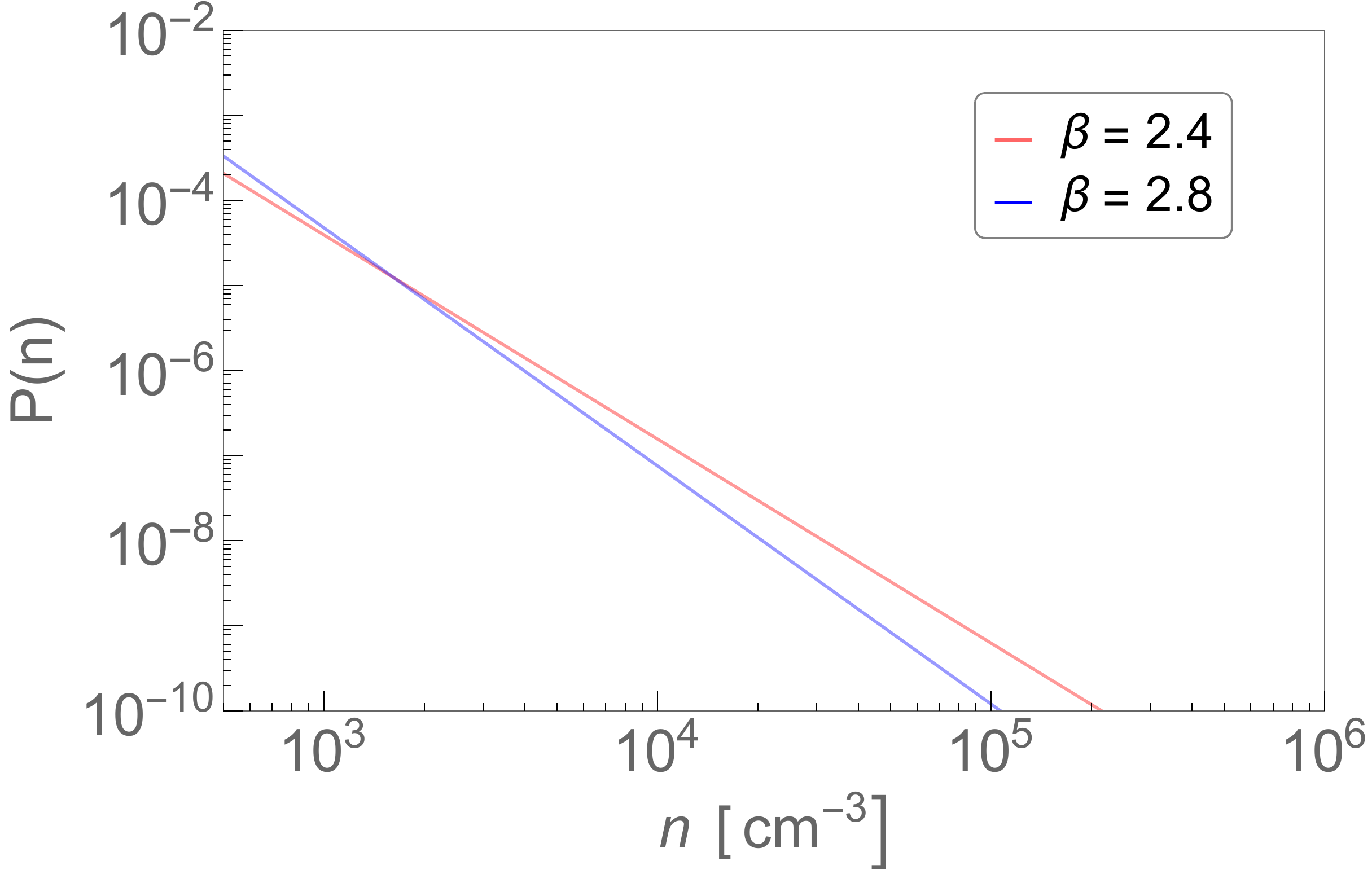}
	\caption{ The power-law profile assumed to describe the distribution of clouds of different density, shown for two choices of the power-law index $\beta$. For $\beta = 2.4$, slightly more dense clumps are present than for $\beta = 2.8$. 
	}
	\label{fig:Clouds}
\end{figure}

A fundamental ingredient is the description of the molecular clouds of the CMZ.
We employ here a slightly simplified model of the cloud density with respect to that discussed in \cref{sec:accretion:MilkyWay:clouds}. We describe the distribution of the hydrogen number density in the clouds with a single power law, 
\begin{equation}
\label{eq:gasdensityPBH}	
	P(n) \propto n^{- \beta}
\end{equation}
To include in this distribution both the less dense, warm clouds and the denser clumps, we set the minimum density to $n= 10^{2.5} \, \mathrm{ cm }^{-3}$. The distribution is normalized such that the overall average molecular hydrogen  number density of the CMZ region is $n_\mathrm{avg} = 150\, \mathrm {cm}^{-3}$\citep{Ferriere2007}. We treat index of the power-law $\beta$ as a free parameter. However, its value is quite well constrained by the measurement of the clouds filling factor. For the reference value of  $\beta =2.4$ \citep{HeRG:2019,HeRG:2020} we obtain a filling factor of $\approx 15 \%$ and an average cloud density of $\bar{n} \approx  \times 10^4 $. Setting $\beta =2.8$ results in a filling factor of $\approx 20 \%$ and an average cloud density of $\bar{n} \approx  \times 7 \times10^3 $. We will consider these as our reference values. The distribution of cloud densities is shown in \cref{fig:Clouds}.The temperature of the clouds enters the accretion rate through the sound speed $c_\mathrm{s}$. As in \cref{sec:accretion:ABH} we set $c_\mathrm{s} = 1$ km/s.  

\paragraph{PBH speed distribution}

We model the velocities of the PBH population with a Maxwell-Boltzmann distribution. The mean value of the velocity is an important parameter, given the strong dependence on it of the accretion rate. Using Eddington's inversion method results in values of about 50 km/s, corresponding to near-peak accretion rates. On the other hand, as mentioned in \cref{sec:accretion:ABH:CMZ:setup}, observations of stars near the GC show a higher velocity dispersion, around 130 km/s \cite{Sanders_2019}.  As in \cref{sec:accretion:ABH}, we treat the mean velocity  $\mu_\mathrm{BH}$ as a free parameter.

\paragraph{PBH spatial distribution}

\begin{figure}[h!]
	\centering
	\includegraphics[width=.6\linewidth]{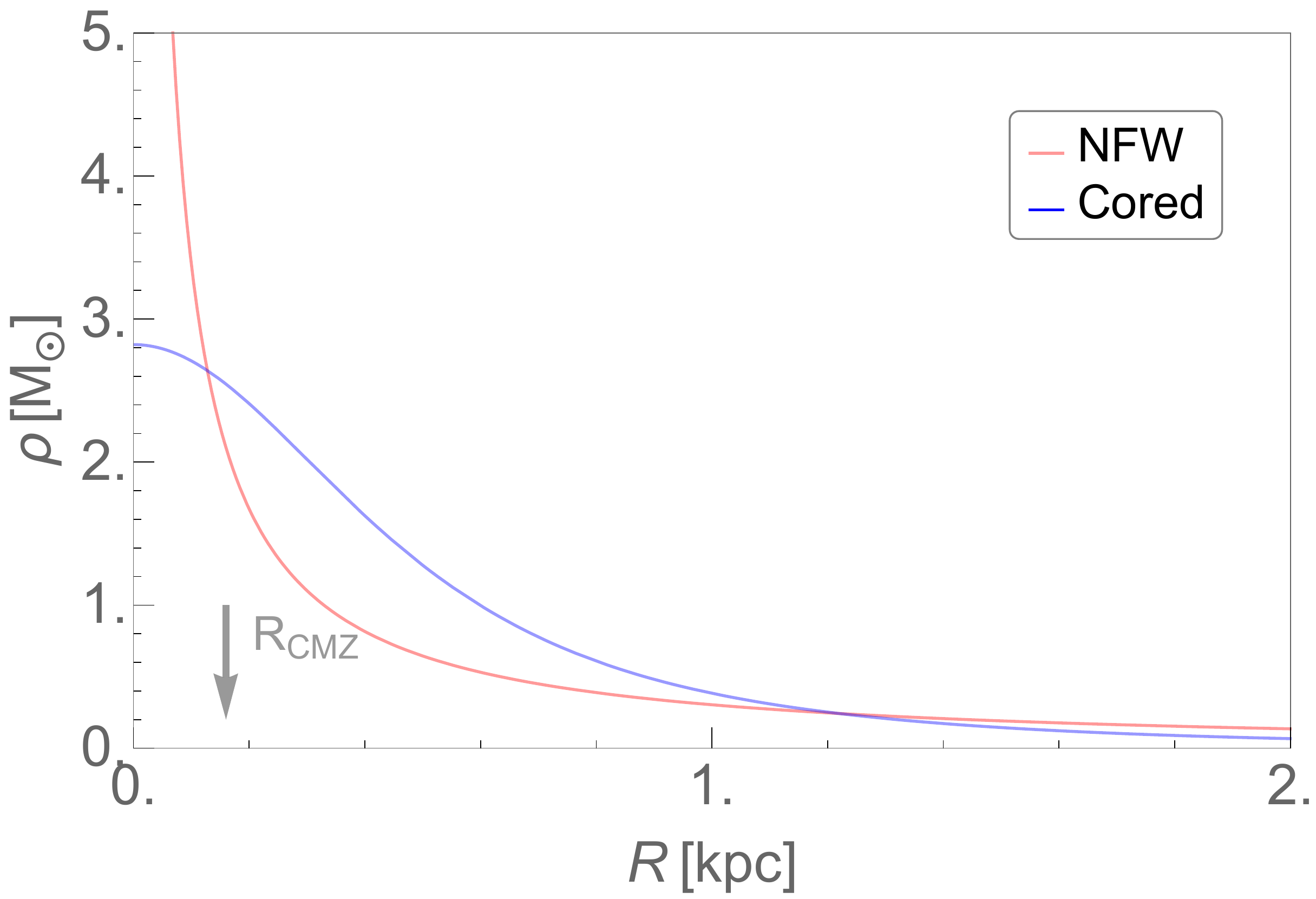}
	\caption{We compare the radial density of the Galactic DM halo described by the NFW profile (red) and a profile showing a central core (blue). The two descriptions are in agreement at large radii, but differ significantly within the CMZ (whose radius is indicated with a grey arrow).}
	\label{fig:DMprofiles}
\end{figure}

Whether primordial black holes constitute all of the DM, or only a fraction of it, we expect them to follow the profile of the Galactic DM halo. 
Here we are interested in the innermost region of the Galaxy: we recall that the CMZ is an approximately cylindrical cloud complex centred in the galactic centre, extending for a radius of around 160 pc and an overall height of about 30 pc. In this central region, the profile of the DM halo is poorly constrained (see eg \cite{Salucci:2018hqu} and references therein). \\
In figure \cref{fig:DMprofiles}, we show a Navarro-Frenk-White (NFW) profile \cite{Navarro:1996gj} and a cored \cite{Governato:2009bg, Binney:2001wu} density profile.
These two classes of profiles differ significantly in the inner region corresponding to the CMZ, whose radius is indicated in figure with a grey arrow. 
Given these uncertainties, we choose to ``smooth out'' the density profile, computing the expected total DM mass enclosed in the CMZ volume. \\
Despite the very different behaviour of the two profiles, the total enclosed mass is similar: assuming an NFW profile we obtain $M_\mathrm{CMZ} \sim 9 \times 10^6$ \Msun , while the cored profile gives $M_\mathrm{CMZ} \sim 6 \times 10^6$ \Msun.

The total enclosed mass in the form of PBHs is then $\fPBH M_\mathrm{CMZ}$, from which one can obtain the total number of PBHs assuming a mass function. The relevant quantity for the problem at hand is the number of BHs that are found within a molecular cloud. To compute this quantity we make, as in \cref{sec:accretion:ABH}, the assumption that both PBH and clouds are distributed uniformly within the region, with no significant correlation/anti-correlation of their locations. Hence we identify the probability of a PBH being in a cloud with the cloud filling factor $f_\mathrm{clouds}$.\\
We remark that this is a simplifying assumption and that some correlation/anti-correlation might exist between the two distributions.
Furthermore, one could expect that on these small scales the distribution of PBHs presents some particular behaviour. In particular, one could expect PBH to be distributed in clusters. We leave the assessment of the impact of these aspects for future work.
\subsection{Searching for primordial black holes in the central molecular zone}
\label{sec:accretion:PBH:CMZ}

\subsubsection{Monochromatic mass function}

We start by assuming a monochromatic mass distribution. We can compute the number of X-ray sources above the NuSTAR threshold, from \cref{eq:Nsources}
\begin{equation}
	\label{eq:NPBHNustar1}
	N^\mathrm{sources} \, = \, N^\mathrm{ clouds} (M)  \int_{\phi > \phi^*} P(\mathrm{ v}_\mathrm{PBH})P(n) \,\diff \mathrm{ v}_\mathrm{PBH}\,\diff n  \; ,
\end{equation}
where the expected number of primordial black holes in the clouds is obtained as
\begin{equation}
	\label{eq:NPBHclouds}
	N^\mathrm{ clouds} (M) = f_\mathrm{clouds} \, \fPBH \dfrac{M_\mathrm{CMZ}}{M} \; ,
\end{equation}
where $f_\mathrm{clouds}$ is the associated filling factor. We stress once again that we are assuming, for simplicity, that no type of spatial correlation or anti-correlation exists between clouds and PBHs.\\
\begin{figure}[h]
	\centering
	\includegraphics[width=.7\linewidth]{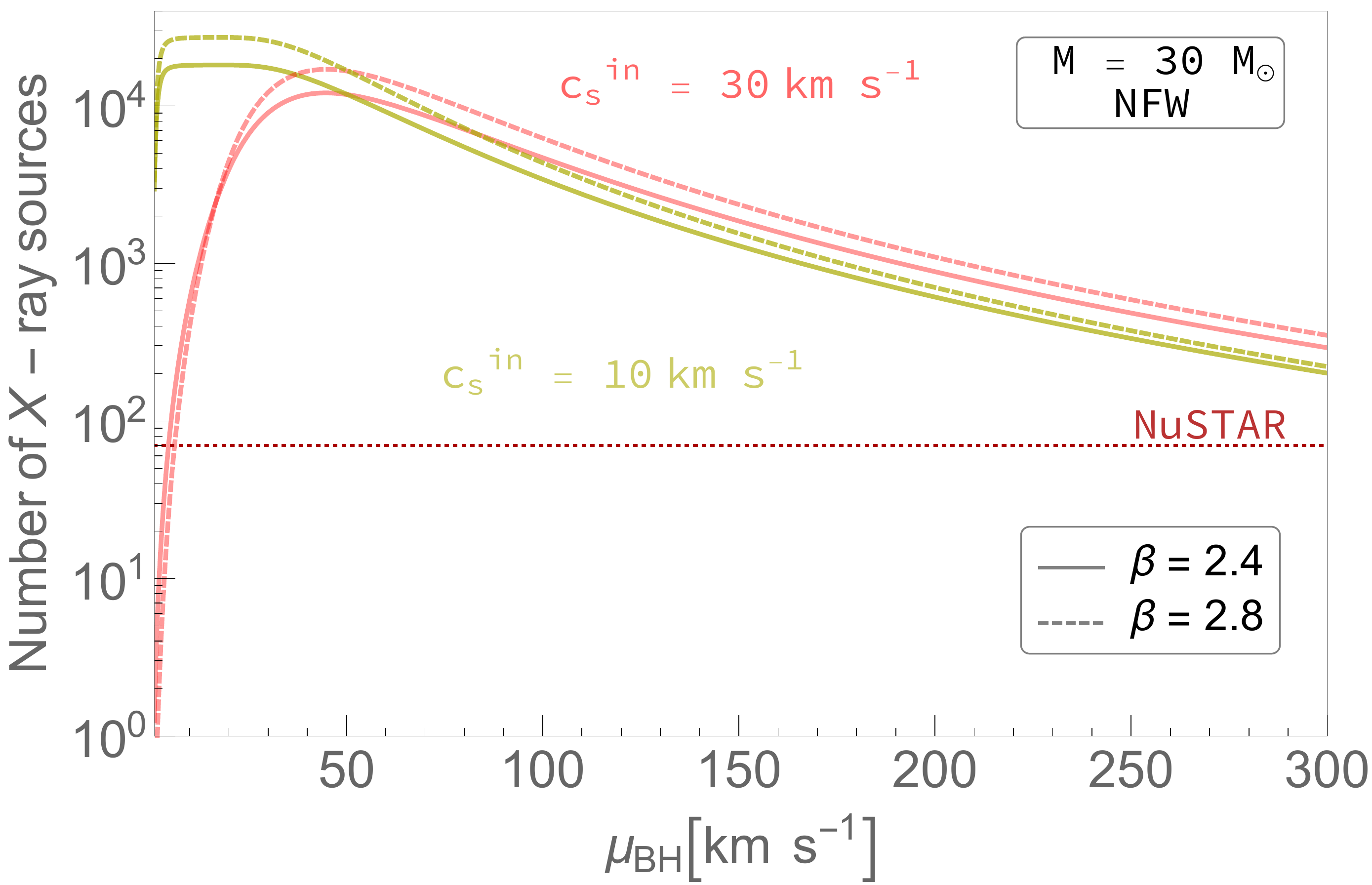}
	\caption{Number of sources above the NuSTAR threshold, assuming an NFW profile and a monochromatic mass function with $M = 30 ~\Msun$. The prediction i shown for different values of the cloud density power-law index $\beta$ and of the sound speed of the ionized gas.}
	\label{fig:Nsources30} 
\end{figure}

In \cref{fig:Nsources30} we show the expected number of X-ray sources detected by NuSTAR assuming a PBH mass of $M = 30 ~\Msun$ and an NFW DM profile, as a function of the average PBH speed $\mu_\mathrm{BH}$. We show this prediction for different choices of 
two other relevant parameters, relative to the description of the gas:
\begin{itemize}
	\item the power-law index $\beta$ describing the cloud density distribution \cref{eq:gasdensityPBH} 
	\item the sound speed $c_\mathrm{s}^\mathrm{in}$ of the ionised gas around the black hole
\end{itemize}
The number of sources scales significantly with the PBH average speed. On the other hand, the prediction is solid with respect to variations in the other two parameters (with the exception of unrealistically small velocities $\mu_\mathrm{BH} \lesssim 20$ km/h). 

The red dashed line in  \cref{fig:Nsources30}  shows the number of sources detected by the NuSTAR survey \cite{Hong:2016qjq}. By comparison with this observation, we obtain a conservative bound on \fPBH as a function of the PBH mass, assuming all detected sources are PBHs.

\begin{figure}
	\centering
	\begin{subfigure}
		\centering
		\includegraphics[width=.49\linewidth]{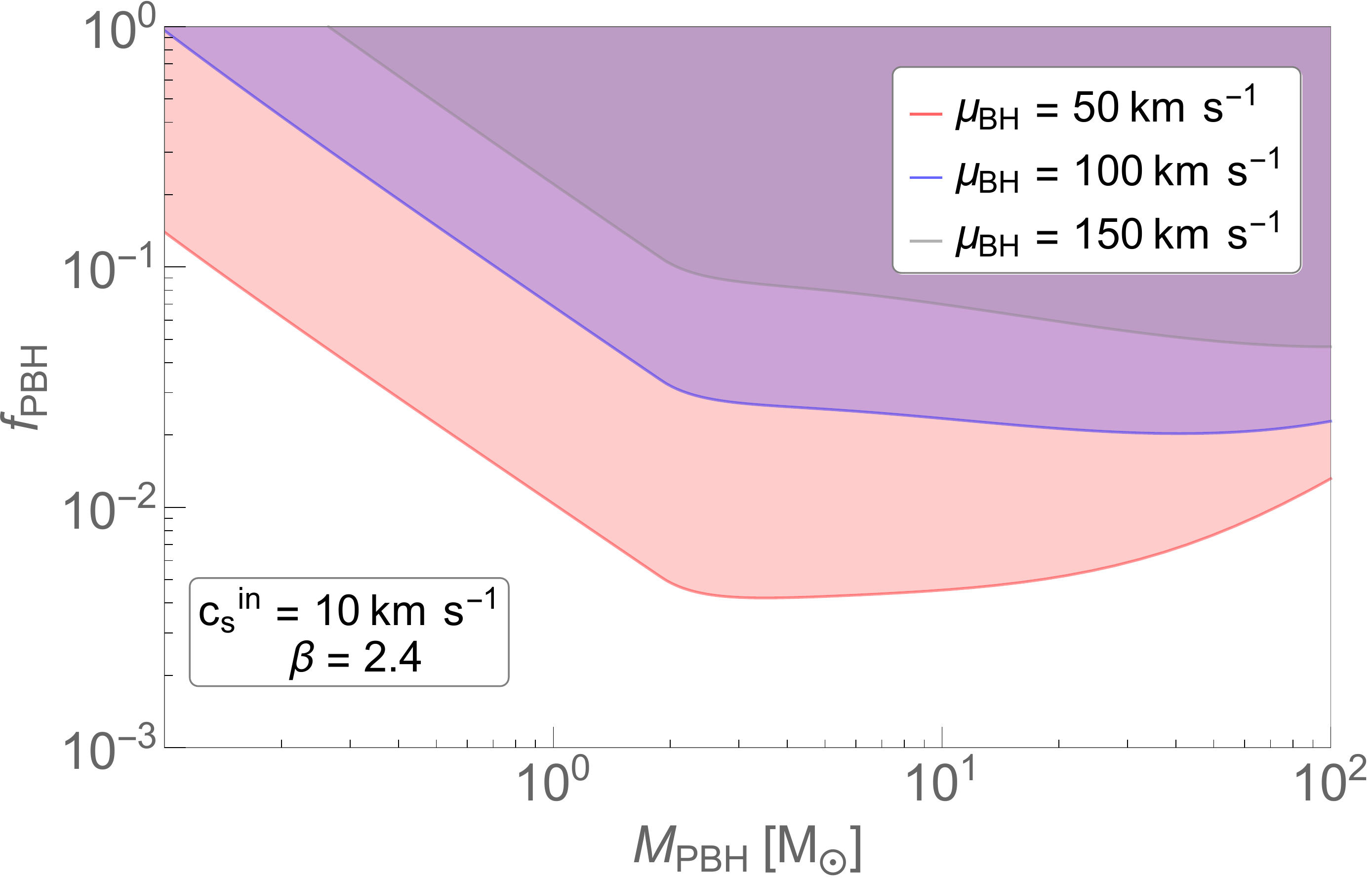}
	\end{subfigure}
	\begin{subfigure}
		\centering
		\includegraphics[width=.49\linewidth]{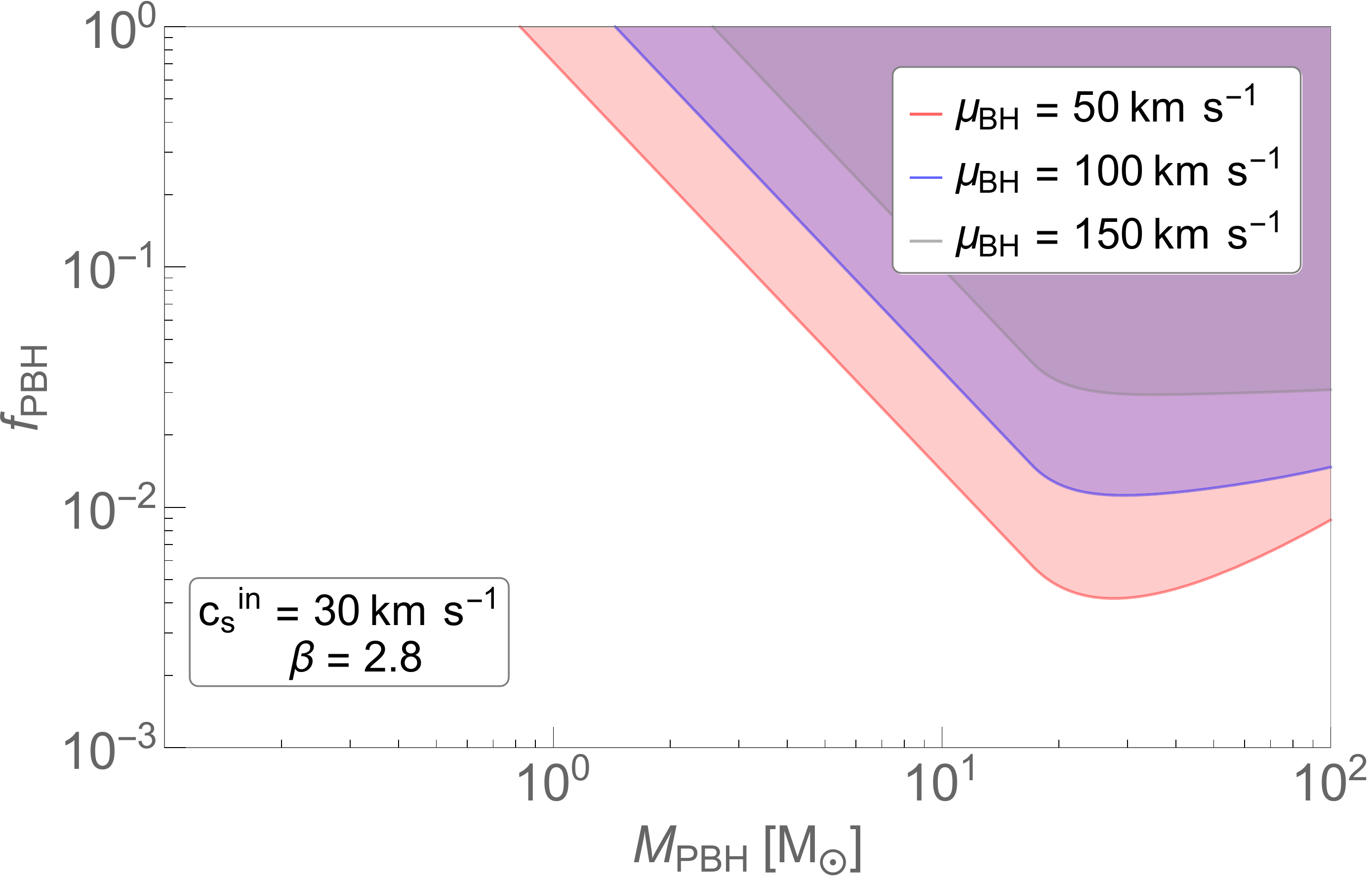}
	\end{subfigure}
	\caption{Bounds on $\fPBH$ obtained by comparison with the catalogue of the NuSTAR survey \cite{Hong:2016qjq}, assuming a monochromatic mass function. The two panels correspond to an optimistic and a conservative case with respect to the description of the molecular clouds. Different colours correspond to different values of the average PBH speed. }
	\label{fig:boundsmono} 
\end{figure}

We present these constraints in \cref{fig:boundsmono}. The two panels correspond to different values of the parameters $\beta$ and $c_\mathrm{s}^\mathrm{in}$.
In different colours we show different values for the average PBH speed.  $\mu_\mathrm{BH}$. 
%
For masses  above $ \sim 10 \Msun $, the two panels are in agreement, with the constraint on \fPBH ranging between  $ 10^{-1} $ and $10^{-3}$. In this mass range, the main uncertainty on the bound  comes from the value of velocity dispersion. The strongest bound are obtained when the PBH speeds are close to the value corresponding to the peak in the accretion rate (peak accretion is attained for $\vBH\simeq 2 c_\mathrm{s}^\mathrm{in} $).

For lower PBH masses, we can notice instead -- comparing the two panels -- that the results show a strong variability with the choice of the parameters that describe the gas. \\

\begin{figure}
	\centering
	\includegraphics[width=.49\linewidth]{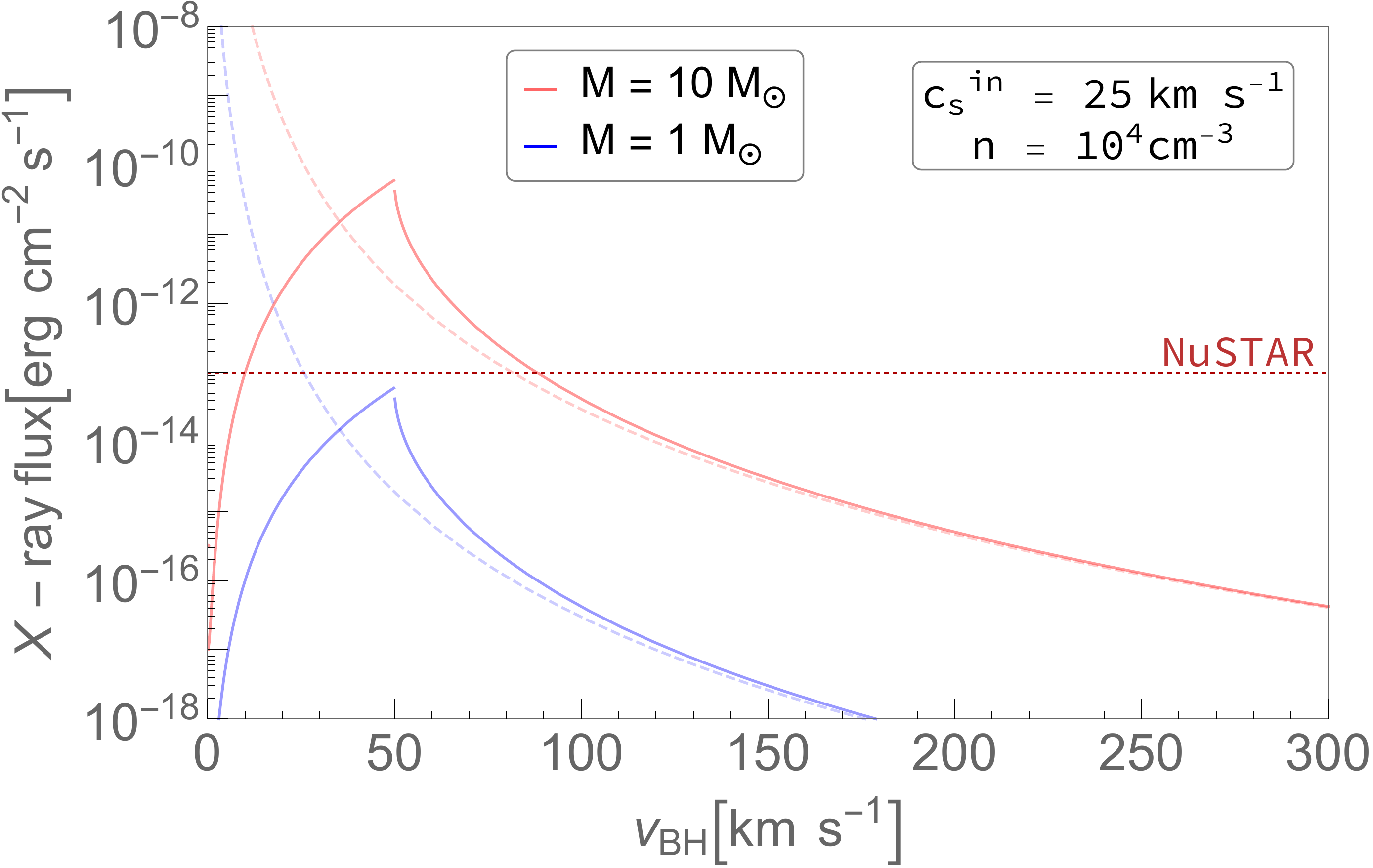}
	\includegraphics[width=.49\linewidth]{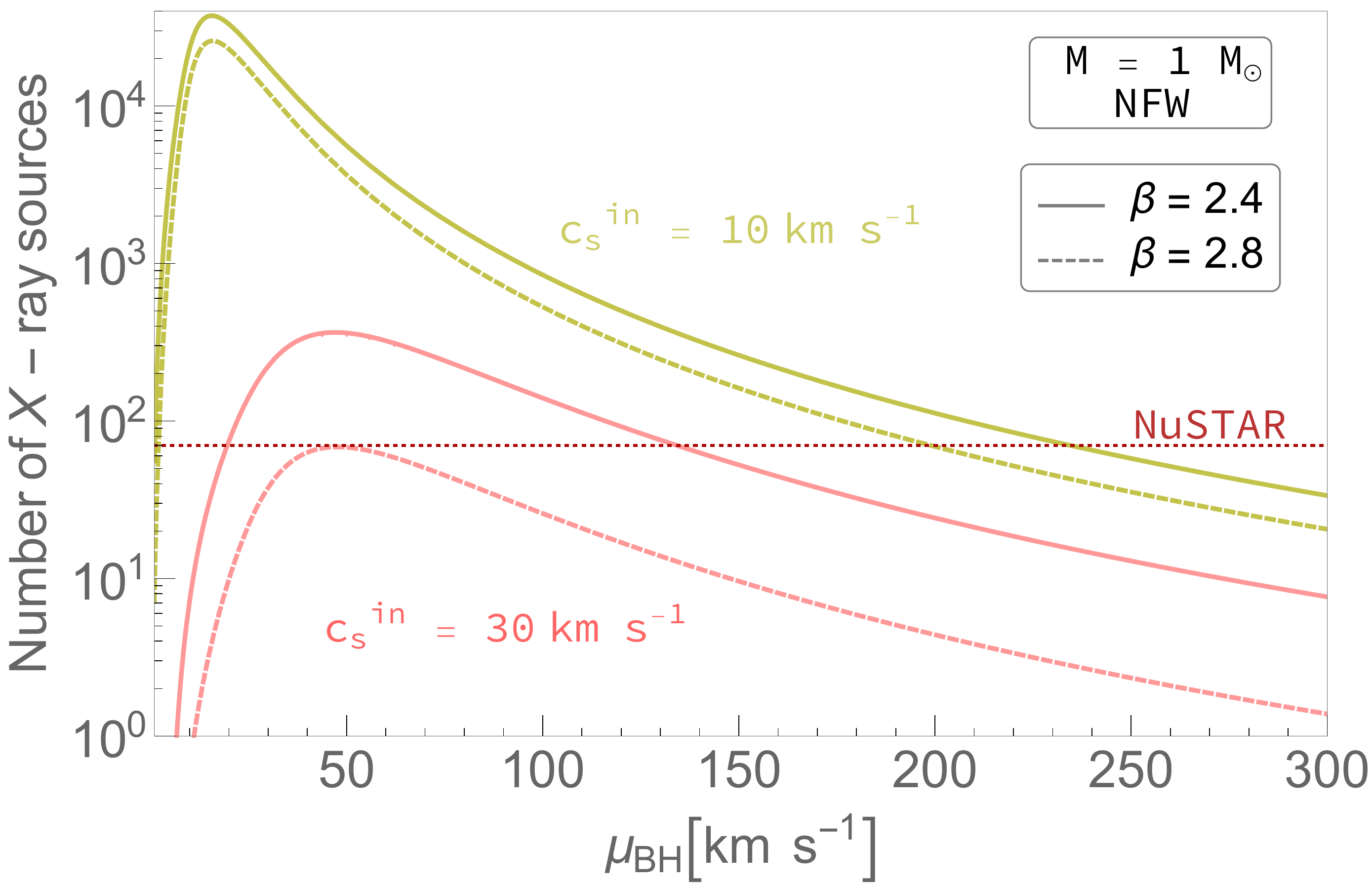}
	\caption{\textbf{Left:} X-ray flux at Earth for two BHs of different mass accreting from a typical molecular cloud ($n = 10^4\, \mathrm{cm^{-3}}$) located at the Galactic centre (GC), as a function of their velocity relative to the cloud. The sound speed of the ionized gas is set to 25 km/s. The BH of $1 ~\Msun$ is below the detection threshold of the NuSTAR survey.  \textbf{Right}: Number of sources above the NuSTAR threshold, assuming an NFW profile and a monochromatic mass function with $M = 1 ~\Msun$. The prediction is shown for different values of the cloud density power-law index $\beta$ and of the sound speed of the ionized gas.}
	\label{fig:NsourcesThreshold} 
\end{figure}

This is due to a presence of a threshold effect, exemplified in \cref{fig:NsourcesThreshold}.  In the left panel, we show the X-ray flux at Earth for two BHs of different mass accreting from a typical molecular cloud ($n = 10^4\, \mathrm{cm^{-3}}$) located at the Galactic centre (GC), as a function of their velocity relative to the cloud. The sound speed of the ionized gas is set to 25 km/s. Comparing the predicted flux to the threshold of the NuSTAR survey, we can notice that the BH of $\sim$ 1 $\Msun$ or less is not detectable for any value of the relative speed. 

The threshold mass can vary by a factor of a few depending on the cloud density and on the the ionized sound speed $c_\mathrm{s}^\mathrm{in}$. For instance, one finds  a threshold mass of approximately $ 0.2 ~\Msun$ setting $c_\mathrm{s}^\mathrm{in}=10$ km/s (in this case, the location of the peak is shifted to the left, so only very slow PBHs are visible, see \cref{fig:accretion2}).
While the density assumed in this figure is that of an average cloud, rare but very dense clumps can be present, such that the threshold becomes blurred: even light PBHs can become visible if the gas is dense enough.  Notice also how the BHL model (dashed lines \footnote{We do not introduce the usual suppression factor of the BHL rate, to allow for a more direct comparison of the models}) fails to capture this threshold effect.\\

In the right panel of \cref{fig:NsourcesThreshold}, we show the expected number of X-ray sources assuming a PBH mass of $M= 1 ~\Msun$. As in \cref{fig:Nsources30}, the prediction is shown as a function of the average PBH speed $\mu_\mathrm{BH}$ and for different choices of the power-law index $\beta$ and the sound speed of the ionized gas $c_\mathrm{s}^\mathrm{in}$.We can notice a striking difference with respect to \cref{fig:Nsources30}. While in that case, i.e. for $M = 30 ~\Msun$, the prediction is quite independent of the parameters $\beta$ and $c_\mathrm{s}^\mathrm{in}$, for small masses their impact becomes very relevant.

\subsubsection{Multimodal mass function}

As we discussed in \cref{sec:intro:pbhs:formation}, variations of the equation of state parameter in the radiation era lead to an enhancement of the probability of collapse for some particular values of the horizon size. The resulting multi-modal PBH mass function presents its most prominent peak at around $1 ~\Msun$ and a second, less pronounced peak at around $30 ~\Msun$. Here we apply our analysis to the study of a population described by this well-motivated mass spectrum.

\begin{figure}[h]
	\centering
	\includegraphics[width=.6\linewidth]{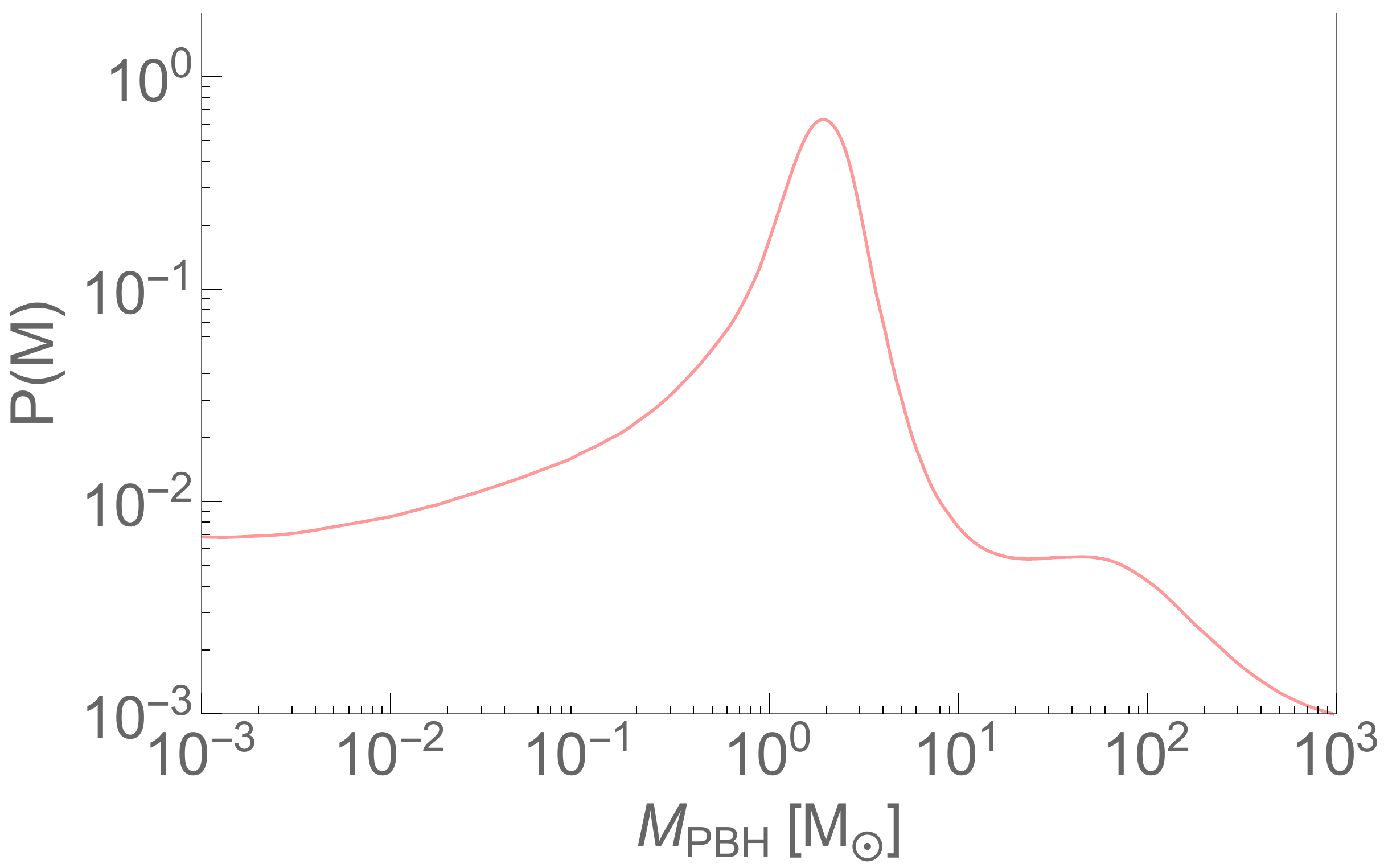}
	\caption{The mass spectrum of PBHs considered in this analysis. From \cite{Carr:2019kxo}. }
	\label{Fig:MassDist}
\end{figure}

We compute the number of X-ray sources above the NuSTAR threshold as
\begin{equation}
	\label{eq:NPBHNustar2}
	N^\mathrm{sources} \, = \, f_\mathrm{clouds} \, \fPBH \, M_\mathrm{CMZ}   \int_{\phi > \phi^*} P(\mathrm{ v}_\mathrm{PBH})P(M)P(n) \,\diff \mathrm{ v}_\mathrm{PBH}\,\diff M\,\diff n
\end{equation}
where the multi-peaked mass distribution $P(M)$ is obtained following~\cite{Carr:2019kxo} and shown in \cref{Fig:MassDist}.

\begin{figure}
	\centering
		\includegraphics[width=.49\linewidth]{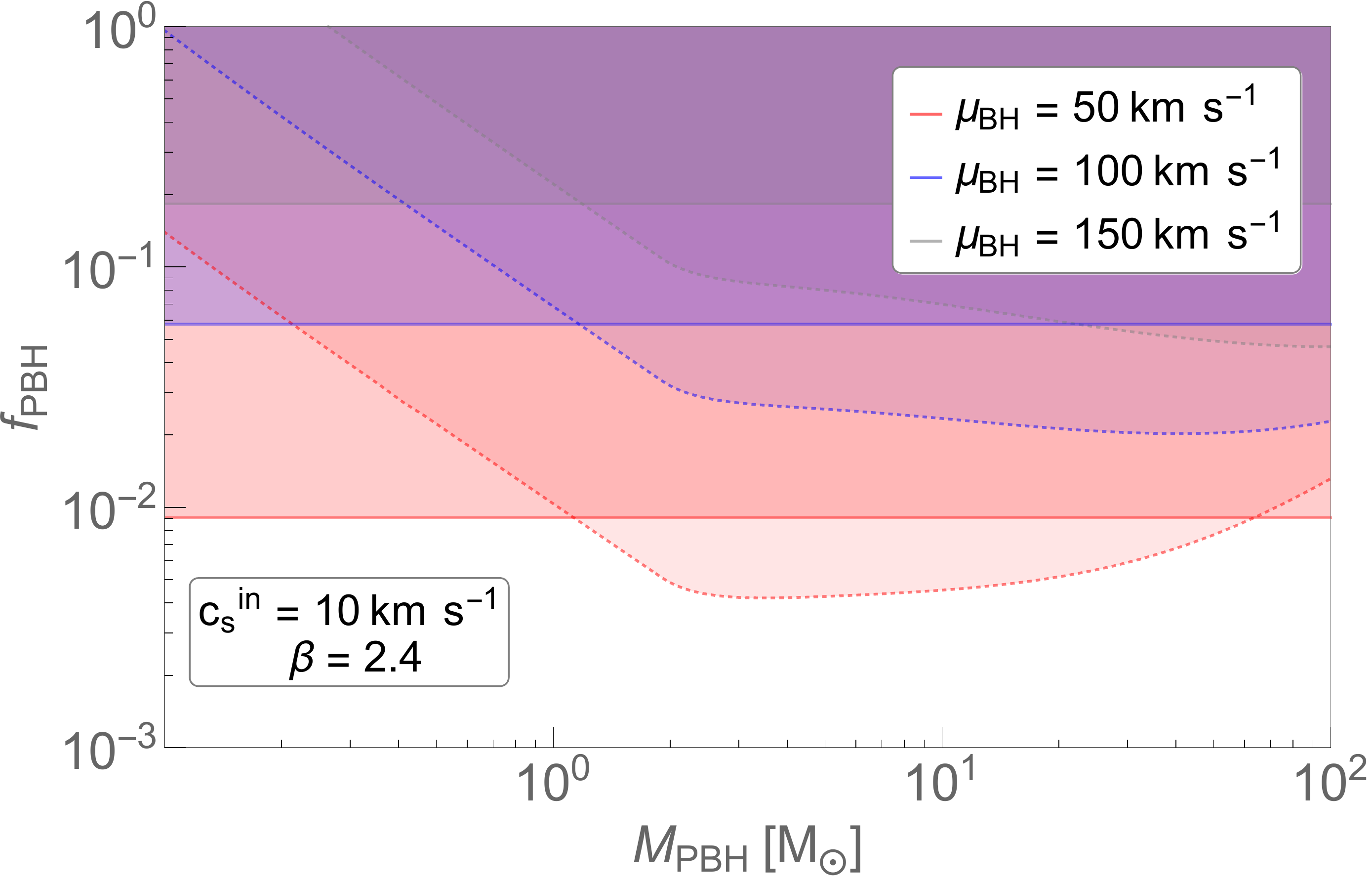}
		\includegraphics[width=.49\linewidth]{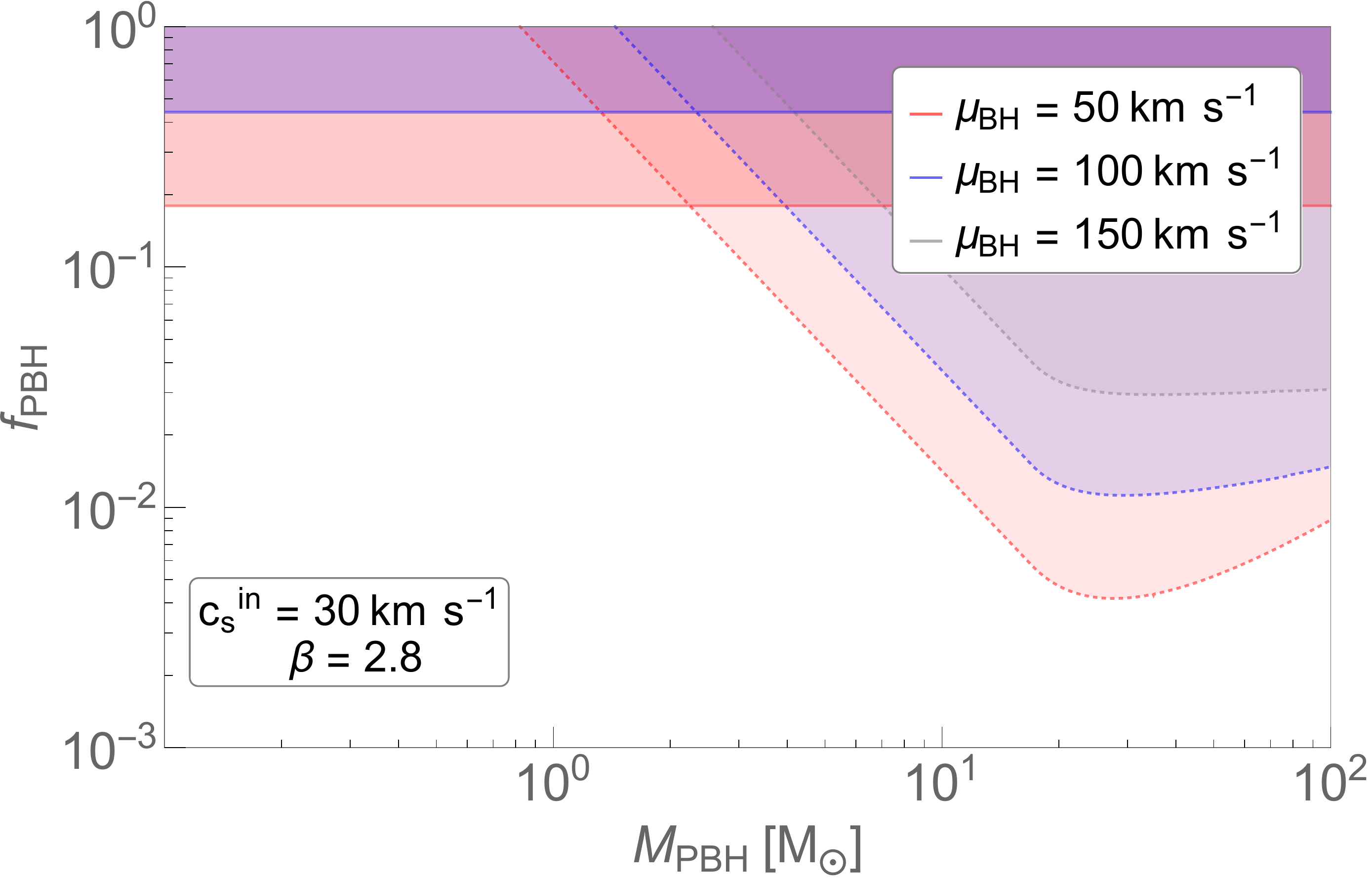}
	\caption{Bounds on $\fPBH$ obtained by comparison with the catalogue of the NuSTAR survey \cite{Hong:2016qjq}. The two panels correspond to an optimistic and a conservative case with respect to the description of the accreted gas. Different colours correspond to different values of the average PBH speed. In solid lines, the results for the multimodal mass distribution scenario. In dashed lines, the monochromatic case.}
	\label{fig:bounds} 
\end{figure}

As for the monochromatic case, we obtain a bound on \fPBH by comparison with the result of the NuSTAR catalogue. The results are shown in \cref{fig:bounds}. Once again, the two panels correspond to different values of the parameters $\beta$ and $c_\mathrm{s}^\mathrm{in}$, while in different colours we show different values for the average PBH speed. The dashed lines depict the bound obtained in the monochromatic case, as a function of the PBH mass.
The horizontal solid lines refer to the multi-modal mass distribution. \\
We can notice that the result is dominated by the peak of the mass distribution, at M $ \sim 1 ~\Msun $. As expected from the discussion above, the bound shows an important variability depending on the choice of parameters. In some regions of the parameter space, and in particular when velocities are assumed to be low, we predict a large number of sources. For other combinations of the parameters, the number of sources is lower than the number of objects in the NuSTAR catalogue and no bound can be placed.\\

In conclusion, we have found that the constraints on the fraction of DM in the form of PBH of masses~$ \sim 10 ~\Msun$ or more are relatively solid. The most relevant uncertainties in this mass range are related to the PBH velocity dispersion.

However, the picture changes when considering lower PBH masses. In this case the bound is much less stable. 
In the case of the multi-modal mass function, the bound is dominated by the peak at~$\sim 1  ~\Msun $ and is therefore affected by the same large variability.
This calls for a more in depth assessment of the uncertainties related with this bound. An analysis within a Bayesian framework is currently underway~\cite{Scarcella:2022pbh}.

On the other hand, we remark that the second peak of the mass function ($M \sim 30 ~\Msun $) offers a very interesting opportunity of detection, in particular considering the significant advances in radio astronomy that will come with SKA (see~\cref{sec:accretion:ABH:SKA}) in the near future.

\subsection{Discussion}

For what concerns the accretion scenario and the emission mechanism, the same discussion presented in \cref{sec:accretion:ABH:Comments} can be applied here.\\ 
Likewise, it is important to underline that the results scale linearly with $N^\mathrm{ clouds}$, the number of PBHs expected to be found in a molecular cloud (see \cref{eq:NPBHclouds}).  
In consideration of this, we also underline the importance of assessing the role of the possible clustering of PBHs. If PBHs are clustered, the number of accreting PBHs would depend on how the locations of clusters correlate with the distribution of the clouds of the CMZ complex.\\

\section{Conclusions}
\label{sec:accretion:Conclusions}

In this Chapter, we have discussed the possibility of detecting isolated black holes in the Milky Way exploiting the radiation emitted during the process of gas accretion from a dense interstellar environment.\\

In \cref{sec:accretion:ABH}, we presented a comprehensive study of the detection prospects for a population of isolated astrophysical black holes.
We adopted the Park-Ricotti (PR13) accretion model that takes into account radiation feedback and is backed up by hydrodynamic numerical simulations, and compared the results obtained within this scenario with the ones obtained by exploiting the well-known Bondi-Hoyle-Littleton formalism.

We first considered the solar vicinity and found that, within the PR13 model, a {few bright X-ray sources associated to isolated accreting black holes } are expected in the region ($R < 250$ pc). A fraction of the sources listed in current catalogues could be associated to such objects.

We then performed an extended parametric study about the number of detectable sources in the vicinity of the Galactic Centre, with particular focus to the promising Central Molecular Zone region. In this region of interest we found that, despite the large uncertainty associated to the free parameters involved in the calculation, the prediction of $\mathcal{O}(10)$  X-ray sources above the detection threshold associated to the NuSTAR catalogue is solid for any reasonable choice of the parameters under scrutiny. In particular, a number of bright X-ray sources comprised between $10$ and $20$ is expected for our reference value of the ionized sound speed $c_\mathrm{s}^\mathrm{ in}= 30$ km/s and an average black hole speed $\mu_\mathrm{BH}= 150$ km/s. 
A multi-wavelength analysis will be needed to clearly identify such a population. In this context, we pointed out promising prospects of detection for future generation experiments in the radio band, with particular reference to the Square Kilometre Array.\\

In \cref{sec:accretion:PBH}, we have turned to PBHs, discussing the possibility of detection in the central region of the Galaxy. We have studied in particular the X-ray emission and compared the predicted fluxes with the NuSTAR survey sensitivity. This comparison can be used to set a constrain on the abundance of PBHs in the Universe;  we have performed a complete parametric study of the uncertainties associated with this bound, assuming a monochromatic mass function. We have found that the bounds present in the literature are relatively solid for masses of the PBHs of order $\mathcal{O}(10 -100) ~\Msun$. At smaller masses, however, the presence of a threshold effect -- such that light PBH are only visible if accreting from very dense clouds and for a small range of velocities -- causes the predictions to vary significantly with the values of the relevant physical quantities (cloud density, average PBH speed, temperature of the ionized gas),  reaching compatibility with $\fPBH = 1$ in some points of parameter space.

Going beyond the the monochromatic assumption, we have then applied our analysis to a PBH population described by a specific multi-modal mass function, motivated by the thermal history of the early Universe. 
This mass distribution presents a peak at $\sim 1 ~\Msun$, which dominates the prediction. As a consequence, we found that it is difficult to place a solid constraint on the fraction of PBH in DM within this scenario. A complete assessment of the uncertainty of the constrain, within a Bayesian framework, is underway and will be included in a future work.

On the other hand, accreting BHs belonging to the subdominant, more massive component (tens of solar masses), are expected to emit significant radiation for a wide range of velocities. An interesting window exists, therefore, for the detection of these objects.
In particular, our result calls for an accurate analysis of existing data to determine whether such objects have already been detected. In this context, the results of Sec. 2.4 would characterize the irreducible "astrophysical floor" associated to the PBH quest. The characterization of both expected signal and background, presented in this chapter, will pave the way to future data analyses and searches in the radio and X-ray channel.


%
\chapter{Gravitational waves: a primordial melody?}
\label{sec:GWPBH}

In 1916, Einstein predicted~\cite{1916SPAW.......688E} that an accelerating mass would generate gravitational waves, vibrations of space-time itself.
Many years later, the first indirect proof of GW existence came following the discovery, in 1974, of the first binary pulsar by R.~Hulse and his Ph.D advisor J.~ Taylor~\cite{Hulse:1974eb}  (both obtained the Nobel prize in 1993). Over the following years, precise measurements of the period of the system~\cite{1982ApJ...253..908T}, determined that it was decaying: the binary was loosing energy at a rate in excellent agreement with the one expected from the emission of gravitational radiation.

Following this discovery, efforts begun to plan and build the first generation of gravitational wave detectors. These were conceived as interferometers to overcome the challenge of measuring extremely small distance variations. The first generation of interferometers already active by the 1990s, did not achieve a detection. The highly improved, second generation detectors of LIGO  Hanford and LIGO Livingston started running in 2015 and observed the first event, GW150914, shortly thereafter~\cite{LIGOScientific:2016aoc}. The signal, lasting around 0.2 seconds, was emitted by the coalescence of a pair of black holes of around $30 \, \Msun$ each, at a distance of around 400 Mpc from Earth. 
For their role in the LIGO collaboration and their decisive contribution to this discovery, the Nobel prize was awarded to R.~Weiss, B.~Barish and K.~Thorne in 2017.

This first GW detection marked the beginning of a new era for the study of our Universe.  In the few years that have followed, the observation of gravitational wave signals has become routine, with around 90 events having been identified at the time of writing. All the signals have been generated by the merging of massive compact objects, the majority of them binary black holes systems. 

The nature of the detected merger events carried some surprises. Firstly, it was thought that binary neutron stars would be the most common sources, above binary black holes, while it turned out to be quite the opposite. The masses of the binary black hole systems were higher than expected and strikingly larger than those observed in the Galactic population  of X-ray binaries. 
As more data was gathered, it was found that the mass distribution of black holes had a large high mass tail, and presented a significant substructure, with a prominent peak around $30 \,  \Msun$.

.  

The idea that primordial black holes could be behind some -- or all -- of these events, was first put forward in the paper ``Did LIGO detect dark matter?''~\cite{Bird:2016dcv}. Many studies have followed up on this idea, on the one hand using the observations to constrain the fraction \fPBH of dark matter in the form of PBH \cite{Ali-Haimoud:2017rtz,Kavanagh:2018ggo,Hall:2020daa}, on the other hand suggesting that the data carried evidence for the presence of a component of primordial origin \cite{Franciolini:2021tla,Franciolini:2021xbq,Hutsi:2020sol}.

Due to the intrinsically simple nature of black holes, single-event based identification of a primordial source is very difficult. The statistical approach also suffers from significant modelling uncertainties.

Amongst the most promising avenues for the identification of a primordial source, is the detection of a sub-solar mass black hole. On the one hand, electron degeneracy pressure prohibits the formation of black holes below the Chandrasekhar limit $ \sim 1.4 ~ \Msun $ \cite{1931ApJ....74...81C}. Searches for such objects have returned no observation~\cite{LIGOScientific:2019kan, LIGOScientific:2021job}. However, the limits that can be placed with current detectors are weak\footnote{One the one hand, the amplitude of these low-mass signals is low. On the other hand, they are present in the detector for along number of cycles, allowing in principle to accumulate SNR. However, this requires very accurate matching with the filter and unfeasibly large template banks. }.  Furthermore, the absence of events can constrain PBH abundance only in this specific mass range.

An unmistakable sign of the primordial nature of a binary black hole, would be its detection at high redshifts, before the formation of the first stars. Conversely, the non-observation of high-redshift events can constrain the PBH abundance across all the range of masses probed. In fact, as we will discuss in \cref{sec:GWPBH:PBHbinaries}, the rate of primordial events is expected to \emph{increase} with redshift. Current facilities can only resolve events up to $z \lesssim 1$, but this type of observations will become a reality with the next generation of detectors.

In this chapter, we start by summarizing some general concepts regarding gravitational waves and their detection, in \cref{sec:GWPBH:gw,sec:GWPBH:detection}. We then turn to discussing the formation of primordial and astrophysical binaries, and their expected merger rates, in \cref{sec:GWPBH:PBHbinaries,sec:GWPBH:PBHbinaries}. \Cref{sec:GWPBH:ET} discusses the prospects for detecting primordial black holes through high-redshift gravitational waves observations, and is based on \cite{Martinelli:2022elq}.

\section{Gravitational wave emission}
\label{sec:GWPBH:gw}

In this section, we summarize some basic properties of gravitational waves, in particular of those emitted by the coalescence of binary systems.

\subsection {Gravitational waves }
\label{sec:GWPBH:gw:intro}

Small perturbations of the flat space-time geometry can by described at linear order by the metric:
\begin{equation}
	\label{eq:linearmetric}
	g_{\mu\nu}\, = \, \eta_{\mu\nu} \, + h_{\mu\nu} , \quad \quad  |h_{\mu\nu} | \ll 1  \; , 
\end{equation}
where $\eta_{\mu\nu} $ is the background metric and $h_{\mu\nu} $ describes the perturbation. In this regime, the Einstein equations can be linearised, substituting \cref{eq:linearmetric} for the metric and discarding higher orders in $h_{\mu\nu} $. This is best done in therms of the trace-reversed metric perturbation
\begin{equation}
	\label{eq:barh}
	\bar{h}_{\mu\nu}\, \equiv \,  h_{\mu\nu} - \frac{1}{2}  \eta_{\mu\nu} h  \; , 
\end{equation}
where $h \equiv  \eta_{\mu\nu}h_{\mu\nu} \equiv h_{\mu}^{\mu} $. Notice that $\bar{h} = h $, and \cref{eq:barh} can be inverted to get
\begin{equation}
	\label{eq:barh2}
	h_{\mu\nu}\, \equiv \,  \bar{h}_{\mu\nu} - \frac{1}{2}  \eta_{\mu\nu} \bar{h}  \; .
\end{equation}
In the harmonic gauge, $\partial ^{\nu} \bar{h}_{\mu \nu}\, =\, 0$, the linear Einstein equations read
\begin{equation}
	\label{eq:EinsteinLin}
	\Box \bar{h}_{\mu\nu}\, = - \dfrac{16 \pi G}{ c^4} \, T_{\mu\nu} \; ,
\end{equation}
where $\Box \, = \, \partial_{\rho}\partial^{\rho}$ is the D'Alambert operator. \Cref{eq:EinsteinLin} describes a curvature perturbation propagating a the speed of light, sourced by the energy-momentum tensor: a gravitational wave.

\subsubsection {Propagation and effect on test masses}
\begin{figure}
	\centering
	\includegraphics[width=.95\linewidth]{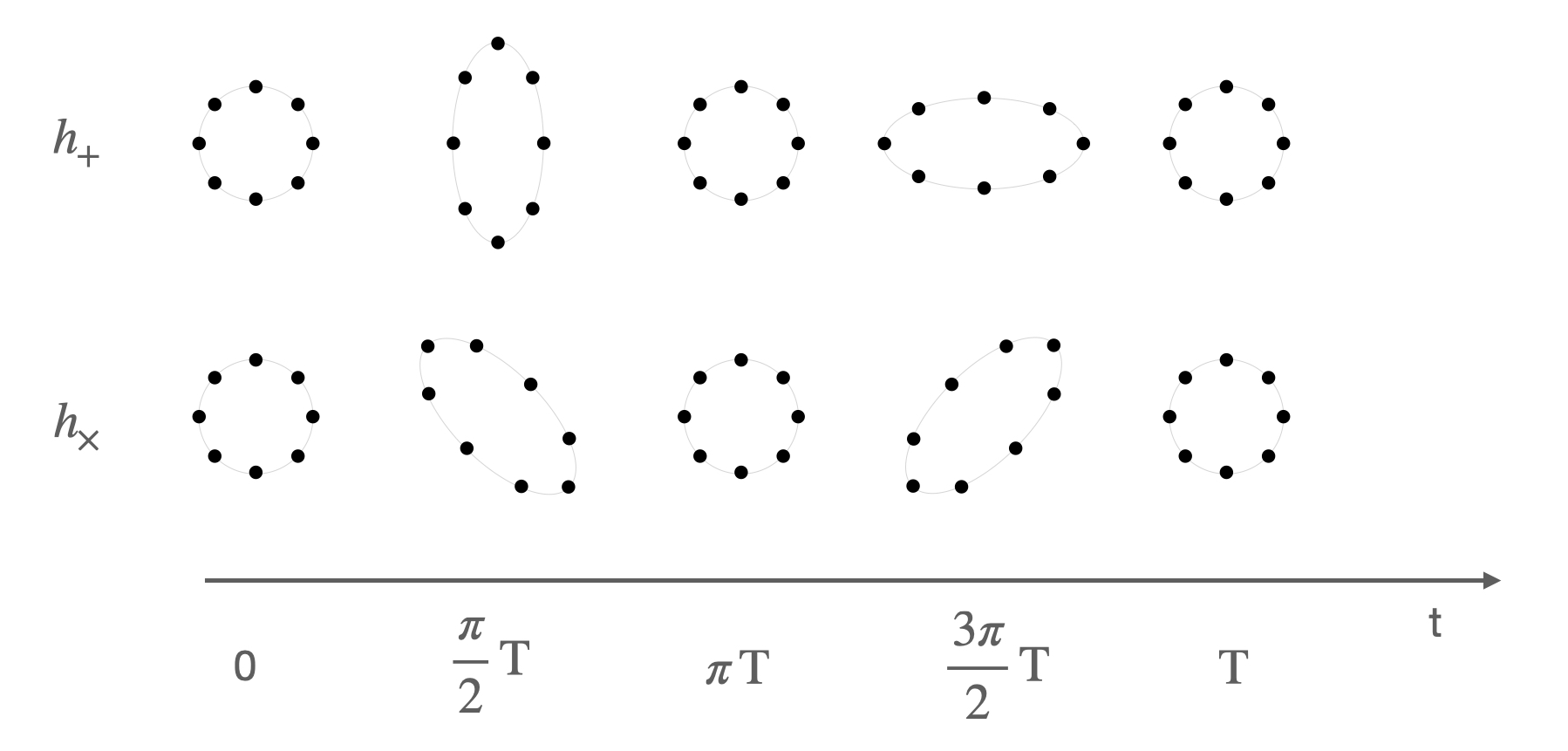}
	\caption{Effect of a gravitational wave on a ring of test masses. The wave is propagating along the axis perpendicular to the figure and T indicates its period. The two rows show the  two independent polarization modes.}
	\label{fig:GWrings}
\end{figure}

We can study the propagation of GWs in vacuum by setting the source term to zero. \Cref{eq:EinsteinLin} becomes
\begin{equation}
	\label{eq:EinsteinLinVac}
	\Box \bar{h}_{\mu\nu}\, = 0 \; .
\end{equation}
This equation has a residual gauge symmetry that can be used to write the metric in the transverse traceless~(TT) gauge, where the properties of the wave become apparent. The gauge conditions that define the TT gauge are
\begin{equation}
	h_{00} =0 \, , \quad h_{0i} =0 \, , \quad   \partial^i h_{ij} =0 \, , \quad  h_i^i =0 \, . \; 
\end{equation}
In the TT gauge, the trace vanishes (hence $\bar{h}_{\mu\nu} \, = \, h_{\mu\nu}$) as do the spatial components parallel to the direction of propagation (on a plane wave propagating in direction $\hat{n}= \vec{k}/k$, the transversality condition $\partial^i h_{ij} =0 $ becomes $n^i h_{ij} =0 $).
There are only two residual non-zero degrees of freedom, which correspond to plane waves oscillating in the plane transverse to the direction of propagation. These are known as the ``plus'' and ``cross''  polarization modes of the gravitational wave ($h_{+}$ and $h_\mathrm{x}$, respectively). Assuming the plane wave propagates along the $z$ axis, we find
\begin{equation}
	h_{ij}^\mathrm{TT} (t, z) \, = \, 
	\begin{pmatrix}
		h_{+} & h_\mathrm{x} & 0 \\
		h_\mathrm{x}	 & - h_{+} & 0 \\
		0 & 0 & 0 
	\end{pmatrix}
	\, \cos \left[\omega(t - z/c) \right] \; .
\end{equation}
%
 \Cref{fig:GWrings} shows how the relative positions of a ring of test masses, disposed on the plane orthogonal to the direction of propagation of the GW, are affected by each of the two polarization modes. The fractional change in lengths produced by the passing of a GW is know as \emph{strain}-- or, equivalently, time travel times -- between test masses.
\subsubsection {Sources of gravitational waves}

\Cref{eq:EinsteinLin} is solved, in the TT gauge, by
\begin{equation}
	h^\mathrm{TT}_{ij} (\vec{x}, t) \, = \;  \frac{4 G}{c^4} \,  \Lambda_{ij, kl} \int  \diff^3 x^\prime\,  \dfrac{T^{kl} \left( t-  \frac{| \vec{x} -\vec{x}^\prime |}{c}, \, \vec{x}^\prime\right)}{| \vec{x} - \vec{x}^\prime |}\; , 
\end{equation}
where $  \Lambda_{ij, kl}(\hat{n}) $ is the projector on the TT gauge\footnote{$  \Lambda_{ij, kl}  \equiv  P_{ik}P_{jl} - \frac{1}{2}P_{1j}{kl} $, with $P_{ij}(\hat{n}) \equiv  \delta_{ij} -n_in_j$} and $\hat{n}$ is the direction of propagation of the wave.

We consider large distances from the source $ r \gg d$, where $d$ is the size of the source and $r$ the distance from the observer. 
At lowest order in $v/c$, where $v$ is the internal linear velocity of the source\footnote{If the frequency of the source motion is $\omega_\mathrm{s} / 2 \pi $, we can write the internal linear velocity of the source as $v = \omega_\mathrm{s} d $}, we have
%
\begin{equation}
	h^\mathrm{TT}_{ij} (\vec{x}, t) \, = \;  \frac{4 G}{c^4} \, \Lambda_{ij, kl}  \, \frac{1}{r}  \int  \diff^3 x^\prime\, T^{kl} \left( t-  \frac{r}{c}, \, \vec{x}^\prime\right)\; , 
\end{equation}
Using the continuity equation $\partial_\mu T^{\mu\nu} =0$, valid in the linear regime, we can express the energy-momentum tensor components as

\begin{equation}
	T^{ij} \, = \, \dfrac{1}{2} \, \partial_0^2 \left( T^{00}\,  x^i x^j  \right) \, + \, \mathrm{spatial \; divergence}.
\end{equation}
Upon integration and writing $T^{00}=\rho c^2 $ (true for $v \ll  c$), we find that the GWs are generated by variations in the quadrupole term of the mass expansion:
\begin{equation}
	h_\mathrm{TT} ^{ij} (\vec{x}, t) \, = \, \frac{2 G}{c^4} \, \Lambda_{ij, kl}  \,  \frac{1}{ r} \, \partial_t^2  \int_{V} \,\diff^3 x^\prime\, \rho\left( \vec{x}^\prime, t - \frac{r}{c} \right)\, x^{\prime i} x^{\prime j}  \; .
\end{equation}
When contracted with the lambda operator $ \Lambda_{ij, kl}$,  only the traceless component of the quadrupole momentum contributes to $h_\mathrm{TT} $. We denote this as $Q^{ij}$
\begin{equation}
	Q^{ij} \, \equiv \,\int_{V} \,\diff^3 x\, \rho( \vec{x}, t ) \, \left(  x^{ i} x^{ j} - \frac{1}{3}|\vec{x}|^2 \delta^{ij} \right) \; ,
\end{equation}

We can then write, in a more compact notation
\begin{equation}
	\label{eq:quadrupoleformula}
	h_\mathrm{TT} ^{ij} (\vec{x}, t) \, = \, \frac{2 G}{c^4}  \,  \frac{1}{ r} \,  \ddot{Q}^{ij} \left( t - \frac{r}{c}\right)\; .
\end{equation}
This is known as the Einstein quadrupole formula. It is valid in the linear regime, for low internal velocities of the source ($v \ll c$) and at large distances from it.
Contrary to electromagnetic waves, which are sourced at first order by variations in the charge dipole, the lowest order emission of GW comes from the quadrupole term. 

This is  a consequence of the conservation of momentum, which forbids isolated mass configurations with a varying dipole. 
A non-zero dipole implies that the centre of mass of the system lies outside the rotation axis; then, a varying dipole requires the centre of mass to be accelerated (as an example of a source with vanishing dipole but non-zero quadrupole, one can think of oblate object spinning around an axis different from its symmetry axis but which runs through its centre of mass).

An order of magnitude expression for the strain $h$ can be obtained writing the quadrupole as  $ Q \sim M d^2$ and estimating the derivative as $\ddot{d^2} \sim v^2$. The typical amplitudes associated to a gravitational wave are extremely small

\begin{equation}
	\label{eq:quadrupoleapprox}
	h\, \approx \, \frac{ G \, M \, v^2}{ r \, c^4}  \; \simeq \, 5 \times 10^{-21 } \left( \dfrac{M}{100 \, M_\odot} \right)  \left( \dfrac{r}{\mathrm{Gpc }}\right)^{-1}  \left( \dfrac{v}{c}  \right)^2
	\end{equation}
Such are the minuscule strains nowadays routinely measured by gravitational wave observatories. \\

\begin{figure}[h!]
	\centering
	\includegraphics[width=\linewidth]{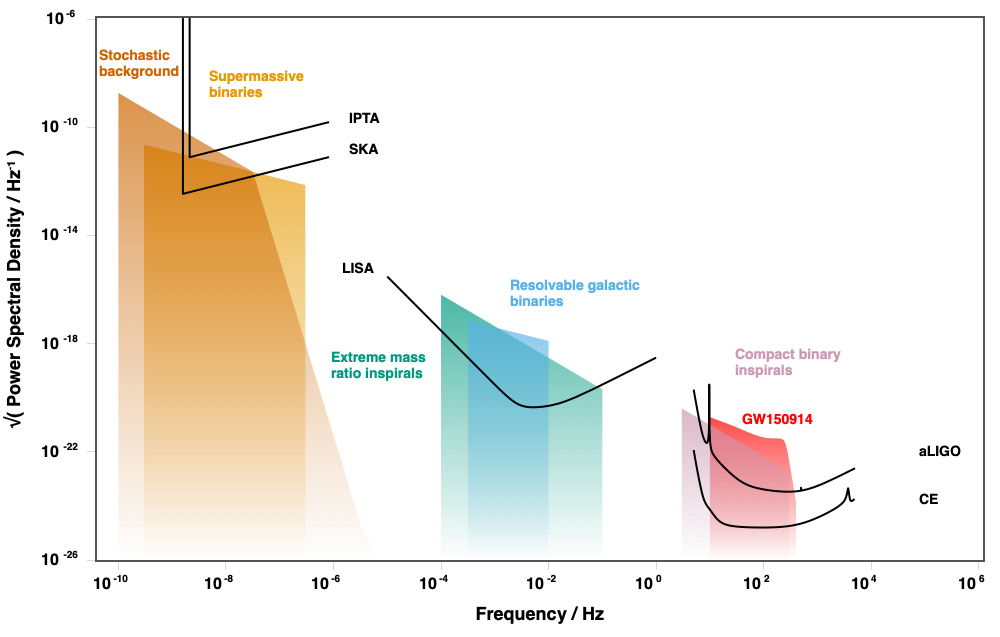}
	\caption{Noise power spectral density for present and future observational facilities. The typical frequencies and amplitudes of different classes of events are shown for comparison. Credit: gwplotter.com}
	\label{fig:allsensitivities}
\end{figure}

Gravitational waves can be generated by a wide range of astrophysical sources, across very different frequencies \cite{Lasky:2015lej}.  The main frequency at which the GW signal is emitted is of the same order as $\omega_\mathrm{s} \sim v / d $, the frequency of the internal motion of the source. In general, we expect larger systems to have longer periods and thus emit at lower frequency (see \cref{fig:allsensitivities}). Notable sources of GW include:

 \begin{itemize}
	 	\item Inspiralling or merging binaries. These emit large amounts of energy and can be modelled extremely well. They are the only sources that have been experimentally identified so far; we will discuss them in detail in the following section
	 	\item  Violent events, such as supernova explosions -- these are known as ``burst'' sources and cannot be modelled precisely. This class of events includes hypothetical ones such as bubble collisions in first order phase transition in the early Universe \cite{PhysRevD.30.272, Kamionkowski:1993fg}, or cosmic strings \cite{LIGOScientific:2021nrg}
	 	\item Oblate compact objects , 
	 	such as non perfectly symmetric neutron stars. If the angular momentum of the object is constant, the gravitational wave will also be emitted at a constant frequency, giving rise to a continuous signal.
	 	\item Scattering processes, or hyperbolic encounters between compact objects. In general, we expect the emitted radiation to be very faint (see however~\cite{Morras:2021atg,Garcia-Bellido:2021jlq}). This radiation plays a role in the dynamical formation of binaries, which we discuss in \cref{sec:GWPBH:PBHbinaries:late} 
	 \end{itemize}
 
\noindent
The superposition of many sources too faint to be resolved individually, can be detected as a stochastic gravitational wave background. A stochastic signal can also be generated by an intrinsically random process; an example are the tensor perturbations in the early Universe~\cite{Caprini:2018mtu, Maggiore:1999vm}.

In the following sections, we focus on GW emission from compact binary mergers. These are the only type of events that have been detected so far (evidence for a stochastic signal has been reported by NANOGrav~\cite{NANOGrav:2020bcs}, but its gravitational origin is questioned), and are the focus of the work presented in this Chapter.

\subsection{Compact binary systems}
\label{sec:GWPBH:gw:compact_bin}

Consider a bound binary system of compact objects, at large enough separation such that the rate of emission of gravitational radiation is initially small. The system loses energy slowly and it slowly becomes more compact. This is known as the \emph{inspiral phase}. As the system shrinks, its orbital period decreases and the acceleration becomes larger, which in turn leads to a higher rate of GW emission. This is a runaway process which eventually, over cosmological times, can lead to the merger of the two objects.

In the \emph{merger phase}, when the system runs out of stable orbits, the frequency of the inspiral increases dramatically, velocities approach the speed of light and the objects plunge into each other. 

\begin{figure}[h!]
	\centering
	\includegraphics[width=.7\linewidth]{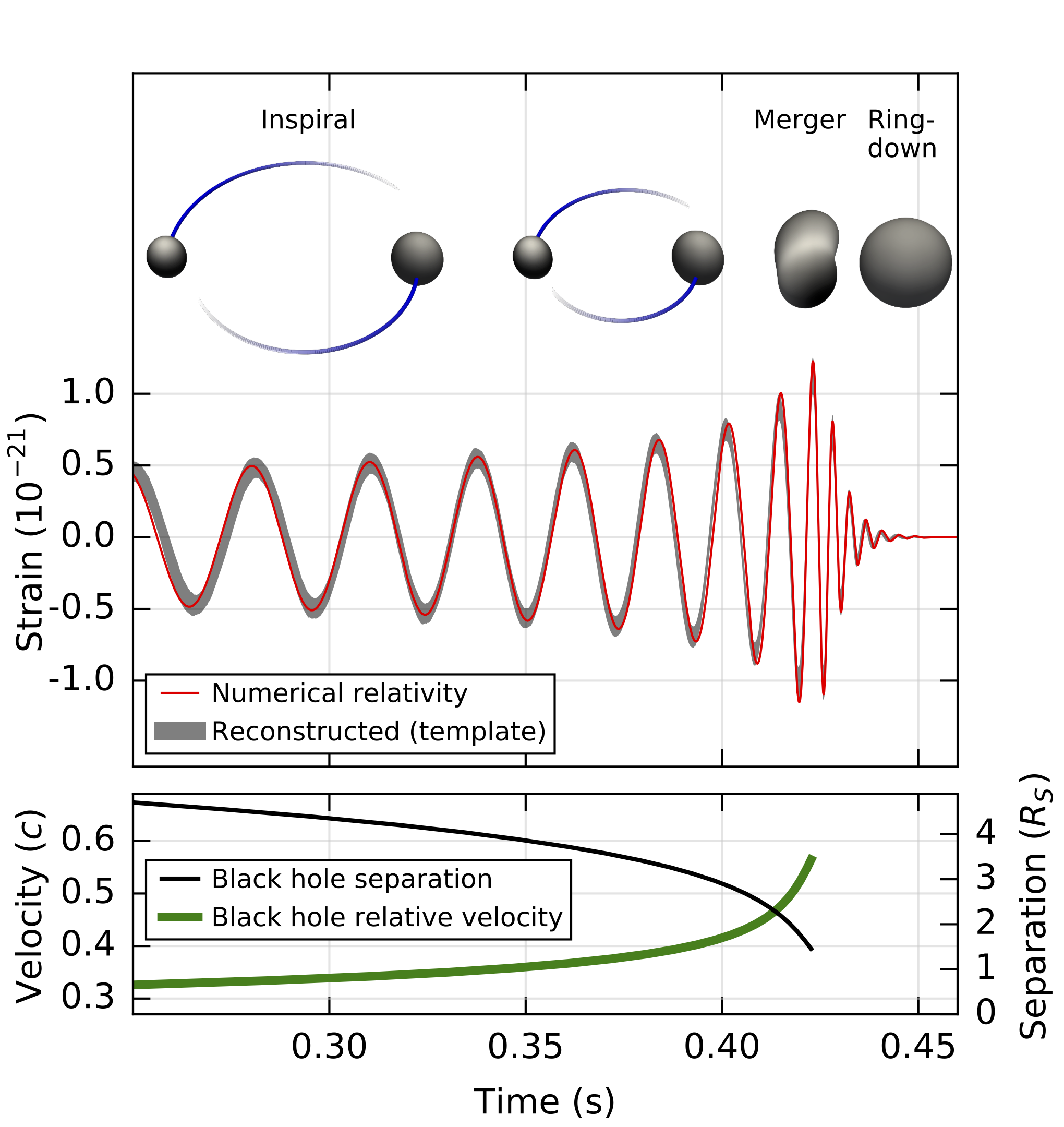}
	\caption{Top:  Reconstructed gravitational-wave strain amplitude of the GW150914 event as observed by LIGO Hanford. The inset images show numerical relativity models of the black hole horizons as the black holes coalesce. Bottom: The Keplerian effective black hole separation in units of Schwarzschild radii $R_\mathrm{s} $ and the effective relative velocity i.e. the post-Newtonian parameter $v/c$. From~\cite{LIGOScientific:2016aoc}.
	}
	\label{fig:GW150914}
\end{figure}

Finally, in what is known as the \emph{ringdown phase}, the merged object settles into its final equilibrium in the form of a Kerr black hole.
\cref{fig:GW150914} shows the gravitational waveform corresponding the merger and ringdown phases for the first detected GW event, GW150914.

\subsubsection{Variation of orbital parameters} We first discuss how the emission of GWs affects the orbit of a binary system during the inspiral phase.
As long as the distances between the two objects are large, we can work in the linear theory.
In this regime, the orbital motion of a binary system can be described by Newtonian physics. This is a classical two body problem, and it can be expressed in terms of an elliptical orbit.
An ellipse is characterised its eccentricity $0 < e < 1$ and semi-major axis $a> 0$. The distance of closest approach is the peri-apsis radius\footnote{other nice names to be used when the central object is a BH: peri-melasma (from greek melos, black), peri-nigricon, peri-bothron (from greek bothros, hole) \cite{enwiki:1084886995}. }, given by $r_\mathrm{p} = a (1-e)$. The semi-minor axis $b$ is related to the above parameters through the relation $e =\sqrt{1-(b/a)^2}$.\\
The total energy of a binary depends on its semi-major axis and masses:
\begin{equation}
	\label{eq:bin_energy}
	E_\mathrm{tot} \, = \,  -\dfrac{ \mu \, G  \, M }{2 a}  \; ,
\end{equation}
where $\mu = m_1 m_2/(m_1 + m_2)$ is the reduced mass, $M = m_1 + m_2$ is the total mass and $m_1, m_2$ are the masses of the two bodies. The angular momentum is 
\begin{equation}
	L  \, = \,  \mu \, \sqrt{a \left(1- e^2\right) \, G M  }\; ;
\end{equation}
it is sometimes convenient to introduce the dimensionless angular momentum $j$, defined as
\begin{equation}
	j \equiv \dfrac{L/ \mu}{\sqrt{ G M a}} \, = \, \sqrt{1- e^2} \,  = \, \dfrac{b}{a} \; ,
\end{equation}
where $L$ is the angular momentum, $M$ is the total mass. 
It is also useful to define the symmetric mass ratio $\eta$, which varies between 0 -- when the mass ratio tends to infinity -- and $1/4$ -- when the masses are equal,
\begin{equation}
	\eta \equiv \dfrac{\mu}{M}  \,   \in  \,  ( 0 , 1/4 ]\; .
\end{equation}
\begin{figure}[t]
	\centering
	\includegraphics[width=.5\linewidth]{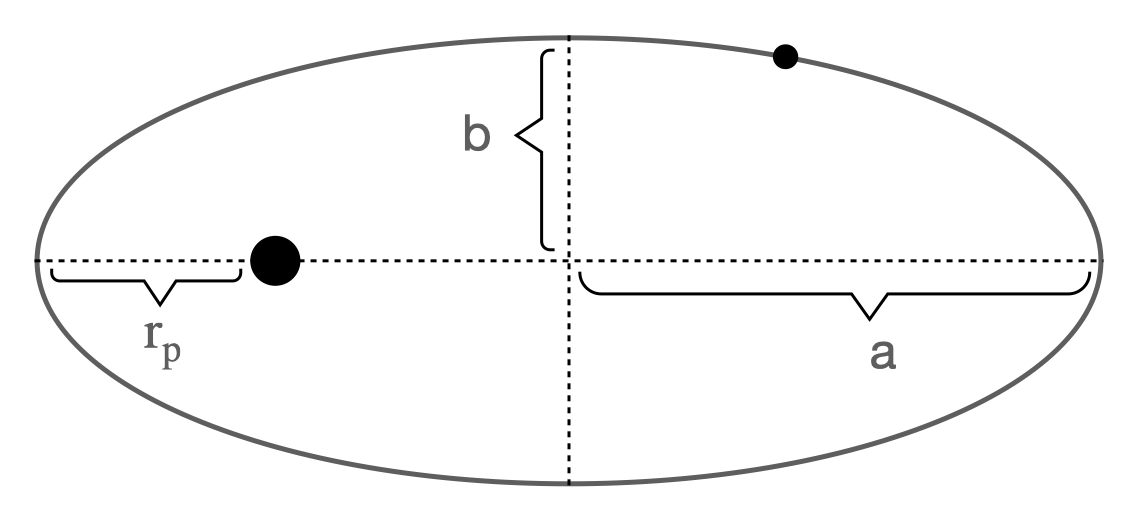}
	\caption{An elliptic orbit. The semi-major and semi-minor axes are indicated with $a$ and $b$ respectively, while $r_\mathrm{p}$ is the peri-apsis radius.  }
	\label{fig:ellipse}
\end{figure}
Gravitational waves carry away both energy and angular momentum from the binary. Averaged over one orbital period, we have: 
\begin{align}
	\label{eq:dEdt}
	&\frac{\diff E}{\diff t} \, = \,  - \frac{32}{5}  \, \dfrac{G^4 \, \eta^2 M^5}{ c^5 \,  a^5}  \, \frac{1}{j^7}  \left( 1 +  \frac{73}{24} e^2+ \frac{37}{96} e^2\right) \; , \\[5pt]
	\label{eq:dLdt}
	&\frac{\diff L}{\diff t} \, = \,  - \frac{32}{5}  \, \dfrac{G^{7/2} \, \eta^2 M^{9/2}}{ c^5 \, a^{7/2}}  \, \frac{e}{j^4}  \left( 1 +  \frac{7}{8} e^2 \right) \; . 
\end{align}
The loss of energy and angular momentum is reflected in the variation of the binary's orbital parameters:
\begin{align}
	\label{eq:dadt}
	&\frac{\diff a}{\diff t} \, = \,  - \frac{64}{5}  \, \frac{G^3 \, \eta M^3}{ c^5 \, a^3}  \, \frac{1}{j^7}  \left( 1 +  \frac{73}{24} e^2+ \frac{37}{96} e^2\right) \; , \\[5pt]
	\label{eq:dedt}
	&\frac{\diff e}{\diff t} \, = \,  - \frac{304}{15}  \, \frac{G^3\,  \eta M^3}{ c^5\,   a^4}  \, \frac{e}{j^5}  \left( 1 +  \frac{121}{304} e^2 \right) \; . 
\end{align}
The emission of GW results in a hardening of the binary, \textit{i.e.}a reduction of its semi-major axis. This is expected, since the energy of an elliptic orbit is $E \propto -1/a $, ( \cref{eq:bin_energy}).  \\ 
For a circular orbit, we have  $\diff e/ \diff t =0$: that is, a circular orbit remains circular. A generic elliptic orbit, on the other hand, is always circularised (this implies that the dimensionless angular momentum $j$ becomes larger with time, despite the fact that the physical angular momentum $L$ decreases).
We can also see that, at fixed $a$, in the limit $e \rightarrow 1$ \cref{eq:dedt} diverges. This corresponds to $r_\mathrm{p} \rightarrow 0$, where indeed the Newtonian acceleration diverges. However, in this limit the linear gravity approximation breaks down and these expressions are no longer valid.

By combining \cref{eq:dadt} and \cref{eq:dedt} we can obtain an expression for the time of merger $\tau_\mathrm{m}$, given the initial values of the orbital parameters:
\begin{equation}
	\label{eq:tmerger}
	\tau_\mathrm{m} (j, a)\, = \, \dfrac{5}{256}  \, \dfrac{c^5}{G^3}  \, \dfrac{  a^4 \, j^7}{M^3 \eta} \, G(e)  \; ,
\end{equation}
where $G(e)$ is a function which is everywhere close to 1, $G(e) \approx 1$ for $e \approx 0$ and $G(e) \rightarrow 768/425 \approx  1.8 $ for $e \rightarrow 1$. This gives
\begin{equation}
	\label{eq:tmergerastimate}
	\tau_\mathrm{m} (j, a)\, \approx \, 10^3 \, \mathrm{Gyr}\;  \left(\dfrac{ M }{ 100 \,  M_\odot} \right)^{-3} \, \left(\dfrac{ \eta }{ 1 /4} \right) \,  \left(\dfrac{  a }{ 1 \, \mathrm{AU}} \right)^4 \,  j^7 \;,
\end{equation}
where the reference is a circular, symmetric binary system. We can see that coalescence times easily exceed the age of the Universe by orders of magnitude. However, the time to merger decreases rapidly for highly eccentric binaries.  For example, if instead of considering a circular binary, we set $j\approx 0.4$ in \cref{eq:tmergerastimate} -- which corresponds to eccentricity $e \approx 0.9$ -- we find $ \tau_\mathrm{m} \, \approx \, 2 $ Gyr.

\subsubsection{Circular binaries}

As we have seen, the emission of GW results in a reduction of the ellipticity of the binary orbit, \cref{eq:dadt}. 

Current detectors only observe these very final stages of the coalescence process, where the binary can safely be considered circular. In this paragraph we describe the shape of the waveform emitted by a circular binary system\footnote{ Future detectors, such as LISA, will be able to observe the binary system starting at larger separation, when the time to coalescence is large. In that case, taking into account the orbital ellipticity becomes important.}.

We should mention that, as the two objects approach each other, the linear approximation breaks down. On the one hand, as the orbital velocity increases, the quadrupole formula is no longer a good approximation and higher order multipoles must be considered. On the other hand, the effects of curvature also become non-negligible. In fact, for a binary system, and more generally for a self gravitating system, the intensity of the gravitational field is related to the velocity of the internal motion. From the virial theorem, we have $v^2 \approx GM/d$. Then we can compare the orbital distance $d$ with the Schwarzchild radius $r_\mathrm{s} = 2GM/c^2$, finding $\,   r_\mathrm{s}  \approx \left(  v/c \right)^2  \, d$. 
When $ v/c $ is no longer negligible, the size of the orbit approaches the Schwarzchild radius and curvature effects become important.
In this situation, one can no longer compute the orbit assuming Newtonian dynamics; the most common approach is to apply a post-Newtonian (PN) expansion of Einstein's equations, that is an expansion in powers of $ v/c $, while consistently computing higher order multipole terms of the GW source. 
State-of-the-art computations allow to obtain the waveforms up to post-Newtonian order 4.5, that is $\propto (v/c)^9$. Predictions for the last phases of the merger rely on the comparison with numerical models, while the GW emission in the ringdown phase is calculated with BH perturbation theory.

Despite its limitations, the linear theory is nevertheless useful to discuss the qualitative features of the waveform, and we rely on it in the following.\\
As long as  $\dot{\omega_\mathrm{s}} \ll \omega_\mathrm{s}^2$, the inspiral motion can be approximated with a  circular one with a slowly varying radius. This is a good approximation up to the final phase of the inspiral. In this regime the (quadrupole) GW emission is peaked at a frequency double of that of the source, $f =   \omega_\mathrm{s}/ \pi$. \\
In the linear theory and for quasi-circular motion, the variation of frequency is related to the frequency itself by
\begin{equation} 
	\label{eq:dotfreq}
	\dot{f}\, =\, \dfrac{96}{5} \, \pi^{8/3}\, \left(\dfrac{G M_\mathrm{c}}{ c^3}\right)^{5/2}\, f^{11/3}
\end{equation}
where 
we have introduced the \emph{chirp mass}:
\begin{equation}
	M_\mathrm{c} =\dfrac{ (m_1 m_2)^{3/5}}{ (m_1+m_2)^{1/5}} \; .
\end{equation}
Precisely due to its role in determining the variation of the GW frequency, the chirp mass is the best measured parameter among those that characterise the binary. The accuracy of its measurement increases with the number of cycles observed by the detector.\\
\Cref{eq:dotfreq} can be solved  for $f $ as a function of the  time to coalescence $ \tau = t_\mathrm{coal} -t $, giving
\begin{equation} 
	f (\tau) \, = \, \dfrac{1}{\pi} \, \left( \dfrac{5}{256} \dfrac{1}{\tau }\right)^{3/8}\, \left(\dfrac{G M_\mathrm{c}}{ c^3}\right)^{-5/8} \; .
\end{equation}
From here we can read off typical frequencies for different masses and times to merger. Considering that the lowest frequencies probed by current detectors are $\mathcal{O}$~(10 Hz), a typical system characterised by $M_\mathrm{c} \approx  50 \, \Msun$ enters the detection band around two seconds before merger
%
%
\begin{equation} 
	\tau( f ) \, =  2 \, \mathrm{s} \,  \left( \dfrac{f}{10 \,  \mathrm{Hz}} \right)^{8/3} \, \left( \dfrac{ M_\mathrm{c} }{ 50 \, M_\odot} \right)^{-5/3}\,  \; .
\end{equation}
Systems characterized by smaller masses enter the detection band at larger time to coalescence and hence can be observed for more cycles (provided that the strain is large enough to allow a detection).  For instance, for $M_\mathrm{c} \approx 2 \, \Msun$, we find that $f \approx 10$ Hz at $\sim$ 7 minutes to merger. 
We can also estimate the corresponding orbital separation, given that $\omega_\mathrm{s}^2= GM/R^3$
\begin{equation} 
	R \, \approx  500 \,  \mathrm{km} \, \left( \dfrac{ M}{ M_\odot} \right)^{1/3}\,  \left( \dfrac{ f }{10 \, \mathrm{Hz}} \right)^{-2/3} \; .
\end{equation}
Comparing this with the solar radius ($\sim 10^6 $ km), we realize that only compact objects can orbit each other at separations small enough that their gravitational radiation is detected by current observatories (future surveys such as LISA will be sensitive to lower frequencies and be able to detect radiation from galactic binary white dwarfs).

The variation of the time-domain amplitude with frequency\footnote{here $h_\mathrm{c} (f)$ does not indicate a Fourier transform. Rather, at each time is associate only one frequency (the main one, given by $\omega_\mathrm{GW}= 2 \omega_\mathrm{s}$)} is given by the relations
\begin{align} 
	\label{eq:modes}
	& h_+ \,  = \, h_\mathrm{c} (f[\tau]) \, \dfrac{1 + \cos^2 {\iota}}{2} \, \cos{\psi}\; ,\\
	& h_\times \,  = \, h_\mathrm{c} (f[\tau]) \, \cos{\iota} \, \sin{\psi} \; ,
\end{align}
where $\iota$ is the inclination (the angle of the orbital plane of the binary with the line of sight) and $\psi$ is the GW phase;  the ``characteristic strain'' $h_\mathrm{c}$ depends on the distance, on the chirp mass  and on the frequency as:
\begin{equation}
	\label{eq:strainfreq}
	h_\mathrm{c} (f)  \, = \,   \dfrac{4}{r}\ \left( \dfrac{G M_\mathrm{c}}{c^2}\right)^{5/3}\, \left(\dfrac{\pi f}{c}\right)^{2/3} \; .
\end{equation}
We notice that the strain increases with the chirp mass as well as with the frequency: the amplitude gets larger and simultaneously raises in pitch, what is known as a \emph{chirp}.

For events taking place at large cosmological distances, the expansion of the Universe must be taken into account. It has the following effects: 
\begin{itemize}
	\item the observed frequencies are redshifted with respect to those emitted in the source frame: $f_\mathrm{obs} = f_\mathrm{s}/(1+z) $
	\item the dependence of the strain on the coordinate distance $r$ is replaced by a dependence on the luminosity distance
	\begin{equation} 
		D_\mathrm{L}(z) \, =  \,   c (1+z) \, \int_0^z \,  \dfrac{\diff z^\prime}{H(z^\prime)} \; ;
	\end{equation}
	\item the redshifting of the frequency results in a redefinition of the chirp mass $M_\mathrm{c}   \rightarrow  \mathcal{M}_c \equiv (1+z)M_\mathrm{c}  $. 
\end{itemize}
In summary, the previous equations take the form:
\begin{align} 
	\label{eq:dotfreq2}
	&\dot{f} \, =\, \dfrac{96}{5} \, \pi^{8/3}\, \left(\dfrac{G \mathcal{M}_c}{ c^3}\right)^{5/2}\, f^{11/3} \; , \\[5pt]
	\label{eq:taufreq2}
	&	f (\tau) \, = \, \dfrac{1}{\pi} \, \left( \dfrac{5}{256} \dfrac{1}{\tau }\right)^{3/8}\, \left(\dfrac{G \mathcal{M}_c}{ c^3}\right)^{-5/8} \; , \\[5pt]
	\label{eq:strainfreq2}
	&h_\mathrm{c}  (f) \, = \,   \dfrac{4}{D_L}\ \left( \dfrac{G \mathcal{M}_c}{c^2}\right)^{5/3}\, \left(\dfrac{\pi f}{c}\right)^{2/3} \; .
\end{align}
Notice there is a degeneracy between redshift and mass, as the mass only appears through the quantity $\mathcal{M}_c$. Breaking this degeneracy requires knowledge of the source redshift. In the absence of a luminous counterpart, this cannot be measured directly.
However, if the amplitudes of the two polarization modes can be measured separately, their ratio allows to infer the inclination and the characteristic strain. Then, knowing  $\mathcal{M}_c$ and $h_\mathrm{c} (f)$, the luminosity distance can be obtained from \cref{eq:strainfreq2}. Finally, the redshift is obtained assuming a cosmological model.

In practice, as we will discuss in the next section, detectors measure a combination of the two polarization modes and resolving them separately is not trivial.

On the other hand, if an electromagnetic counterpart is present, one can locate the galaxy where the merger took place and measure directly its redshift.
In this case, redshift and luminosity distance can be measured independently, providing what is known as a \textit{standard siren} (by analogy with standard candles). These are precious objects, as they allow to test the expansion history of the Universe. Only one such event has been detected so far~\cite{BNSmerger}, but more are expected to be observed by next-generation experiments in the future decades~\cite{Belgacem:2019tbw}.

\vspace{40pt}

 \section{Detection of gravitational waves}
 \label{sec:GWPBH:detection}

 \subsection{Identification of signals}

The output of a detector is a scalar, time-dependent quantity. It is the combination of the detector noise $n(t)$ and (possibly) a gravitational wave signal $h(t)$ 
\begin{equation}
	s(t) = h(t) + n(t)
\end{equation}
The signal $h(t)$ can be expressed as a linear combination of the two GW polarization modes, with coefficients that depend on the direction of propagation of the wave
 \begin{equation}
 	\label{eq:hpatterns}
	 	h(t) = F_+ (\theta, \phi) \, h_+(t) + F_\times (\theta, \phi) \, h_\times (t) \; .
\end{equation}
The coefficients $F_{+,\times}$ are known as the \emph{antenna patterns}, and describe the angular dependence of the sensitivity of the detector. The angles $(\theta,\phi)$ define the position of the GW source on the sky. 
The antenna patterns functions also depend on the angle $\psi$, on the plane orthogonal to the direction of propagation, which defines the two polarization modes. This angle defines an arbitrary coordinate system, and the detector output $h(t)$  is independent of its choice.\\
 Notice that four independent quantities combine to produce the signal: the amplitudes of the two polarization modes and the two angles $\theta, \phi$.  One detector is then not able to differentiate the two modes or infer the sky localization.
Two detectors can measure two independent strains; if they are at a non-negligible distance from each other, the time delay provides an additional measurement (this is the case, for instance, for the two LIGO interferometers, located in the Livingston and Hanford observatories). This is still not enough information to break the degeneracy and measure the sky position and the two polarization modes.  Recalling the discussion in \cref{sec:GWPBH:gw:compact_bin}, resolving the two modes independently is also necessary to measure the luminosity distance. For this, three detectors are needed: if located at different sites, they provide five independent measurements: three signals and two time delays~\cite{Maggiore2007}. 

The noise $n(t)$ is usually expressed in terms of the noise spectral density, or noise power spectrum. Assuming that the noise can be described by a Gaussian stochastic process, the noise power spectral density is defined by 
\begin{align}
	S_ n(f) &= 2 \int^\infty_{-\infty} \mathrm{d}\tau \; R(\tau) \, \mathrm{e}^{\mathrm{i} 2 \pi f \tau},
	\intertext{where}
	R(\tau) &= \langle n(t + \tau)\; n(t) \rangle  
\end{align}
is the autocorrelation of the noise (angle brackets indicate the ensemble average\footnote{ In practice, the ensemble average is computed comparing the noise signal over many equivalent time periods }). Inverting this expression and setting $\tau =0$ we find
 \begin{equation}
	 \langle n^2 (t) \rangle = \dfrac{1}{2} \int_{-\infty}^\infty \mathrm{d} f  \;S_n(f) = \int_0 ^\infty \mathrm{d} f  \;S_n(f) 
\end{equation}
In the absence of correlation, $R(\tau)  = \delta (\tau) $ and $S_n(f) $ is independent of the frequency (white noise).

It is possible to exploit the knowledge of the expected waveform of the GW signal to identify signals which are lower in amplitude than the detector noise. This is done through the \emph{matched-filtering} procedure, in which the measured strain $s(t) $ is multiplied with a filter $K(t)$ - corresponding to a template waveform -  and integrated over time.
 \begin{equation}
	\hat{s} \equiv \int_{- \infty}^{\infty} \diff t \,   s(t) K(t) = \int_{- \infty}^{\infty} \diff t \,   h(t) K(t)  + \int_{- \infty}^{\infty} \diff t \,   n(t) K(t) \; .
\end{equation}
 If a signal matching the filter is present, the first integrand is definite positive and its integral grows with time; the second integrand, containing the noise component, oscillates and its time average tends zero. It is essential, particularly in the case of events consisting of many cycles, that the filter matches the signal with extreme accuracy; the presence of a mismatch or \emph{dephasing} between the two, can easily result in a destructive interference and in the loss of the signal. 
 
 The signal-to-noise ratio (SNR) is defined as $\rho \equiv S/N$, where $S$ is the value of $\hat{s}$ when a signal is present
  \begin{equation}
 	S = \int_{- \infty}^{\infty} \diff t  \,   h(t) K(t) =  \int_{- \infty}^{\infty} \diff f \,  \tilde{h}(f) \tilde{K}^*(f) \; ,
 \end{equation}
while $N$ is given by $N^2 =  \langle \hat{s}^2 \rangle$, and can be expressed in terms of the spectral noise density  as
  \begin{equation}
	N^2 =  \int_{- \infty}^{\infty} \diff f \,  \frac{1}{2} S_n(f) |\tilde{K}^*(f) |^2 \; .
\end{equation}
Thesignal-to-noise ratio of an event is then given by 
\begin{equation}
	\label{eq:snrgeneral}
	\rho = \frac{\int^\infty_{-\infty} \mathrm{d}f \;  \tilde{h}(f)\, K^*(f)}{\sqrt{\int^\infty_{-\infty} \mathrm{d}f \; \frac12 \, S_n(f) \, |\tilde{K}^*(f)|^2}},
\end{equation}
where $K(f)$ is the filter which maximises the signal-to-noise ratio. The optimal filter is given by the Wiener filter
\begin{equation}
	\tilde{K} (f) \propto \dfrac{ \tilde{h}(f)}{S_\mathrm{ n}(f)} \; .
\end{equation}
This is simply proportional to the signal in the case of white noise, otherwise it is weighted with the noise spectral density. Substituting this expression in \cref{eq:snrgeneral} gives
\begin{equation}
	\rho^2 = 4 \int^\infty_{0} \mathrm{d}f \; \frac{|\tilde{h}(f)|^2 }{S_n(f)}\, = 4 \int^\infty_{-\infty} \mathrm{d}\log{f}\; \frac{f |\tilde{h}(f)|^2  }{S_n(f)}\,
\end{equation}

In GW searches, the signal waveform is not known beforehand, and the matched-filtering procedure is applied using large templates banks to search for a match in the detector output.

 The matched-filtering procedure is efficient, as we have seen, in eliminating Gaussian noise: this can be done by setting a threshold on the signal-to-noise ratio ($\rho > 8$ is often used as a reference value). However, non-Gaussian noise can instead present high SNR tails; a more effective way to remove these is to compare the output of different detectors looking for coincidences.

\subsection{Gravitational wave detectors}

GW detectors can be divided in three main categories: ground-based interferometers, space based detectors and pulsar timing arrays. We review here some of the main characteristics of present and future observatories (see \cite{2021NatRP...3..344B} for an updated  review).

\begin{figure}[h!]
	\centering
	\includegraphics[width=.8\linewidth]{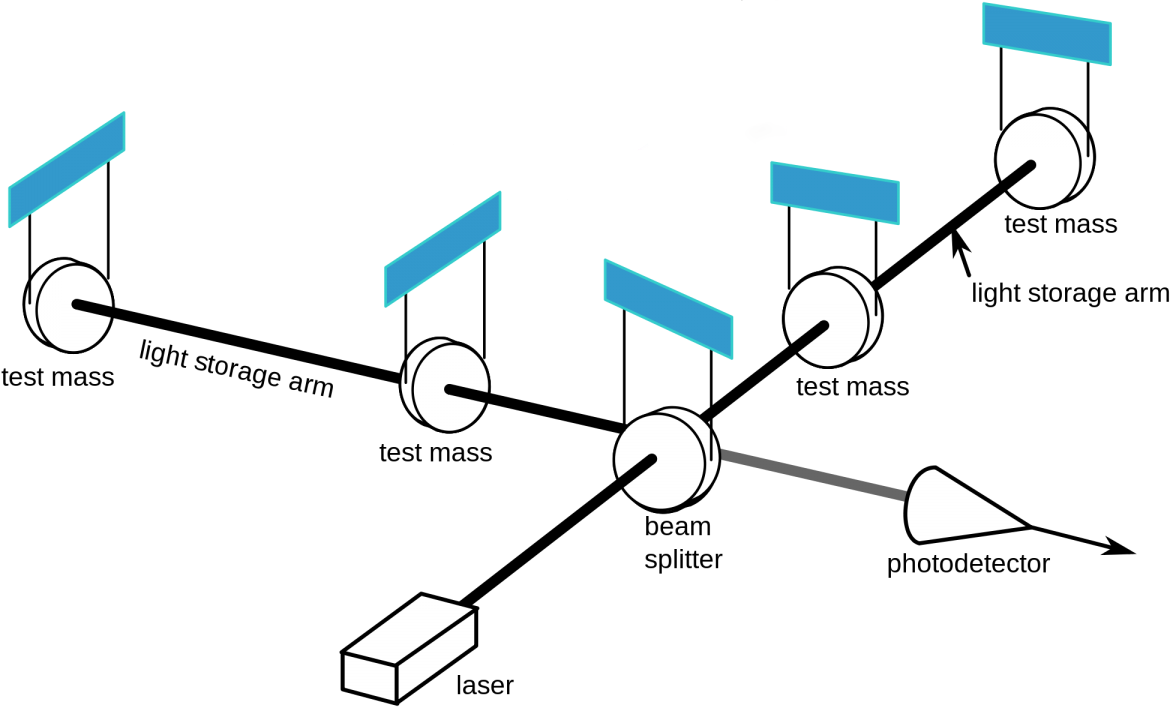}
	\caption{Basic scheme of an enhanced laser interferometer with suspended mirrors. Credit: ligo.org}
	\label{fig:interferometer}
\end{figure}

\paragraph{Ground-based interferometers}
These detectors are based on enhanced Michelson laser interferometry (see \cref{fig:interferometer}). The role of test masses is played by the mirrors, which are suspended with pendulum systems in such way that each approximates a local inertial frame (in the direction of light propagation). 
These instruments are sensitive to the frequency range $\mathcal O$ (10 Hz -- 10 kHz ).
At low frequencies, ground based detectors are limited by seismic and Newtonian noise; the main source of noise in the detection range are thermal noise in the suspension systems and in the mirror themselves; in the high frequency range, the shot-noise due to the quantum nature of light comes to dominate. \\
The currently operating network of ground-based detectors includes the US-based interferometers of LIGO-Livingston and LIGO-Hanford~\cite{2015CQGra..32g4001L}, each with 4 km arms, the Italy-based Virgo~\cite{VIRGO:2014yos} and the Japan-based KAGRA~\cite{KAGRA:2018plz}, both with 3 km arms. A smaller interferometer GEO 600, with arms of 600 m, is operating in Germany~\cite{Dooley:2014nga}.\\
The third generation (3G) of ground-based interferometers is planned for the next decade. The European Einstein Telescope~\cite{Punturo:2010zz}, will consist of three 10 km arms placed in triangular shape and will be built underground to minimize seismic noise. Two separate laser beams will travel through each arm, and two interferometers will be placed at each of the three joints, probing different frequency ranges. The US-based Cosmic Explorer~\cite{LIGOScientific:2016wof}, will instead maintain the current design -- two L-shape interferometers -- but the arm length will be increased to 40 km. \\
3G instruments will have a higher overall sensitivity, allowing to detect events at very large cosmological distances. Furthermore, the frequency range will be increased significantly. High frequencies will allow to better resolve the last stages of the mergers, including possible tidal effects, and to probe merger events of smaller mass. The low frequency range allows the detection of systems of larger total mass, but also to see lighter systems at larger separation and hence follow their inspiral for a greater number of cycles.

\paragraph{Space-based interferometers}
The Laser Interferometer Space Antenna (LISA)~\cite{Klein:2015hvg, Amaro-Seoane2017}, is also planned to be launched in the 2030s. It will consist of three satellites in an Earth-trailing orbit, separated by a distance of  $2.5 \times 10^9$ m. LISA will be free from seismic and Newtonian noise, and will be sensitive to low frequency GW, in the range $\mathcal{O}$(\SI{100}{\micro\hertz} -- \SI{100}{\milli\hertz} ).  
It will be able to detect coalescing black hole systems in the mass range $10^2 - 10^7 \Msun$, allowing it to trace the formation of supermassive black holes; furthermore, the detection of extreme mass ratio inspirals will allow to probe the extreme environments close to the event horizon of these supermassive objects.
Other proposed space-based observatories include DECIGO~\cite{Kawamura:2020pcg}, TianQin and Taiji~\cite{Gong:2021gvw}.

\paragraph{Pulsar timing arrays}	 
Pulsar timing arrays (PTA), rather than using laser beams to measure length variations, use variations in the radio frequency pulse arrival times from an array of millisecond pulsars~\cite{1978SvA....22...36S, 1979ApJ...234.1100D}. These pulsars are characterized by an extremely stable period (comparable to that of an atomic clock). PTAs look for deviations from the predicted arrival times of pulses and for spatial correlation of these across the array.
These experiments probe the very low frequency range,  $\mathcal{O}$($10^{-9}$ -- $10^{-6}$  Hz). 
Today, there are three major operating PTAs: the Parkes PTA in Australia~\cite{Hobbs:2013aka}, the European PTA Consortium~\cite{Kramer:2013kea} and the NANOGrav consortium in North America~\cite{Brazier:2019mmu}. These arrays monitor over 100 millisecond pulsars (MSPs), which collectively form the International Pulsar Timing Array (IPTA)~\cite{Manchester:2013ndt}.

 \subsection{Present status of observations}
 \label{sec:GWPBH:gw:data}

Since the first detection of a binary BH (BBH) merger in 2015, the LIGO/Virgo Collaboration (LVC) has performed three runs of observation with increasing instrumental sensitivities. In 2020, the KAGRA observatory in Japan~\cite{KAGRA:2018plz} joined the global network of gravitational wave detectors, reporting its first observations in conjunction with the GEO600 instrument in March 2022~\cite{LIGOScientific:2022myk}. The last observation run O3 was completed in March 2020 and the latest catalogue GWTC-3~\cite{LIGOScientific:2021psn} published by the LIGO, Virgo and Kagra (LVK) Collaboration includes a total of 90 events, the great majority originating from binary black hole mergers.

\begin{figure}
 	\centering
 	\includegraphics[width=.9\linewidth]{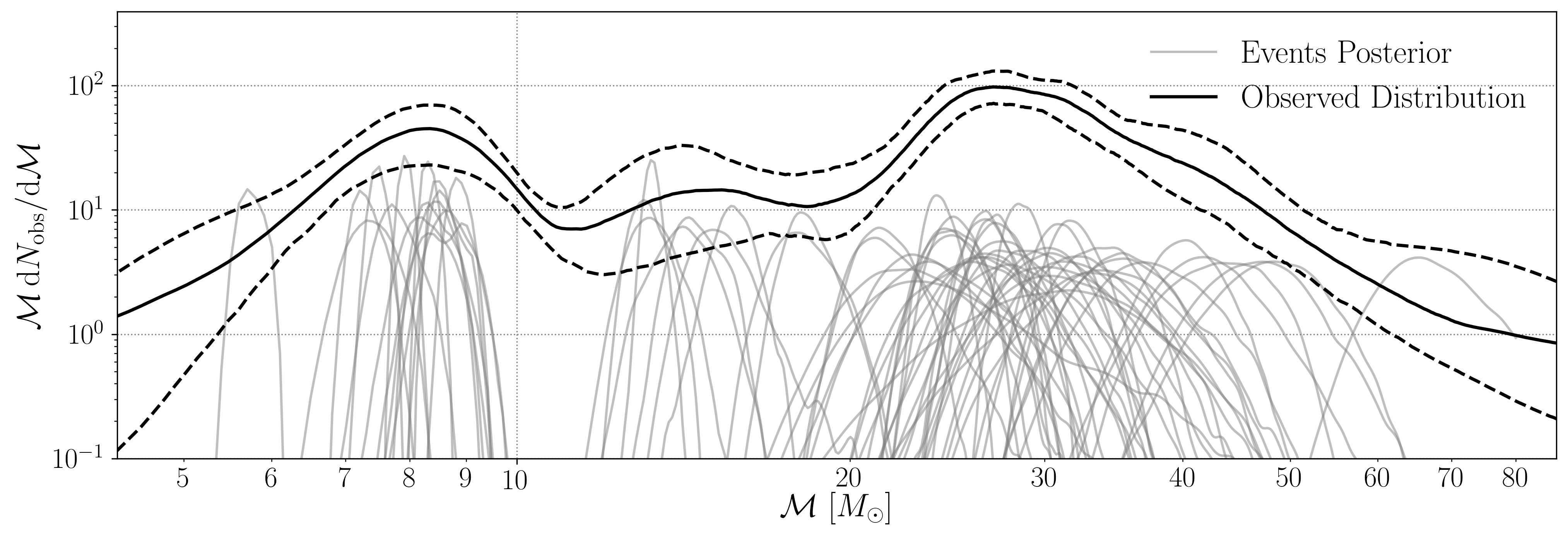}
 	\caption{ Chirp mass of BBH events from te GWTC-3 catalogue~\cite{LIGOScientific:2021psn}. The figure shows individual event chirp masses (grey) and the observed chirp mass distribution (black solid) obtained using an adaptive kernel density estimator. 
 	}
 	\label{fig:O3events}
 \end{figure}

 In November 2021, the LVK Collaboration released ~\cite{LIGOScientific:2021psn} the analysis of the properties of the compact binary population observed by LIGO and Virgo since 2015. The binary black hole (BBH) population analysis is based on 69 confident events with false alarm rate below $\SI{1}{\per\year}$. These events and their inferred chirp masses are shown in \cref{fig:O3events}.
 
 The collaboration reported an inferred BBH merger rate between $\SI{17.9}{\per\giga\parsec\cubed\per\year} $ and $\SI{44}{\per\giga\parsec\cubed\per\year} $ at redshift $z =0.2$ with $90 \%$ CL. While all events are observed at low redshifts $z \lesssim 1$, they find that the rate increase as $(1+z)^\kappa$ with $\kappa \sim 3$. 
 
\begin{figure}[tb]
	\centering
	\includegraphics[width=0.59\textwidth]{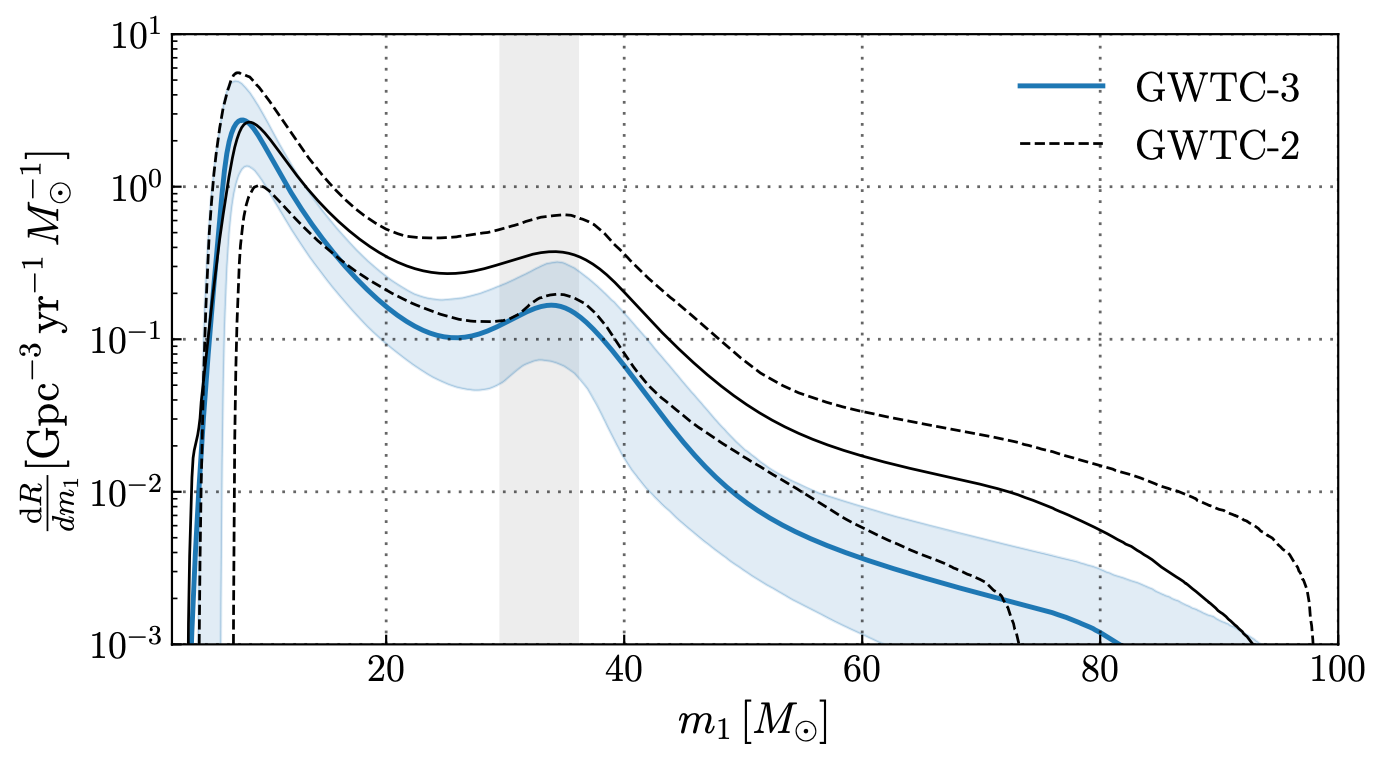}
	\hfill
	\includegraphics[width=0.39\textwidth]{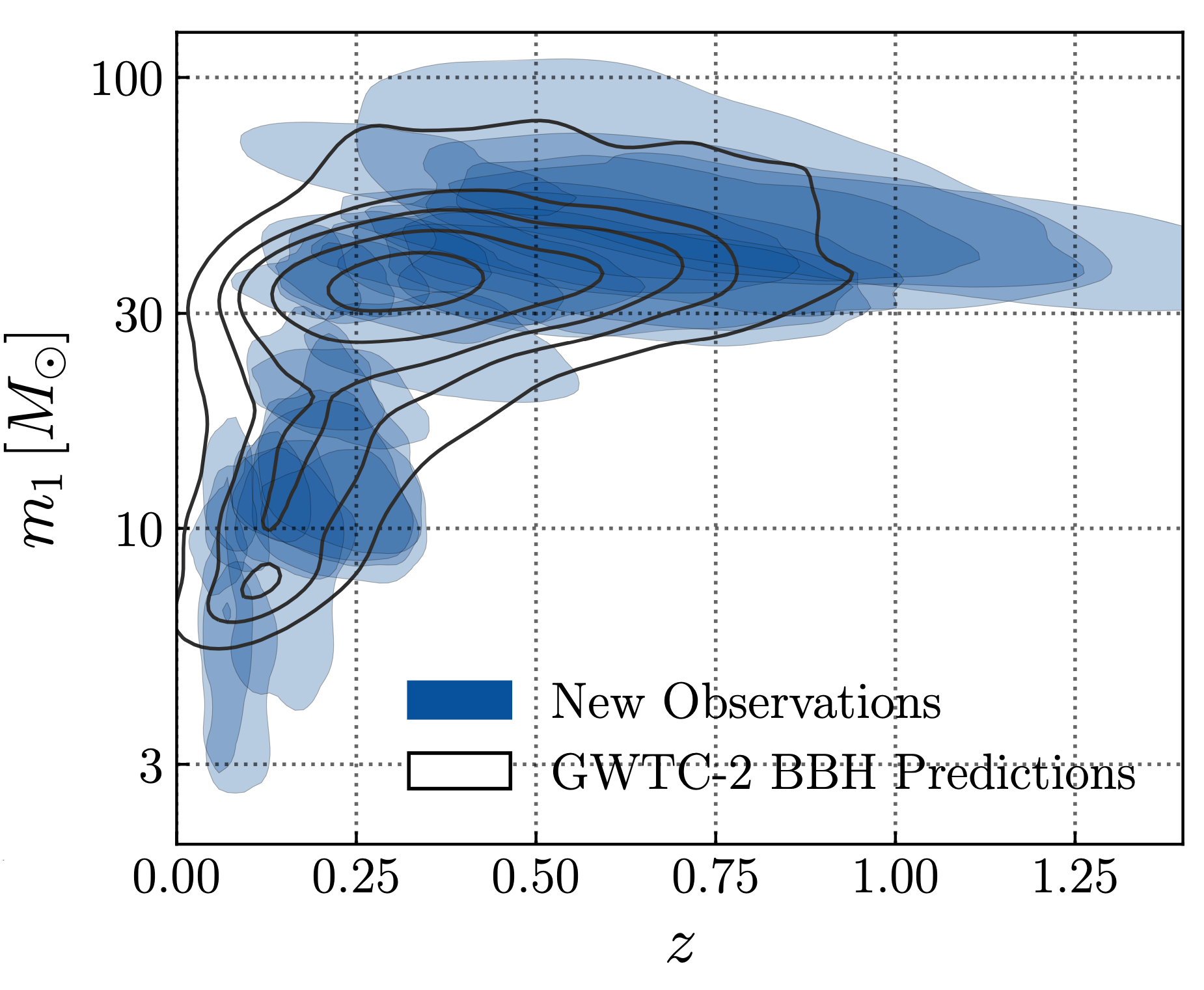}
	\caption{Left: the BBH primary mass distribution for the power law+peak model, showing the differential merger rate as a function of primary mass or mass ratio. The solid blue curve shows the posterior population distribution with the shaded region showing the 90\% credible interval. The black solid and dashed lines show the results obtained analysing GWTC-2. The vertical grey band shows 90\% credible intervals on the location of the mean of the Gaussian peak. Right: Primary mass versus redshift. Reported parameters (blue shaded regions) for O3 events, compared to the expected population of detected BBHs (black contours) as inferred from past analysis of GWTC-2. Figures from~\cite{LIGOScientific:2021psn}.}
	\label{fig:O3massdist}
\end{figure}

 The rate of mergers displays a significant dependence with the mass of the compact objects. Taking into account the selection effects (heavy BBH systems can be observed at larger distance, see \cref{fig:O3massdist}, right panel), the true mass distribution of the BBH population can be inferred (\cref{fig:O3massdist}, left panel).  The mass distribution has its global maximum at~$\sim 10 \, \Msun$, compatible with the mass function inferred from galactic X-ray binaries. However, it also presents a large high mass tail and a significant second peak at~$\sim 35 \,  \Msun$.  Confirming the result of the analysis of the previous GWTC-2 catalogue~\cite{LIGOScientific:2020kqk}, the distribution is found to be inconsistent with a simple power law model. \Cref{fig:O3massdist} shows the inferred differential merger rate as a function of the primary BH mass (the mass of the largest BH in the binary), fitted using a model composed by the combination of a truncated power-law and a Gaussian peak.

 At high masses, standard stellar evolution models predict a lack of black holes in the range $\sim 50 - 120 \, \Msun$, known as the pair-instability supernova gap. The observations are instead consistent with a monotonically decreasing mass distribution at high masses, lacking evidence for an upper mas gap. Of notable importance is the observation of the event GW190521~\cite{LIGOScientific:2020iuh, LIGOScientific:2020ufj}, with primary mass $85_{- 14}^{+ 21}  \, \Msun  $ at $90 \%$ CL. Overall the merger rate for primary mass in the bin $50 - 100 \Msun$ is reported to be between $\SI{0.01}{\per\giga\parsec\cubed\per\year} $ and $\SI{0.4}{\per\giga\parsec\cubed\per\year} $ at redshift $z =0.2$ with $90 \%$ CL (lower than the GWTC-2 estimate, $0.71_{-0.37}^{+0.65} \,  \textrm{Gpc}^{-3}  \textrm{yr}^{-1} $).
 
 The existence of a lower mass gap at $ 3 - 5 \, \Msun$ between the heaviest neutron stars and the lightest BHs, is also challenged by the observations. In particular, the high mass ratio event GW190814~\cite{LIGOScientific:2020zkf} involved a secondary compact object whose mass was measured with high precision to be $2.59_{-0.09}^{+0.08} \, \Msun$, possibly representing the lightest BH ever observed. Tidal effects, which would identify the object as a NS, have not been observed; this is not sufficient however to rule out the NS hypothesis,  in particular due to the large mass asymmetry between the merging objects (a highly asymmetric NS-BH system is expected to merge before tidal deformations can impact the waveform). Nevertheless, its mass is above the maximum mass that the NS equation of state is expected to support. Similar considerations apply to the event GW200210.
 
 With few exceptions, such as the aforementioned GW190814, most binaries present small mass ratios~\cite{LIGOScientific:2021psn}, see \cref{fig:O3massdist}. Observed effective spins\footnote{The effective spin is defined as the projection of the two spins $\vec{S_1}, \vec{S_2}$ on the total angular momentum~$\vec{L}$ of the binary: $\chi_\mathrm{eff}  \equiv \left( m_1 \vec{S_1} +m_2\vec{S_1}\right)/ M  \cdot \vec{L}$} are also small, concentrated around $\chi_\mathrm{eff} \approx 0$. Finally, a correlation between mass asymmetry and spin magnitude is observed.

The substructure detected in the mass distribution has triggered a debate relative to the existence and the nature of two different populations of BBH. On the one hand, different channels for the formation of astrophysical black hole pairs have been proposed \cite{Zevin:2020gbd,Mapelli:2021gyv}. On the other hand, the suggestive possibility exists that a population of primordial black holes has been detected by LVK~\cite{Franciolini:2021tla,Franciolini:2021xbq,Hutsi:2020sol}.

However, significant uncertainties and complexities are associated with the modelling of both the astrophysical background and the primordial component. The way these populations are modelled, and the corresponding predictions of the merger rates, are the subject of the next sections.

 \clearpage

\section{Primordial black hole binaries}
\label{sec:GWPBH:PBHbinaries}

A pair of free streaming black holes that undergo a close encounter within a cluster can form a binary system by gravitational capture. This requires a very small impact parameter, since typical kinetic energies are large. It is usually referred to as the \emph{late-time} binary formation channel. \\
However, the formation of PBH binaries is much more efficient in the early Universe. 
Furthermore, an important fraction of early-Universe binaries are formed in configurations that correspond to short (i.e. within the age of the Universe) merger times.
Ref.~\cite{Nakamura:1997sm} carried out the the first estimate of the merger rate due to early-Universe binaries, which was later refined following the first LIGO detection ~\cite{Ali-Haimoud:2017rtz, Raidal:2018bbj}.

In this chapter, we review these two formation channels and the estimated merger rates for the two populations.
As we have seen in \cref{sec:GWPBH:gw:compact_bin}, the time of merger is strongly dependent on $a$ and $j$, $\tau_\mathrm{m} \propto a^4 j^7$. Therefore, obtaining the merger rate of binaries requires a careful estimate of the expected distribution of these parameters among the binary population.

\subsection{Early Universe binary formation}
\label{sec:GWPBH:PBHbinaries:formation}

We start by considering binary formation in the Early Universe. PBH are considered to be initially at rest with respect to the expanding Universe, subject only to the Hubble flow\footnote{This is due to the fact that, while PBH can be born with peculiar velocities (due to inhomogeneities in the primordial fluid in which they form), these velocities are expected be diluted by the Hubble flow to negligible values, long before the binary is able to decouple}. 
Eventually, the mutual gravitational  attraction between PBH becomes dominant and pairs that are initially close decouple from the expansion. After decoupling, given the absence of peculiar velocities, a pair of black holes would coalesce to a single black hole on the free fall time scale. However, a torque is provided to the system by the anisotropy in the surrounding gravitational field, allowing the formation of a binary system. \\

\subsubsection{Pair decoupling and semi-major axis}

\paragraph{Time of decoupling} A first estimate of the time of decoupling can be obtained, following Ref.~\cite{Nakamura:1997sm}, assuming that the binary decouples from the Hubble flow when $\rho_{\mathrm {bin}} \approx \rho_\mathrm{rad}$, where $\rho_{\mathrm {bin}}$ is the energy density associated to a PBH pair of mass $M$. If the pair is initially at an initial comoving separation $x$, this can be estimated as
\begin{equation}
	\rho_{\mathrm {bin}}=  \dfrac{3 M}{4 \pi \, x^3}  s^{-3}\; ,
\end{equation}
where $s$ indicates the scale factor. Then, normalizing the scale factor to $s_{\mathrm eq}= 1$, we have $s_{\mathrm {dec}}\, \approx \, \rho_{\mathrm {m}}/ \rho_{\mathrm {bin}} $.
 \\
In terms of the initial comoving separation of the pair $x$, this can be expressed as
\begin{equation}
	\label{eq:sdec1}
	s_{\mathrm {dec}}\, \approx \,\left(  \dfrac{x}{x_{\mathrm {m}}} \right)^3 ,
\end{equation}
where we have defined
\begin{equation}
	x_{\mathrm{m}}(M) \equiv \left( \dfrac{3 M}{4 \pi \, \rho_{\mathrm{eq}}} \right)^{-\frac{1}{3}}  \; .
\end{equation}
The quantity $x_{\mathrm{m}}$ can be interpreted as the comoving separation associated to the total matter, assuming it was entirely made of objects of mass M.
While \cref{eq:sdec1} stresses that the time of decoupling depends only on the pair's separation $x$ and mass $M$, it is more convenient to express this quantity in terms of the average PBH separation $\bar{x} =\, \left( \fPBH \, f_\mathrm{DM} \right) ^{\frac{1}{3}} x_{\mathrm{m}}$,
where $f_\mathrm{DM} \approx 0.85$ is the dark matter fraction. Defining  $f \equiv \fPBH \, f_\mathrm{DM}$, and substituting in \cref{eq:sdec1}, we obtain
\begin{equation}
	\label{eq:sdec2} 
	s_{\mathrm {dec}} \, \approx \,  \dfrac{1}{f} \, \left( \dfrac{x}{\bar{x} } \right)^3	\; .
\end{equation}
As we mentioned, in the case of binary formation we are interested in pairs whose separation is smaller than average, $x < \bar{x}$. While $f \lesssim 1 $, we can see from \cref{eq:sdec2} that all binaries decouple in the radiation era. If instead the density of PBH is very low, binary formation can happen later. However, such late binaries will be characterised  by large physical separations and, as we will discuss later, their time to merger typically exceeds the age of the Universe. Thus, when calculating the merger rate, we are interested in binaries that decouple early, deep in the radiation era.\\
Ref.~\cite{Ali-Haimoud:2017rtz} performed a more refined estimate of the decoupling process, comparing the Hubble flow and the gravitational attraction between the pair in the Newtonian approximation. They found $s_{\mathrm {dec}} \, \approx \,  1/(3 f) \, ( x/\bar{x} )^3$, in good agreement with \cref{eq:sdec2}.
%
%
%
%

\paragraph{Semi-major axis} Assuming that the decoupling is instantaneous, the physical separation of the pair at decoupling is approximately $s_{\mathrm {dec}} x$ and the semi-major axis $a$ of the binary can be estimated from \cref{eq:sdec2} as 
\begin{equation}
	\label{eq:semimajorx1}
	a \,  \approx  \, \dfrac{1}{2} \, s_{\mathrm {dec}} \, x \; = \dfrac{1}{2 f} \,  \dfrac{x^4}{\bar{x}^3 } .
\end{equation}
The estimate obtained in~\cite{Ali-Haimoud:2017rtz} is somewhat smaller, $a \, \approx 0.1 \, f^{-1} \,  x^4 /\bar{x}^3$.

One can now obtain the distribution of semi-major axes of PBH binaries. Assuming the PBHs are initially distributed à la Poisson, the comoving separation follows $x \sim  x^2 \,  e^{-(x/\bar{x})^3}$. As the semi-major axis is $a \propto x^4$, we have
 \begin{equation}
 	\label{eq:Pofa}
 	\mathcal{P} (a) \,   \propto \, a^{- 1/4} \, e^{  - \left(a/{\bar{a}} \right)^{3/4}} \;  ,
 \end{equation}
 where $\bar{a} \sim \bar{x}/ 10 f  $. This distribution peaks at zero separation and has average $\langle a \rangle = \bar{a}$.

\subsubsection{Angular momentum and eccentricity}

In the absence of torques, a pair of PBHs would free-fall onto each other after decoupling. 
Head on collisions are avoided thanks to the torque provided by the inhomogeneities in the surrounding gravitational field. This torque provides an angular momentum to the system, allowing a binary to form.\\ For large enough \fPBH, the inhomogeneities are dominated by the PBH shot noise.
\begin{figure}[tb]
	\centering
	\includegraphics[width=0.6\textwidth]{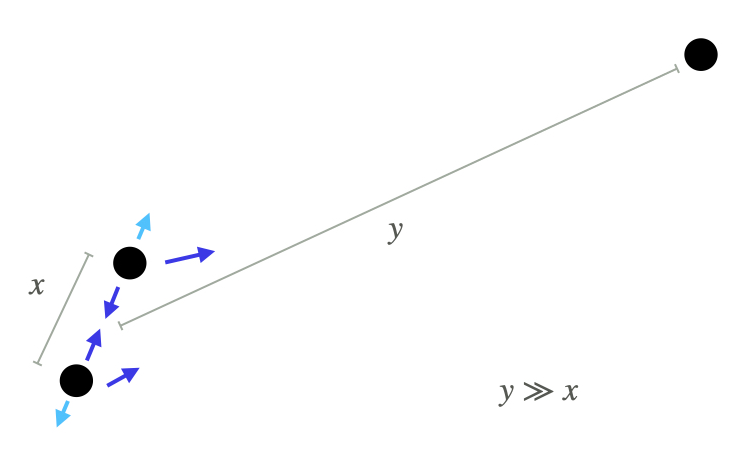}
	\caption{Schematic representation of the formation of a binary in the early Universe.}
	\label{fig:earlybinform}
\end{figure}
As a first approximation, we consider the torque provided by the nearest neighbour to the BH pair, which we assume to be located at comoving separation $y \gg x$ from the centre of the pair. The gravitational attraction of this object generates a tidal force, which in general will have a component perpendicular to the axis of the binary. 
As an order of magnitude estimate, the tidal force is~\cite{Nakamura:1997sm}
\begin{equation}
	F_\mathrm{tid} \,   = \,2 \, \dfrac{G M^2 \, (s_{\mathrm {dec}} x)}{(s_{\mathrm {dec}} y)^3}\; .
\end{equation}
Assuming that this force acts perpendicularly to the initial separation axis with constant magnitude for a time approximately equal to the free fall time, $t_\mathrm{ff} \approx(s_\mathrm {dec} x)^3/ (GM)$, we obtain an estimate for the semi-minor axis $b$ 
\begin{equation}
	b \,   \approx \, \dfrac{1}{2} \, a_\mathrm{tid} \, t_\mathrm{ff}^2 \, =\,  \left( \dfrac{x}{y} \right)^3 a  \;  .
\end{equation}
This corresponds to a binary eccentricity of
\begin{equation}
	\label{eq:bin_e1}
	e \,   \approx  \, \sqrt{  1 - (x/y)^6} \;  .
\end{equation}
Expressing this result in terms of the dimensionless angular momentum, one has $j \approx \left( x/y\right) ^3$. Considering the separation to the third binary to be approximately equal to the average PBH separation $ y  \approx \bar{x}$, we can estimate the typical value of the latter to be $j \,   \approx  \, \left( x /\bar{x}\right) ^3 $.

A more careful computation of the angular momentum due to the torque produced by a third black hole at distance $y \gg x$ gives~\cite{Ali-Haimoud:2017rtz},
\begin{equation}
	\label{eq:bin_e2}
	\vec{j}\,   \approx  \,   1.5 \,    \left( \hat{x} \cdot \hat{y }\right) \left( \hat{x} \times \hat{y }\right) \,  (x/y)^3 \, \;  ,
\end{equation}
corresponding to a magnitude in good agreement with \cref{eq:bin_e1}
\begin{equation}
	j \,   \approx  \,   0.8 \,  \sin (2 \theta)  (x/y)^3 \, \;  ,
\end{equation}
From \cref{eq:bin_e1,eq:bin_e2} we can notice an important property of early time binaries: very tight binaries, formed from pairs that start at very small initial separations $x \ll \bar{x}$, are also expected to have very large eccentricities $j \ll 1$. This follows, intuitively, from the fact that tidal forces become smaller as the separation of the PBH pair decreases.
\begin{figure}[tb]
	\centering
	\includegraphics[width=0.49\textwidth]{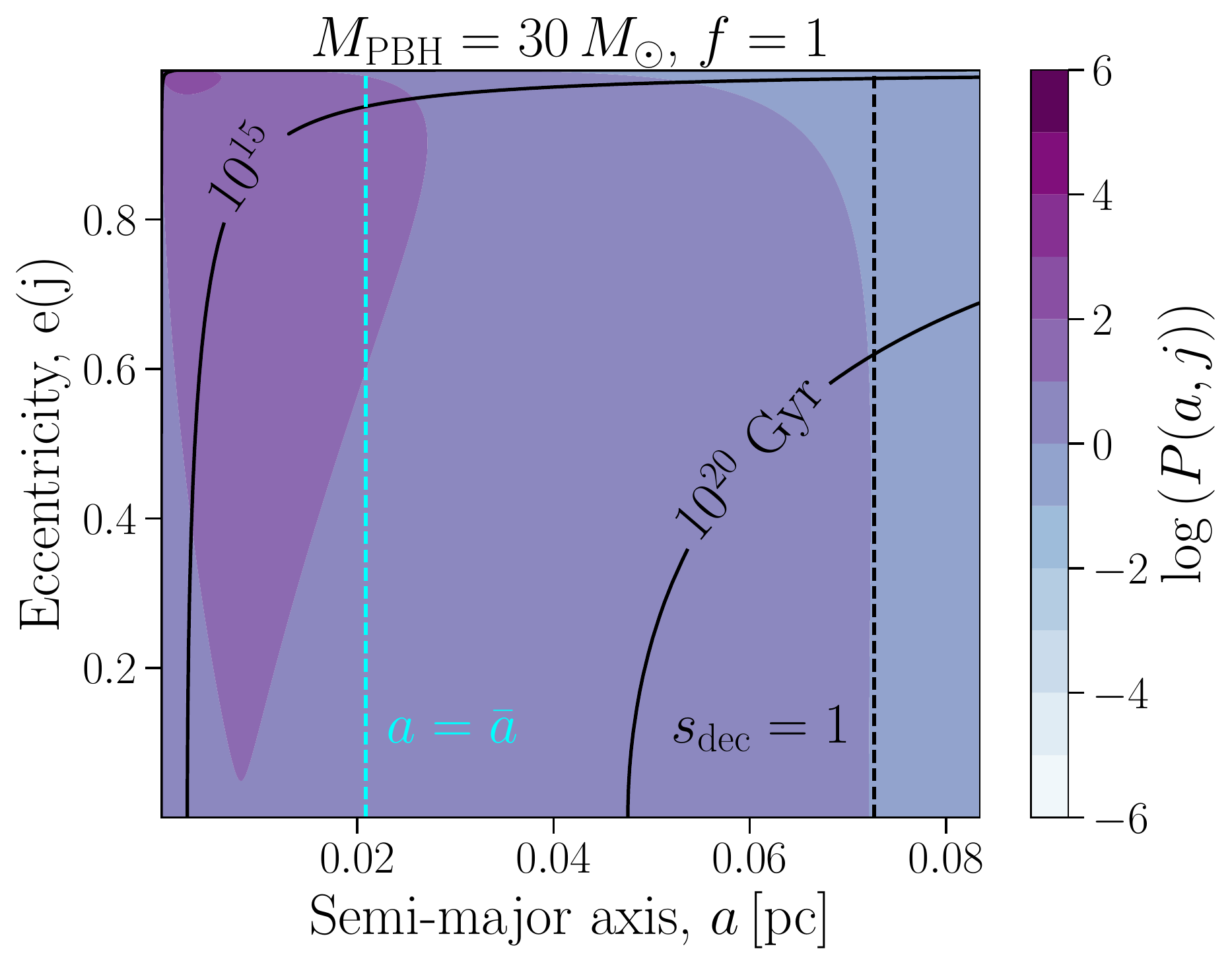}
	\hfill
	\includegraphics[width=0.5\textwidth]{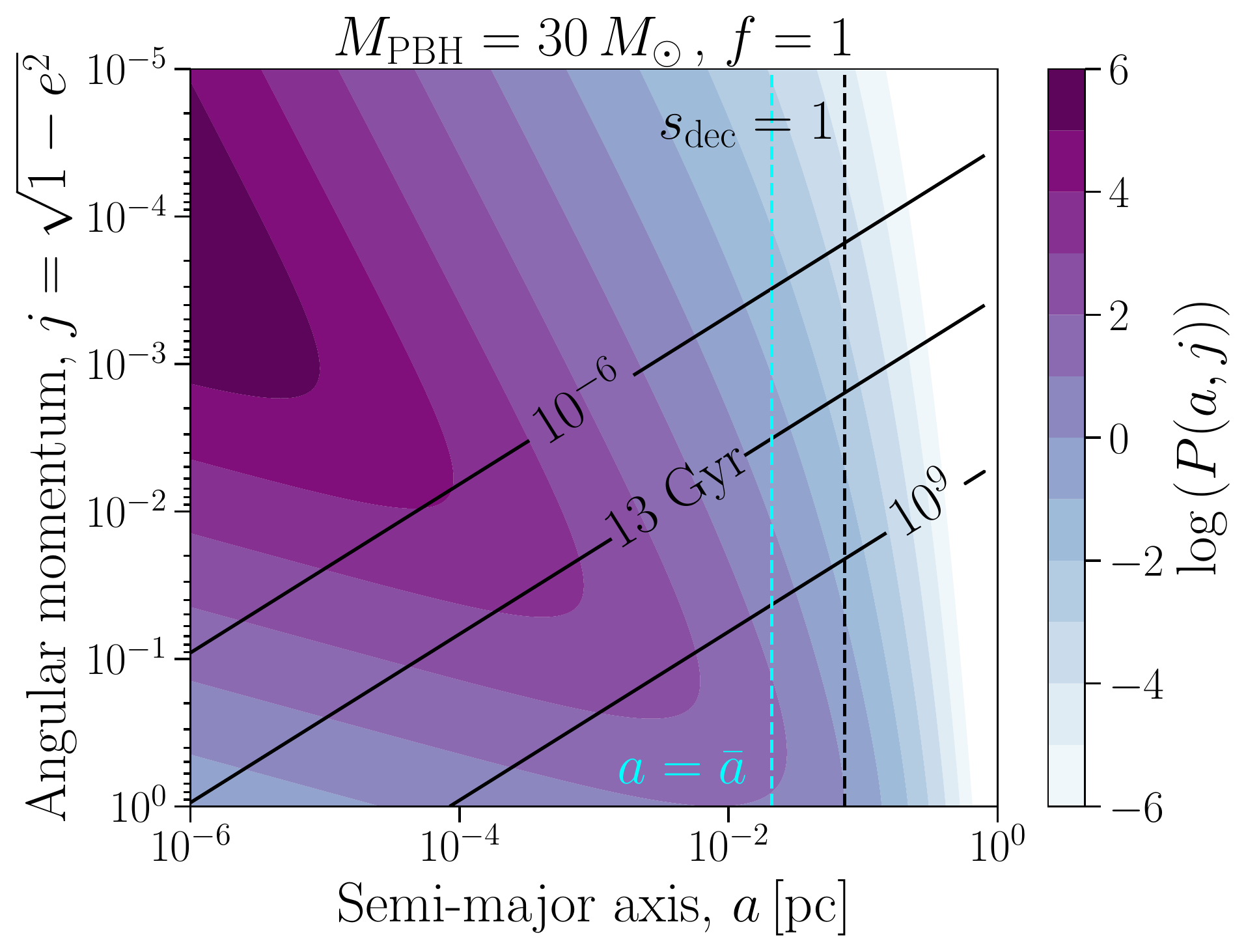}
	\caption{Joint PDF for the orbital binary parameters $a, j$ assuming $\fPBH =1 $, $M=30 \Msun$ and an initial Poisson distribution. The right panel zooms in at small values of $a, j$, responsible for the mergers that occur within the age of the Universe. The black contour lines indicate equal merger times. Notice how the time to merger decreases towards the peak of the distribution.}
	\label{fig:ae_distribution1}
\end{figure}
\begin{figure}[tb]
	\centering
	\includegraphics[width=0.49\textwidth]{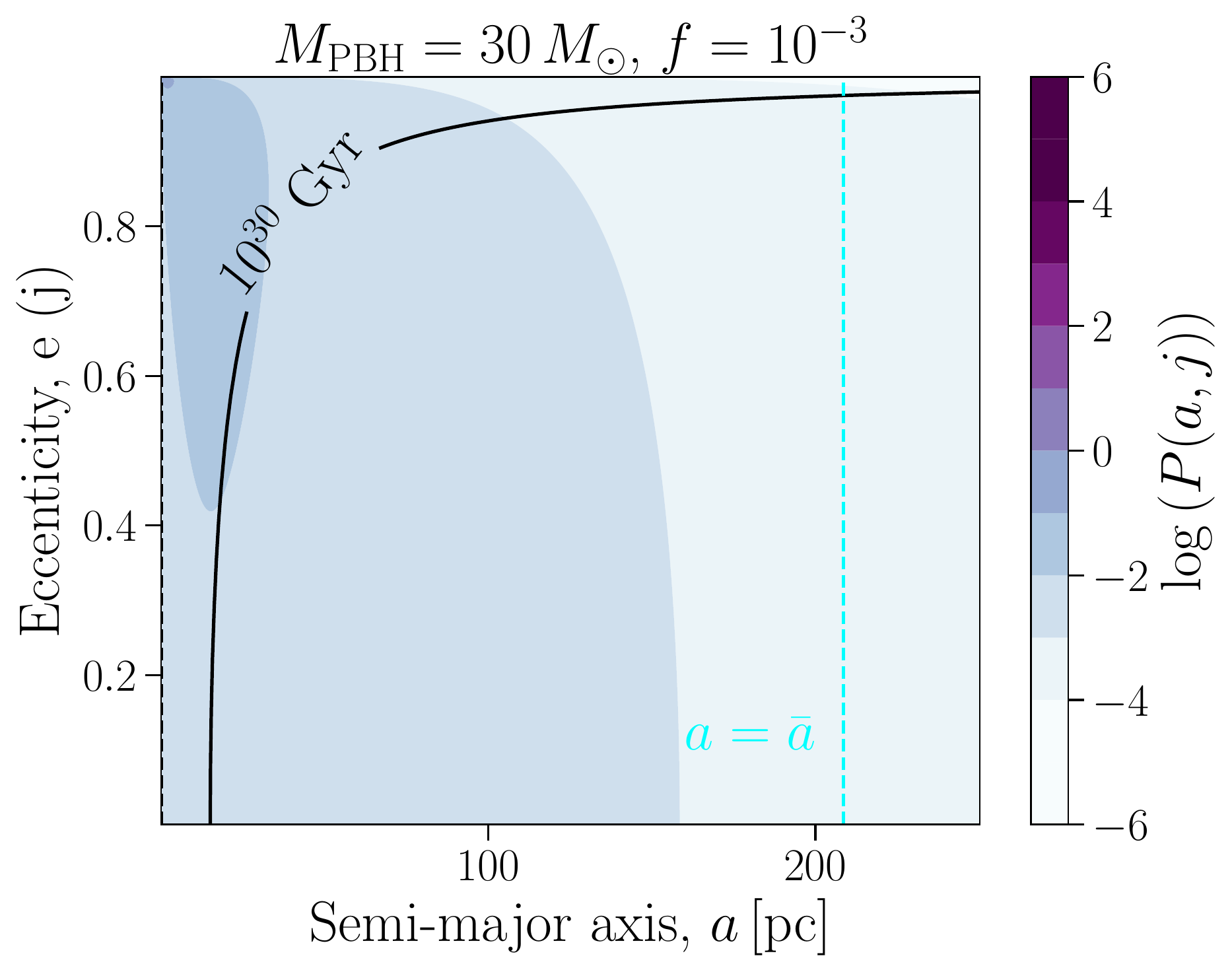}
	\hfill
	\includegraphics[width=0.5\textwidth]{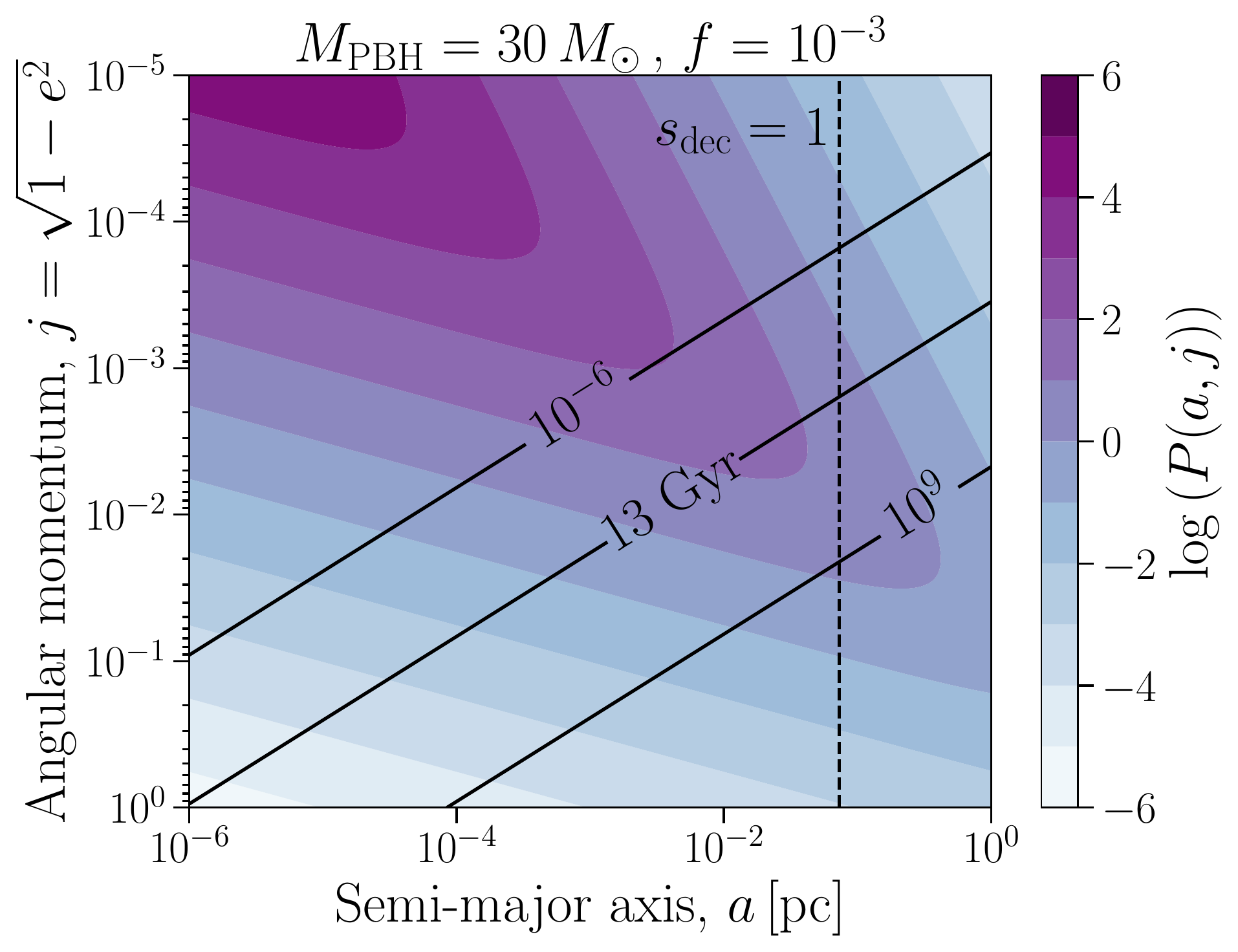}
	\caption{Same as \cref{fig:ae_distribution1}, but for $\fPBH = 10^{-3}$. }
	\label{fig:ae_distribution001}
\end{figure}

Ref.~\cite{Ali-Haimoud:2017rtz} further extended the computation to take into account the torques by \emph{all} other PBHs (assuming $y \gg x$ is true for all). In this case the eccentricity depends on the particular configuration of the PBH in space. Under the assumption that the PBHs are Poisson distributed, one finds the following probability distribution for the dimensionless angular momentum $j$ given a pair separation~$x$ \cite{Ali-Haimoud:2017rtz, Raidal:2018bbj}
\begin{equation}
	\label{eq:Pofj}
	\mathcal{P} (j \, | \, x) \,   =  \,  \frac{1}{j}\, \dfrac{(j/j_x)^2}{(1+(j/j_x)^2 )^{3/2}}\;  ,
\end{equation}
where
\begin{equation}
	j_x \, \equiv \, \dfrac{1}{2} \left( \dfrac{x}{\bar{x}} \right)^3
\end{equation}

For small fractions \fPBH, the contribution to tidal torques from adiabatic perturbations in the smooth matter component also becomes important. If these are Gaussian with variance $\sigma_\mathrm{M}$, the resulting angular momentum is distributed normally in the plane perpendicular to $\vec{x}$, with average magnitude~\cite{Ali-Haimoud:2017rtz} 
%
\begin{equation}
	\langle j \rangle \, \approx \,   \langle j^2 \rangle^{1/2} \, \approx \,  \dfrac{1}{2}  \dfrac{\sigma_\mathrm{M}}{f} \left( \dfrac{x}{\bar{x}} \right)^3 \; ,
\end{equation}
The above contribution can be taken into account by substituting in \cref{eq:Pofj}
\begin{equation}
	j_x \, = \, \dfrac{1}{2} \, \left(    1 +   \dfrac{\sigma_\mathrm{eq}^2}{f^2}    \right)^{1/2}  \, \left( \dfrac{x}{\bar{x}} \right)^3 \; .
\end{equation}

The joint PDF for the orbital parameters $P(a, j)$ is shown in \cref{fig:ae_distribution1,fig:ae_distribution001} assuming an initial Poisson distribution and $M = 30 \, \Msun$. In \cref{fig:ae_distribution1} \fPBH is set to 1, while in \cref{fig:ae_distribution1} $ \fPBH=10^{-3} $. The black contour lines indicate equal merger times. The right panels zoom in at small values of $a$ and $j$, where the population responsible for the mergers that occur within the age of the Universe resides. One can notice how the joint PDF time to merger decreases when moving in the direction of the peak of the distribution. As a consequence, the PDF for the time to merger increases towards short times, and hence towards high redshifts. 

 \paragraph{Early-time disruption of binaries by neighbouring PBHs} 
We have assumed that the pair composing the binary is at a separation much smaller than average~$x\ll\bar{x}$, and that other PBHs are at much larger distances~$y \gg x$. The first assumption, as we have seen, is justified by the fact that we are interested in binaries which merge within a Hubble time. Nevertheless, the second assumption can still be false, if a third BH also happens to be placed at short separation from the pair ~$y \sim x \ll\bar{x}$. \\ 
 This situation was considered in Ref.~\cite{Raidal:2018bbj}, where, through numerical simulations, it was shown that if the third PBH is close enough, the binary will be disrupted. Hence, when computing the probability of forming a binary, one should only consider pairs that are sufficiently isolated from all other PBHs. This results in a suppression of the number of binaries that are expected to form, and hence of the merger rate. As one can expect, the suppression effect is important when the average PBH density is high; it becomes negligible for lower densities~\cite{Hutsi:2020sol}.

\paragraph{Initial clustering} 
The assumption of initially Poisson distributed PBHs is expected to hold if PBH are formed from Gaussian distributed density perturbations~\cite{Ali-Haimoud:2018dau,Desjacques:2018wuu,Ballesteros:2018swv}.
However, if PBH are formed from a non-Gaussian power spectrum, one expects them to form in clusters~\cite{Tada:2015noa,Young:2015kda,Suyama:2019cst}. While at first one would expect the initial clustering to increase the merger rate, see e.g. \cite{PhysRevD.99.063532}, the disruption of binaries in very  high density environments might instead result in a reduction of the merger rate~\cite{Young:2019gfc,Raidal:2018bbj}. We do not consider the impact of initial clustering in the following.

\subsection{Evolution of early binaries}
\label{sec:GWPBH:PBHbinaries:evolution}

Given an initial distribution $P(a, j)$ and the expression for the time to merger given by \cref{eq:tmerger}, one can estimate the probability distribution of binary merger times (see \cref{sec:GWPBH:PBHbinaries:MR}).
However, \cref{eq:tmerger} is obtained considering the evolution of orbital parameters exclusively through the emission of gravitational waves. In fact, these can be modified by a number of other phenomena, altering the expected time to merger and hence affecting the merger rate. We briefly list here the most relevant.
 
\textbf{Accretion} Matter accretion can increase the BH masses and consequently shrink the binary~\cite{Caputo:2020irr, DeLuca:2020bjf, DeLuca:2020qqa}. This effect becomes more relevant for high BH mass. In particular, the effect of accretion is an enhancement of the high mass tail of the mass distribution. Assuming adiabatic accretion, the eccentricity is unaffected while the semi-major axis varies with the mass as~\cite{Ali-Haimoud:2017rtz,DeLuca:2020qqa}:
\begin{equation}
	\dfrac{\dot{a}}{a} \, \simeq \,  - 3 \, \dfrac{\dot{M}}{M}  \; .
\end{equation}

\textbf{DM spikes} If PBHs exist along with particle DM, the latter forms mini-halos (spikes) around the PBHs. In this case, another perturbation to the binary can come from the dynamical friction that these spikes induce on the PBHs when they approach each other~\cite{Kavanagh:2018ggo}. While the binaries are dramatically perturbed by this interaction, the shrinking of the binary compensates its increase in angular momentum in such way that the coalescence time is almost unaffected.

\textbf{Clustering} Under the action of gravity PBHs form bound structures (clusters), where complex $N$-body interactions take place. Possibly the greatest uncertainties in the theoretical estimation of the PBH merger rate come from the difficulty of modelling these interactions. Various works~\cite{DeLuca:2020jug, Vaskonen:2019jpv, Raidal:2018bbj, Jedamzik:2020ypm, Jedamzik:2020omx, Tkachev:2020uin} have been dedicated to studying the effect of clustering on the evolution of binaries, showing that it can have a significant impact the merger rate. On the one hand, new binaries can be formed in these dense environments \cite{DeLuca:2020jug}. On the other hand, we have seen that binaries need to be extremely eccentric to merge within a Hubble time; given the strong dependence of the time of merger on the angular momentum, $t_{\mathrm{merger}} \propto j^7$, even a small increase in $j$ is sufficient to delay the merger beyond our time. Ref.~\cite{Jedamzik:2020ypm} carried out Monte-Carlo simulations of three bodies interactions in clusters, finding these have a significant impact on the distribution of orbital parameters (in particular, producing a shift towards harder binaries with lower eccentricities). In turn, this resulted on a reduction, by orders of magnitude, of the expected rate of binary mergers.

A semi-analytical modelling of the impact of clustering was proposed in~\cite{Vaskonen:2019jpv}, based on the clustering formation model of \cite{Inman:2019wvr}. This calculation is based on estimating the fraction of clusters and sub-clusters that undergo core collapse following gravo-thermal instability. It is assumed that all binary systems within these structures do not contribute to the merger rate: they end up being perturbed in the high density cores in such way that their coalescence time exceeds the age of the Universe. In this sense, the calculation can be considered an over-estimation of the effect (while it is very likely that a binary is perturbed in a dense environment, it is not necessarily so; furthermore, new binaries can be created in the cores). However, the estimate is conservative in that it does not consider perturbations to binaries in stable clusters -- those not affected by core collapse.
This suppression factor increases going towards low redshifts, as clusters of larger size undergo core collapse. As expected, it also increases with \fPBH. For small PBH fractions, $\fPBH \lesssim 10^{-3}$, the effect becomes negligible.

\subsection{Late-time binary formation }
\label{sec:GWPBH:PBHbinaries:late}

PBH binaries can also form at later time, through gravitational capture within clusters. In a close encounter (scattering) between two black holes, the acceleration of the bodies results in the emission of gravitational waves. If enough energy is lost, the system can become gravitationally bound. 
This formation channel was first discussed, in the context of PBHs, in~\cite{Bird:2016dcv}.

We consider two BHs of masses $m_1 , m_2$ approaching each other with relative velocity $\mathrm{v}_\mathrm{rel}$ and impact parameter $\beta$.
For the pair of PBHs to form a binary, the energy of the system after the scattering event must be negative. Then the energy loss must be $ \Delta E  \gtrsim  K_\infty  = 1/2 \, \mu \, \mathrm{v}_\mathrm{rel} $.  To favour the capture process, we assume that the system is initially in a near-to parabolic orbit ($E \gtrsim 0$).
\begin{figure}[t]
	\centering
	\includegraphics[width=0.7\textwidth]{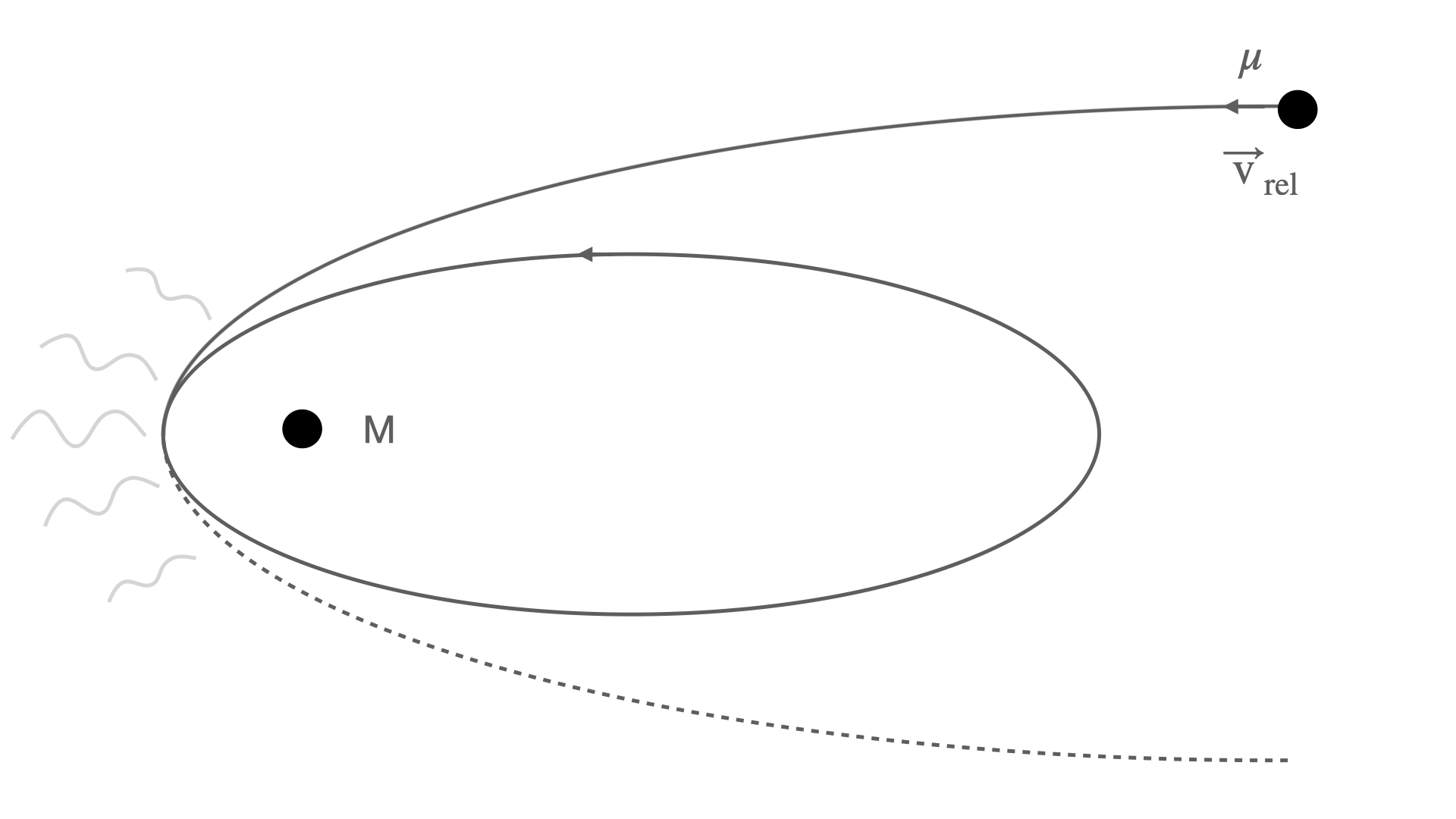}
	\caption{Schematic representation of the formation of a  binary by dynamical capture. The system, initially in a hyperbolic orbit, looses energy through the emission of gravitational waves when the two BHs approach each other. If the energy loss is sufficiently large, the system may become bound. }
	\label{fig:capture}
\end{figure}

For non relativistic velocities, the energy loss in a parabolic or hyperbolic encounter is given by~\cite{Mouri:2002mc, 1989ApJ...343..725Q}
\begin{equation}
	\Delta E ( r_\mathrm{p})  \, = \,  \dfrac{8}{15} \dfrac{G^{7/2}}{c^5}  \, \dfrac{(m_1 +m_2)^{1/2} \, m_1^2 m_2^2}{r_\mathrm{p}^{7/2}}  \, g(e)\; ,
\end{equation}
where $e \geq 1$ is the eccentricity of the orbit, $r_\mathrm{p}$ is the distance of closest approach and $g(e)= 425 \pi / (32 \sqrt(2))$ at $e=1$ (corresponding to a parabolic orbit). 
Imposing  $ \Delta E  \gtrsim 1/2 \, \mu \, \mathrm{v}_\mathrm{rel} $, where $\mathrm{v}_\mathrm{rel}$ is the relative velocity at the closest approach, we find
\begin{equation}
	r_\mathrm{p} \, \lesssim \,  G \,  \left( \dfrac{16 }{ 15 \, c^5}   \right)^{2/7} \, \dfrac{\eta^{2/7} \, M}{\mathrm{v}_\mathrm{rel}^{1/7}}     \; .
\end{equation}
The minimum approach distance is related to the impact parameter $\beta$ by
\begin{equation}
	r_\mathrm{p}\, =  \, \dfrac{\beta^2 \, \mathrm{v}_\mathrm{rel}}{ 2 \, G \, M}     \; .
\end{equation}
Hence at given relative velocity $\mathrm{v}_\mathrm{rel}$, a binary will form if the impact parameter is be smaller than
\begin{equation}
	\label{eq:impactparam}
	\beta^2 \, \lesssim \, \, \left( \dfrac{2 \, G  M}{c^2} \right)^2  \,  \left(  \dfrac{\mathrm{v}_\mathrm{rel} }{c} \right) ^{-18/7}  \, \eta^{2/7}  \;.
\end{equation}
Assuming $m_1 = m_2 = \, 50 \, \Msun$ and $\mathrm{v}_\mathrm{rel} = 200 $ km/s, we find an impact parameter of $\beta \sim 10^6$ km (comparable to the Earth -Moon distance).\\
The cross section $\sigma = \pi \beta^2$ associated to the process is 
%
\begin{equation}
	\sigma \, \approx \,  10^{-15} \, \mathrm{pc}^2 \,  \left( \dfrac{M}{100 \, M_\odot}  \right)^2 \, \left( \dfrac{\mathrm{v}}{200 \, \mathrm{km}/ \mathrm{s}}  \right)^{-18/7} \left(\dfrac{1/4}{\eta} \right)^{2/7}  \, \; .
\end{equation}	
The largest cross sections correspond to final orbital energies $E \lesssim 0$.
Since $E_\mathrm{orbit}\propto~- 1/a~=~-(1-e)/ r_\mathrm{p} $, having $E_\mathrm{orbit} \approx 0$ at small $r_\mathrm{p}$ requires $e \approx 1$. These binaries are expected to have very high eccentricities  merge in short times.

The rate of binary formation in a halo of volume $V_\mathrm{halo}$ with a uniform PBH density of $n_\mathrm{PBH}$ is
\begin{equation}
	\mathcal{R}_\mathrm{halo}  \, = \, V_\mathrm{halo}\frac{1}{2} \, n_\mathrm{PBH}^2 \, \sigma \, \langle v \rangle \;  ,
\end{equation}	
where $\langle v \rangle$ is the average PBH speed.  Comparing with \cref{eq:impactparam} one can see that the rate scales as $\mathcal{R} \, \propto \, v^{-11/7}$. Low velocity dispersion produces higher rates, as the capture process is favoured by low kinetic energies.  We can also notice that $ \mathcal{R} \, \propto \, n_\mathrm{PBH}^2$, in contrast with the linear dependence of the early-time formation (see \cref{eq:PBHrate2}). We can expect the contribution of the late-time mergers to acquire relative importance for increasing values of $\fPBH$. 

An accurate estimate of the total rate of late binary formation and merger requires to take into account the halo mass distribution, their density profiles and velocity dispersion. The rate is largely dominated by small halos, due to their high concentrations and low velocity dispersion. This makes the computation highly uncertain, since the halo mass function is poorly constrained at small scales.\\
 In Ref.~\cite{Bird:2016dcv}, the estimate was carried out assuming NFW profiles for the halo densities, a Maxwell-Boltzmann distribution for the velocities and different halo mass functions. For $\fPBH=1$ they obtained, roughly,
\begin{equation}
	\mathcal{R}_\mathrm{late} \, \lesssim \, \SI{1 }{\per\giga\parsec\cubed\per\year }\;  ,
\end{equation}	
in agreement with the rate inferred by observations. 
However, it was was later realized that this rate was subdominant with respect to the one from early-time binaries~\cite{Ali-Haimoud:2017rtz}. For this reasons, this contribution is often neglected.
(See also \cite{Korol:2019jud}, which suggests that mergers from late time encounters are one order magnitude less). We notice, however, that it has been recently argued that three body interactions in clusters can provide an efficient mechanism of late-time PBH binary formation~\cite{Franciolini:2022ewd}

As mentioned, the late-time formation channel acquires relative importance for large values of \fPBH, since it scales quadratically with the PBH density. Furthermore, large values of \fPBH correspond, as we have see, to a high probability of disruption of binaries formed in the early Universe. In the following, we will be interested in small fractions of PBHs, $\fPBH \lesssim 10^-3$: we will therefore neglect the contribution of late-time binaries to the merger rate .

 \subsection{Merger rate}
 \label{sec:GWPBH:PBHbinaries:MR}

Given a statistical distribution of the orbital parameters and formation times of  a population of binaries, we can estimate the expected merger rate density per unit time and 
volume as
\begin{equation}
	\label{eq:PBHrate}
	\mathcal{R}(t)\, = \, \frac{1}{2} \, n_\mathrm{PBH} \, \int  \diff t_\mathrm{f} \mathrm{d} a \,  \mathrm{d} j   \, P(a, j, t_\mathrm{f} ) \, \delta \left[  t - (t_\mathrm{f} +\tau_{\mathrm{m}}(a, j)) \right] ,
\end{equation}
where $P(a, j, t)$ is the probability of forming a binary with orbital parameters $a, j$ at time $t$, $n_\mathrm{PBH}$ is the comoving number density of PBHs, assumed constant and uniform, $t_\mathrm{f}$ indicates the time of formation, $tau_{\mathrm{m}}$ is the time to merger and the factor $\frac{1}{2}$ avoids over-counting.

To estimate the merger rate of PBHs, we neglect the late-time formation channel and consider only the dominant contribution from binaries formed in the early Universe.
As we have discussed, the binaries that merge within a Hubble time form deep in the radiation era $t_\mathrm{f} \ll \tau_{\mathrm{m}} $. Then, we can perform the integration over $t_\mathrm{f}$ and identify the coalescence time $\tau_{\mathrm{m}}$ with the time of merger $t_\mathrm{f} +\tau_{\mathrm{m}} $. \\
We obtain the merger rate density at given redshift, per unit time and volume as
\begin{equation}
	\label{eq:PBHrate2}
	\mathcal{R}[z(t)]\, = \, n_\mathrm{PBH} \, \frac{1}{2} \int \mathrm{d} a \,  \mathrm{d} j   \, P(j | a)P(a)  \, \delta \left[  t - \tau_{\mathrm{m}}(a, j) \right] .
\end{equation}
Where $P(a) $ and $P(j | a)$ are the PDFs for the orbital parameters, given by \cref{eq:Pofa,eq:Pofj} for a Poisson distribution. \\

Notice that this formulation relies the underlying assumption that \emph{all} PBH form binaries, while this is only true for pairs formed with small initial separation. This can be taken into account by integrating over the semi-major axis $a$ only up to some cutoff value. This is usually set by requiring $s_\mathrm{dec} < s_\mathrm{dec} $ \footnote{ In practice, the integral is not sensitive to the value of the cutoff , sice the time to merger rapidly exceeds the age of the Universe for large separations (in the limit $j \rightarrow 0$, large separations can be compatible with small merger times. However, as we can see from \cref{fig:ae_distribution1}, the contribution of such binaries to the merger rate is negligible)}.\\
A more relevant caveat is that even PBH pairs which form at small separations from each other may not be able to form a binary, if a third PBH is formed within a small distance. As discussed above, this third object can perturb the pair and prevent a binary from forming and taking this into account results in a suppression of the merger rate. However, as in the following we will restrict our calculations to low PBH densities $\fPBH \lesssim 10^{-3}$, we neglect this factor.

Furthermore, it is important to underline once again that \cref{eq:PBHrate2} assumes that the orbital parameters evolve exclusively through GW emission. In fact, as we have discussed, these can be altered by a number of additional phenomena. The most relevant of these is the suppression due to interactions in clusters, which we model following Ref.~\cite{Vaskonen:2019jpv}, as detailed in \cref{sec:appendix:clustering}.  

\begin{figure}[tb]
	\centering
	\includegraphics[width=0.49\textwidth]{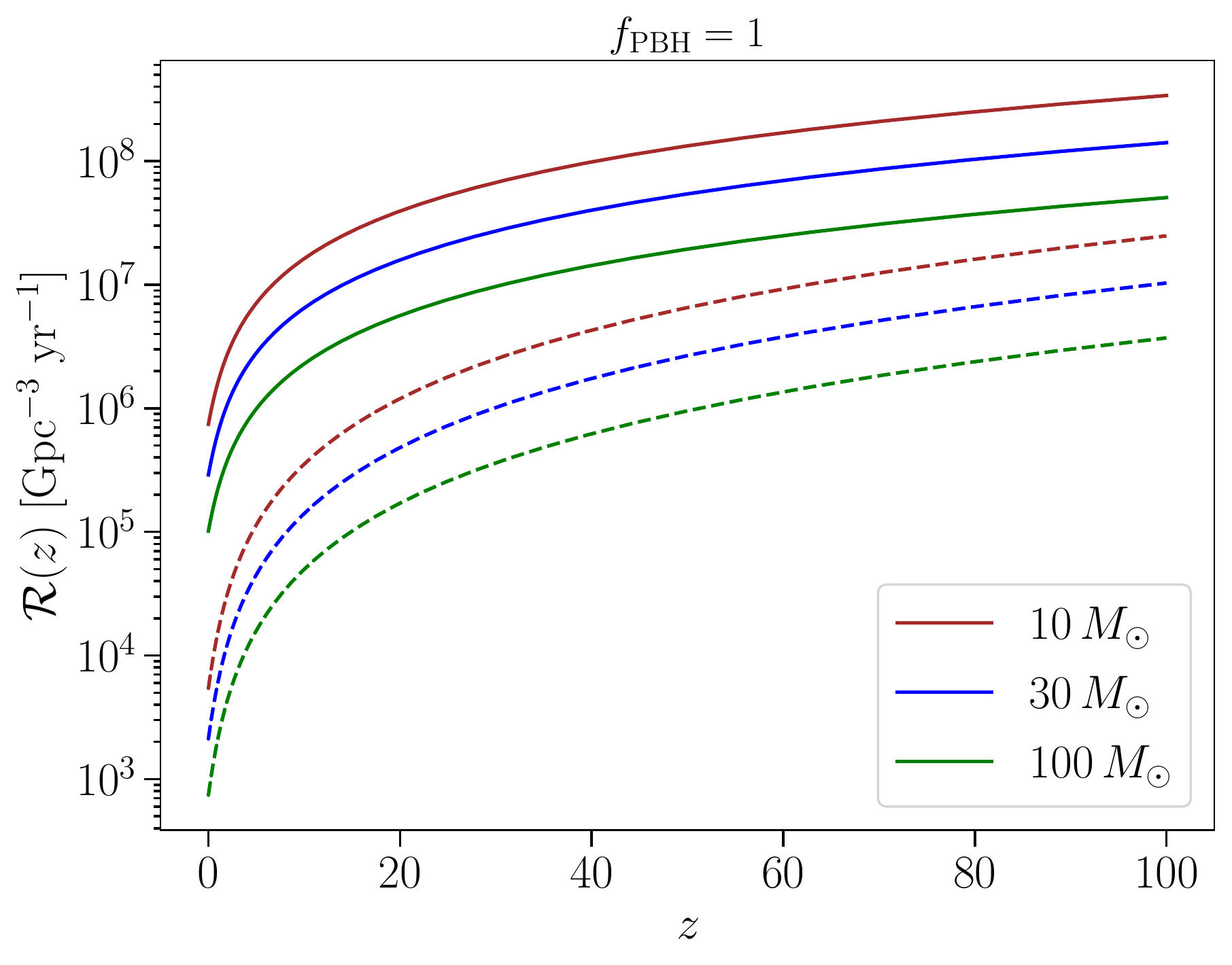}
	\hfill
	\includegraphics[width=0.5\textwidth]{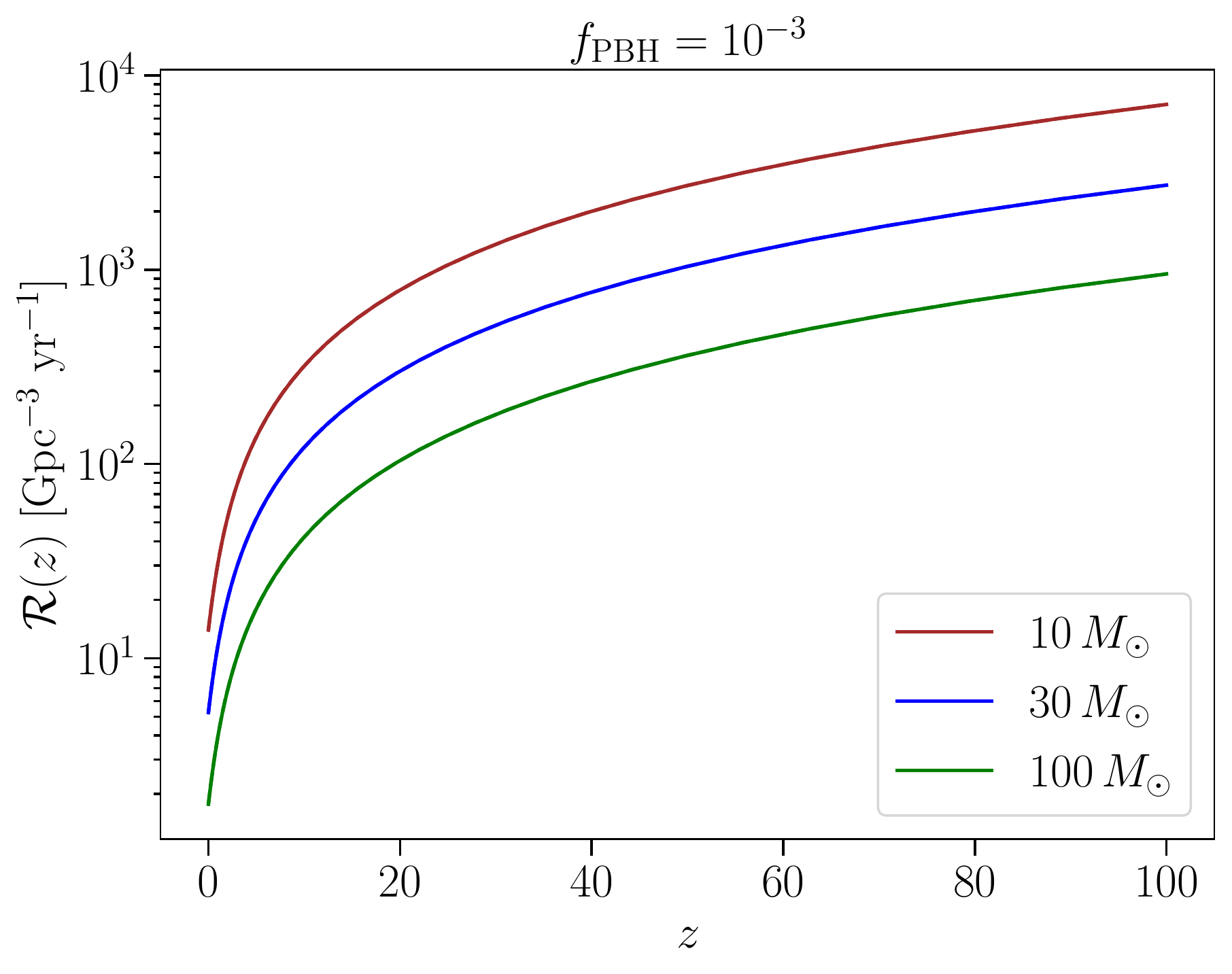}
	\caption{PBH merger rate  density as a function of redshift, for different values of the PBH mass. The dashed lines show the merger rate suppression due to late-time clustering (see \cref{sec:appendix:clustering}).}
   \label{fig:MRclusteringz}
	\end{figure}

\Cref{fig:MRclusteringz} shows the PBH merger rate  density as a function of redshift, for different values of the PBH mass (we assume a monochromatic mass function for simplicity). The left panel shows the prediction for \fPBH=1, while in the right panel we have set $\fPBH=10^{-3}$.  We notice that the merger rate is an increasing function of redshift, a consequence of the distribution of orbital parameters shown in \cref{fig:ae_distribution1,fig:ae_distribution001}. \\
The dashed lines show the merger rate suppression due to late-time clustering according to the prescription of Ref.~\cite{Vaskonen:2019jpv}. As the rate of disruption in clusters increases with time, overall it causes an enhancement of the slope of the function $\mathcal{R}(z)$, making small values at $z=0 $ compatible with larger overall PBH abundances~\cite{Chisholm:2005vm,Chisholm:2011kn,Raidal:2018bbj}.
While this effect is very significant for large PBH densities, for $\fPBH = 10^{-3}$ its impact is negligible.

\subsection{Predicting the observed merger rate}
\label{sec:GWPBH:PBHbinaries:Obs_rate}

In the previous section, we have computed the merger rate density~$\mathcal{R}(z)$, that is, the number of mergers per unit comoving volume and per unit time in the rest frame of the source, as a function of redshift~$z$.

From an observational point of view,  the relevant quantity is rather the number of mergers that would be observed by an ideal detector per unit redshift and per unit time (in the rest frame of the detector)
%
\begin{equation}
	R_\mathrm{ideal}(z) = \frac{\mathcal{R}(z)}{1+z}\,\frac{\diff V_\mathrm{ c}}{\diff z} \, .
	\label{eq:MergerRate}
\end{equation}
$V_\mathrm{ c}$ denotes the comoving volume; $\diff V_\mathrm{ c}/\diff z$ then represents the comoving volume of a spherical shell between $z$ and $z+\diff z$ around the detector,
\begin{equation}
	\frac{\diff V_\mathrm{ c}}{\diff z} = 4\pi r^2(z)\,\frac{c}{H(z)}\, ,
	\qquad \text{with} \quad
	r(z) \equiv \int_0^z \diff\zeta \; \frac{c}{H(\zeta)}\,,
\end{equation}
the comoving distance at redshift $z$.  The comoving volume $\diff V_\mathrm{ c}/\diff z$ increases with redshift, as the differential sphere gets larger.
The factor of $(1+z)^{-1}$ converts the source-frame time to the detector-frame time: the frequency of events in the detector frame is reduced by a factor $(1+ z)$ with respect to the source frame. This effect eventually comes to dominate over the increase in volume, suppressing the merger rate high redshifts.

The redshift distribution of events, still in the case of an ideal detector, is given by
\begin{equation}
	\label{eq:prob_redshift}
	p(z)
	=
	\frac{R(z)}{\int_{z_\mathrm{min}}^{z_\mathrm{max}}
		\diff\zeta \; R(\zeta)} \, .
\end{equation}
where we limit ourselves to mergers in a redshift range $z \in [z_\mathrm{min}, z_\mathrm{max}]$.

\begin{figure}[h!]
	\centering
	\includegraphics[width=0.6\textwidth]{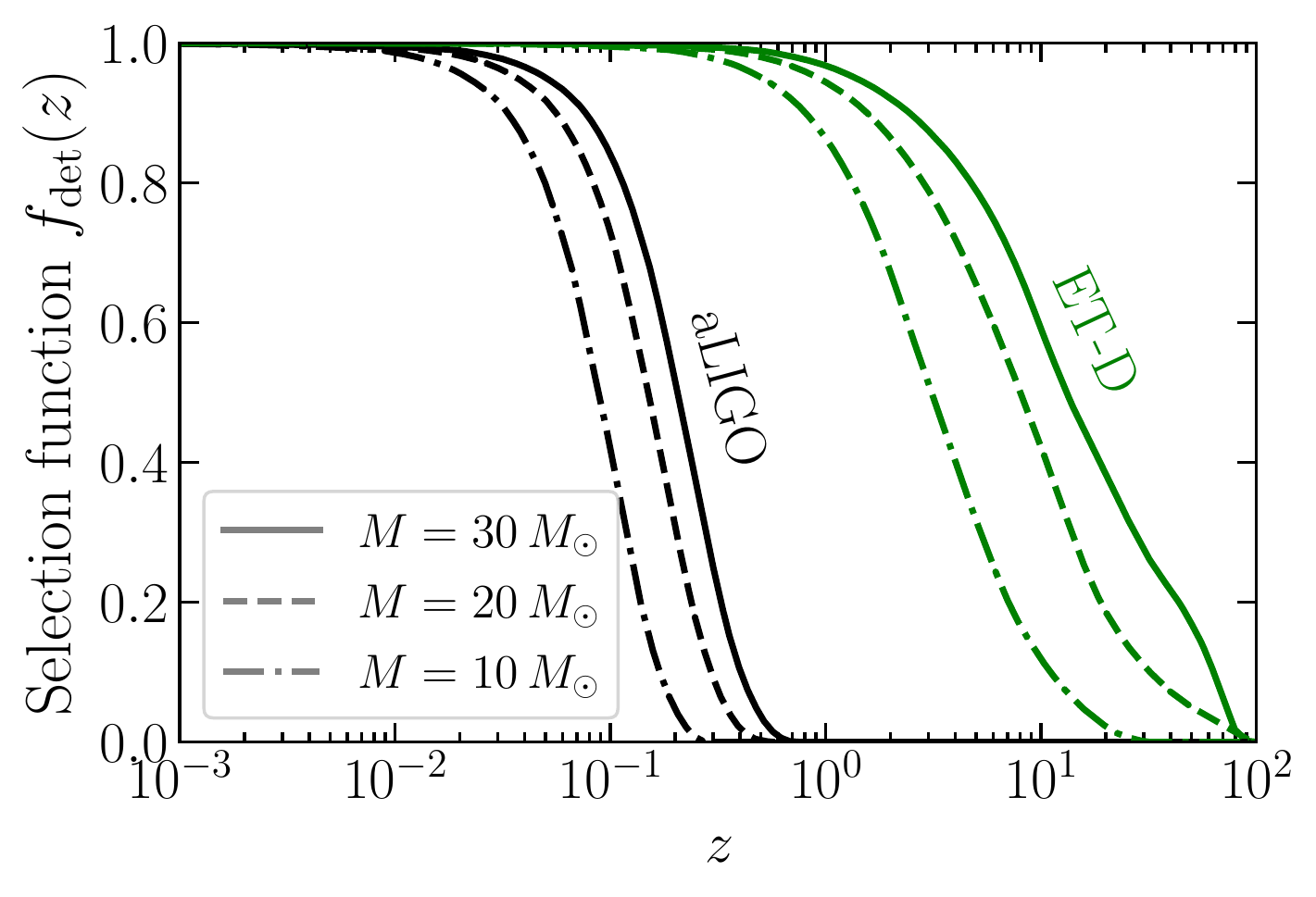}
	\caption{The selection functions $f_\mathrm{ det}(z)$ for aLIGO (black) and ET-D (green) for three different total masses $M$: \SI{10}{\solarmass} (dot-dashed lines), \SI{20}{\solarmass} (dashed lines) and \SI{30}{\solarmass} (solid lines). The selection function is defined as the fraction of mergers which are detectable at a given redshift.}
	\label{fig:ET_selection_functions}
\end{figure}

Not all  mergers will have a signal-to-noise ratio (SNR) large enough to be detectable (see \cref{sec:GWPBH:detection}). We introduce the detector selection function $f_\mathrm{det}(z)$, which encodes the fraction of mergers which are detectable at a given redshift. 
This can be obtained by generating mock merger events with different parameters and computing the detector's response to each, then averaging over the non relevant variables such as position and binary orientation.\\
\Cref{fig:ET_selection_functions} shows the selection functions for aLIGO and Einstein Telescope computed with the code \texttt{darksirens}.\footnote{\url{https://darksirens.readthedocs.io}}These are obtained sampling over binary orientations and sky positions -- following the approach of Ref.~\cite{gw-horizon-plot} -- and requiring a threshold of $\mathrm{SNR}_\mathrm{min} = 8$. We illustrate these selection functions for a range of BH masses.

Therefore, given a detector and its selection function $f_\mathrm{det}(z)$, the rate of \textit{detectable} events is given by
\begin{equation}
	\label{eq:detectable_rate}
	R_\mathrm{det}(z) = f_\mathrm{ det}(z) \, R(z) \, .
\end{equation}
The number of expected events in a given redshift range for an observation time $T$ is then
\begin{equation}
	\label{eq:N_bar_tot}
	N = T
	\int_{z_\mathrm{min}}^{z_\mathrm{max}} \diff z \; f_\mathrm{ det}(z) \, R(z) \; .
\end{equation}
In term of the merger rate density $\mathcal{R} (z)$ that is equivalent to  
\begin{equation}
	\label{eq:N_bar_tot2}
	N = T
	\int_{z_\mathrm{min}}^{z_\mathrm{max}} \diff z \; \dfrac{\diff \langle V T \rangle }{\diff z}\, \mathcal{R}(z) \ ,
\end{equation}
where we have introduced the differential \textit{time-volume sensitivity}
\begin{equation}
	\label{eq:dVTdz}
	\dfrac{\diff \langle V T \rangle }{\diff z}\,\equiv \, \dfrac{f_\mathrm{ det}(z)}{1+z}\,\frac{\diff V_\mathrm{ c}}{\diff z}
\end{equation}
If the rate can be considered constant in redshift, the number of events is simply given by $N =  \langle V T \rangle  \, \mathcal{R}(z)$.

\subsection{Comparison with present data}
\label{sec:GWPBH:PBHbinaries:data}

We are now able to compare the predicted PBH merger rate to the observational data.
We refer to the  merger rate density estimates reported in Table 4 of \cite{LIGOScientific:2020kqk} for GWTC-2 and Table 4 of \cite{LIGOScientific:2021psn} for GWTC-3. These are reported assuming a non-evolving rate, while our models for the ABH and PBH merger rates evolve with redshift. We therefore compute the average merger rate density $\langle \mathcal{R} \rangle$, defined as
\begin{equation}
	\label{eq:rate_avg}
	\langle \mathcal{R} \rangle \, =\, \dfrac{\int_0^{z_\mathrm{ max}}  R_\mathrm{det}(z) \,\mathrm{d}z}{\int_0^{z_\mathrm{ max}} \frac{f_\mathrm{det}(z)}{1+z}\frac{\mathrm{d}V_c}{\mathrm{d}z} \,\mathrm{d}z  } \, .
\end{equation}
This corresponds to the non-evolving merger rate density which would give rise to the same number of expected detectable events as our redshift-dependent model, over the redshift range $z \in [0, z_\mathrm{max}]$. For $f _\mathrm{ det}(z)$ we use the selection function for aLIGO shown in \cref{fig:ET_selection_functions} and we fix $z_\mathrm{ max}=2$.\footnote{In practice, this integral is largely insensitive to the precise value of $z_\mathrm{max}$, as the selection function drops rapidly above $z \sim 0.1$.} For each mass bin considered in these catalogues, we make the conservative choice to exclude the values of \fPBH that would correspond to an average merger rate density $\langle \mathcal{R} \rangle$ that exceeds the measured one by 2$\sigma$.

\begin{figure}
	\centering
	\includegraphics[width=0.6\textwidth]{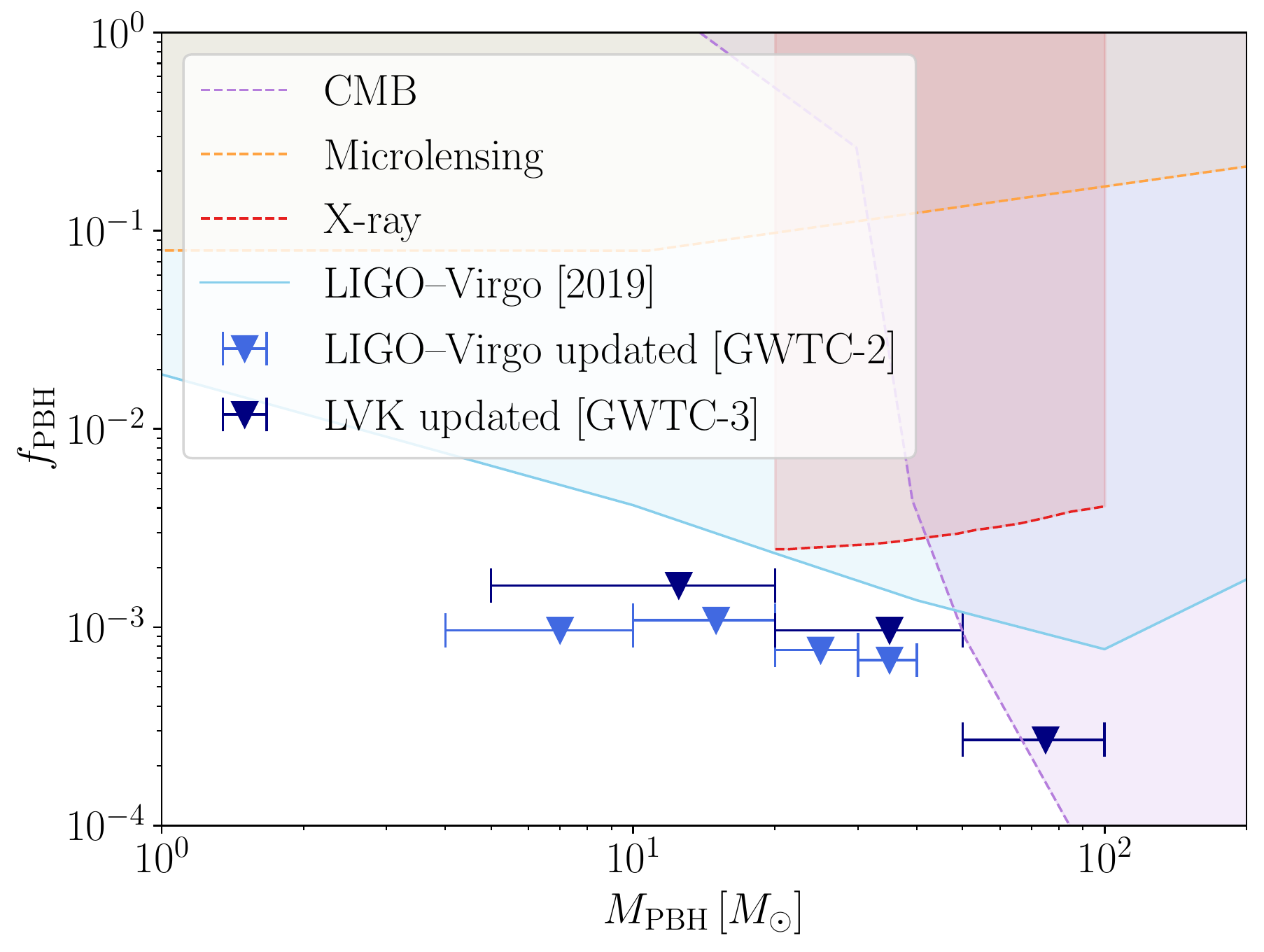}
	\caption{Updated LIGO--Virgo and LVK bounds on $f_\mathrm{PBH}$, based on the observed low redshift merger rate reported in recent GW transient catalogues. The various bins in $M_\mathrm{PBH}$ correspond to the reported upper limits on the merger rate across different mass bins. For comparison, we also show bounds from the CMB  due to PBH accretion~\cite{Serpico:2020ehh}, microlensing constraints from the high redshift star Icarus~\cite{Oguri:2017ock}, and X-ray observations of the Milky Way~\cite{Manshanden:2018tze}. We obtained these bounds from the \href{github.com/bradkav/PBHbounds}{PBHbounds repository}~\cite{PBHbounds}.}
	\label{fig:LIGObound}
\end{figure}

We obtain the constraint shown in \cref{fig:LIGObound}, setting an upper limit of $\fPBH \lesssim 10^{-3}$, in agreement with previous results.

\begin{figure}
	\centering
	\includegraphics[width=.6\linewidth]{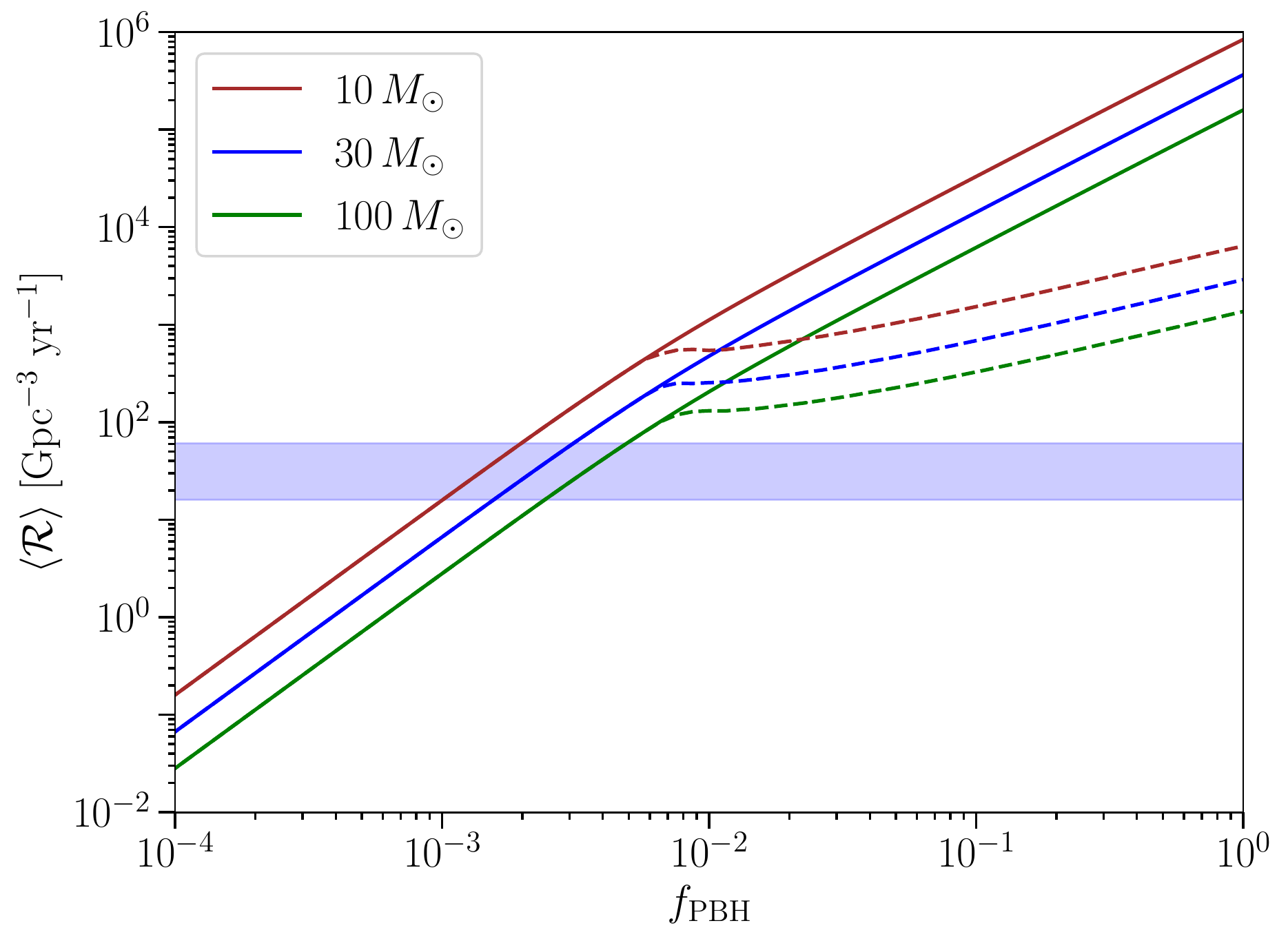}
	\caption{Merger rate of PBH binaries formed in the early Universe as would be measured by LIGO  as a function of \fPBH,  for $M_\mathrm{PBH}=10\,M_{\odot}$, $M_\mathrm{PBH}=30\,M_{\odot}$ and $M_\mathrm{PBH}=100\,M_{\odot}$. The dashed lines are obtained considering the disruption in clusters. The coloured band indicates the latest measure of the total binary black hole rate as reported in~\cite{LIGOScientific:2021psn}.}
	\label{fig:MRclust}
\end{figure}

We find that the suppression due to late-time clustering does not impact this upper limit. Its effect is relevant only for values of \fPBH larger than $\mathcal{O}(10^{-2})$, as shown in \cref{fig:MRclust}. Though the rate is significantly suppressed by clustering for large values of \fPBH, the predicted PBH merger rate would still exceed the merger rate observed by LVK, and large values of \fPBH remain excluded. Around our reported bound of $f_\mathrm{PBH} \lesssim 10^{-3}$, the suppression becomes negligible. 

Even so, we emphasise that this upper limit has to be taken \emph{cum grano salis}. Disruption effects due to late-time clustering would benefit from more careful scrutiny. Furthermore, scenarios featuring non-trivial mass functions and/or non-negligible initial clustering may potentially evade the bound.

\section{Astrophysical black hole binaries}
\label{sec:GWPBH:ABHbinaries}

In this section, we briefly discuss the models of formation of astrophysical black hole binaries and the associated rate of merger events. 

Around half of stars, and more importantly, 70\% of stars with $M> 10 \Msun$, are found in binaries \cite{2012Sci...337..444S}.  Although the violent phases of the BH formation are likely to disrupt the binary system, a fraction of them is expected to survive. The difficulty however, lies in modelling the formation of binaries that are hard enough to merge within a Hubble time. We first discuss this problem and how it can be overcome; we then examine the expected dependence of the merger rate with redshift.

\subsection{Formation channels}
\label{sec:GWPBH:ABHbinaries:formation}

During their evolution, before turning into compact objects, massive stars undergo a giant phase where their size increases to become of order 1 -10 au.  If a pair of stars is initially in a binary system with separation smaller than that, they will engulf each other during the giant phase, rather than forming a compact binary system.
From \cref{eq:tmergerastimate} for the time of merger, we see that a separation of $\sim 10 $ au corresponds, for a circular binary, to a time to merger of $10^7$ Gyr -- that is, around $ 10^6$ Hubble times. A time to merger of order $ 1 $ Gyr  at this separation requires an extremely high eccentricity, $e  \gtrsim 0.995$. This is neither viable, as it corresponds to a periastron of $r_\mathrm{p} \sim 0.05$ au. 
 
Two main formation channels have been proposed to overcome this difficulty and form hard astrophysical binary systems: one applies to the common evolution of isolated binaries, the other involves dynamical processes in dense environments. We summarize them briefly here, and refer to~\cite{2020FrASS...7...38M, Celoria:2018mzr, Barack:2018yly} for recent reviews.

\begin{figure}
	\centering
	\includegraphics[width=.8\linewidth]{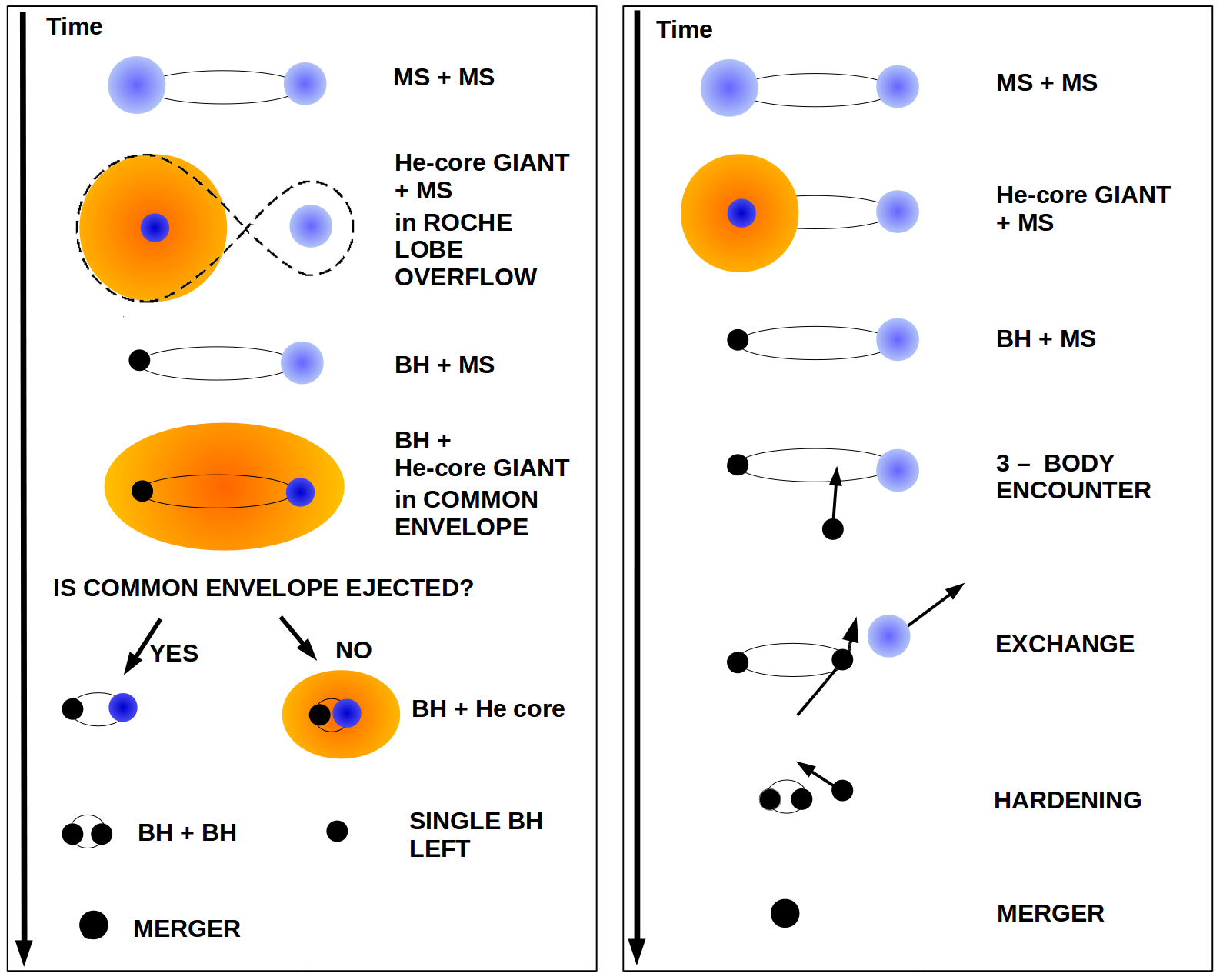}
	\caption{The two main astrophysical binary formation channels. Credit:~\cite{2020FrASS...7...38M} }
	\label{fig:astrobin}
\end{figure}

\paragraph{Common evolution} 
The evolution of an astrophysical binary is complex and undergoes different phases. As one of the stars leaves the main sequence to become a giant, the binary can undergo Roche lobe overflow, leading to mass transfer and possibly to the giant involving the other object in its hydrogen envelope (which separates from the denser helium core).  If this envelope is not ejected, the system merges prematurely;  however, if an ejection occurs, the binary system survives and shrinks by a factor $10^2-10^3$, leading to a separation of around $a \sim R_\odot$. The process is shown in the left panel of figure \cref{fig:astrobin}. We refer to ~\cite{2013A&ARv..21...59I} for a review on the status of the common-envelope models and~\cite{Fragos:2019box} for the results of recent hydrodynamical simulations. 

\paragraph{Dynamical channel}   Binaries formed with large separations can be can be hardened through three-body interactions in dense environments such as globular clusters, or galaxy cores \cite{Samsing:2013kua, Rodriguez:2016kxx, Antonini:2016gqe}. These interactions  involve \emph{dynamical hardening} (\cite{1975MNRAS.173..729H,1975AJ.....80..809H}) and \emph{dynamical exchanges}, where one member of a binary is replaced by a more massive object (see \cref{fig:astrobin}, right panel). These processes are more efficient if velocity dispersions are low. \\
Furthermore, it is argued that dense environments can favour hierarchical mergers, possibly providing an explanation to the existence of a high mass BH population \cite{DiCarlo:2019pmf}.
The estimation of binary formation rates requires an accurate modelling of cluster formation and evolution. For results based on recent simulations see ~\cite{Mapelli:2021gyv, Zevin:2020gbd}.

\subsection{Star formation rate}
\label{sec:GWPBH:ABHbinaries:SFR}

Despite the modelling uncertainties, the astrophysical merger rate --  and in particular, its dependence with redshift -- is expected to be strongly related to the the rate of star formation (SFR). \\
The instantaneous star-formation rate density $\mathcal{R}_\mathrm{ SF}(z)$ is directly measured by galaxy surveys performed in the ultraviolet band and in the far-infrared band. The ultraviolet wavelengths directly trace the presence of short-lived massive stars; the infrared ones are a signature of UV light emitted by the same population of stars and subsequently absorbed and re-emitted by dust.
In-depth analyses of these data allow to model the cosmic history of the SFR. Importantly, they unambiguously highlight a rising trend at low redshift up to a peak of very intense star formation in the range $1 \leq z \leq 2$, followed by a fall-off at large redshift~\cite{SFR_MadauReview}.

The high-redshift behaviour of the SFR may be affected by significant uncertainties and biases, mainly due to dust obscuration and to the  fact that early star formation took place in very faint galaxies, typically missed in existing surveys.
An alternative, indirect tracer of the high-redshift SFR is the rate of gamma ray bursts (GRBs). 
In this case, the main uncertainty is the relation between the rate of GRBs and the star formation rate itself. The SFR rate models obtained from these tracers peaks at slightly higher redshifts, but shows the same overall behaviour described above (see e.g.~\cite{Vangioni:2014axa} and references therein).

The SFR can be parametrized as~\cite{Nagamine:2003bd} 
\begin{equation}
	\label{eq:SFR}
	\mathcal{R}_\mathrm{ SF}(z)
	=
	k \,
	\frac{a \, \mathrm{e}^{b(z - z_\mathrm{ m})}}
	{a - b + b \, \mathrm{e}^{a(z - z_\mathrm{ m})}} \ .
\end{equation}
The free parameters $k, a, b, z_\mathrm{ m}$ in this expression are fitted by comparison to existing catalogues the different tracers of star formation.

\subsection{Merger rate}
\label{sec:GWPBH:ABHbinaries:rate}

The astrophysical merger rate, as we discussed, carries significant of modelling uncertainties. It can be obtained through numerical simulations as in, for example, \cite{Mapelli:2018wys, Belczynski:2017gds}.
A more phenomenological approach is to estimate the redshift behaviour of the merger rate density $\mathcal{R}_\mathrm{ ABH}$ starting from the star formation rate as~\cite{Dvorkin:2016wac,Mukherjee:2021ags} 
\begin{equation}
	\mathcal{R}_\mathrm{ ABH}(z[t]) = \mathcal{N}  \int_{t-  \Delta t_\mathrm{ min}}^{t- \Delta t_\mathrm{ max}}
	\diff t_\mathrm{f} \; P(t- t_\mathrm{f} ) \, \mathcal{R}_\mathrm{SF}(t_\mathrm{f} )\ ,
	\label{eq:ABHrate}
\end{equation}
where $ t_\mathrm{f} $ is the formation time of the binary stars and  $P(\Delta t)$ is the distribution function of the time delay between the formation of the stars and the BBH system merger. As stars that form BHs need to be very massive and therefore short lived, the time delay is dominated by the time to merger (this is more complicated in the case of dynamical formation channels).

The fraction of stars that will form hard binary black hole systems, despite the recent progresses in numerical simulations we mentioned above, remains highly uncertain. Assuming that this fraction is only mildly varying with cosmological  time, the uncertainty can be absorbed in the normalization factor $ \mathcal{N}  $  (which we expect to be mass dependent).\\
In this approach, the normalization is not set \emph{a priori}. It can instead be fixed by comparison with the observed low-redshift BBH merger rate (notice that if one considers hybrid scenarios where a portion of the events is ascribed to a PBH population, each value of \fPBH actually corresponds to a different normalization factor for the rate associated to the ABH population).
\begin{figure}[tb]
	\centering
	\includegraphics[width=0.49\textwidth]{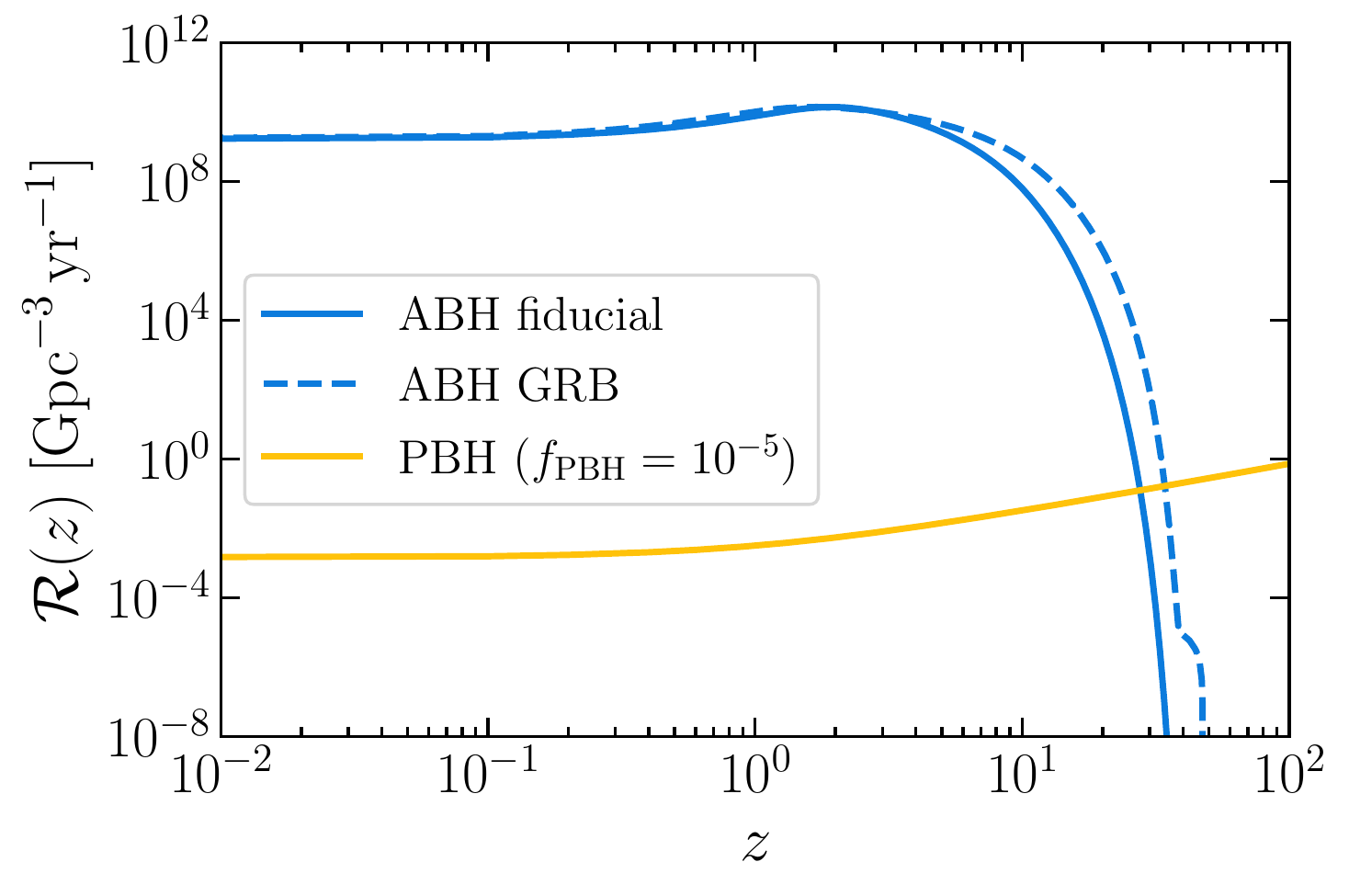}
	\hfill
	\includegraphics[width=0.48\textwidth]{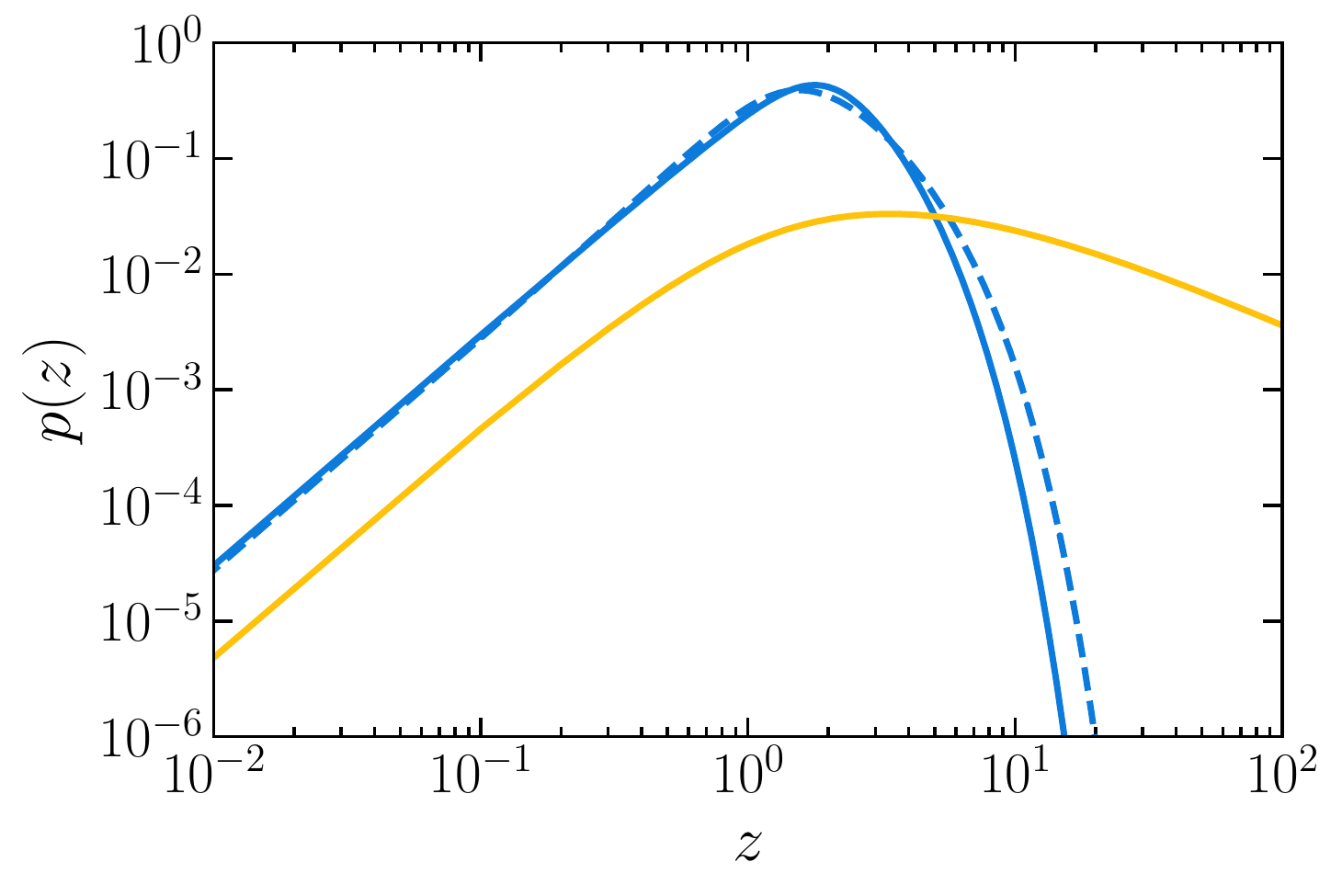}
	\caption{Distribution of merger events as a function of redshift for ABHs and PBHs. We show results for our fiducial ABH model (solid blue), the more optimistic ABH model (dashed blue) based on GRB data, and our PBH model with $f_\mathrm{PBH} = 10^{-5}$ (yellow).
		\textbf{Left:} Comoving merger rate density per unit volume per unit time $\mathcal{R}(z)$. \textbf{Right:} Redshift probability distribution $p(z)$, defined in \cref{eq:prob_redshift}. This corresponds to the distribution of observed event redshifts for an ideal detector, ignoring the effects of a selection function $f_\mathrm{det}(z)$.}
	\label{fig:MergerRateAstro}
\end{figure}

\subsection{Primordial vs astrophysical merger rates}

We obtain the astrophysical rate trough \cref{eq:ABHrate} and modelling time delay distribution, following Ref.~\cite{Dvorkin:2016wac}, as $P(\Delta t) \propto 1/\Delta t$ for $\Delta t_\mathrm{ min} \leq \Delta t \leq \Delta t_\mathrm{ max}$ with $\Delta t_\mathrm{ min} = \SI{50}{\mega\year}$ and $\Delta t_\mathrm{ max}=H_0^{-1}$. This expression is motivated for instance by the high-precision numerical simulations of binary-black-hole formation via the evolution of isolated binary stars presented in~\cite{Belczynski:2016obo}.\\
We use expression \cref{eq:SFR} for the star formation rate and consider two different choices for the parameters, aiming to bracket the measurement uncertainty. The reference set of parameters corresponds to the \emph{fiducial model} quoted in \cite{Chen:2019irf} based on a fit to observations of bright galaxies. The values are
$k = \SI{0.178}{\solarmass\per\year\per\mega\parsec\cubed}$,
$z_\mathrm{m} = 2$,
$a = 2.37$,
$b = 1.8$.
We also consider a GRB-based fit as a \emph{maximal} model, with the values taken from~\cite{Vangioni:2014axa}:
$k = \SI{0.146}{\solarmass\per\year\per\mega\parsec\cubed}$,
$z_\mathrm{m}  = 1.72$,
$a = 2.8$,
$b = 2.46$.

In figure \cref{fig:MergerRateAstro} we compare the merger rates as a function of redshift for an astrophysical and a primordial population. 
We can appreciate the crucial difference between the \emph{redshift evolution} of the ABH and PBH merger rates. The ABH merger, following the SFR, peaks at $1 \leq z \leq 2$ and decays towards high redshift. Instead, the PBH merger rate is expected, as we discussed in the previous sections, to increase monotonically with redshift. Hence, even a small population of PBHs is expected to dominate the merger rate at early times.

This naturally suggests a focus on high redshift GWs for the purposes of detecting PBHs. Current gravitational wave observatories only reach up to $z \sim 1$, but third generation detectors are planned to reach deep into the dark ages, where the astrophysical background is expected to be negligible, see \cref{fig:ET_selection_functions}. \\
The extremely large distances involved call for a careful assessment of the expected statistics and of the errors associated to redshift measurements, in order to determining the potential of PBH searches based on high-redshift observations.

In the following section, based on~\cite{Martinelli:2022elq}, we present a forecast of the capability of the Einstein Telescope to identify and measure the abundance of a subdominant population of distant PBHs. The study is based on the generation and analysis of realistic mock catalogues of the luminosity distances and errors that would be obtained from GW signals observed by the Einstein Telescope.\\

\section{Prospects for the Einstein Telescope\footnote{Adapted from \cite{Martinelli:2022elq}}}
\label{sec:GWPBH:ET}

The most up-to-date configuration of the planned ET facility is known as ET-D~\cite{Hild:2011np}, which proposes a three-armed observatory consisting of three interferometers arranged in an equilateral triangle. In the ET-D configuration, each detector is in fact made up of a \textit{pair} of detectors -- one sensitive to a lower frequency range and the other to a higher frequency range -- thus greatly increasing the overall sensitivity of the instrument with respect to the current generation of GW observatories. 
Furthermore, the triangular shape of the observatory will enable improved sky localisation of GW events~\cite{Vitale:2016icu}. Lastly, the interferometers and detectors will all be constructed underground, in an effort to reduce seismic noise~\cite{2011GReGr..43..623B}. 

\begin{figure}
	\centering
	\includegraphics[width=0.6\textwidth]{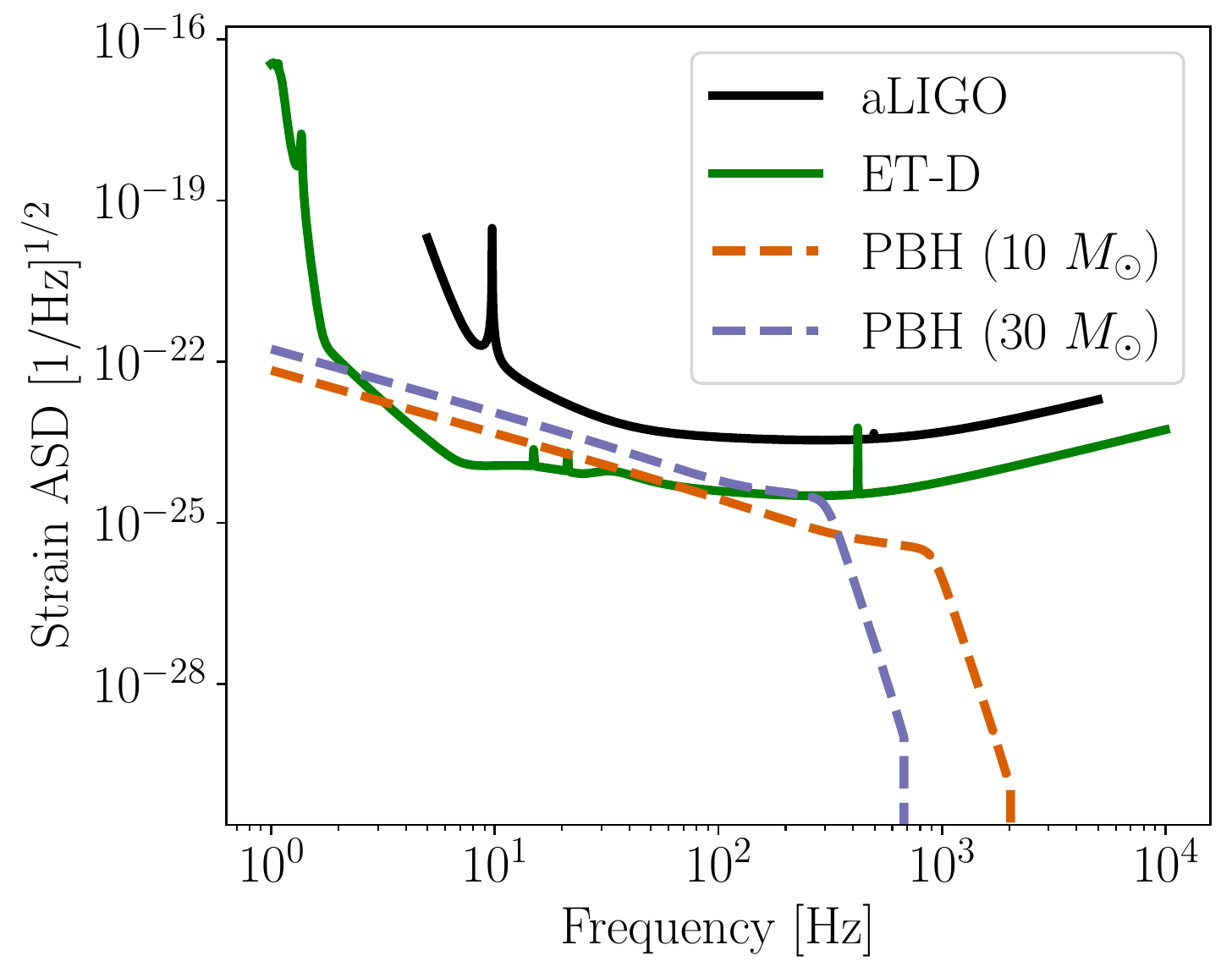}
	\caption{The strain amplitude spectral densities (ASD) of advanced LIGO (black) and the ET-D configuration of the Einstein Telescope (green), along with the characteristic strain of two PBH merger events, with component masses of \SI{10}{\solarmass} (orange) and \SI{30}{\solarmass} (purple) taking place at $z=1$, that is, with a luminosity distance $D\approx\SI{6.8}{\giga\parsec}$. } 
	\label{fig:ET_sensitivity}
\end{figure}

The less noise in the detector, the smaller the amplitude of GWs --~or \textit{strain}~-- that can be detected.  This noise is typically quantified by the strain amplitude spectral density (ASD), which we show for advanced LIGO\footnote{\url{https://dcc.ligo.org/LIGO-T1800042/public}.} (aLIGO) and ET-D\footnote{\url{http://www.et-gw.eu/index.php/etsensitivities}.} in \cref{fig:ET_sensitivity}. These ASD curves effectively show the lowest GW strain that can be detected and we show for comparison the characteristic strains of two PBH merger events with the example masses we use in the rest of this work: \SI{10}{\solarmass} and \SI{30}{\solarmass}. From this plot, we can see that ET-D will be sensitive to strains around two orders of magnitude smaller than what aLIGO can currently detect, and will also probe a much broader range of frequencies. 

With all these factors taken together, the ET-D design is expected to yield many more observations of GWs at ever-greater cosmic distances (or redshifts) than the current generation of terrestrial detectors. 

In the previuos section, we highlighted the crucial difference between the \emph{redshift evolution} of the ABH and PBH merger rates. While there is broad consensus on a steeply decreasing rate for ABHs beyond $z \simeq 2$ (see e.g. \cite{Dvorkin:2016wac} and references therein), the PBH merger rate is  instead expected, to be a \emph{monotonically increasing} function of redshift: the distribution of the orbital parameters for these primordial binaries peaks at low values of both semi-major axis and angular momentum, and therefore low values of the merger time.

This naturally suggests a focus on high redshift GWs for the purposes of detecting PBHs~\cite{Koushiappas:2017kqm}. They may manifest as resolved individual events~\cite{Chen:2019irf,DeLuca:2021wjr,DeLuca:2021hde,Ng:2021sqn}, but also as a measurable contribution to the stochastic GW background, as recently pointed out in \cite{Atal:2022zux,Braglia:2022icu}. Here we focus on the former case: the prospects for detecting an anomalously large number of \emph{individual resolved events} at large cosmic distances. 

Specifically, we propose (i) a careful theoretical modelling of both the PBH and ABH merger rates; and (ii) a realistic assessment of the ability of future observatories -- specifically, the ET -- to disentangle the two populations.

As far as modelling is concerned, a useful starting point is the formalism for the PBH merger rate developed in \cite{Nakamura:1997sm,Ioka:1998nz,Ali-Haimoud:2017rtz}, based on the assumption of an initially uniform and isotropic distribution of PBHs all having the same mass (in other words, a ``monochromatic'' mass function). In this scenario, neighbouring PBHs may decouple from the Hubble flow before matter--radiation equality, forming binaries which merge with an ever decreasing rate throughout the age of the Universe. We extend this formalism to include the effects of the early-time formation of PBH clusters~\cite{Chisholm:2005vm,Chisholm:2011kn,Raidal:2018bbj} on the evolution of the binaries. The ``background'' to these PBH merger events is provided by the mergers of ABHs. The ABH merger rate traces the binary ABH birth rate, with some delay, which in turn is expected to trace the star formation rate, again with some delay~\cite{Dvorkin:2016wac}. A careful description of PBH and ABH merger rates as a function of redshift is therefore crucial in understanding how well future GW observatories can distinguish these two populations. 

In this analysis, we make the conservative imposition of monochromatic mass functions for both the ABH and PBH populations -- in other words, all the BHs in a given population have the same mass -- and focus on the redshift dependence of the merger rates as the only discriminant between the two. 

The uncertainty in the measured luminosity distances of events at high redshift is also expected to play an important role in separating the two populations of BHs. The authors of~\cite{Ng:2021sqn} simulate the response of different detector networks located in Europe and the USA, and assess whether a single-event-based PBH identification can be unambiguous, if the redshift is large enough. The authors conclude that the typical redshift measurement is not precise enough to conclude with certainty that a single source is of primordial origin.

Motivated by these results, we present a framework that implements a statistically sound assessment of the capability of ET to:
\begin{itemize}
	\item \emph{ Detect} an excess of merger events at high redshift with respect to the astrophysical expectation;
	\item \emph{ Measure} the $f_\mathrm{ PBH}$ associated with the detection, if present, or constrain $f_\mathrm{ PBH}$ to lie within some range.
\end{itemize}

To this aim, we generate a mock data set associated to the null hypothesis of $f_\mathrm{ PBH} = 0$, and several data sets associated to different PBH fractions. We analyse these data using two different methods. 
First, we use an intuitive two-bin approach to determine the minimum abundance of PBHs that can be detected as a significant ``excess'' with respect to the astrophysical background, identifying the optimal binning in redshift.
Next, we develop a parameter estimation pipeline, in which we estimate the posterior distribution of $f_\mathrm{ PBH}$, by comparing the mock data generated for different fiducial values of $f_\mathrm{ PBH}$ with the theoretical distribution of event distances.

The code associated with this work~\cite{Martinelli:2022elq}, \texttt{darksirens},\footnote{\url{https://darksirens.readthedocs.io}} is publicly available on GitLab \href{https://gitlab.com/matmartinelli/darksirens}{\faGitlab}.

\paragraph{Conventions, notation, cosmology.} We assume a spatially flat \LCDM cosmology throughout, using $H_0 = \SI{67.4}{\kilo\meter\per\second\per\mega\parsec}$ and $\Omega_\mathrm{ m} = 0.315$ as reported by the \emph{Planck} collaboration~\cite{Aghanim:2018eyx}. A three-bar equality sign ($\equiv$) indicates a definition; an upper-case $P$ indicates a probability, while a lower-case $p$ indicates a probability density function (PDF); bold symbols ($\vect{x}, \vect{\theta}, \vect{\mathcal{A}}$, \ldots) stand for vectors or matrices. The luminosity distance is simply denoted with a $D$ throughout, because we use no other notion of distance. We endeavour with all heroism to keep explicit factors of $G$ and $c$ where relevant.

\subsection{Normalisation of the astrophysical merger rate}
\label{sec:GWPBH:ET:bound}

Given the models for the redshift evolution of the ABH and PBH rate density described respectively in \cref{sec:GWPBH:ABHbinaries:rate} and \cref{sec:GWPBH:PBHbinaries:MR}, we now aim to: \emph{ (i)} obtain an expression for the actual merger rate by convolving the theoretical rate estimates with the detector space-time sensitivity; \emph{ (ii)} compare the predicted merger rate for each value of \fPBH with the recent estimates provided by LVK. In this way, we can set the normalisation of the astrophysical component for each value of \fPBH, and obtain an upper limit on this quantity.

We compare the predicted PBH merger rate to the observational data from the latest LVK Gravitational-Wave Transient Catalog (GWTC-3), for different PBH mass intervals and values of \fPBH. In particular, we adopt the  merger rate density estimates reported in Table 4 of \cite{LIGOScientific:2021psn} for GWTC-3. The procedure is discussed in \cref{sec:GWPBH:PBHbinaries:data}. 

For values of \fPBH smaller than the upper limits obtained in~\cref{sec:GWPBH:PBHbinaries:data}, we adjust the ABH merger rate normalisation $\mathcal{N}_\mathrm{ABH}$ so that the sum of the averaged rates corresponds to the reported one, that is
\begin{equation}
	\langle \mathcal{R}_\mathrm{ PBH}(\fPBH)\rangle
	+ \langle \mathcal{R}_\mathrm{ ABH} (\mathcal{N}_\mathrm{ ABH})\rangle
	= \mathcal{R}_\mathrm{ obs}^\mathrm{ LVK} \,. 
\end{equation}

\subsection{Generating mock catalogues}
\label{sec:GWPBH:ET:generating_mock}

This section presents our method to produce mock data for the ET, accounting for the instrumental response and other observational effects such as lensing of the signal. Since we focus on the information contained in the redshift distribution of the GW events -- or, more accurately, their luminosity distance distribution which is the actual observable -- in the remainder of this section, a \emph{data set}~$\mathcal{D}$ will refer to $N_\mathrm{ det}$ luminosity distance measurements~$D_i$ with their uncertainty~$\sigma_i$, $\mathcal{D}=\{(D_i, \sigma_i)\}_{i=1,\ldots, N_\mathrm{ det}}$. Our mock generation code, \texttt{darksirens} \href{https://gitlab.com/matmartinelli/darksirens}{\faGitlab}, is publicly available and could be easily modified to include other observables.

\subsubsection{Parameters}

The parameters that must be set to produce a mock catalogue of GW distance measurements can be divided into four categories:
\begin{itemize}
	\item \textbf{Cosmology.} We assume a spatially flat homogeneous and isotropic \LCDM cosmological background. The parameters to be set are the Hubble--Lema\^{i}tre constant $H_0$ and the total matter density parameter~$\Omega_\mathrm{ m}$.
	\item \textbf{Primordial black holes} The PBH population is characterised by the masses of the objects, $M_\mathrm{ PBH}$, the fraction of DM made up of PBHs, $f_\mathrm{ PBH}$, and whether or not clustering of PBHs is considered (we remark once again that the effect of clustering is negligible within the current implementation). 
	\item \textbf{Astrophysical black holes.} The ABH population is characterised by the masses of the objects~$M_\mathrm{ ABH}$ and the SFR parameters $z_\mathrm{ m}$, $a$ and $b$ which enter into \cref{eq:SFR}.
	\item \textbf{Specifications.} This class of parameters allows us to set the observational time $T_\mathrm{ obs}$ of the survey and the SNR threshold, SNR$_\mathrm{ min}$, which determines if a candidate event is detected or not. It also allows us to customise details of the mock generation, i.e.~to specify the redshift range over which calculations are done ($z\in\left[z_\mathrm{ min},z_\mathrm{ max}\right]$) and whether or not to include the effect of lensing.
\end{itemize}

For the results presented in \cref{sec:GWPBH:ET:cut_and_count,sec:GWPBH:ET:likelihood_based_method}, we fix the parameters used to generate the mock data to the fiducial values listed in \cref{tab:fiducials}, unless otherwise specified in the text. We report the survey specifications assumed for the ET observations in  \cref{tab:specs}.

\begin{table}[ht]
	\centering{}%
	\begin{tabular}{|c|c||c|c|c|c||c|c|}
		\hline
		\multicolumn{2}{|c||}{Cosmology} & \multicolumn{4}{c||}{ABH parameters} & \multicolumn{2}{c|}{PBH parameters} \\
		\hline
		$\Omega_\mathrm{ m}$  & $H_0$ [\si{\kilo\meter\per\second\per\mega\parsec}] & $M_\mathrm{ ABH}$ [$M_\odot$]  & $z_\mathrm{ m}$  & $a$    & $b$ & $M_\mathrm{ PBH}$ [$M_\odot$] & clustering \\
		\hline
		$0.315$ & $67.4$ & $7$ & $2$ & $2.37$ & $1.8$ & $10$ & yes \\
		\hline 
	\end{tabular}\protect
	\caption{Cosmological and BH related parameters accessible in \texttt{darksirens} \href{https://gitlab.com/matmartinelli/darksirens}{\faGitlab}. The values shown here represent our baseline settings used throughout the analysis. We do not report a value of $f_\mathrm{ PBH}$ as this will be changed depending on the analysis done with the data.}\label{tab:fiducials}
\end{table}

\begin{table}[ht]
	\centering{}%
	\begin{tabular}{|c|c|c|c|c|}
		\hline
		\multicolumn{5}{|c|}{Specifications} \\
		\hline
		$T_\mathrm{ obs}$ [yrs]  & SNR$_\mathrm{ min}$  & lensing & $z_\mathrm{ min}$  & $z_\mathrm{ max}$\\
		\hline 
		$1$ & $8$ & yes & $0.001$ & $100$\\
		\hline
		
	\end{tabular}\protect
	\caption{Parameters available in \texttt{darksirens} \href{https://gitlab.com/matmartinelli/darksirens}{\faGitlab} to specify the characteristics of the survey and the redshift range over which the mock is constructed. The values shown here represent our baseline settings used throughout the analysis.}\label{tab:specs}
\end{table}

\subsubsection{Sketch of the generation algorithm}

Once the parameters are chosen, \texttt{darksirens} produces a mock catalogue~$\mathcal{D}$ of GW distances with their uncertainties as they would be measured by ET. For $\fPBH \neq 0$, the data is a mix of ABH and PBH mergers, which takes into account the potential clustering of the latter, the effect of lensing on distance measurements, and ET's instrumental uncertainties. We now explain how the mock data are generated.

The very first step consists in computing the total number $N_\mathrm{ tot}$ of mergers occurring in the redshift range $[z_\mathrm{min}, z_\mathrm{max}]$. We draw a random number from a Poisson distribution whose mean, $\bar{N}_\mathrm{ tot}$, is given by \cref{eq:N_bar_tot}, in which we consider the total merger rate~$R(z)=R_\mathrm{ ABH}(z)+R_\mathrm{ PBH}(z)$ as described in \cref{sec:GWPBH:PBHbinaries:Obs_rate}. \\
Each data point~$i = 1,\ldots,N_\mathrm{ tot}$ is then produced as follows:
\begin{enumerate}
	\item Randomly draw the ``true'' redshift~$z_i$ of the merger from the redshift distribution $p(z)$ given in \cref{eq:prob_redshift}, once again considering the total merger rate.
	\item Convert $z_i$ into the ``true'' unlensed luminosity distance~$\bar{D}_i\equiv \bar{D}(z_i)$ of the merger, using the background cosmological model and \texttt{CAMB}.
	\item Compute the unlensed signal-to-noise ratio (SNR)~$\bar{\rho}_i$ of the event. The position, polarisation and inclination of the event are drawn randomly and the SNR computed based on these quantities along with the specifications of the ET. 
	The full details of this calculation are given in \cref{sec:appendix:mockdata}.
	\item Determine the weak gravitational lensing magnification~$\mu_i$ of the event by randomly drawing it from the theoretical PDF $p(\mu)$ -- see \cref{sec:appendix:mockdata:lensing} for details on lensing and its statistics. The lensing magnification enhances (or reduces) the SNR as $\rho_i = \sqrt{\mu_i} \bar{\rho}_i$, and reduces (or increases) the luminosity distance as $\tilde{D}_i = \bar{D}_i/\sqrt{\mu_i}$.
	\item Determine whether the SNR is large enough for the event to be properly detected: if $\rho_i < 8$, then the event is considered to be too faint to be a true GW candidate and is removed from the catalogue. The choice of discarding events with an SNR smaller than eight follows the approach taken by the LIGO collaboration as an estimate for the detection threshold~\cite{Abbott:2016xvh}.
	\label{step:SNR_cut}
	\item Compute the measured luminosity distance,
	$
	D_i = \tilde{D}_i + \Delta D_i,
	$
	where $\Delta D_i$ represents the instrumental error on the measurement. The latter is drawn from a Gaussian distribution $\mathcal{N}(0, \sigma^\mathrm{ inst}_i)$, with error inversely proportional to the SNR, $\sigma^\mathrm{ inst}_i = 2 \tilde{D}_i/\rho_i$~\cite{Li:2013lza}. 
	\item Compute the total uncertainty~$\sigma_i$ on this data point as the quadratic sum of the instrumental and lensing uncertainties,
	\begin{equation}
		\sigma_i^2
		= \left(\sigma_i^\mathrm{ inst}\right)^2 + \left(\sigma_i^\mathrm{ lens}\right)^2
		= \left( \frac{2\tilde{D}_i}{\rho_i} \right)^2
		+ \sigma_\kappa^2(z_i) \, \bar{D}_i^2 \ ,
		\label{eq:distance_error}
	\end{equation}
	where $\sigma_\kappa^2(z)$ is the variance of the weak-lensing convergence -- see \cref{sec:appendix:mockdata:lensing} for details.
\end{enumerate}

\begin{figure}[ht!]
	\centering
	\includegraphics[width=0.8\columnwidth]{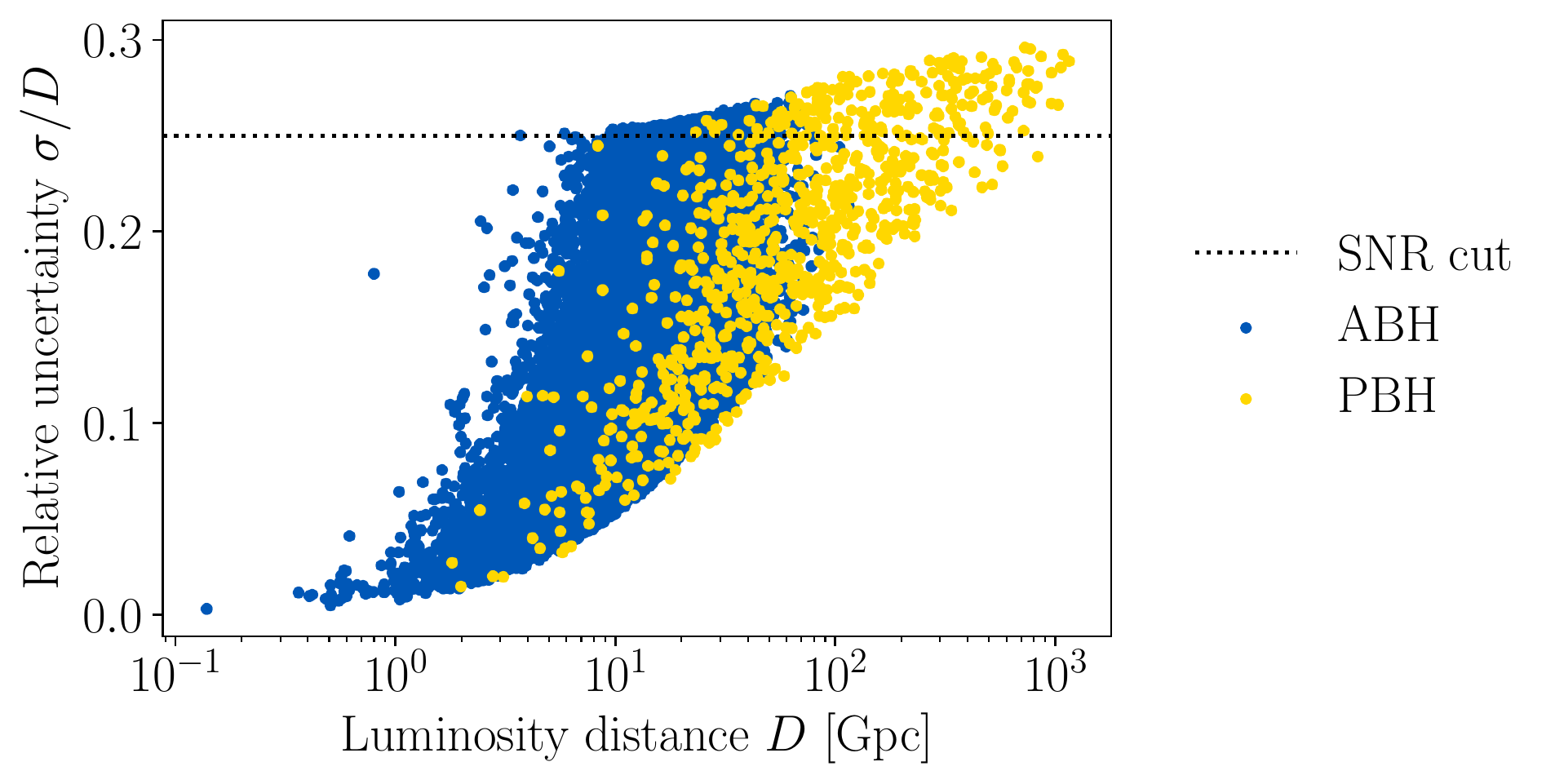}
	\caption{Data set generated using \texttt{darksirens} \href{https://gitlab.com/matmartinelli/darksirens}{\faGitlab} with the baseline parameters of \cref{tab:fiducials} and \cref{tab:specs}, with $f_\mathrm{ PBH}=10^{-5}$. The blue dots show the observed ABH events, while the yellow dots represent the observed PBH events. At a fixed distance~$D$, the typical uncertainty is smaller for PBHs than for ABHs due to our choice of a larger PBH mass ($M_\mathrm{PBH} = 10\,M_\odot$, $M_\mathrm{ABH} = 7\,M_\odot$). The dotted horizontal line shows the level of instrumental uncertainty corresponding to our SNR cut of $\rho_i = 8$. The events that lie above this line still satisfy the SNR condition we impose, 
		but receive an extra contribution to the error from lensing.}
	\label{fig:mockplot}
\end{figure}

\noindent
The end product is a catalogue of distance measurements with their uncertainties, $\mathcal{D}=\{(D_i, \sigma_i)\}_{i=1,\ldots, N_\mathrm{ det}}$, where $N_\mathrm{ det}\leq N_\mathrm{ tot}$ is the number of events that survive the SNR cut of step~\ref{step:SNR_cut}. We have checked that $N_\mathrm{ det}$ agrees on average with the theoretical expectation~$\bar{N}_\mathrm{ det}$. \Cref{fig:mockplot} shows an example of a catalogue that can be obtained using this approach. The figure shows the distance $D_i$ of the events that survive the SNR cut,\footnote{A plot including the events which would not survive the SNR cut is shown in \cref{fig:SNRplot}.} together with their relative error, for both ABH (blue) and PBH (yellow). The data shown here are obtained setting $f_\mathrm{ PBH}=10^{-5}$.

\subsubsection{Distance uncertainty}

Correctly modelling uncertainties in the luminosity distance of the GW events is crucial to disentangling high redshift and low redshift populations of mergers. A full analysis pipeline based on real GW data would provide as an output a probability distribution for the true luminosity distance $P(\bar{D})$ of the merger. Instead, in our catalogues, the distance uncertainty is described by only a single number $\sigma_i$, given in \cref{eq:distance_error}. We therefore model the probability distribution for the luminosity distance as a Gaussian, given explicitly as:\\
\begin{equation}
	p(\bar{D}|D_i) = \frac{1}{\sqrt{2 \pi} \sigma_i} \exp\left[- \frac{(\bar{D} - D_i)^2}{2\sigma_i^2}\right]\,.
	\label{eq:Gaussian_error}
\end{equation}

\medskip

We will now present two different analysis methods of the mock data we have generated:
an intuitive binned approach, and a more complete Bayesian analysis of the data, in \cref{sec:GWPBH:ET:cut_and_count} and \cref{sec:GWPBH:ET:likelihood_based_method} respectively. With the first method we aim to estimate the detectability threshold of a PBH population, while with the second method we also assess the capability of the ET to measure the PBH fraction. We refer to these methods as ``cut-and-count'' and ``likelihood-based'' respectively.

\subsection{Detecting PBHs: the cut-and-count method}
\label{sec:GWPBH:ET:cut_and_count}

In this section, we present the ``cut-and-count'' method which we use to determine the smallest PBH fraction that could be detected by the ET. The idea is simple and intuitive: since we expect no ABHs to be formed beyond some high redshift (as there is a significant delay between the beginning of the Universe, the birth and death of the first stars and hence the formation and merging of ABHs), any sufficiently high redshift GW event produced by a BH merger should be the result of merging \textit{primordial} BHs. 

In practice, however, large distances are also the most uncertain ones, since the GW signal is typically much fainter than closer events. Hence the question we aim to answer is better phrased as: what is the lowest value of \fPBH that would still produce a sufficient number of large-distance events so that they would be statistically distinguishable from $\fPBH=0$ when observing with the ET? We hence \textit{cut} the data set into two subsets containing the small- and large-distance events, \textit{count} the number of events in the large-distance subset, and compare to the expected number for $\fPBH=0$. This method only exploits a fraction of the available information in the data set, but it is nevertheless a natural starting point.

\subsubsection{Description of the method}
\label{sec:GWPBH:ET:cut_and_count:method}

Let $D_* \equiv \bar{D}(z_*)$ be an arbitrary distance threshold ($z_*$ is the corresponding background redshift threshold). Given a data set~$\mathcal{D}$, we divide it into two subsets: a small-distance subset $\mathcal{D}_\leq$ on the one hand, made of the events with distances beneath the threshold ($D\leq D_*$); and a large-distance subset~$\mathcal{D}_>$ on the other hand, with events beyond the threshold ($D> D_*$). We call $N_>$ the cardinal of $\mathcal{D}_>$, i.e., the number of events above the threshold~$D_*$; the larger \fPBH, the larger the expected~$N_>$.

To be more specific, the computation of $N_>$ does not simply consist in counting the number of events whose best-fit distance is above the threshold. Due the measurement errors, it may happen that an event truly lies beneath the distance threshold ($\bar{D}_i < D_*$) but is actually measured beyond it ($D_i > D_*$), or vice versa. In order to account for this, we calculate the probability that the true distance lies beyond the arbitrary distance threshold as
\begin{equation}\label{eq:probdens}
	P_i
	=
	\frac{1}{\sqrt{2\pi}\sigma_i}
	\int_{D_*}^\infty{\mathrm{ d}\bar{D}_i \; \exp\left[-\frac{(\bar{D}_i-D_i)^2}{2\sigma_i^2}\right]}\, ,
\end{equation}
using the definition of the uncertainty from \cref{eq:Gaussian_error}.
We can therefore obtain the value of $N_>$ by summing over all the events the probability of falling within the bin being considered,
\begin{equation}\label{eq:stat_counts}
	N_> \equiv \sum_{i=1}^{N_\mathrm{ det}}{P_i}\,.
\end{equation}

The uncertainty on $N_>$ is twofold. First, since the data set is discrete, it inevitably comes with a Poisson uncertainty with variance
\begin{equation}
	\sigma^2_\mathrm{ P} = \bar{N}_> \ ,
\end{equation}
where $\bar{N}_{>}$ is the theoretical expectation value of $N_>$.\footnote{In practice,
	we estimate $\bar{N}_>$ by generating a number of mock data sets and taking the mean of the $N_>$ values obtained for each.} On the other hand, the number~$N_>$ of large-distance events may be seen as the sum of $N_\mathrm{ det}$ independent Bernoulli variables, the $i^\mathrm{ th}$ one having a probability $P_i$ of being equal to $1$, and a probability $1-P_i$ of being equal to zero. This observational contribution to the variance of $N_>$ is the sum of the individual Bernoulli variances,
\begin{equation}
	\sigma_\mathrm{ B}^2 = \sum_{i=1}^{N_\mathrm{ det}}{P_i(1-P_i)}\ .
\end{equation}
Summing the two sources of uncertainty in quadrature yields the total error on $N_>$,
\begin{equation}\label{eq:binerror}
	\sigma_{>}
	= \sqrt{\sigma^2_\mathrm{ P} + \sigma_\mathrm{ B}^2} \ .
\end{equation}

\begin{figure}[ht!]
	\centering
	\includegraphics[width=0.6\columnwidth]{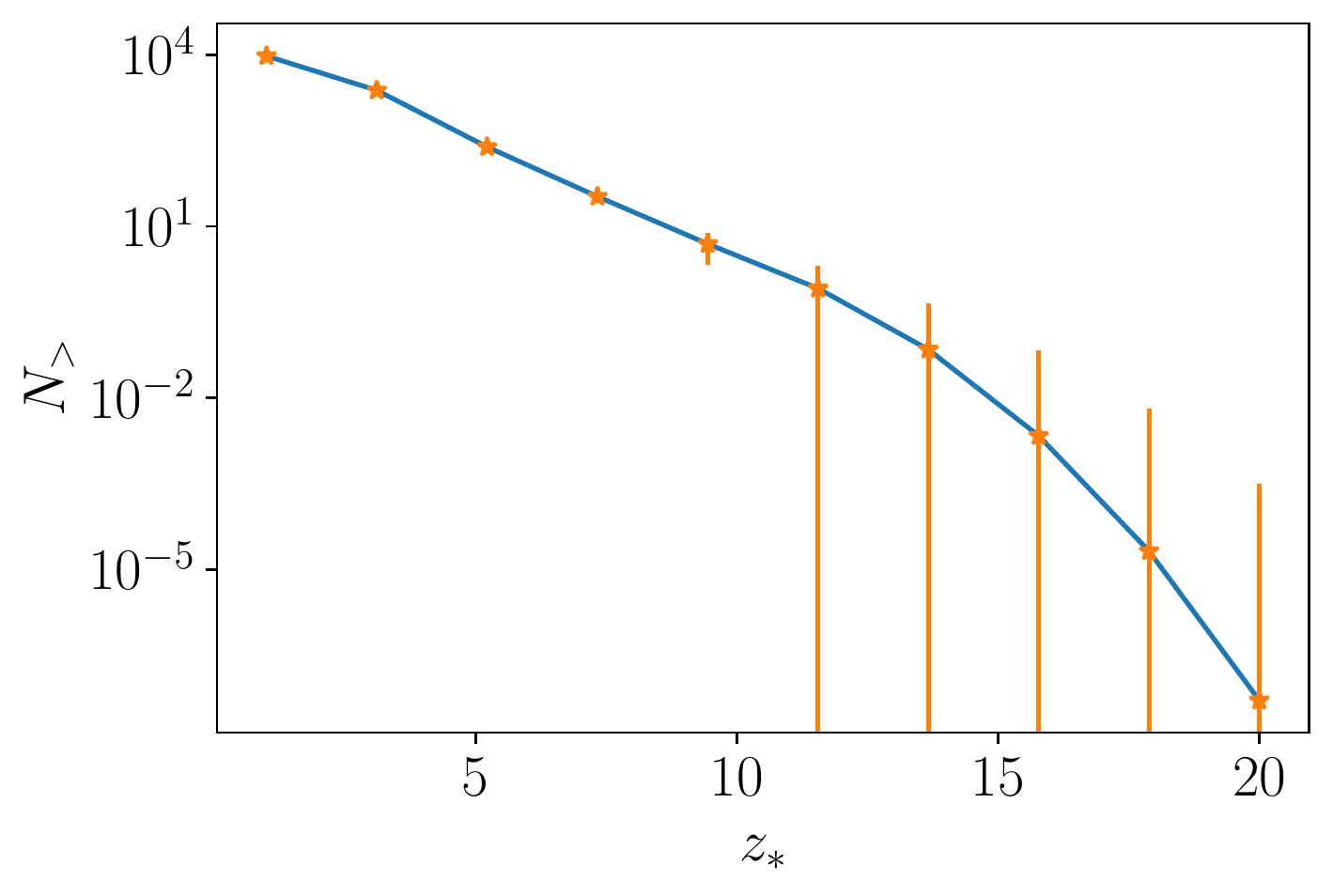}
	\caption{Value of $N_>$ as a function of the redshift threshold~$z_*$, for a data set with a negligible number of PBH. The error bars represent the uncertainty~$\sigma_>$ on $N_>$.}
	\label{fig:N_>_no_PBH}
\end{figure}

As an example, we consider the null case of a data set~$\mathcal{D}_0$ containing no PBH ($\fPBH \to 0$). \Cref{fig:N_>_no_PBH} shows the expected value of $N_>$ as a function of the redshift threshold~$z_*$, to which the $N_>$ of an observed data set should be compared. The error bars of that figure are computed following the approach described aobve, and they represent the uncertainty~$\sigma_>$ on $N_>$. 

\subsubsection{Optimal distance threshold and smallest detectable PBH fraction}
\label{sec:GWPBH:ET:cut_and_count:results}

Let us now determine what is the smallest detectable PBH fraction~\fPBH that could be detected using the cut-and-count method. This will also be the occasion to determine the optimal choice for the arbitrary distance threshold, that is, the value of $D_*$ leading to the maximal sensitivity in \fPBH.

For that purpose, we generate mock data sets~$\mathcal{D}_{\fPBH}$ for $100$ values of \fPBH ranging from $10^{-6}$ to $10^{-2}$. For each of these data sets, we apply the cut-and-count method for 10 different values of the redshift threshold~$z_*$, taken as equispaced in the range $[5, 50]$. This yields 
a number $N_>(\mathcal{D}_{\fPBH}, z_*)$ for each mock data set and each value of $z_*$. Those numbers come with their own uncertainty~$\sigma_>(\mathcal{D}_{\fPBH}, z_*)$, computed as described above for the null case~$\mathcal{D}_0$. We estimate the statistical significance of the detection of a non-zero \fPBH as
\begin{equation}\label{eq:detfac}
	\mathcal{S}(\mathcal{D}_{\fPBH}, z_*)
	\equiv
	\frac{\left| N_>(\mathcal{D}_{\fPBH}, z_*) - N_>(\mathcal{D}_0, z_*) \right|}
	{\sqrt{\sigma^2_{>}(\mathcal{D}_{\fPBH}, z_*) + \sigma^2_{>}(\mathcal{D}_0, z_*)}}\ ,
\end{equation}
which represents the ``number of sigmas'' with which we could claim a detection.

\begin{figure}[ht!]
	\centering
	\begin{tabular}{cc}
		\includegraphics[height=5.6cm]{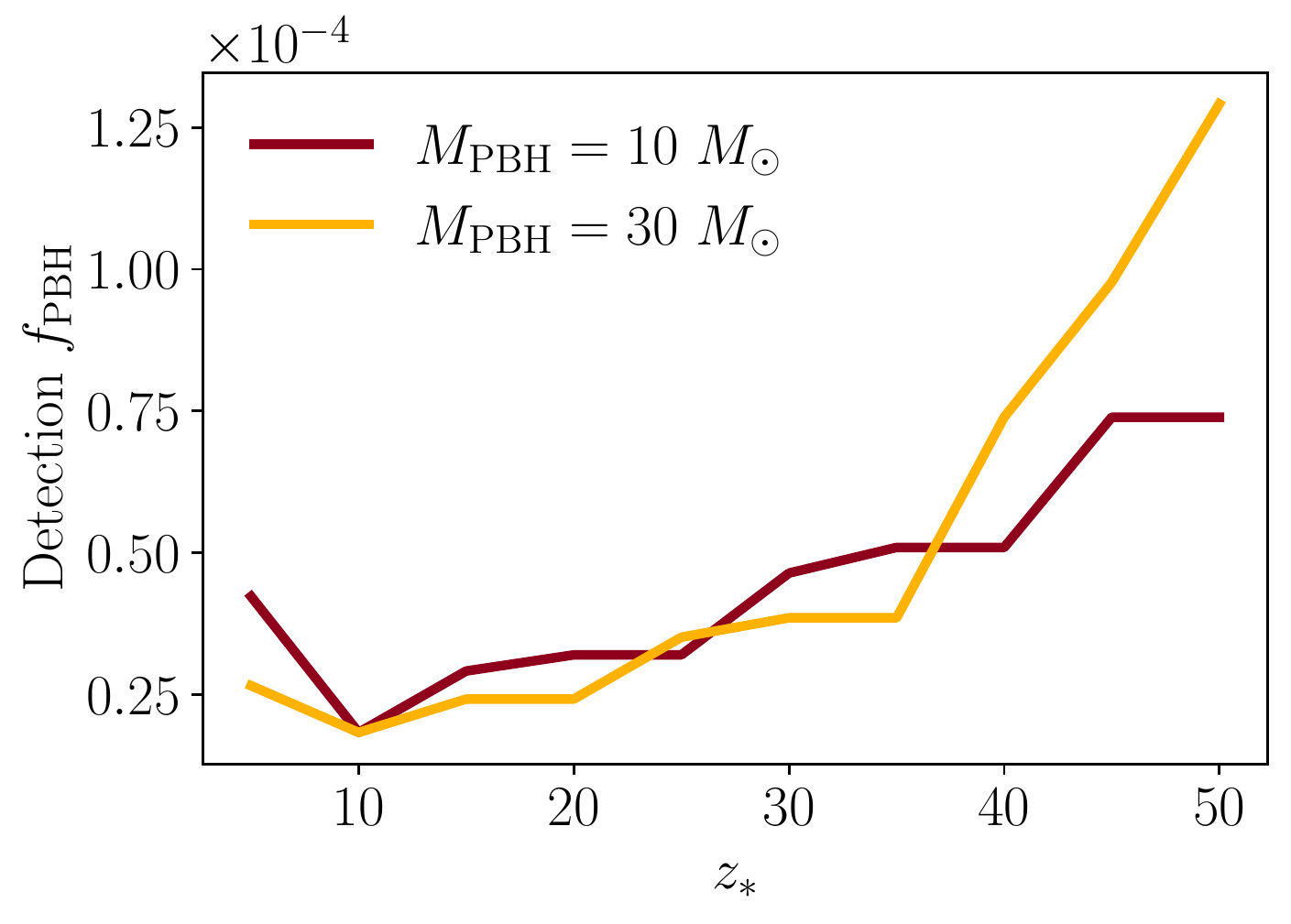} &  
		\includegraphics[height=5.6cm]{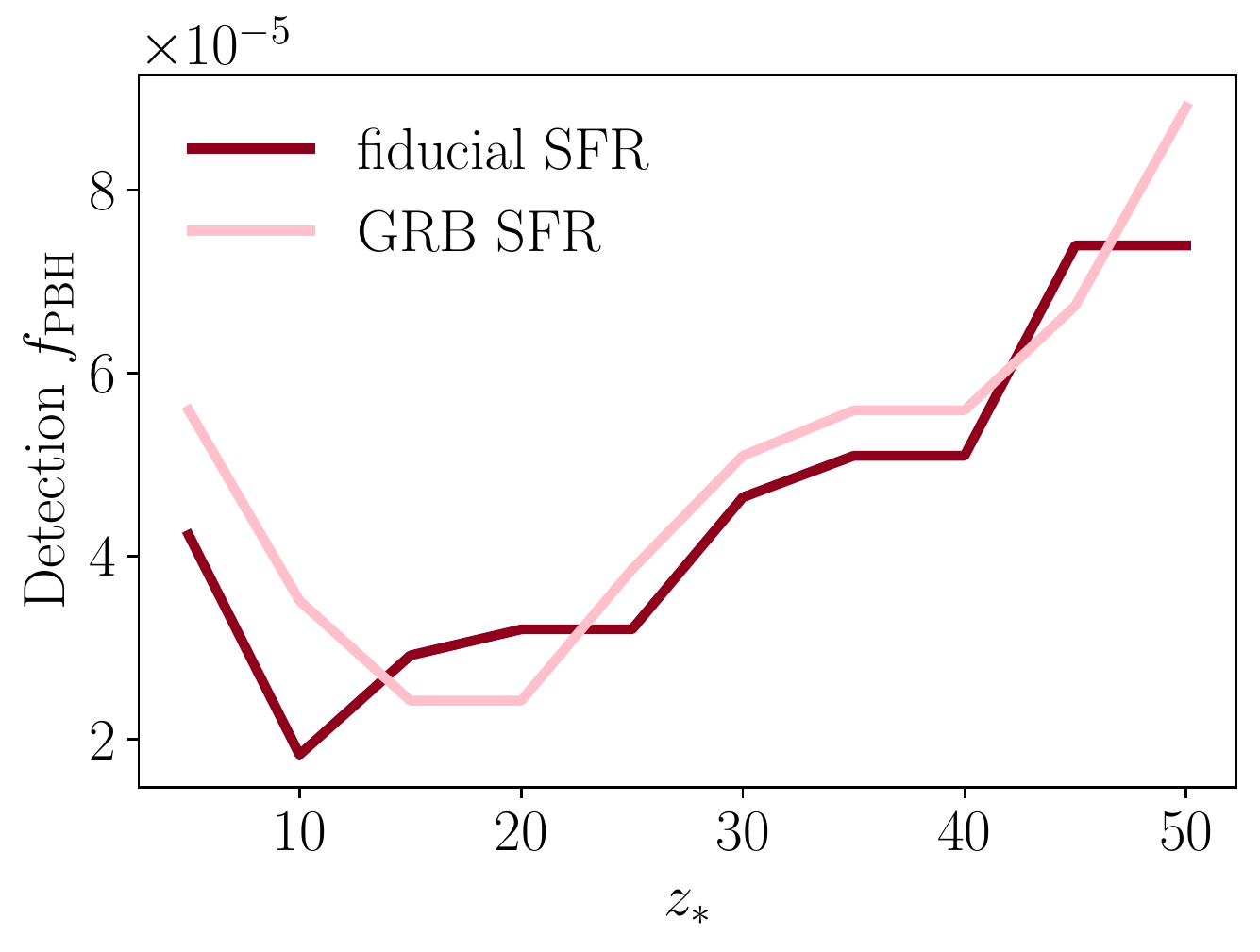}\\
		\multicolumn{2}{c}{\includegraphics[height=5.6cm]{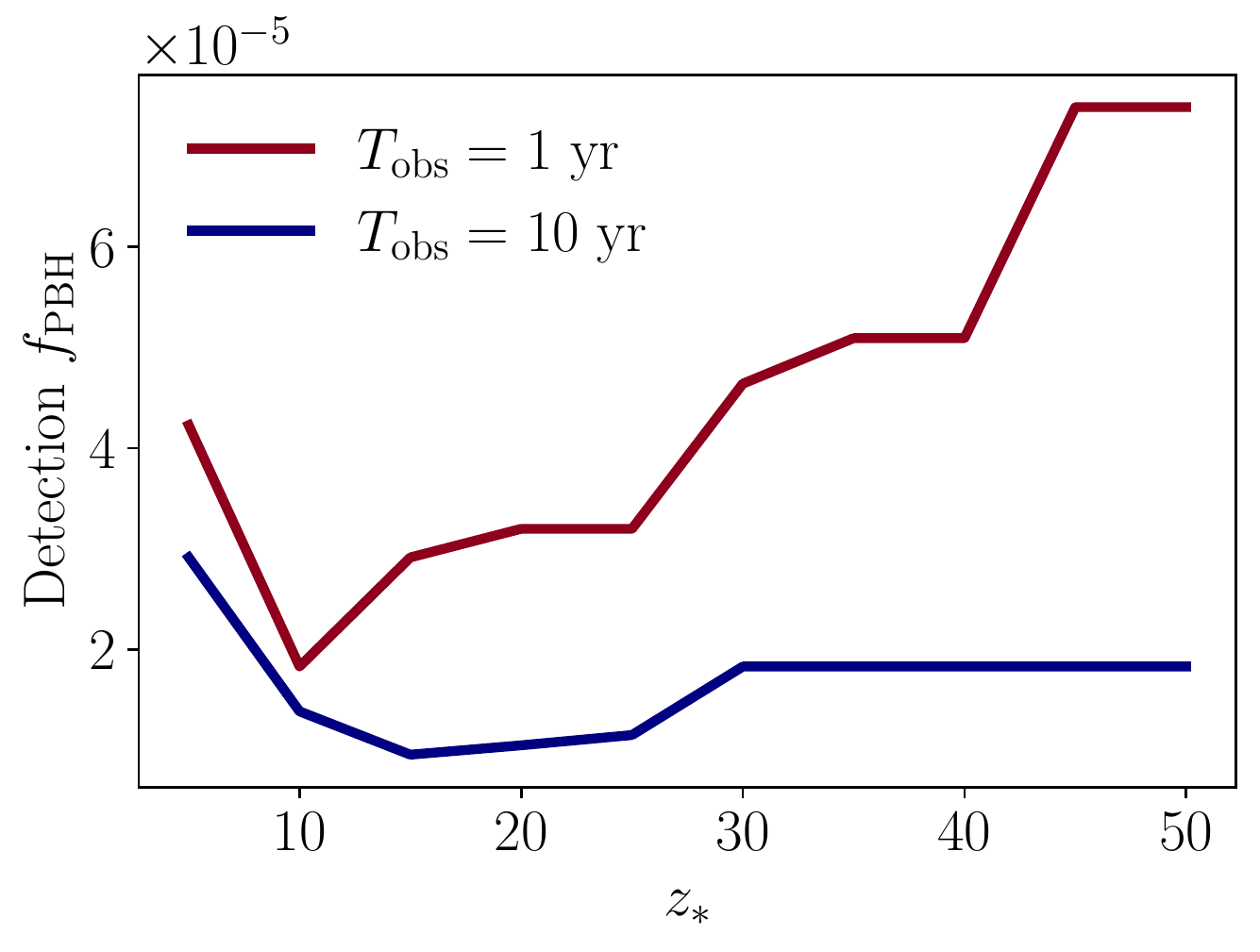}}
	\end{tabular}
	\caption{
		\textbf{ Top left}: Lowest $3\sigma$-detectable $f_\mathrm{ PBH}$ as a function of the redshift~$z_*$ discriminating between the low redshift and high redshift subsets, for $M_\mathrm{ PBH}=10\,M_\odot$ (red lines) and $M_\mathrm{ PBH}=30\,M_\odot$ (yellow lines). 
		\textbf{ Top right}: Lowest $3\sigma$-detectable $f_\mathrm{ PBH}$ as a function of $z_*$, for $M_\mathrm{ PBH}=10\,M_\odot$ for the fiducial ABH merger rate (red line) and for the one obtained using the GRB SFR (pink line).
		\textbf{ Bottom}: Lowest $3\sigma$-detectable $f_\mathrm{ PBH}$ as a function of the chosen $z_*$, for $M_\mathrm{ PBH}=10\,M_\odot$ for an observation time $T_\mathrm{ obs}=\SI{1}{\year}$ (red line) and for $T_\mathrm{ obs}=\SI{10}{\year}$ (blue line).}
	\label{fig:detection_zT}
\end{figure}

In the following, we consider a ``significant'' detection to be one made at $3\sigma$; in other words, for each value of $z_*$, the smallest detectable \fPBH is determined by finding the data set $\mathcal{D}_{\fPBH}$ such that $\mathcal{S}(\mathcal{D}_{\fPBH}, z_*)>3$. In the top left panel of \cref{fig:detection_zT}, both curves (for $M_\mathrm{ PBH} = \SI{10}{\solarmass}$ in red and $M_\mathrm{ PBH} = \SI{30}{\solarmass}$ in yellow) show the evolution of the smallest detectable \fPBH with $z_*$. The optimal value for the redshift threshold, leading to the best sensitivity in \fPBH, is found to be $z_*\approx 10$, corresponding to a distance of $D_*\approx\SI{106}{\giga\parsec}$.

The existence of an optimal value for $z_*$ arises from a trade-off between two effects: on the one hand, increasing $z_*$ reduces the contamination due to ABH events in the high redshift subset; on the other hand, reducing $z_*$ increases the number of events in that high redshift subset and hence the significance of the detection. 
Our results suggests that choosing $z_*<10$ implies too much ABH contamination to easily assess whether or not PBHs are present in the data set, while $z_*>10$ worsens the statistics and only provides a detection for large values of \fPBH.
These conclusions are valid for both $M_\mathrm{ PBH}=\SI{10}{\solarmass}$ and $M_\mathrm{ PBH}=\SI{30}{\solarmass}$, although the exact value of the optimal $z_*$ does change when the mass of the PBHs is changed.

We expect the SFR model to affect the results, as it determines how fast the probability of ABH mergers vanishes with redshift, see \cref{fig:MergerRateAstro}. We verify the impact of this uncertainty by repeating our analysis using the merger rate obtained assuming the GRB SFR model.
In the top right panel of \cref{fig:detection_zT}, we compare the GRB SFR results with the baseline case ($M_\mathrm{ PBH}=10\ M_\odot$, fiducial SFR and $T_\mathrm{ obs}=1$ yr). We can see how in the GRB case, the optimal threshold redshift~$z_*$ increases due to the ABH merger probability being non-vanishing up to higher redshifts than in the fiducial case. We also explore the impact of the observation time of the survey on these results; we compare our baseline results with an extended survey time for ET, setting $T_\mathrm{ obs}=10$ yr. As can be seen in the bottom panel of \cref{fig:detection_zT}, increasing the time of the survey generally leads to a lower detection threshold for $f_\mathrm{  PBH}$, while also decreasing the dependency of such a threshold on the choice of the binning strategy. With a greater survey time, more high redshift events are observed, meaning that a large detection significance is still achievable even as the high redshift bin is moved to larger values of $z_*$. 

Overall, we find that a fraction $\fPBH\approx2\times10^{-5}$ is the lowest value detectable by the ET with this method for $T_\mathrm{ obs}=1$ year, with a weak dependency on the mass of the progenitor systems and on the SFR model chosen. In our baseline case, for the optimal $z_*\approx10$, such a fraction of PBH corresponds to $N_>=16\pm4.6$, while the no PBH case yields $N_>=1\pm1.7$, a result that puts this \fPBH over the $3\sigma$ threshold we consider for detection.

\begin{figure}[ht!]
	\centering
	\includegraphics[width=0.49\textwidth]{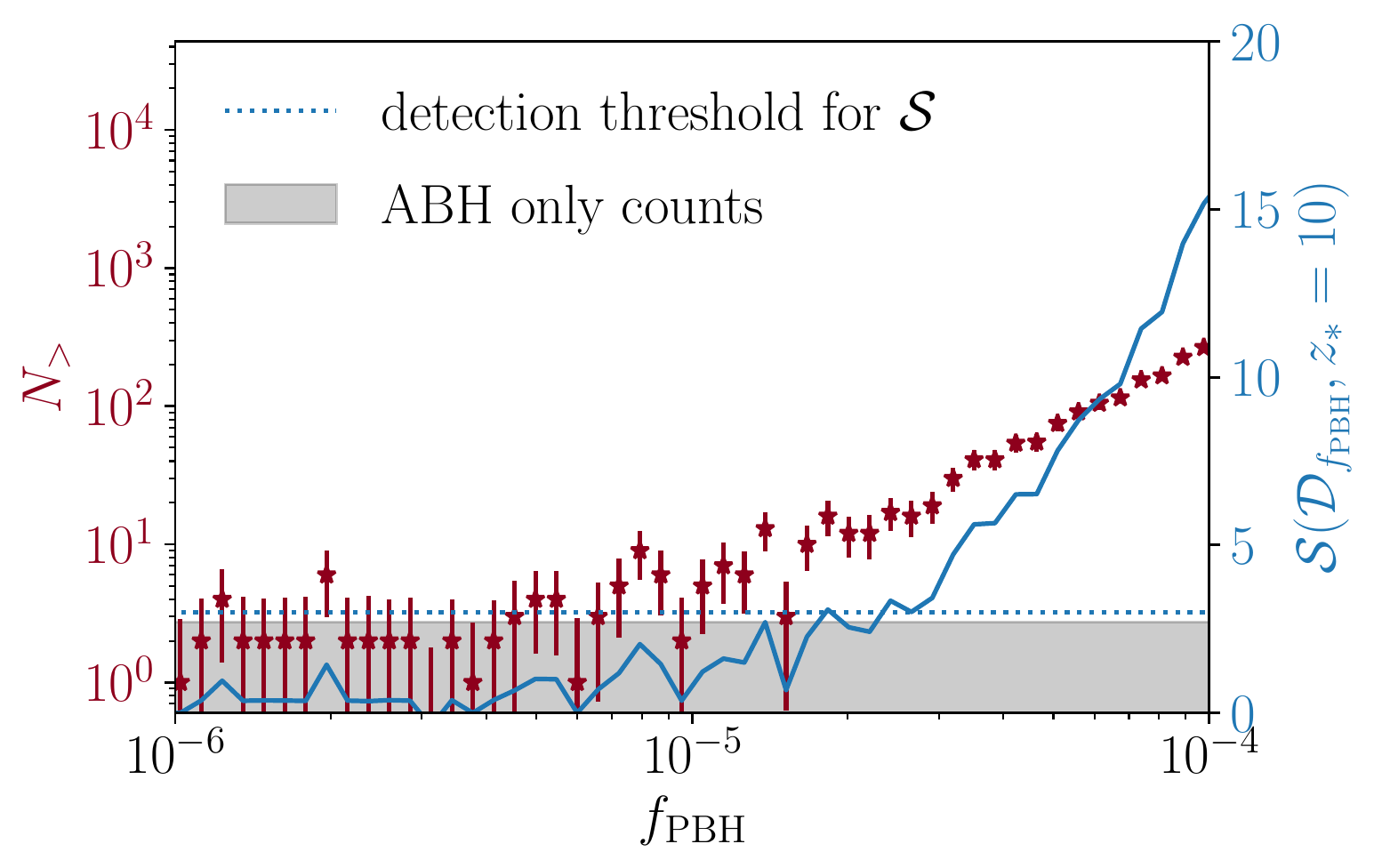}
	\hfill
	\includegraphics[width=0.49\textwidth]{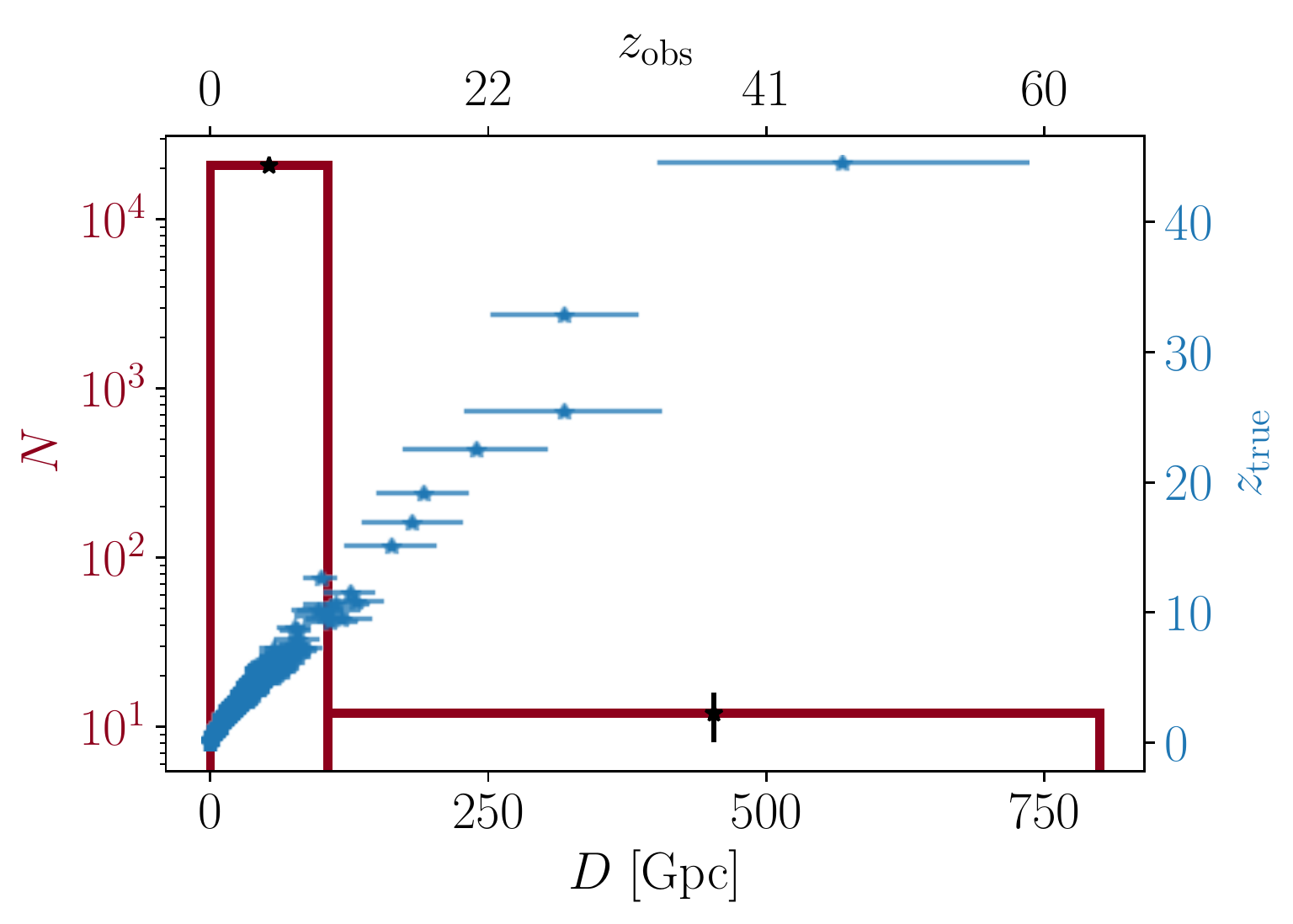}
	\caption{\textbf{ Left}: Number $N_>$ of events in the high redshift subset, $z>z_*=10$, for each value of $f_\mathrm{ PBH}$ in the baseline case (see \cref{tab:fiducials} and \cref{tab:specs}) (red points). The grey band shows the uncertainty for $N_>$ in the no-PBH case. The blue line shows instead the trend of the detection significance for a departure from the no-PBH case as a function of $f_\mathrm{ PBH}$. 
		\textbf{ Right}: Counts in the small- and large-distance bins are shown by the red bar plot for the baseline case, with $z_*=10$ and $f_\mathrm{ PBH}=2\times10^{-5}$, together with the errors we determine on the value of the counts (left $y$-axis). In blue, the distances and their errors for all the events contained in the data set are shown as a function of the true redshift $z_\mathrm{ true}$ (right $y$-axis), while the top $x$-axis shows the redshift $z_\mathrm{ obs}$ that could potentially be inferred from the distance measurements.}\label{fig:best_baseline}
\end{figure}

\Cref{fig:best_baseline} shows more details for the results obtained in our baseline case, when taking $z_*=10$. The left panel shows in red the results for $N_>$ for the sampled values of \fPBH, together with their error bars and the uncertainty on $N_>$ in the no PBH case (grey band). In this plot, we also show the trend of the detection significance $\mathcal{S}$ with \fPBH (blue line). The right panel shows the counts in the two redshift bins for $\fPBH=2\times10^{-5}$ in red, i.e. the first value for which the counts in the large-distance bin result in $\mathcal{S}\geq3$. Together with the counts we also show  the distance and its error for the events contained in this mock data set, as a function of the event redshifts (which would not be observed in reality).

The conclusions we have found for our baseline case using this cut-and-count approach are compatible with those that can be obtained using the approach of \cite{Ng:2021sqn}, where the discriminant for the presence of PBH in the observed data set is the observation of events at $z>30$, where the contribution of ABHs is negligible. With our simulated data we indeed find that $f_\mathrm{ PBH}\approx2\times10^{-5}$ is the lowest value in the baseline case for which we have at least one event above $z=30$ with $99.7\%$ confidence level.

\subsection{Quantifying the PBH fraction: a likelihood-based method}
\label{sec:GWPBH:ET:likelihood_based_method}

Having followed our simple cut-and-count method to quantify the lowest value of \fPBH that would be statistically distinguishable from \fPBH$=0$ (i.e. the ability of the ET to \textit{detect} PBHs), we now utilise a more powerful, likelihood-based method to assess the ET's potential capacity to both detect PBHs \textit{and} measure the \fPBH associated with a detection.

The analysis that follows assumes fixed values of the cosmological parameters $\Omega_\mathrm{m}$ and $H_0$, considering them as priors set from independent experiments. We thus neglect the potential degeneracies that could exists between \fPBH and the cosmological model. This approach is justified by the fact that even future GW experiments are expected to yield much looser cosmological constraints than established probes such as \emph{Planck}.

\subsubsection{Description of the method}

We describe an unbinned likelihood-based approach to comparing the data to a given model for the merger rate, which may include a contribution from both ABH and PBH mergers. To do this, we must compute the likelihood $\mathcal{L}(\fPBH)$, which is the probability to observe the data set $\mathcal{D}$ given the PBH fraction $f_\mathrm{PBH}$.

We write the probability distribution for the \textit{true} luminosity distances of ABH and PBH mergers as  $p_\mathrm{ABH}(\bar{D})$ and  $p_\mathrm{PBH}(\bar{D}|\fPBH)$ respectively, where we have indicated explicitly that the distribution for PBH mergers depends on \fPBH. The probability that a BH merger has a true luminosity distance in the range $[\bar{D}, \bar{D} + \diff \bar{D}]$ is given by
\begin{equation}
	p(\bar{D}|\fPBH) \, \diff \bar{D}
	= \frac{N_\mathrm{ABH}}{\bar{N}_\mathrm{obs}} \, p_\mathrm{ABH}(\bar{D}) \, \diff \bar{D} + \frac{N_\mathrm{PBH}}{\bar{N}_\mathrm{obs}} \, p_\mathrm{PBH}(\bar{D}|\fPBH) \, \diff \bar{D} \,, 
	\label{eq:P_full}
\end{equation}
where $\bar{N}_\mathrm{obs}(\fPBH) = N_\mathrm{ABH} + N_\mathrm{PBH}(\fPBH)$ is the total \textit{expected} number GW events and $N_\mathrm{ABH}$ and $N_\mathrm{PBH}$ are the \textit{expected} numbers of ABH and PBH mergers respectively. These are the numbers of events expected to be observed above the SNR threshold, taking into account the detection efficiency of the observatory, as detailed in \cref{subsec:preliminary_definitions}. 

We do not observe the true distance $\bar{D}$ but instead an estimate of the luminosity distance $D$. The probability that we observe an event with estimated luminosity distance $D$ can be written as
\begin{equation}
	p(D|f_\mathrm{PBH}) = \int p(D|\bar{D})\,p(\bar{D}|f_\mathrm{PBH})\,\mathrm{d}\bar{D}\,.
	\label{eq:measurementerror}
\end{equation}
The distribution of ``true'' luminosity distance $p(\bar{D}|f_\mathrm{PBH})$ is given by the theoretical expectation in \cref{eq:P_full}. The term $p(D|\bar{D})$ can be obtained from the measurement uncertainty in \cref{eq:Gaussian_error} via Bayes' theorem~\cite{Bayes:1764vd},
\begin{equation}
	p(D|\bar{D}) = \frac{p(\bar{D}|D)}{\tilde{p}(\bar{D})}\,
	\tilde{p}(D)\,.
\end{equation}
Here, $\tilde{p}(D)$ and $\tilde{p}(\bar{D})$ are the \textit{overall} probability distributions of $\bar{D}$ and $D$, where by \textit{overall} we mean that they are marginalised over all theory parameters (which in our case consists of only $f_\mathrm{PBH}$).
The term $\tilde{p}(D)$ enters as an overall normalisation which can be pulled out of the integral in \cref{eq:measurementerror} and does not depend on the theory parameters. This can therefore be safely neglected in the likelihood. The final task is then to compute $\tilde{p}(\bar{D})$, for which we need a prior on the PBH fraction $\mathrm{Pr}(f_\mathrm{PBH})$:
\begin{equation}
	\tilde{p}(\bar{D}) = \int_{0}^{1} p(\bar{D}|f_\mathrm{PBH}) \, \mathrm{Pr}(f_\mathrm{PBH})\,\mathrm{d}f_\mathrm{PBH}\,.
\end{equation}
We assume an uninformative log-flat prior,  $\mathrm{Pr}(f_\mathrm{PBH}) \propto 1/f_\mathrm{PBH}$, with $f_\mathrm{PBH}\in[10^{-9}, 10^{-3}]$.

For a sample of $N_\mathrm{obs}$ observed merger events, the likelihood can then be written as
\begin{equation}
	\mathcal{L}(\mathcal{D}|\fPBH) = \frac{\bar{N}_\mathrm{obs}(f_\mathrm{PBH})^{N_\mathrm{obs}} \mathrm{e}^{-\bar{N}_\mathrm{obs}(f_\mathrm{PBH})}}{N_\mathrm{obs}!} \times \prod_{i = 1, N_\mathrm{obs}} p(D_i|\fPBH)\,.
	\label{eq:likelihood}
\end{equation}
The first term is the Poisson probability to observe $N_\mathrm{obs}$ merger events, given that we expect to observe $\bar{N}_\mathrm{obs}(f_\mathrm{PBH})$. The second term accounts for the contribution of each observed event to the likelihood, where $D_i$ are the estimated luminosity distances. Combining the results above, the probability of observing a merger at an estimated distance $D_i$ is given by
\begin{equation}
	p(D_i|f_\mathrm{PBH}) \propto \int \frac{p(\bar{D}_i|D_i)}{\tilde{p}(\bar{D}_i)}\,p(\bar{D}_i|f_\mathrm{PBH})\,\mathrm{d}\bar{D}_i\,,
\end{equation}
which depends on the uncertainty $\sigma_i$ through $p(\bar{D}_i|D_i)$ in \cref{eq:Gaussian_error}. 

We then adopt a Bayesian approach in order to construct the posterior distribution function $p(f_\mathrm{PBH}|\mathcal{D})$ and thus determine projected constraints on the PBH fraction,
\begin{equation}
	\label{eq:PDF}
	p(f_\mathrm{PBH}|\mathcal{D}) \propto \mathcal{L}(\mathcal{D}|f_\mathrm{PBH}) \mathrm{Pr}(f_\mathrm{PBH})\,.
\end{equation}
We sample $\log_{10}f_\mathrm{ PBH}$ from a uniform prior $\log_{10}f_\mathrm{ PBH}\in\left[-9,-3\right]$ using the public cosmological sampling code \texttt{Cobaya} \cite{Torrado:2020dgo}, while fixing all other parameters to the values used to generate mock data.
We analyse the results using \texttt{GetDist} \cite{Lewis:2019xzd} to obtain the bounds achievable on $f_\mathrm{ PBH}$ using this approach.

\subsubsection{Future constraints on \fPBH}
\label{sec:GWPBH:ET:likelihood_based_method:results}

\begin{figure}[th!]
	\centering
	\begin{tabular}{cc}
		\includegraphics[height=5.cm]{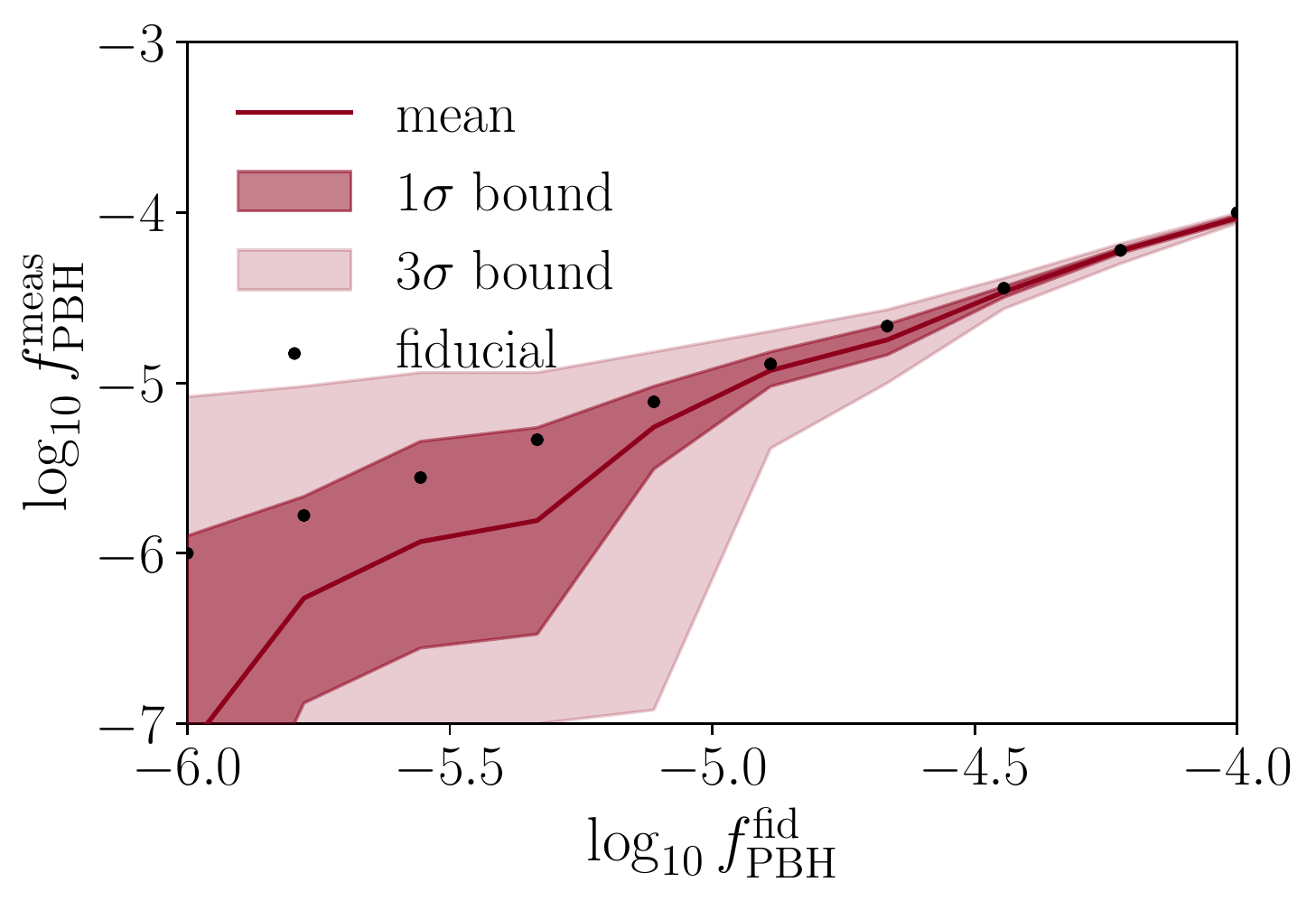} &
		\includegraphics[height=5.cm]{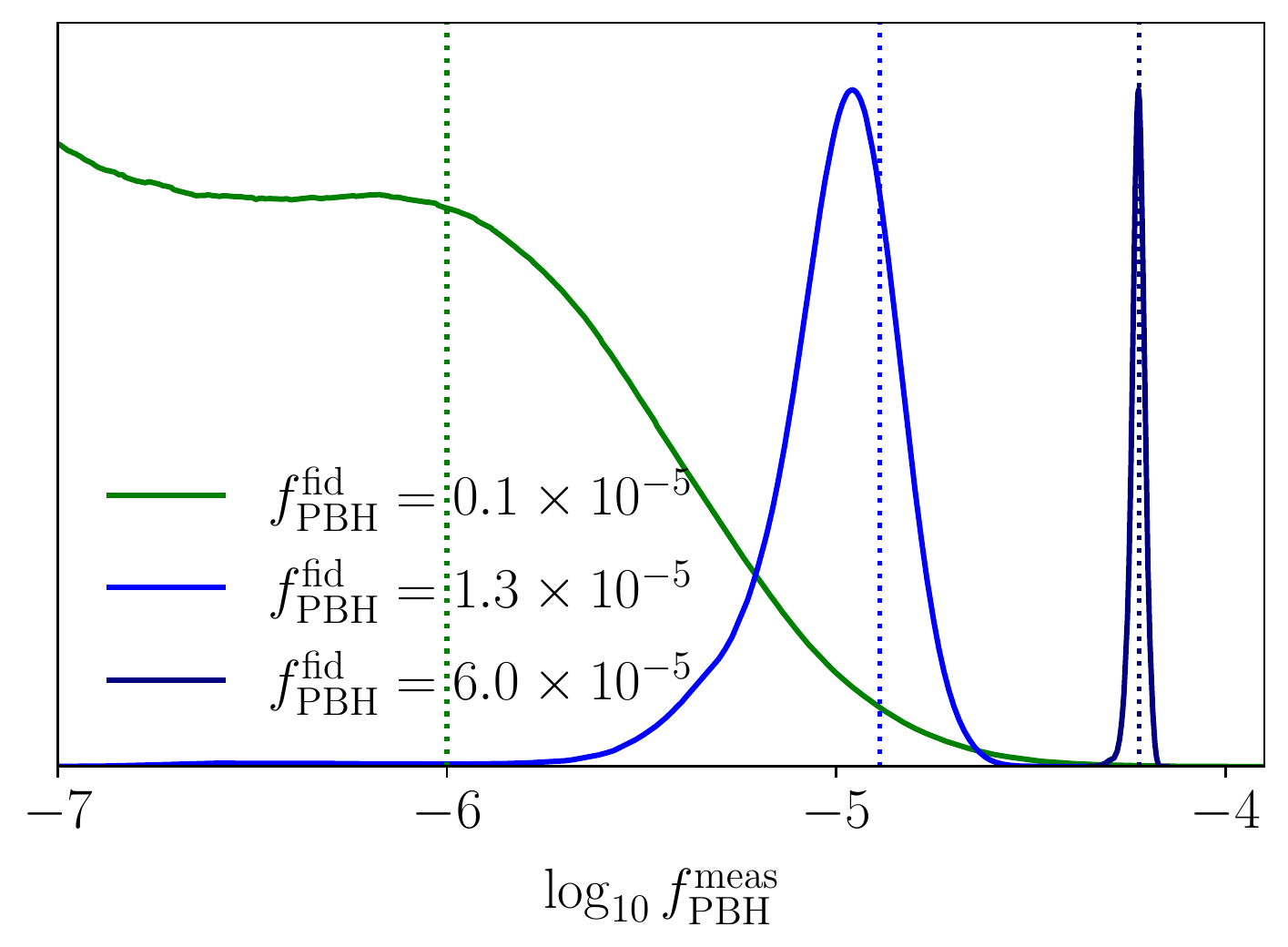}\\
		\multicolumn{2}{c}{\includegraphics[height=5.cm]{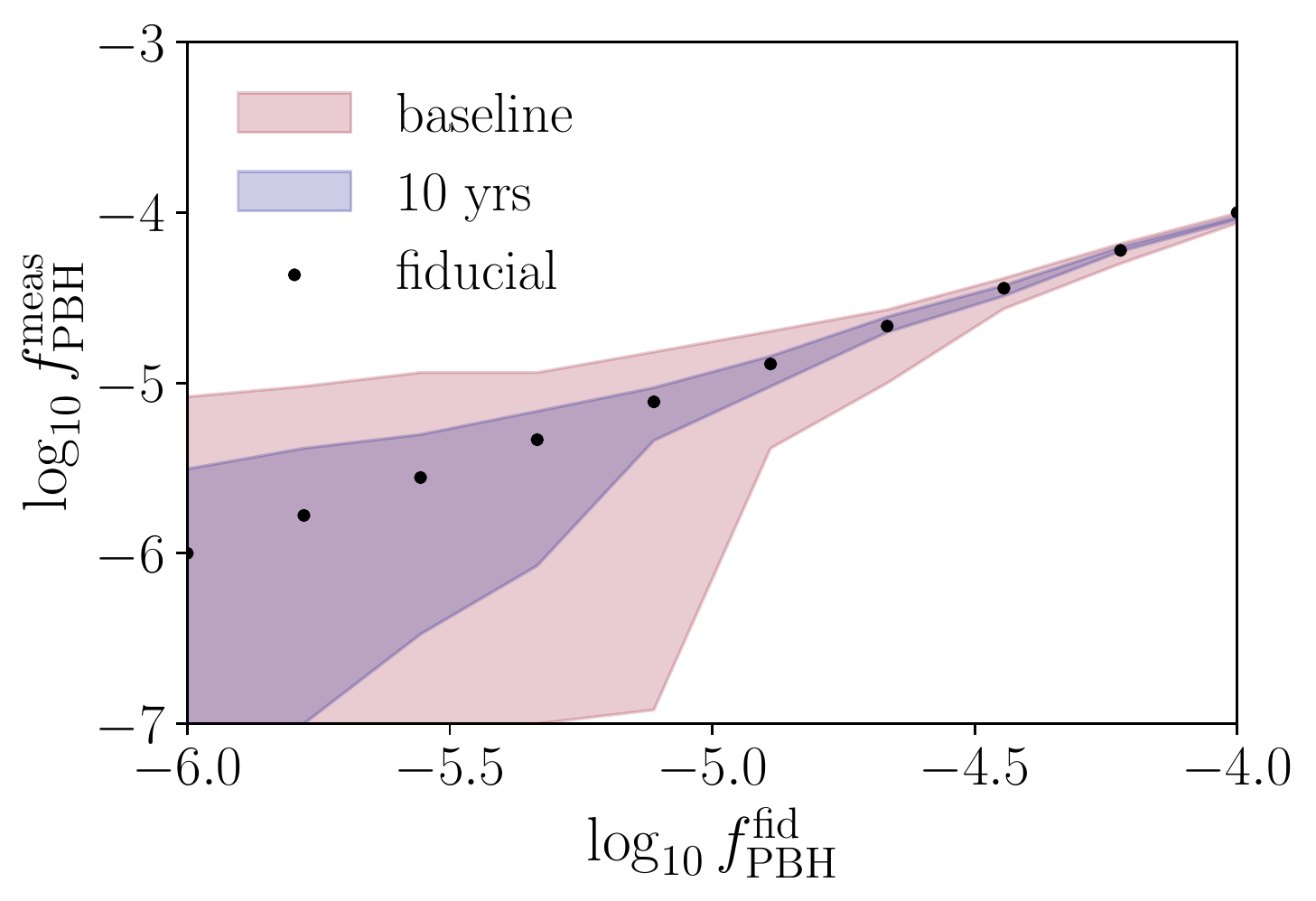}}
	\end{tabular}
	\caption{\textbf{ Left:} recovered mean value $f_\mathrm{ PBH}^\mathrm{ meas}$ (solid line) and $68\%$ and $99.7\%$ confidence level limits (red bands) as a function of the fiducial $f_\mathrm{ PBH}^\mathrm{ fid}$ (black dots). \textbf{ Right:} posterior distributions obtained on $f_\mathrm{ PBH}^\mathrm{ meas}$ for low (green), intermediate (blue) and high (navy) values of $f_\mathrm{ PBH}^\mathrm{ fid}$, with the vertical dotted line showing the value of $f_\mathrm{ PBH}^\mathrm{ fid}$ corresponding to each color. \textbf{ Bottom:} Comparison between the $99.7\%$ ($3\sigma$) confidence regions for the baseline case (outer band) and for an observation time of $T_\mathrm{ obs}=10$ yrs (inner band).}\label{fig:bradleyplot}
\end{figure}

In this section we discuss  the bounds on $f_\mathrm{ PBH}$ that ET can potentially obtain, as a function of the ``true'' value of this parameter, using the likelihood-based method.
In order to estimate the constraints, we generate 10 mock data sets with fiducial values $f^\mathrm{ fid}_\mathrm{ PBH}$ logarithmically distributed in the range $\left[10^{-6},10^{-4}\right]$. 
For each fiducial value of $f_\mathrm{ PBH}^\mathrm{ fid}$, we compute the likelihood associated to each data set, and obtain the posterior distribution function applying \cref{eq:PDF}. We define the mean of the distribution as the ``measured value'' $f^\mathrm{ meas}_\mathrm{ PBH}$. We show the results of this analysis in \cref{fig:bradleyplot}.
In the top left panel, we visualise the $68\%$ and $99.7\%$ confidence level bounds (i.e. $1\sigma$ and $3\sigma$) on the measured PBH fraction for each of the chosen $f^\mathrm{ fid}_\mathrm{ PBH}$ in our baseline settings, together with the mean value obtained for $f_\mathrm{ PBH}^\mathrm{ meas}$. We interpolate between the results obtained for our ten values of $f_\mathrm{ PBH}^\mathrm{ fid}$, in order to visualise the trend of these bounds.

We notice how the qualitative behaviour of these results is similar to that of the previous method; the analysis highlights how for low fiducial values ($f_\mathrm{ PBH}\lesssim10^{-5}$) the ET is not able to detect the presence of PBHs, and only an upper bound can be placed on $f_\mathrm{ PBH}$. Choosing a threshold of $3\sigma$ for a non-vanishing value of $f_\mathrm{ PBH}^\mathrm{ meas}$ to be considered a detection, as we did in \cref{sec:GWPBH:ET:cut_and_count}, we find that ET will be able to detect the presence of PBHs for $f_\mathrm{ PBH}^\mathrm{ fid}\approx7\times10^{-6}$. Such a value is the result of interpolating between the $99.7\%$ confidence level lower bounds obtained for the different data sets, and then finding for which value of $f_\mathrm{ PBH}^\mathrm{ fid}$ this function would exclude the no PBHs case (set to be at $f_\mathrm{ PBH}=10^{-7}$). Such a value is roughly three times lower than what we found with the cut-and-count method, highlighting how using the full amount of information present in the data set helps to boost the survey sensitivity.

For higher values of $f^\mathrm{ fid}_\mathrm{ PBH}$, a detection is possible and the constraining power of ET increases with higher fiducial values.
In the right panel of \cref{fig:bradleyplot}, we show the posterior distribution of $f_\mathrm{ PBH}^\mathrm{ meas}$ for three different $f_\mathrm{ PBH}^\mathrm{ fid}$; here we highlight how for low $f_\mathrm{ PBH}^\mathrm{ fid}$, the posterior distribution is extremely flat at low $f_\mathrm{ PBH}$, as the ET cannot distinguish between fractions of PBHs that produce a very low number of events. Moving towards higher $f_\mathrm{ PBH}^\mathrm{ fid}$, the posterior becomes increasingly peaked, showing how these PBH abundances could be measured with very high precision.

These results highlight how, should the Universe contain a high enough number of PBHs, the method proposed in this work will be able to constrain the value of $f_\mathrm{ PBH}$. For example, considering $f_\mathrm{ PBH}^\mathrm{ fid}=1.3 \times 10^{-5}$, i.e. the first generated data set above the detection threshold of $f_\mathrm{ PBH}\approx7\times10^{-6}$ found above, we find that one can obtain a measurement of $f_\mathrm{ PBH}^\mathrm{ meas}$ with a precision of $\SI{35}{\percent}$ at $\SI{68}{\percent}$ confidence level.

Finally, we compare the baseline results with what can be achieved increasing the observation time to $T_\mathrm{ obs}=\SI{10}{yrs}$. This comparison is depicted in the bottom panel of \cref{fig:bradleyplot}, where the baseline bound is shown in red, while the result for the extended observation time is shown in navy. Our results confirm that extending the survey time allows the detection of lower fractions of PBH; specifically, the smallest detectable $f_\mathrm{ PBH}^\mathrm{ fid}$ in the ten year case is is about five times smaller than in the one year case. For a given detectable value of $\fPBH$, the uncertainty on its measurement is also reduced; for instance, with $f_\mathrm{ PBH}^\mathrm{ fid}=1.3\times10^{-5}$ we find that $f_\mathrm{ PBH}^\mathrm{ meas}$ can be obtained with a precision of $\SI{11}{\percent}$ with $\SI{68}{\percent}$ confidence in the ten year case.

\subsection{Discussion}
\label{sec:GWPBH:ET:discussion}

The rationale of the two statistical frameworks presented here is to present complementary approaches, and emphasise different aspects of the analysis of GWs from BH merger events as a tool to detect and measure a PBH population. In particular, the cut-and-count method is especially designed to intuitively capture and visualise the essential aspect of the problem, namely the potential discovery of an anomalous excess of high redshift events beyond some reference distance. On the other hand, the likelihood-based approach exploits the capability of a Bayesian framework while using all the available information encoded in the redshift distribution of the events. 
In both cases, the characterisation of the instrument response plays a crucial role, and the main results we obtained revealed the capacity of the ET to disentangle astrophysical and primordial BH merger events on solid statistical grounds.

However, when searching for signatures of new physics against a highly uncertain astrophysical background, a key question is whether such signatures remain undetected simply due to the limitations of instrumental sensitivity or instead due to poor understanding of either the background or the signal itself, and the \emph{ systematic} uncertainties associated with each. 
In the search for PBHs we are considering here, the modelling of both the background and the signal poses a clear challenge.

Beginning with the signal, our model relies on a set of simplifying assumptions which includes: \emph{(i)} a statistically \emph{uniform} initial spatial distribution of PBHs (i.e. no initial clustering); and \emph{(ii)} a \emph{monochromatic} mass function (i.e. all the PBHs have the same mass). Regarding \emph{(i)}, as pointed out in \cite{DeLuca:2021hde}, initial clustering may significantly change the picture: the authors of that study aimed to find the minimum PBH abundance testable by future detectors and concluded that abundances as small as $\fPBH \sim 10^{-10}$ can be probed, if PBHs are highly clustered at formation; in the case of no initial clustering, they estimated that values of $\fPBH \sim 10^{-5}$ can provide at least one event per year at the ET. Our results based on the mock data generation and careful statistical analysis are compatible with the latter estimate. We leave the assessment of the role of the initial clustering, as well as a more refined analysis of the late-time clustering, to future work.

Regarding \emph{(ii)}, realistic PBH production scenarios generally predict an extended mass function rather than a monochromatic one as assumed here. Assessing the role of the mass function requires us to understand the role of the PBH mass itself in the merger rate. On the one hand, the larger the chirp mass of a PBH binary, the higher the amplitude of the corresponding GW signal,
which may result in a higher SNR, and hence an increased detectability of the PBH population. However, changing the chirp mass also affects the frequency range of the GW emitted by the merger. Our choice for the PBH mass, $M_\mathrm{PBH}=\mathcal{O}(\SI{10}{\solarmass})$, roughly coincides with an optimal detectability of high-redshift events, given the ET noise curve.\footnote{Note however that the optimal PBH mass depends on the precise value of the event's redshift.} On the other hand, for a fixed \fPBH, the larger the PBH mass, the fewer PBHs, and hence the rarer the merger events. Therefore, if we consider a broad mass function, we expect that a high-mass tail would be associated with rare but (potentially) high-SNR events, while a low-mass tail would correspond to abundant but low-SNR events.
Overall, for a relatively narrow mass function we do not expect our main conclusions to dramatically change, but we leave the study of more complex mass functions for future work.

Related to the above point, we chose to keep the masses of the ABH and PBH populations fixed in the statistical analysis. We have not used any information about the inferred masses of the mergers in this work, instead focusing on how the distance information can disentangle the two populations. Of course, including mass information is likely to enhance our ability to detect a PBH population, since detecting a deviation from the conventional ABH mass distribution model would be a hint at a possible population of PBHs, but that would require further assumptions about the ABH mass distribution.

The possibility to detect a subdominant population of PBHs by analysing the mass function of binary BH mergers has progressively gained momentum with the detection of many BH merger events with masses around $\SI{30}{\solarmass}$ \cite{LIGOScientific:2020kqk, LIGOScientific:2021psn}.
For example, the authors of \cite{DeLuca:2021wjr} claim that there is decisive evidence in favour of a two-population model from these data, with a second, subdominant, population peaked around $\SI{30}{\solarmass}$. This population displays a significant high-mass tail, hence its presence is  further supported by the detection of a BH with mass in the \emph{ pair-instability gap} region \cite{LIGOScientific:2020iuh} (several astrophysical models of BH formation in the pair-instability mass gap also exist, see e.g. \cite{LIGOScientific:2020ufj}). 

Motivated by these considerations, we emphasise that an exploration of the impact of broad mass distributions for the PBH population (possibly with multiple peaks on different mass scales~\cite{Carr:2019kxo}) on the PBH detection prospects with the ET is an important future goal in this research line. 
This would require a more careful treatment of late-time clustering than that presented in this work, given the complex phenomenology that may arise (for instance, mass segregation in clusters \cite{Trashorras:2020mwn}).

We have assumed throughout this work that PBH binaries formed in the early Universe dominate the merger rate. Dense environments, such as PBH clusters, can potentially lead to the disruption of early Universe binaries. We have considered  the effect of late-time clustering (see \cref{sec:appendix:clustering}), which seems to be negligible for the small PBH abundances considered in this work; the impact of initial PBH clustering is left for future work. PBH binaries could also be disrupted by interactions with other objects; while it is unlikely that they will be disrupted by stellar encounters in the disks of typical galaxies~\cite{Sasaki:2016jop}, it seems feasible that they may be disrupted in sufficiently dense environments, such as close to the centre of the host galaxy. This is unlikely to affect the high redshift signals we study here, with the first haloes having little effect on early-formed binaries~\cite{Ali-Haimoud:2017rtz}. On the other hand, dense environments can lead to the formation of a population of late-time PBH binaries. This is typically expected to be subdominant~\cite{Bird:2016dcv,Ali-Haimoud:2017rtz}, in particular for the small values of \fPBH considered in this work. These late-time binaries may be more difficult to distinguish from ABH ones, as they are expected to form only at low redshift, in which case alternative approaches are necessary, such as a measurements of the clustering bias~\cite{Canas-Herrera:2021qxs}.
However, as long as the population of early-time binaries remains relatively unperturbed, high redshift observations would be able to probe it, and hence detect the presence of PBHs, even if this population were to be subdominant with respect to the late-time one. Regarding the measurability discussion, a sizable population of late-time binaries merging at relatively high redshifts could lead us to overestimate the abundance of PBHs. In neglecting this, we rely on the expectation that the merger rate due to such binaries is only relevant at low redshifts and for large PBH abundances.

There are also a number of ways in which the modelling of the astrophysical background could alter the results presented here. In particular, we emphasise that our model for the ABH evolution, characterised by a peak at $z \sim 2$ and a monotonic decrease at higher redshifts, is designed to capture the rate associated to binary systems made of second- and third-generation stars (usually called Population I and II). These systems may form via different channels, such as binary stellar evolution in galactic fields, or  dynamical formation through multi-body interactions in star clusters.

However, a significant contribution to the binary BH merger rate, including a peak at high redshift, $z  \sim \mathcal{O}(10)$, can be expected from the first-generation (Population III) stars, which formed out of the pristine gas left over after cosmological nucleosynthesis, and generated the first heavy elements in the Universe \cite{Vangioni:2014axa,deSouza2011Aaanda}. Taking into account this additional component would certainly require, again, to address the mass dependence in more detail. In fact, this population would be modelled as a high (or intermediate) mass model of star formation at high redshift, to be added to the background model~\cite{Ng:2020qpk}. An even more detailed treatment would imply a marginalisation over the parameters that describe the metallicity dependence of the SFR, treated as nuisance parameters. 
We leave this more detailed treatment to future studies, and emphasise that the numerical and statistical tools presented here are the ideal framework to consistently address uncertainties in both the signal and background models.

As we were finalising this work, Ng et al.~\cite{Ng:2022agi} proposed an independent analysis on the same topic. Ref.~\cite{Ng:2022agi} explores how a network consisting of two third generation detectors (ET and CE) can be used to distinguish between ABH and PBH populations using high redshift mergers. Their analysis differs from ours in that they consider exclusively population~III ABHs and restrict their analysis to events above $z > 8$, as well as working directly with measurement errors on the event redshift, rather than on the luminosity distance, as we do here. 
Their results are broadly consistent with ours, highlighting that a PBH population with $f_\mathrm{PBH} \sim \mathcal{O}(10^{-5})$ should be within reach of third generation detectors.  We find a one year sensitivity estimate for $f_\mathrm{PBH}$ roughly a factor of three stronger, which may be partly due to differences in the distance uncertainties and population models assumed. This emphasises the importance of careful modelling of these details in quantifying the PBH detection and measurement potential of future GW detectors. 

\subsection{Conclusions}
\label{sec:GWPBH:ET:conclusions}

In this work we have assessed the capability of the forthcoming Einstein Telescope (ET) gravitational wave (GW) observatory to detect and measure a subdominant population of primordial black hole (PBH) mergers using a novel statistical framework. 

We have described a procedure that computes the redshift evolution of both the expected background associated with the astrophysical black hole (ABH) merger events, and the signal under consideration associated with the PBH population (assumed to be characterised by masses of the same order of magnitude). In our modelling, we paid particular attention to the impact of late-time clustering of PBHs. We set the relative normalisation of the two contributions by comparison with the latest (low redshift) data released by the LIGO, Virgo and KAGRA collaborations. This procedure naturally provided an updated upper limit on the fraction of dark matter in the form of PBHs (quantified using the parameter \fPBH), based on the third Gravitational Wave Transient Catalog (GWTC-3) -- see \cref{fig:LIGObound}.  

The key feature of our merger rate models is the different behaviour at high redshift of signal (i.e.\ PBH mergers) and background (i.e.\ ABH mergers), with the former monotonically increasing with increasing distance, and the latter steadily decreasing. Motivated by this qualitative aspect, and taking into account the expected high sensitivity of ET, we presented two methods to assess the potential of discovering a distant PBH population, which both accurately take into account the experimental sensitivity. Both methods are based on the generation and analysis of a set of mock data catalogues, associated to the null hypothesis (no PBHs) and to different values of the fraction of DM in the form of PBHs.

The first method (dubbed ``cut-and-count'') relies on a two-redshift-bin data analysis strategy that highlights the PBH contribution as a significant excess in the high redshift bin. We demonstrated that a PBH fraction as low as $f_\mathrm{ PBH}\approx2\times10^{-5}$ can be detected with a 3$\sigma$ significance, given a proper choice of the cut position -- see \cref{fig:detection_zT}.

The second method takes a Bayesian approach to the problem, and fully exploits the information on the redshift dependence of both the signal and background. Within this framework, we provided an assessment of the posterior probability density function for different fiducial values of \fPBH. We found that fractions of DM in the form of PBHs as low as $\mathcal{O}(10^{-5})$ can be measured by ET with one year of data taking, with a precision of $\sim\SI{35}{\percent}$ at $\SI{68}{\percent}$ confidence level -- see \cref{fig:bradleyplot}. These results demonstrate a well-defined avenue towards a discovery or, in the absence of a discovery, the prospect of setting a significantly improved upper limit on the existence of PBHs in this mass window.

Furthermore, in the process of this~ work, we have developed the \texttt{darksirens}~\href{https://gitlab.com/matmartinelli/darksirens}{\faGitlab} code, which allows for the fast generation of realistic mock GW catalogues sourced by ABH and PBH merger events. We have made the code publicly available for community use, with extensive scope for development and application to other investigations.

The final message of this work is clear: the possibility of detecting and measuring a distant population of PBHs is well within the capabilities of future GW observatories such as the ET. The exciting prospect that these exotic objects could constitute a fraction of the dark matter, a substance that despite its fundamental importance in the Universe we still know so little about, makes the pursuit of this goal nothing short of vital. In the rigorous analysis we have here presented, based on the twin pillars of robust statistics and a thorough treatment of the modelling of both signal and noise, we have shown that the ET will be able to measure a fraction of PBHs as low as $\mathcal{O}(10^{-5})$, through observations of their luminosity distances alone. If a high redshift population of PBHs exists in our Universe, the detection of these distant dark sirens is therefore something we no longer need to hope for, but can begin to expect.
 

%
\chapter{Conclusions}
\label{sec:conclusions}

In this thesis, we have considered the possibility that the Dark Matter, or a fraction of it, is composed by Primordial Black Holes (PBHs), compact objects generated in the early Universe. We have discussed some phenomenological consequences of the existence of these objects, focusing on two detection channels: electromagnetic radiation from gas accretion and gravitational waves from binary merger events. These constitute the most powerful observation channels adapted to the detection of black holes in the solar-mass range.\\

\Cref{sec:accretion} has been dedicated to gas-accreting isolated black holes in the Milky Way. These may be detected through the radiation emitted by the heated gas as it is drawn towards the event horizon. 

A key ingredient required to predict the luminosity of such sources is the accretion rate: the analysis presented in this thesis relies on a state-of-the-art accretion model, able to account for radiation feedback effects on the accreted gas. Results obtained based on the more standard Bondi-Hoyle-Littleton model are also discussed for comparison.

We have first focused our attention on astrophysical black holes, in \cref{sec:accretion:ABH}. These are expected to constitute an irreducible background for the search of primordial compact objects, although their detection can also be regarded as relevant astrophysical problem \emph{per se}. 
We have presented a comprehensive study of the prospects of detection of a population of such objects, considering two target observation regions: the vicinity of the Solar System and the Central Molecular Zone, situated at the Galactic Centre and occupied by a giant molecular cloud complex.\\
Concerning the vicinities of the Solar System, our findings suggest that a few tens of X-ray sources associated to isolated accreting black holes could already be present as unidentified objects in existing catalogues. 
An in-depth analysis has been dedicated to the promising target represented by the Central Molecular Zone. We have performed an extended parametric assessment of the number of X-ray sources in this region of interest, considering the detection threshold associated to the NuSTAR survey. Our study suggests that $\mathcal{O} (10)$ sources should be already present --~as unidentified objects -- in the associated catalogue, a prediction which appears to be solid for any reasonable value of the parameters under scrutiny.

A multi-wavelength analysis will be needed to unambiguously identify such a population; we have hence extended the analysis to emission in the radio band. We have shown that, when considering the sensitivity thresholds of present catalogues, no radio counterpart is expected for these X-ray sources. On  the other hand, we have estimated that the majority of gas-accreting black holes in the Central Molecular Zone should be detectable by the planned Square Kilometre Array observatory, which therefore holds the potential to unveil a very large population of these objects. In summary, our results suggest that the identification of a population of isolated black holes at the Galactic centre may be around the corner.

In \cref{sec:accretion:PBH}, we turned to the study of an hypothetical primordial black hole population located in the CMZ. 
The accretion channel can be used, by comparison with existing observations, to constrain the abundance of PBH in the Universe. We have applied the method developed in the study of the astrophysical background to the assessment of the uncertainties associated with this bound. We found in particular that significant uncertainties are associated to the modelling of the phase space distribution of PBHs in the central region of the Galaxy. Furthermore, any correlation or anti-correlation between the positions of PBH and the clouds will impact the result. 
Overall, we have found that the bounds present in the literature are relatively solid for PBH masses between order ten and one hundred solar masses (assuming a monochromatic mass function). However, a threshold effect comes into play when smaller masses are considered, such that PBHs of around one solar mass or less are unlikely to be visible even when accreting from dense clouds. For such small values of the PBH mass, the maximum allowed fraction dark matter in the form of PBHs varies by orders of magnitude within the parameter space considered, being at some points compatible with unity.

Motivated by this observation, we have turned to the study of a specific, well-motivated mass distribution for the PBHs. This mass function, which arises from the evolution of the equation of state of the primordial plasma, presents a prominent peak at around one solar mass, and a second smaller peak at around thirty solar masses. We have found that the constraint on the abundance of PBHs is dominated by the peak at one solar mass, as expected. As a consequence, it is very difficult to place a solid constraint on the fraction of PBH in dark matter within this scenario. We have performed a parametric study of the impact of the different physical assumptions on the result.  Assessing the global uncertainty on the constraint will require a full Bayesian analysis, which is underway and will be included in a future work.
On the other hand, accreting PBHs belonging to the subdominant, more massive component are expected to emit significant radiation for a wide range of velocities. An interesting window exists, therefore, for the detection of these objects. Careful multi-wavelength analyses of present catalogues might be able to identify such objects if they are indeed present in our Galaxy. \\

The second part of this thesis, \cref{sec:GWPBH}, has been dedicated to the search for primordial black holes through gravitational waves. After discussing the current status of observations, we have turned to future third generation observatories. Highly improved sensitivities will allow these detectors to observe merger events which have occurred before the birth of the first stars. Such high redshift observations will provide an avenue towards the identification of an hypothetical primordial black hole population, above the astrophysical background. 

In the work presented in \cref{sec:GWPBH:ET}, we have assessed the capability of the European third generation observatory, the Einstein Telescope, to detect and measure a sub-dominant population of primordial black holes. This study is based on the generation of realistic mock data catalogues, which include, for each merger event, the estimated luminosity distance and the relative error on its measurement. \\
As a starting point, we have discussed the modelling of the redshift evolution of both the expected background -- consisting of astrophysical merger events -- and the signal associated to a primordial population, paying particular attention to the impact of late-time clustering of PBHs. 
Based on the generation of a set of mock catalogues, associated to varying fractions of dark matter in the form of primordial black holes, we have then performed two distinct statistical analysis aimed at assessing the potential of the Einstein Telescope.

The first analysis -- dubbed the ``cut-and-count'' method -- relies on a two-redshift-bin data strategy, that highlights the PBH contribution as a significant excess in the high-redshift bin. We have demonstrated that a PBH fraction as low as $10^{-5}$ can be \emph{detected} with a 3$\sigma$ significance with this method. The optimal position for the bin separation was found to be at around redshift ten.
The second method takes a Bayesian approach to the problem, and fully exploits all the information regarding the redshift dependence of both signal and background. Within this framework, we have provided an assessment of the posterior probability density function of the fraction of dark matter in the form of PBHs, associated to different fiducial values. Through this analysis, we showed that fractions as low as $10^{-5}$ can be effectively \emph{measured} by the Einstein Telescope with one year of data taking.

In summary, we have shown that the possibility of detecting and measuring a distant population of PBHs is well within the capabilities of future gravitational wave observatories such as the Einstein Telescope, which will be able, through observations of luminosity distances alone, to identify a population of PBHs constituting as little as one part in $\mathcal{O}(10^{5})$ of the dark matter.


\appendix\cleardoublepage
%
\chapter{Clustering}
\label{sec:appendix:clustering}

In this appendix,\footnote{From  Ref. \cite{Martinelli:2022elq}} we discuss the suppression of the PBH merger rate due to PBH clustering. We follow the numerical implementation given in Ref.~\cite{Vaskonen:2019jpv} but here we provide additional details which are useful for understanding why the suppression of the merger rate is negligible for the ranges of $f_\mathrm{PBH}$ allowed by present bounds.

\section{PBH cluster formation}

Even if PBHs are uniformly distributed in the early Universe (i.e.\ according to a Poisson distribution), their discrete nature means that the mean PBH density in a given region can be subject to large fluctuations. Regions with overdensities of PBHs begin to collapse early, forming PBH \textit{clusters}~\cite{Chisholm:2005vm,Chisholm:2011kn,Inman:2019wvr}. Here, we briefly review this mechanism, before commenting on the relevance for the survival of PBH binaries. 

At matter--radiation equality, $t_\mathrm{ i} = t_\mathrm{eq}$, we consider a spherical region of radius $r_\mathrm{ i}$, containing an overdensity $\delta_\mathrm{ i} = (\rho_\mathrm{ i} - \rho_\mathrm{eq})/\rho_\mathrm{eq}$. This overdense region will evolve as an isolated matter-dominated Universe with effective curvature~\cite{Gunn:1972sv,1984ApJ...281....1F,1985ApJS...58...39B},
\begin{equation}
	\mathcal{K}
	= \frac{8 \pi G \rho_\mathrm{eq} \delta_\mathrm{ i} r_\mathrm{ i}^2}{3}
	= \frac{2 G \bar{M}_\mathrm{ i} \delta_\mathrm{ i}}{r_\mathrm{ i}}\,,
\end{equation}
where $\bar{M}_\mathrm{ i} = (4 \pi/3) G \rho_\mathrm{eq} r_\mathrm{ i}^3$ would be the initial mass of the region in the unperturbed background. The evolution of the radius of this region $r(t)$ can be solved parametrically as
\begin{align}
	r(\theta) &= \frac{G M_\mathrm{ i}}{\mathcal{K}}(1-\cos \theta)\,, \\
	t(\theta) &= \frac{G M_\mathrm{ i}}{\mathcal{K}^{3 / 2}}(\theta-\sin \theta)\,.
\end{align}
The region collapses then at
\begin{align}
	t_\mathrm{coll} = t(\theta = 2\pi) = \frac{2\pi G M_\mathrm{ i}}{\mathcal{K}^{3/2}}\,,
\end{align}
or, using the fact that $z \propto t^{-2/3}$ in a matter-dominated Universe:
\begin{align}
	z_\mathrm{coll} = z_\mathrm{eq} \left(\frac{t_\mathrm{eq}}{t_\mathrm{coll}}\right)^{2/3} = z_\mathrm{eq} \mathcal{K} \left(\frac{t_\mathrm{eq}}{2\pi G M_\mathrm{ i}}\right)^{2/3}\,.
\end{align}
We can relate $t_\mathrm{eq}$ and $r_\mathrm{ i}$ by assuming that the size of the region corresponds to its comoving size at matter--radiation equality. With this, we find the redshift of collapse as
\begin{equation}
	z_\mathrm{coll} = \frac{2}{(18\pi^2)^2} \, z_\mathrm{eq} \,
	\frac{\delta_\mathrm{ i}}{(1+\delta_\mathrm{ i})^{2/3}}
	\approx 0.36 \, z_\mathrm{eq} \, \delta_\mathrm{ i}\,,
\end{equation}
for $\delta_\mathrm{ i} \ll 1$. The redshift at which the expansion of the region turns around is a little larger, $z_\mathrm{ta} \approx 0.56 \, z_\mathrm{eq} \, \delta_\mathrm{ i}$.

Consider now a region of the Universe which we expect to contain, on average, $N$ PBHs with total PBH mass $M = N M_\mathrm{PBH}$. If a cluster forms from this collection of PBHs, it is likely to form from a typical overdensity in PBHs of $\delta_\mathrm{PBH} = \sqrt{N}/N$, due to Poisson fluctuations. If PBHs make up only a fraction $f_\mathrm{PBH}$ of the total DM density, then the total overdensity will be $\delta_\mathrm{ i} = f_\mathrm{PBH}/\sqrt{N}$ (assuming that the DM fluctuations are subdominant). Dropping some order one factors, we therefore expect that the PBH cluster to form at a redshift
\begin{equation}
	z_\mathrm{coll} \approx z_\mathrm{eq} f_\mathrm{PBH}/\sqrt{N}\,.
	\label{eq:z_coll}
\end{equation}
While this expression is a good estimate of collapse time for large values of $N$, it fails for smaller values. In our numerical analysis, we rely on the more complete approach discussed in \cite{Inman:2019wvr}.
However, in the following sections we will make use of \cref{eq:z_coll} to estimate the typical size of clusters which are disrupted at late times.

\section{Relaxation time}

We will assume that a PBH binary is completely disrupted if the PBH cluster it resides in undergoes core collapse due to gravo-thermal instability and that all other PBH binaries are unperturbed. We take the characteristic core-collapse timescale to be $t_\mathrm{ cc} \sim 18 \, t_\mathrm{ r}$~\cite{Quinlan:1996bw}, where $t_\mathrm{ r}$ is the relaxation time. 

The relaxation time $t_\mathrm{ r}$ for a PBH cluster containing $N$ PBHs. For a system with density $\rho$ and component masses $m$, this relaxation time can be estimated as~\cite{Quinlan:1996bw}
\begin{equation}
	t_{\mathrm{r}}=0.065 \, \frac{ v^{3}}{ G^2 m \rho \ln \Lambda}\,,
\end{equation}
where $\ln\Lambda \approx \ln(N/f_\mathrm{ PBH})$ is the Coulomb logarithm associated with interactions between component masses. Assuming that the fraction of PBHs in clusters matches the mass fraction in the Universe, then the total mass of each cluster is $M = m N/f_\mathrm{PBH}$.

The mean density of a cluster will be
\begin{equation}
	\bar{\rho}
	\equiv \frac{3 M}{4\pi R^3}
	\approx 18 \pi^2\rho_c a^{-3}
	= 18 \pi^2 \rho_\mathrm{ c} \, a_\mathrm{eq}^{-3}\,\frac{f_\mathrm{PBH}^{3}}{N^{3/2}}\,,
	\label{eq:MeanDensity}
\end{equation}
where the second equality follows from the theory of spherical collapse~\cite{1980lssu.book.....P} and from setting $a = 1/z_\mathrm{coll}$ (the scale factor at which the cluster forms). Here, $\rho_\mathrm{ c}$ is the critical density.

Next, we set the PBH velocity $v$ in the cluster equal to some typical velocity dispersion $v \approx \sigma_v \approx \sqrt{GM/R}$. Using the definition of $\bar{\rho}$ in Eq.~\eqref{eq:MeanDensity}, we then have
\begin{align}
	\sigma_v^3
	= \left(\frac{GM}{R}\right)^{3/2}
	= \sqrt{\frac{4\pi}{3}} G^{3/2} M \sqrt{\bar{\rho}}\,.
\end{align}

Assuming that there is no substantial growth or evaporation of the cluster after formation, the relaxation timescale can then be written as
\begin{align}
	t_r &= 0.065 \, \frac{\sigma_v^3}{G^2 m \bar{\rho}\ln\Lambda}
	= 0.065 \, \sqrt{\frac{4\pi}{3}}
	\frac{M}{\sqrt{G \bar{\rho}} \, m \ln \Lambda}\\
	&= 0.065 \, \sqrt{\frac{2}{27\pi}}
	\frac{1}{\sqrt{G\rho_{\mathrm{m, eq}}}}  \frac{N^{7/4}}{f_\mathrm{PBH}^{5/2}\ln\Lambda}
	= \SI{2.1}{\kilo\year}\,\frac{N^{7/4}}{f_\mathrm{PBH}^{5/2}\ln\Lambda}\,.
	\label{eq:relaxationtime}
\end{align}
Here, we have defined $\rho_{\mathrm{m,eq}} = \rho_\mathrm{ c} a_\mathrm{eq}^{-3}$ (the matter density at matter--radiation equality). We have also taken the numerical values $\rho_\mathrm{ c} = 2.78 \times 10^{11} h^2~M_\odot~\si{\per\mega\parsec\cubed} $~\cite{Zyla:2020zbs} and $H_0 = h\times \SI{100}{\kilo\meter\per\second\per\mega\parsec}$, with $h = 0.673$~\cite{Aghanim:2018eyx}. This value for the  relaxation time $t_\mathrm{ r}$ matches the result given in Eq.~(7) of \cite{Vaskonen:2019jpv}.

\section{Cluster collapse}

\begin{figure}[tb]
	\centering
	\includegraphics[width=0.49\textwidth]{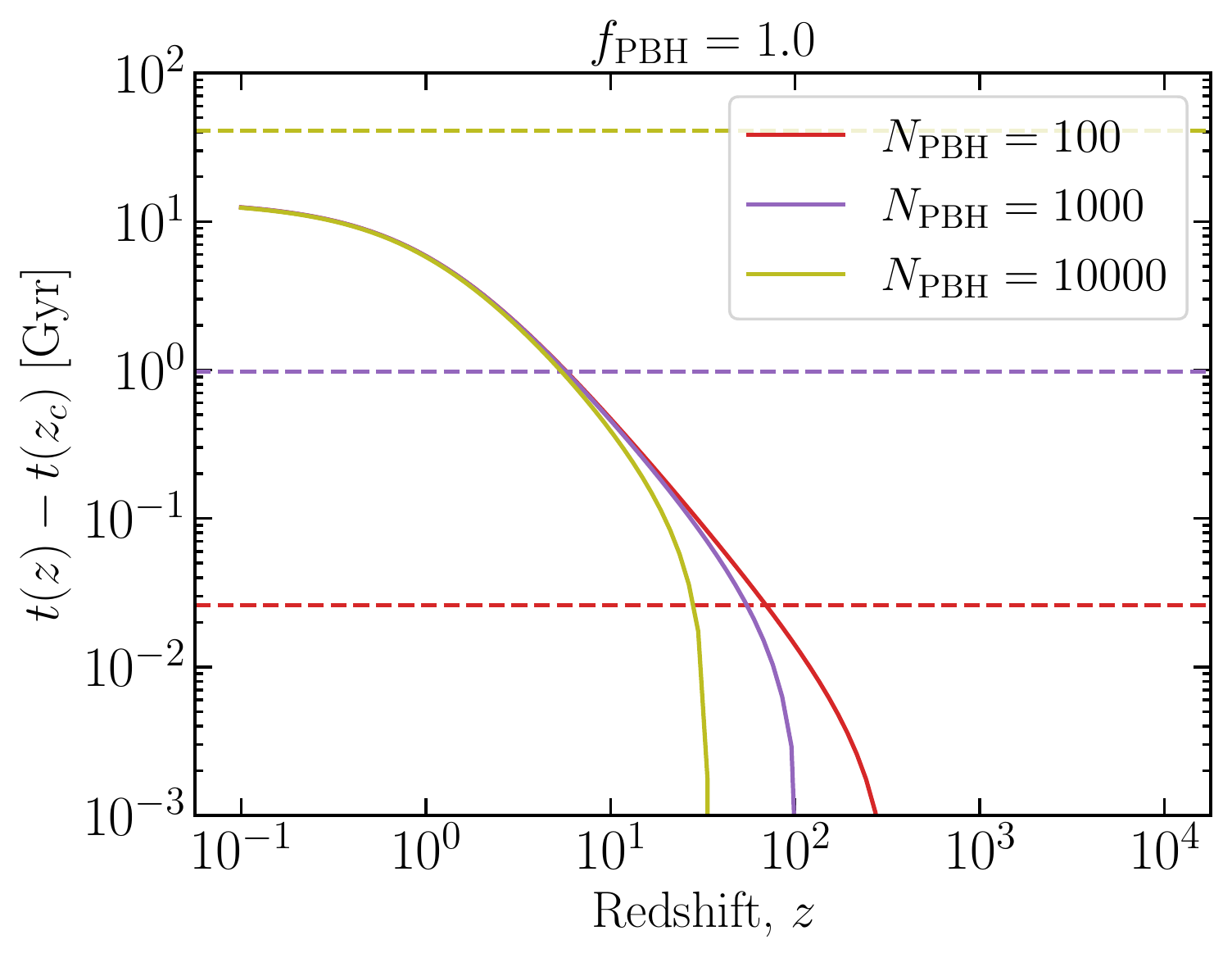}
	\hfill
	\includegraphics[width=0.49\textwidth]{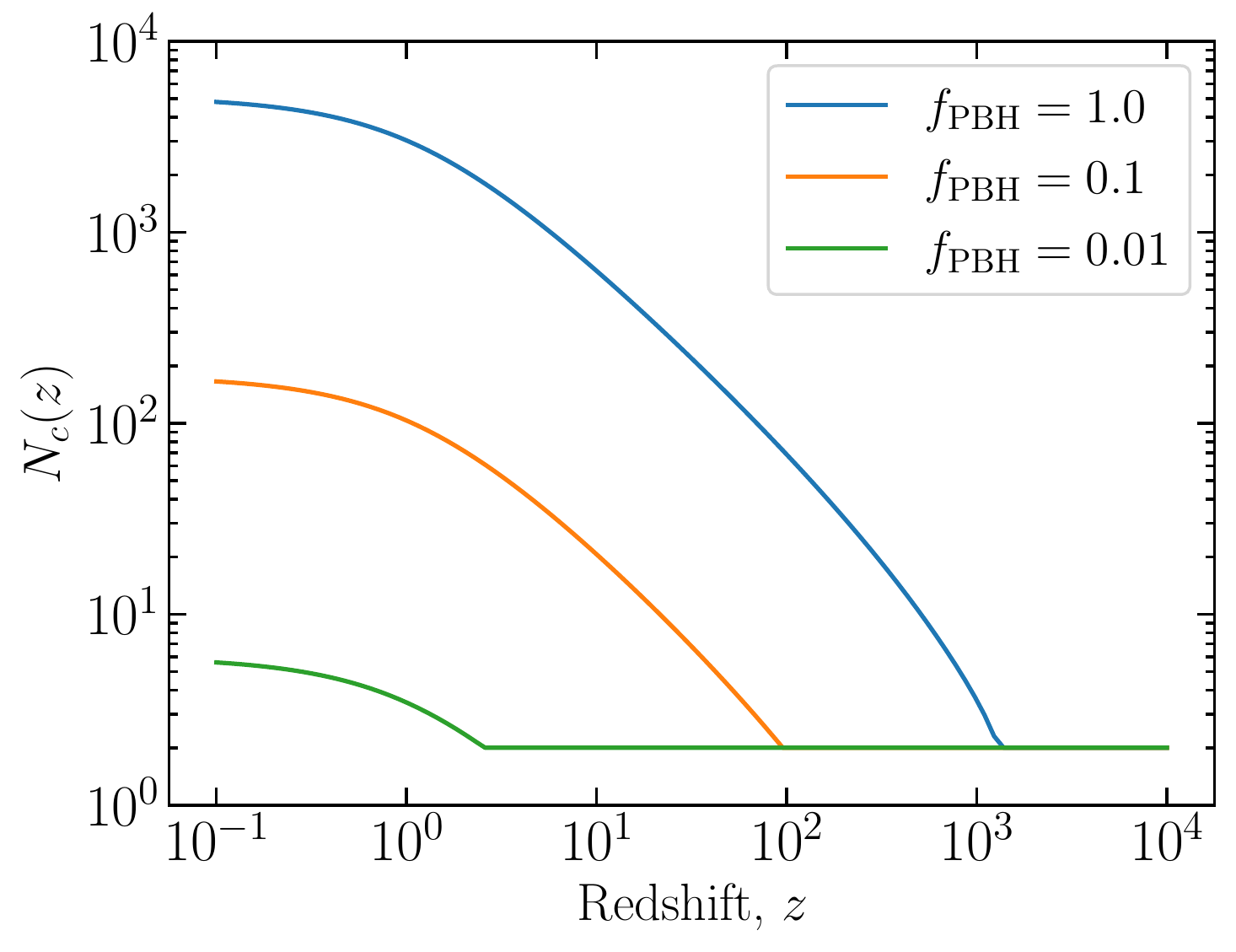}
	\caption{\textbf{Left:} Time-scales associated with PBH clusters. Solid lines show the time between the redshift of cluster formation $z_\mathrm{ c}$ and a given redshift $z$, for clusters containing different numbers of PBHs $N_\mathrm{PBH}$. Horizontal dashed lines show the core-collapse timescale for clusters with $N_\mathrm{PBH}$ members. Where the solid and dashed lines (of a given colour) cross then $N_\mathrm{PBH} = N_\mathrm{ c}(z)$. \textbf{Right:} Critical number of PBHs per cluster $N_\mathrm{ c}$ as a function of redshift $z$. Clusters with $N < N_\mathrm{ c}(z)$ will be disrupted due to core-collapse before redshift $z$.}
	\label{fig:Nc}
\end{figure}

In order to undergo core collapse before redshift $z$, a PBH cluster must be formed with fewer than a critical number of members $N_\mathrm{ c}(z)$. This critical number $N_\mathrm{ c}(z)$ is obtained by equating the core-collapse timescale to the time between formation and collapse,
\begin{equation}
	\label{eq:Nc}
	18 \, t_\mathrm{ r}(N_\mathrm{ c}) = t(z) - t(z_\mathrm{ c})\,,
\end{equation}
where $z_\mathrm{ c}$ is the redshift at which clusters of size $N_\mathrm{ c}$ are formed. Here, $t_\mathrm{ r}(N_\mathrm{ c})$ is the relaxation time for clusters of size $N_\mathrm{ c}$, defined in \cref{eq:relaxationtime}. Recall also from the previous sections that
\begin{equation}
	z_\mathrm{ c} \approx z_\mathrm{eq} \, \frac{f_\mathrm{PBH}}{\sqrt{N_\mathrm{ c}}}\,.
\end{equation}
With these definitions, we can calculate $N_\mathrm{ c}(z)$ from Eq.~\eqref{eq:Nc}. This is illustrated in Fig.~\ref{fig:Nc}. There, we see that for $f_\mathrm{PBH} = 1$, the critical size of clusters is $N_\mathrm{ c} \sim 5000$ at $z = 0$. This decreases with decreasing $f_\mathrm{PBH}$. Below $f_\mathrm{PBH} \sim 0.01$, core-collapse becomes more or less irrelevant, as the critical number of PBHs tends to one.

\section{Merger rate suppression factor}

Following Ref.~\cite{Vaskonen:2019jpv}, the suppression factor is expressed as the probability of a binary not belonging to a cluster that undergoes gravo-thermal collapse. Estimating this factor requires knowledge of the PBH halo mass function, which we describe using the analytical model discussed in \cite{Inman:2019wvr}. We indicate the probability of a PBH belonging to a cluster of $N$ elements at given $z$ and \fPBH with $P(N | z, \fPBH)$. We thus have $P(N | z, \fPBH) \propto  \exp[-N/N^*(z, \fPBH)]$, where $N^*(z, \fPBH)$ is the characteristic number of PBHs in clusters forming at redshift $z$. 

\begin{figure}[tb]
	\centering
	\includegraphics[width=0.49\textwidth]{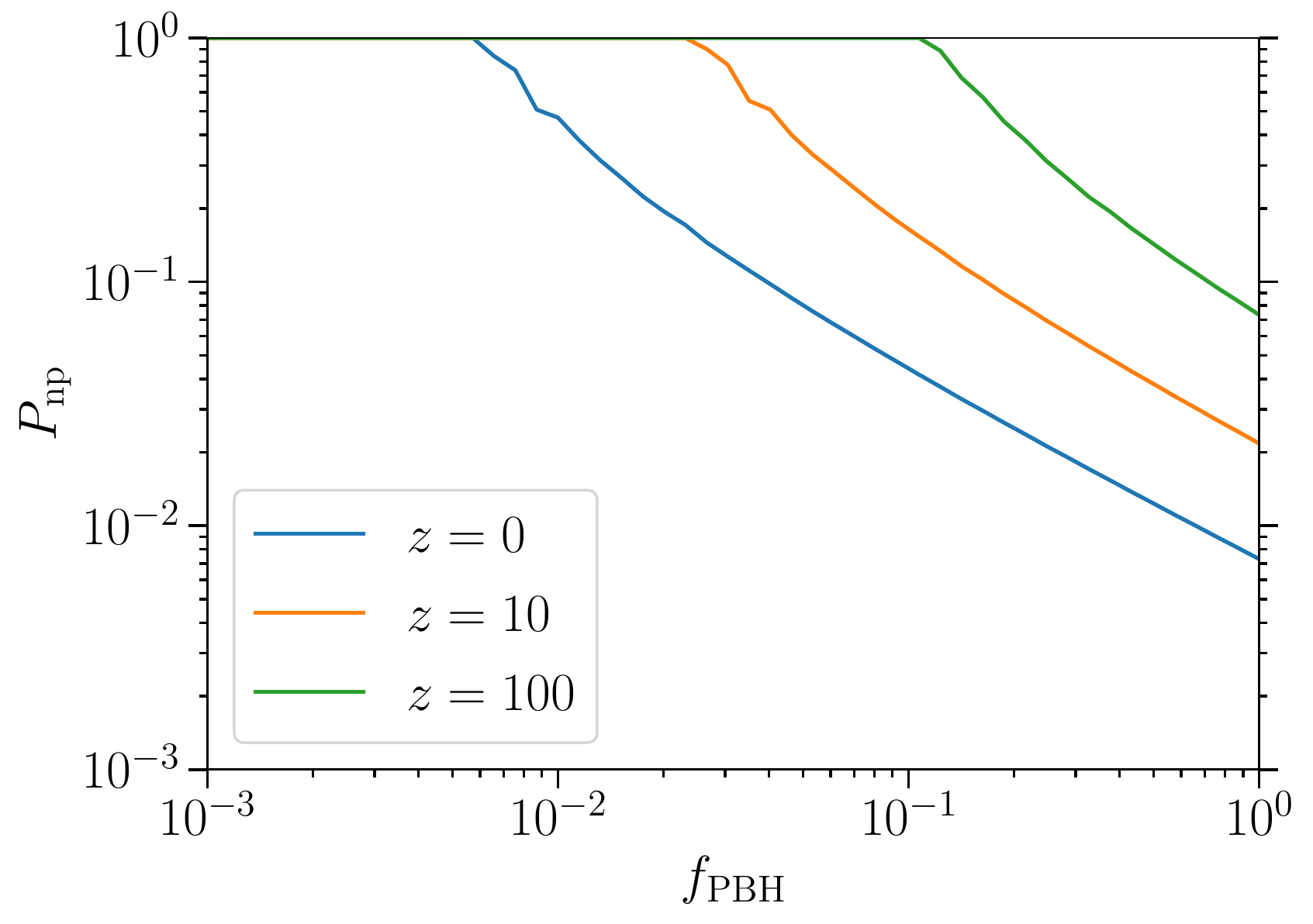}
	\hfill
	\includegraphics[width=0.49\textwidth]{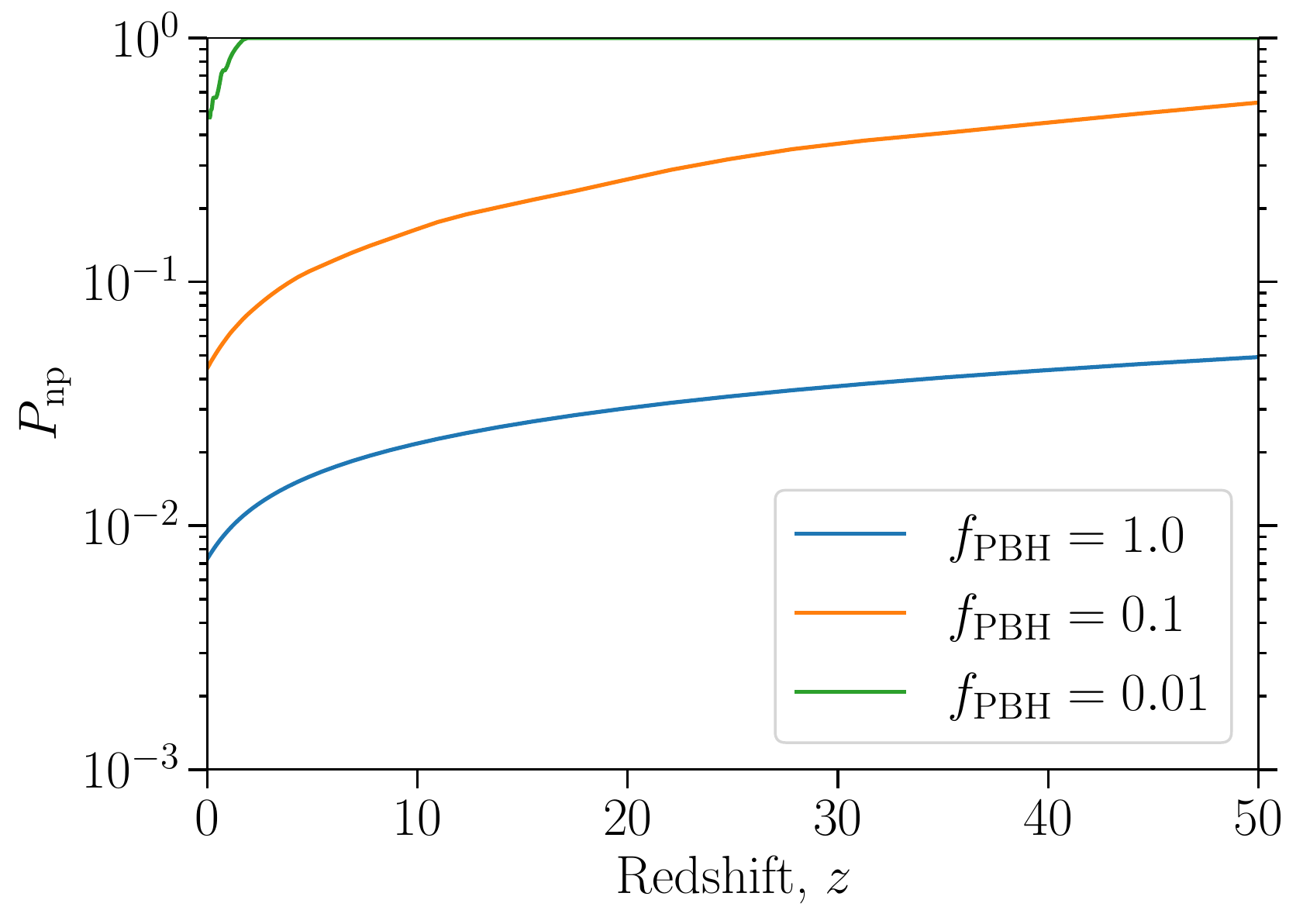}
	\caption{ Merger rate suppression factor due to interactions in gravo-thermally unstable clusters, computed following \cite{Vaskonen:2019jpv}. \textbf{Left:} Probability of  a binary \textit{not} being disrupted as a function of \fPBH for different redshifts. Notice that, depending on the value of $z$, there is a threshold value of \fPBH below which no binaries are perturbed.   \textbf{Right:} Same probability as a function of redshift, for different values of \fPBH.  }
	\label{fig:Pnp}
\end{figure}

The probability of a binary being disrupted by redshift $z$ is given, according to \cite{Vaskonen:2019jpv}, by the sum of two terms:
\begin{enumerate}
	\item The probability of the binary belonging to a cluster which has reached instability before redshift $z$, i.e. a cluster with $N \leq N_\mathrm{ c} (z)$,
	\begin{equation}
		P_\mathrm{ p}^{(1)} = \sum\limits_{N=3}^{N_\mathrm{ c}} P_\mathrm{ bin}(N|z_\mathrm{ c}) \;,
	\end{equation}
	where the probabilities of binaries belonging to a cluster $P_\mathrm{ bin}(N)$ are taken to be $\propto P(N)$ but normalised to exclude $P(N = 1)$:
	\begin{equation}
		\sum\limits_{N=2}^{\infty} P_\mathrm{ bin}(N) =  1 .
	\end{equation}
	
	\item The probability of a binary belonging to a sub-cluster of $N \leq N_\mathrm{ c}$ within a larger cluster ($N > N_\mathrm{ c}$), 
	\begin{equation}
		P_\mathrm{ p}^{(2)} = \sum\limits_{N>N_\mathrm{ c}} \left( \sum\limits_{N'=3}^{N_\mathrm{ c}} P_\mathrm{ sub}(N'|z_\mathrm{ c}) \right) P_\mathrm{ bin}(N|z_\mathrm{ c}) \; ,
	\end{equation}
\end{enumerate}
where the probabilities of belonging to a sub-cluster $P_\mathrm{ sub}(N)$ are taken to be proportional to $P(N)$, but normalised summing up to the size of the cluster that contains them
\begin{equation}
	\sum\limits_{N' = 2}^{N} P_\mathrm{ sub}(N') =  1 .
\end{equation}
All the probabilities in the expressions above are evaluated at $z=z_\mathrm{ c}$, which is the typical redshift of formation of the clusters with $N_\mathrm{ c}$ elements. Then $z_\mathrm{ c}$ is obtained by requiring that $N^*(z_\mathrm{ c}) = N_\mathrm{ c}$. Finally, the probability of a binary not being perturbed is given by $ P_\mathrm{np} = 1 - P_\mathrm{ p}^{(1)}- P_\mathrm{ p}^{(2)} $. This suppression factor is shown in Fig.~\ref{fig:Pnp} for different values of $z$ and \fPBH.

%
\chapter{Einstein Telescope mock data generation}
\label{sec:appendix:mockdata}

This appendix\footnote{From  Ref. \cite{Martinelli:2022elq}} provides a detailed description of how the mock catalogues used for the analysis in \cref{sec:GWPBH:ET} are constructed. 

\section{Antenna patterns}\label{sec:appendix:mockdata:antenna}

We generate mock catalogues of GW events based on the configuration and sensitivity of the ET, a proposed third-generation GW observatory. Specifically, we consider the ET-D configuration, which involves three interferometers arranged in an equilateral triangle shape and two detectors at each vertex of the triangle, one sensitive to higher frequencies and one to lower, i.e. six detectors in total. The antenna patterns for a single interferometer with a $60$ degree opening angle are given by
\begin{align}
	F_+ &= \frac{\sqrt{3}}{2}\left[\frac12 (1 + \cos^2 \theta) \cos(2\phi) \cos (2\psi) - \cos\theta \sin(2\phi) \sin(2\psi) \right],\\
	F_\times &= \frac{\sqrt{3}}{2}\left[\frac12 (1 + \cos^2\theta) \cos(2\phi) \sin(2\psi) + \cos\theta \sin(2\phi) \cos(2\psi) \right].
\end{align}
The antenna patterns for the two other interferometers in the ET-D configuration are then given by $F_{+, \times}(\theta, \phi + 2\pi/3, \psi)$ and $F_{+, \times}(\theta, \phi + 4\pi/3, \psi)$.

We assume that the basic information available from the detection of GWs from merging compact objects by the ET will be the luminosity distance to the merger event, $D$, and the uncertainty on that luminosity distance, $\sigma$, which has contributions from the instrumental noise and from the uncertainty due to weak lensing of the GW.

\section{Luminosity distance} \label{sec:appendix:mockdata:dl}

To obtain the luminosity distance for a given simulated event, we firstly draw a redshift from a probability distribution defined based on the merger rate of the progenitor system being considered (binary PBHs or binary PBHs).

However, unlike binary neutron star mergers which produce an electromagnetic counterpart, the redshift of an individual\footnote{There have been proposals to cross-correlate GW events with galaxy catalogues (see e.g. \cite{Canas-Herrera:2019npr,Mukherjee2021}), allowing the redshift of the event to be estimated, but this introduces a good deal of uncertainty, especially with the current small number of events and relatively poor sky localisation.} binary BH merger is always unknown. We therefore convert the simulated redshift to a luminosity distance in the standard $\Lambda$CDM cosmology using \CAMB. All of the analysis presented in the \cref{sec:GWPBH:ET} is based solely on the luminosity distance information, rather than the redshift information, in order to accurately simulate the actual analysis that could be done with a catalogue which contains only binary BHs.

Finally, we rescale the luminosity distance to account for weak lensing of the signal. The full computation is shown in \Cref{sec:appendix:mockdata:lensing}.

\section{Signal to noise ratio} \label{sec:appendix:mockdata:SNR}

The optimal unlensed signal-to-noise ratio is obtained using a Wiener filter, and is given by
\begin{equation}
	\rho_\mathrm{ opt} =  \left[4 \int^{f_\mathrm{ upper}}_{f_\mathrm{ lower}} \mathrm{d}f \, \frac{h(f) h^*(f)}{S_n(f)} \right]^{\frac12},
\end{equation}
where $f_\mathrm{ upper}$ and $f_\mathrm{ lower}$ are the cutoff frequencies for the strain, beyond which it is assumed to be zero. The waveform $h(f)$ is the Fourier transform of the strain $h(t)$ given by \cref{eq:hpatterns}. We randomly draw the angles $\phi$ and $\psi$ from a uniform distribution between $0$ and $2\pi$. However, the angles $\theta$ and $\iota$ (the inclination of the event i.e. the angle between the source plane and the detector plane) should have their cosine uniformly distributed, meaning that for these quantities we draw their values randomly from a uniform distribution between $-1$ and $1$ and then take the arccosine of the result. The inclination of the event enters into the function used to weight the final signal-to-noise ratio sum, as we will see in a moment.

We use the publicly available package \texttt{PyCBC}\footnote{\url{https://pycbc.org/}.} to generate a mock waveform $h(f)$ for a given event, using the \texttt{IMRPhenomD} waveform model, and by inputting the masses and simulated luminosity distance of the event. We keep the spins fixed to zero. We also generate the frequency range, and hence $f_\mathrm{ upper}$ and $f_\mathrm{ lower}$, using \texttt{PyCBC}.
To describe the power spectral density  we use the publicly available data for ET-D\footnote{\url{http://www.et-gw.eu/index.php/etsensitivities}.}.

With the waveform and power spectral density in hand, the signal-to-noise ratio for an event seen in a single interferometer can be computed. The total signal-to-noise ratio for an event $i$ seen by the three-armed ET-D observatory is obtained by summing the squares of the individual interferometer signal-to-noise ratios, 
\begin{equation}
  	\bar{\rho}_i = \sqrt{w \left(\rho^2_{\text{$i$ opt, \nth{1} arm}} + \rho^2_{\text{$i$ opt, \nth{2} arm}} + \rho^2_{\text{$i$ opt, \nth{3} arm}} \right)},
\end{equation}  	
where the weighting factor $w$ is given by
\begin{equation}
  w = F_+^2(\theta, \phi, \psi) (1+ \cos^2\iota)^2 + 4F_\times^2(\theta, \phi, \psi) \cos^2\iota \, ,
\end{equation} 
where $F_+, F_\times$ are the antenna patterns and $\iota$ is the inclination.

\begin{figure}[h]
	\centering
	\includegraphics[height=5.6cm]{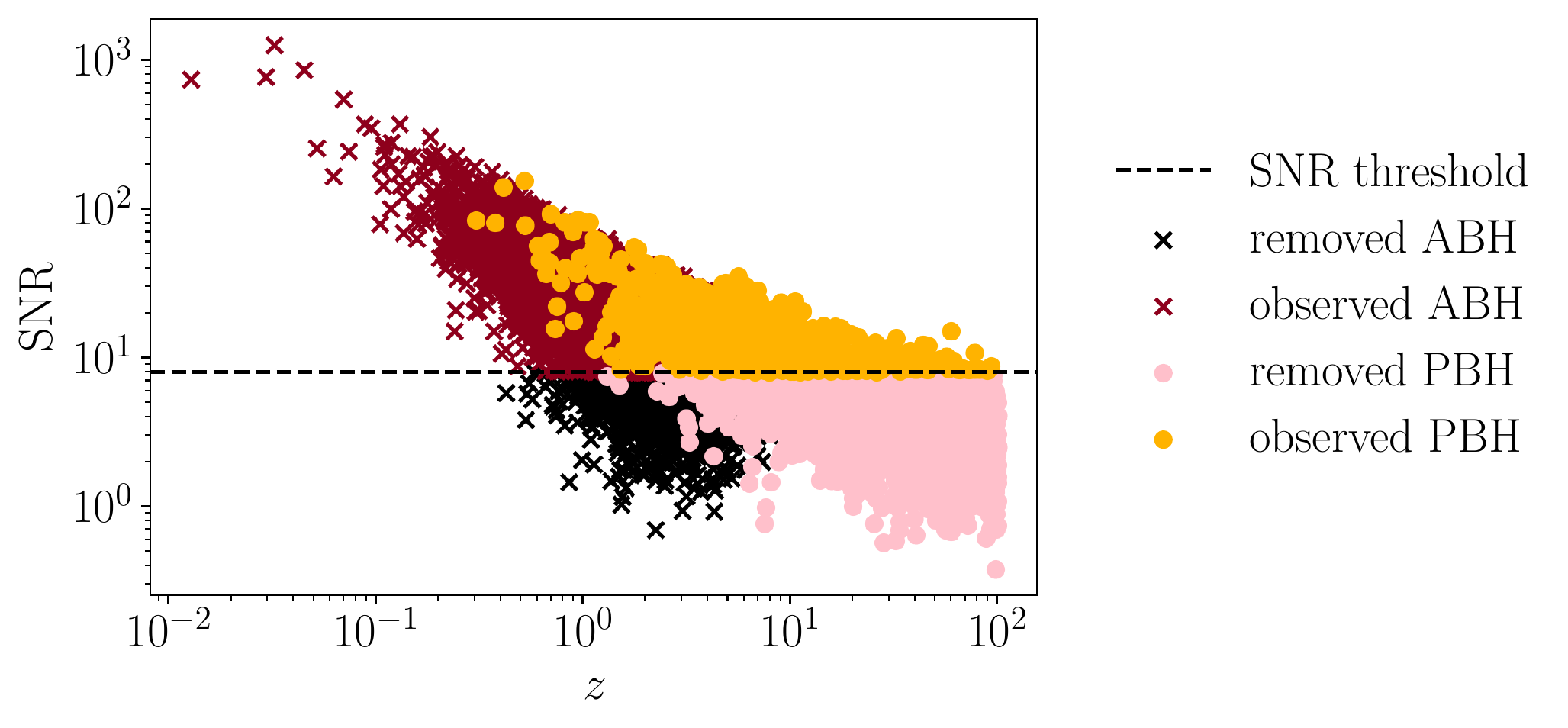}
	\caption{Trend of the signal-to-noise ratio with event redshift for ABHs (crosses) and PBHs (dots). Events that are above the signal-to-noise ratio threshold (black dashed line) are considered to be properly observed GW events (red for ABHs and yellow for PBHs), while the black and pink points are respectively the ABHs and PBHs with signal-to-noise ratios that make them too faint to be observed.}\label{fig:SNRplot}
\end{figure}

Finally, as mentioned in \cref{sec:GWPBH:ET:generating_mock}, after computing the \textit{lensed} signal-to-noise ratio for all the events, we remove all the events with a lensed signal-to-noise ratio of less than eight from the catalogue. This serves to exclude events which may or may not be true detections of GWs. The signal-to-noise ratio threshold that we use follows that of the LIGO collaboration  \cite{Abbott:2016xvh}. In \cref{fig:SNRplot} we show the computed signal-to-noise ratio as a function of redshift for ABH and PBH events, highlighting the significant amount of events which are discarded due to the cut in the signal-to-noise ratio.

\section{Uncertainty on the lensed luminosity distance} \label{sec:appendix:mockdata:sigmadl}

The second step in the generation of the mock catalogue is to compute the uncertainty on the lensed luminosity distance for a given event. As we mentioned, this has two components: the instrumental noise and the noise due to weak lensing of the GW. The instrumental noise depends on the signal-to-noise ratio for a given event.
We approximate the instrumental uncertainty on the lensed luminosity distance $\tilde{D}_i$ for a given event $i$ as
\begin{equation}
	\sigma^\mathrm{ inst}_{i} \approx \frac{2 \tilde{D}_{i}}{\rho_i} \ ,
	\qquad
	\rho_i = \sqrt{\mu_i} \, \bar{\rho}_i \ .
	\label{eq:inst_error}
\end{equation}
Here, $\bar{\rho}_i$ is the unlensed signal-to-noise ratio for the event determined by the observatory in question and $\mu_i$ is the magnification of the signal-to-noise ratio due to weak lensing. 
The factor $2$ in \cref{eq:inst_error} accounts for the contribution of the uncertainty on the inclination $\iota$ to the instrumental noise \cite{Li:2013lza}. Once the rescaling of the unlensed signal-to-noise ratio by the lensing magnification has been carried out, it is then trivial to compute the instrumental uncertainty for the event from \cref{eq:inst_error}. 

However, we note that a precise modelling of the uncertainty on the distance measurement would require access to the full parameter estimation pipeline of the experiment. Furthermore, the uncertainty actually depends on the full network of observatories that may be operating in collaboration with the ET, see \cite{Vitale:2016icu}. Extending the uncertainty computation in this way is beyond the scope of this work.

Our analysis also rests on the assumption that the expression for the instrumental uncertainty given in \cref{eq:inst_error} is correct. The signal-to-noise ratio in this expression depends on the noise being stationary and Gaussian. In a real detector this is not exactly the case. Depending on the specific template used in the matched filtering process, so-called glitches -- artifacts with very high signal-to-noise ratios -- can be seen in the detector \cite{Nuttall2018}. The identification and removal of glitches and other noise artifacts through proper characterisation of the detectors is an important part of  current gravitational wave analysis pipelines \cite{LIGO:2021ppb}. However, since we are working solely with mock data, we generate our catalogues using the assumption of stationary, Gaussian noise. This is effectively equivalent to using a real catalogue of events with glitches removed.

\section{Lensing}
\label{sec:appendix:mockdata:lensing}

This appendix provides details about our modelling of the weak lensing of GW signals. \Cref{sec:appendix:mockdata:lensing:magnification_GWs} is a short theoretical reminder where we state our assumptions and define the relevant quantities to be used. \Cref{sec:appendix:mockdata:lensing:variance_magnification} indicates how the variance of the magnification is estimated. \Cref{sec:appendix:mockdata:lensing:lognormal_PDF_magnification} deals with our model for the magnification PDF in the mocks.

\subsection{Magnification of gravitational waves}
\label{sec:appendix:mockdata:lensing:magnification_GWs}

Consider a GW modelled as a small perturbation~$h_{\mu\nu}$ of an arbitrary background geometry~$g_{\mu\nu}$. If the wave is propagating in vacuum, in the sense that we can neglect its direct interaction with matter, then its equation of motion reads~\cite{Straumann:2013spu}
\begin{equation}
	\label{eq:EoM_GW}
	\nabla_\rho \nabla^\rho h_{\mu\nu} = 0 \ ,
\end{equation}
in the linear regime, harmonic transverse-traceless gauge, and assuming that the GW's wavelength is much shorter than the typical curvature radii of the background spacetime geometry, whose covariant derivative is denoted with $\nabla_\rho$ in \cref{eq:EoM_GW}.

Introducing the wave ansatz $h_{\mu\nu}=H_{\mu\nu}\exp(\rmi w)$, where $H_{\mu\nu}$ and $w$ respectively denote the amplitude and the phase of the GW, \cref{eq:EoM_GW} implies (i) that the wave follows null geodesics of the background spacetime; and (ii) that its amplitude satisfies
\begin{equation}
	(1+z)^{-1} D H_{\mu\nu} = \text{const.} ,
\end{equation}
where $z$ is the source's redshift and $D$ is the electromagnetic luminosity distance -- see, e.g., \cite{Dalang:2019rke} for further details. It follows that $H_{\mu\nu}\propto D^{-1}$, that is, the energy of a GW ``dilutes'' just like the energy of an electromagnetic wave as it propagates. Since this is valid in any background spacetime, we conclude that GWs experience the same gravitational lensing effects as light.

In the following, we only consider \emph{weak-lensing} effects, in the sense that we neglect the possibility that a GW source might be multiply imaged. We also assume that the luminosity distance can be computed from the propagation of an infinitesimal beam of null geodesics. In that framework, the distortions of the beam with respect to the homogeneous-isotropic FLRW case are customarily encoded in the so-called distortion matrix. If $\vect{\theta}$ is the observed incoming direction of GWs and $\vect{\beta}$ the direction in which they would be observed in FLRW, then the distortion matrix is defined as the Jacobian matrix of $\vect{\theta}\mapsto\vect{\beta}(\vect{\theta})$,
\begin{equation}
	\vect{\mathcal{A}}
	\equiv
	\frac{\diff\vect{\beta}}{\diff\vect{\theta}}
	=
	\begin{bmatrix}
		1 - \kappa - \gamma_1 & -\gamma_2 + \omega\\
		-\gamma_2 - \omega & 1 - \kappa + \gamma_1
	\end{bmatrix} ,
\end{equation}
where $\kappa$ is called the convergence of the beam, $\gamma=\gamma_1+\rmi\gamma_2$ its shear distortion, and $\omega=\mathcal{O}(\gamma^2)$ its rotation.

The correction to the observed luminosity distance, $D$, relative to the FLRW case, $\bar{D}$, is quantified by the \emph{magnification}~$\mu$, which is related to the distortion matrix as follows,
\begin{equation}
	\label{eq:magnification}
	\mu^{-1}
	\equiv
	\left(
	\frac{D}{\bar{D}}
	\right)^2
	=
	\frac{\diff^2\vect{\beta}}
	{\diff^2\vect{\theta}}
	=
	\det\vect{\mathcal{A}}
	=
	(1-\kappa)^2 - |\gamma|^2 + \omega^2 .
\end{equation}
In particular, at lowest order in $\kappa, \gamma, \omega$, the magnification only depends on the convergence, $\mu \approx 1 + 2\kappa$. We shall use this approximation for the estimate of the variance of the magnification -- see next subsection. It follows from \cref{eq:magnification} that the absolute uncertainty on the luminosity distance due to lensing may be estimated as
\begin{equation}
	\sigma_\mathrm{ lens}
	= \frac{1}{2} \, \bar{D} \, \sigma_\mu \ ,
\end{equation}
where $\sigma_\mu$ denotes the standard deviation of the magnification.

\subsection{Variance of the magnification in weak lensing}
\label{sec:appendix:mockdata:lensing:variance_magnification}

We aim to model the PDF of the magnification of the observed GW signals in a realistic inhomogeneous universe. As a first step, we shall estimate the variance of that distribution. Previous analyses, both in the context of supernova cosmology~\cite{2014A&A...568A..22B} or GW cosmology~\cite{Sathyaprakash:2009xt, Zhao:2010sz, Cai2016, Du:2018tia, Jin:2020hmc, Hogg:2020ktc}, considered a dispersion of magnitude due to that grows linearly with redshift~\cite{2010MNRAS.405..535J}, $\Delta m = 0.055\,z$, which corresponds to a magnification dispersion of $\sigma_\mu=4\ln(10)\Delta m/5=0.10\,z$. While this linear approximation may be valid at low redshift, we expect it to fail at the high redshifts ($z\sim 10$ to $100$) considered here. The intuition is that at high redshift the Universe is increasingly homogeneous, thereby reducing the growth of the lensing dispersion.

In order to get a more accurate estimate of $\sigma_\mu$ across a wide range of redshifts, we shall use perturbation theory at second order. In that framework, we first notice that
\begin{equation}
	\sigma_\mu^2 
	\equiv
	\langle \mu^2 \rangle - \langle \mu \rangle^2
	=
	4\sigma_\kappa^2
	+ \mathcal{O}(\kappa^4) \ ,
\end{equation}
where $\sigma_\kappa^2$ is the variance of the convergence, because $\langle\kappa\rangle = 0$ at linear order and $\omega\sim\kappa^2\sim\gamma^2$. Hence, we may estimate $\sigma_\mu$ at second order from the linear-order results on $\sigma_\kappa$.

In the flat-sky and Limber approximations, the variance of the convergence for a source at redshift $z$ is related to the convergence power spectrum~$P_\kappa(\ell, z)$ as
\begin{equation}
	\sigma_\kappa^2(z)
	= \int_0^\infty \frac{\ell\diff\ell}{2\pi} \; 
	P_\kappa(\ell, z) \ ,
\end{equation}
which is itself related to the power spectrum~$P_W$ of the Weyl potential $W=(\Phi+\Psi)/2$, $\Phi$ and $\Psi$ being the Bardeen potentials, via
\begin{equation}
	\label{eq:P_kappa}
	P_\kappa(\ell, z)
	= \ell^2(\ell+1)^2 \int_0^r
	\diff r'
	\left(
	\frac{r - r'}{r}
	\right)^2
	\left(
	\frac{1}{\ell+1/2}
	\right)^4
	P_W \left(\eta_0-r', \frac{\ell+1/2}{r'} \right) ,                          
\end{equation}
where $\eta_0$ denotes today's conformal time, and $r=r(z)$ is the comoving distance to the source.

\begin{figure}
	\centering
	\includegraphics[width=0.49\columnwidth]{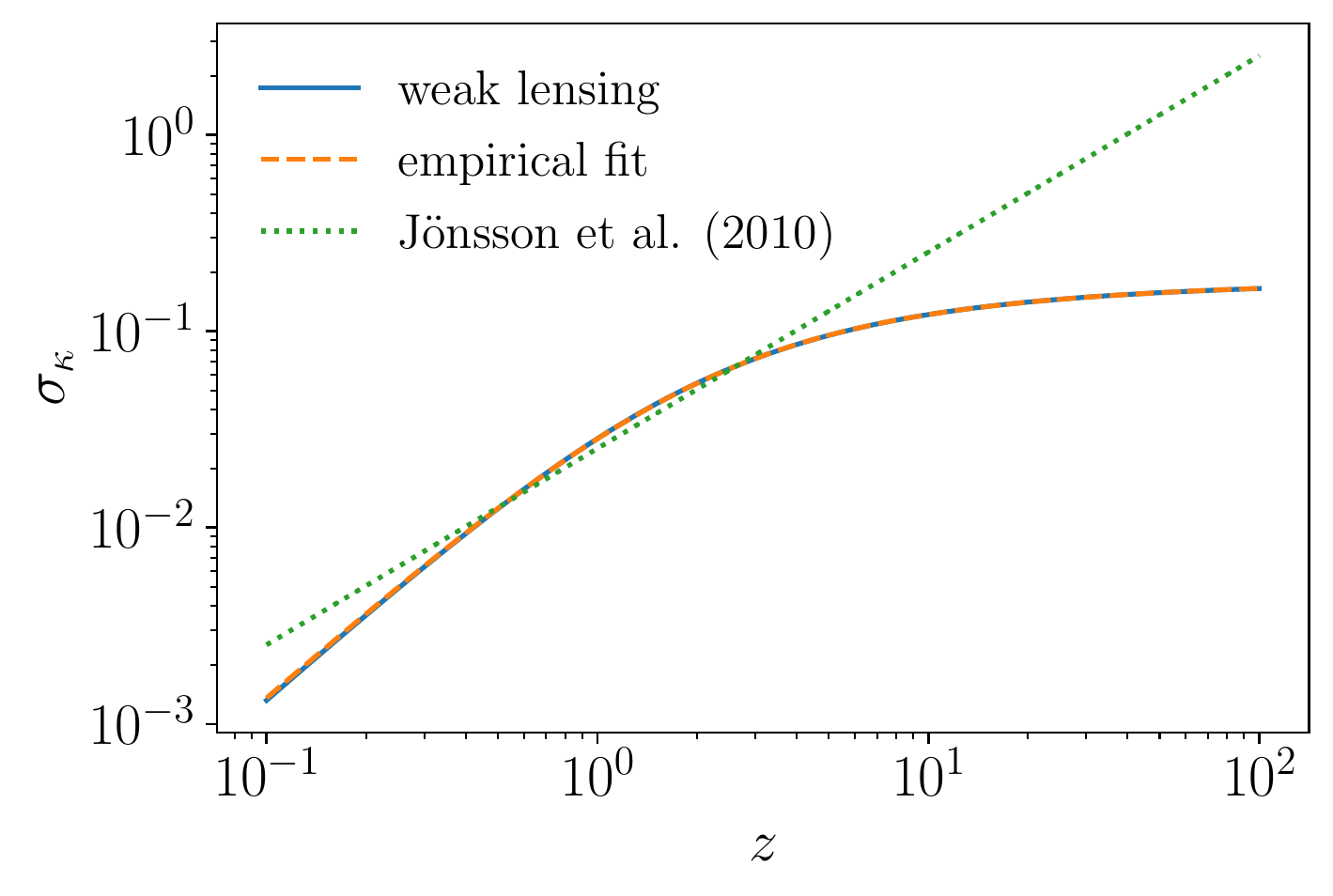}
	\hfill
	\includegraphics[width=0.49\columnwidth]{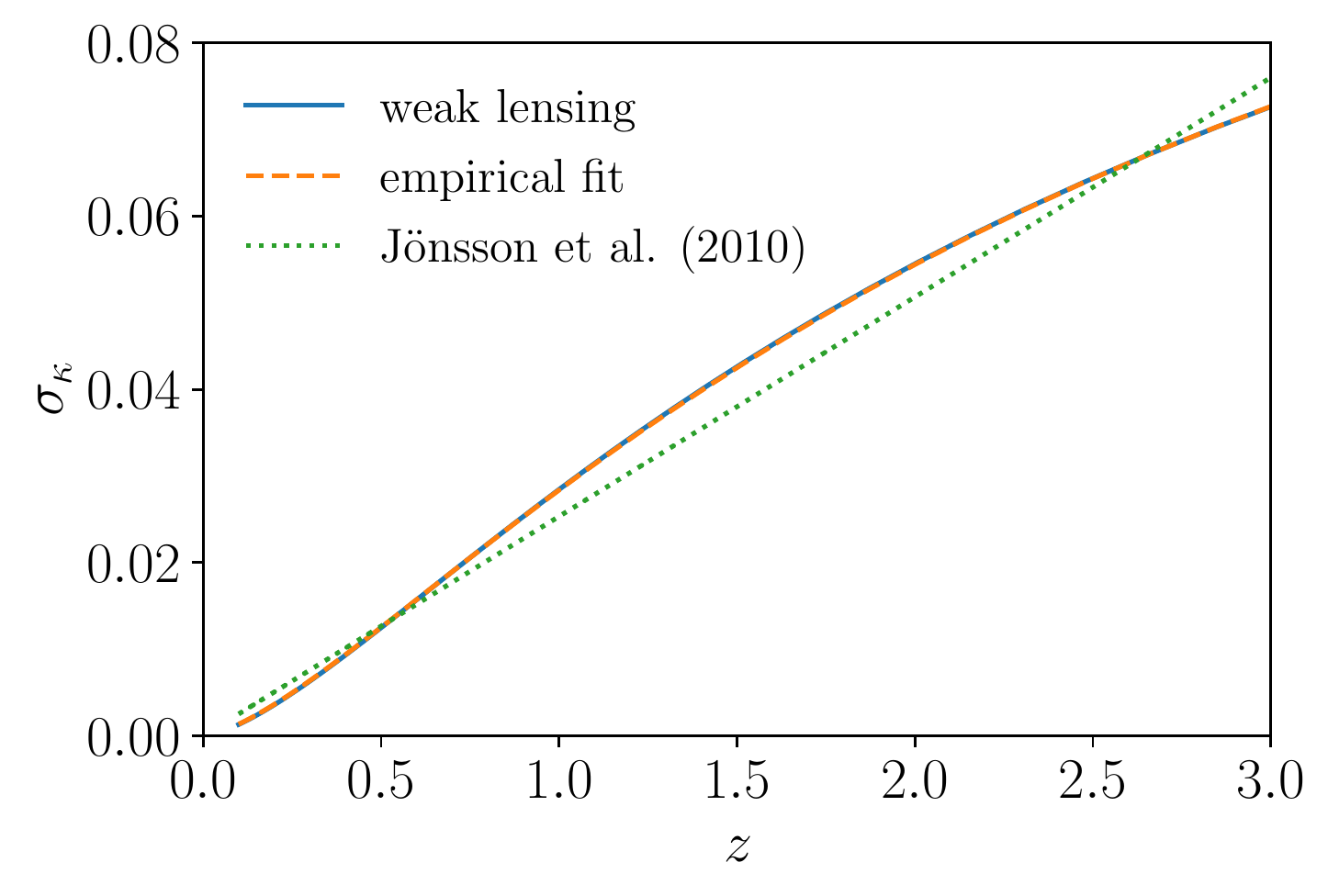}
	\caption{Standard deviation of the convergence~$\sigma_\kappa$ as a function of the redshift~$z$ of the source. Blue solid lines indicate the numerical results from \CAMB; dashed orange lines indicate the empirical fitting function~\eqref{eq:empirical_fit_sigma_kappa}; green dotted lines show the linear prescription of \cite{2010MNRAS.405..535J}. The left panel shows the entire redshift range $z\in[0, 100]$, while the right panel focuses on $z<3$.}
	\label{fig:sigma_kappa}
\end{figure}

We compute $P_W$ and the associated quantities using \CAMB. For a \LCDM cosmology with $H_0=\SI{67.4}{\kilo\meter\per\second\per\mega\parsec}, \Omega_\mathrm{ m} = 0.315, \Omega_\mathrm{ b} = 0.05,  A_\mathrm{ s}=2\times 10^{-9}, n_\mathrm{ s}=0.965$, we find that the standard deviation of the magnification as a function of redshift is very well fit (with sub-percent accuracy) by
\begin{equation}
	\label{eq:empirical_fit_sigma_kappa}
	\sigma_\kappa(z)
	= a \arctan\left[\left(1 + b \, z^c\right)^d - 1\right]
\end{equation}
with $a=0.116, b=1.26, c=1.46, d=0.268$.\footnote{These numerical values depend on the cosmology; our code includes a notebook \href{https://gitlab.com/matmartinelli/darksirens/-/blob/master/notebooks/variance_convergence.ipynb}{\faGitlab} capable of generating such a fitting function for other values of the cosmological parameters.} The result is depicted in \cref{fig:sigma_kappa}, where we also indicate the linear ansatz of \cite{2010MNRAS.405..535J} for comparison. The latter is a reasonably good approximation of $\sigma_\kappa(z)$ up to $z\approx 3$, but it highly overestimates it at high $z$. At $z=100$, we find $\sigma_\kappa \approx 16\%$, which is more than an order of magnitude below the prediction of the linear ansatz. This emphasises the importance of a careful modelling of lensing at high redshift.

\subsection{Lognormal probability distribution of the magnification}
\label{sec:appendix:mockdata:lensing:lognormal_PDF_magnification}

\begin{figure}
	\centering
	\includegraphics[width=0.49\columnwidth]{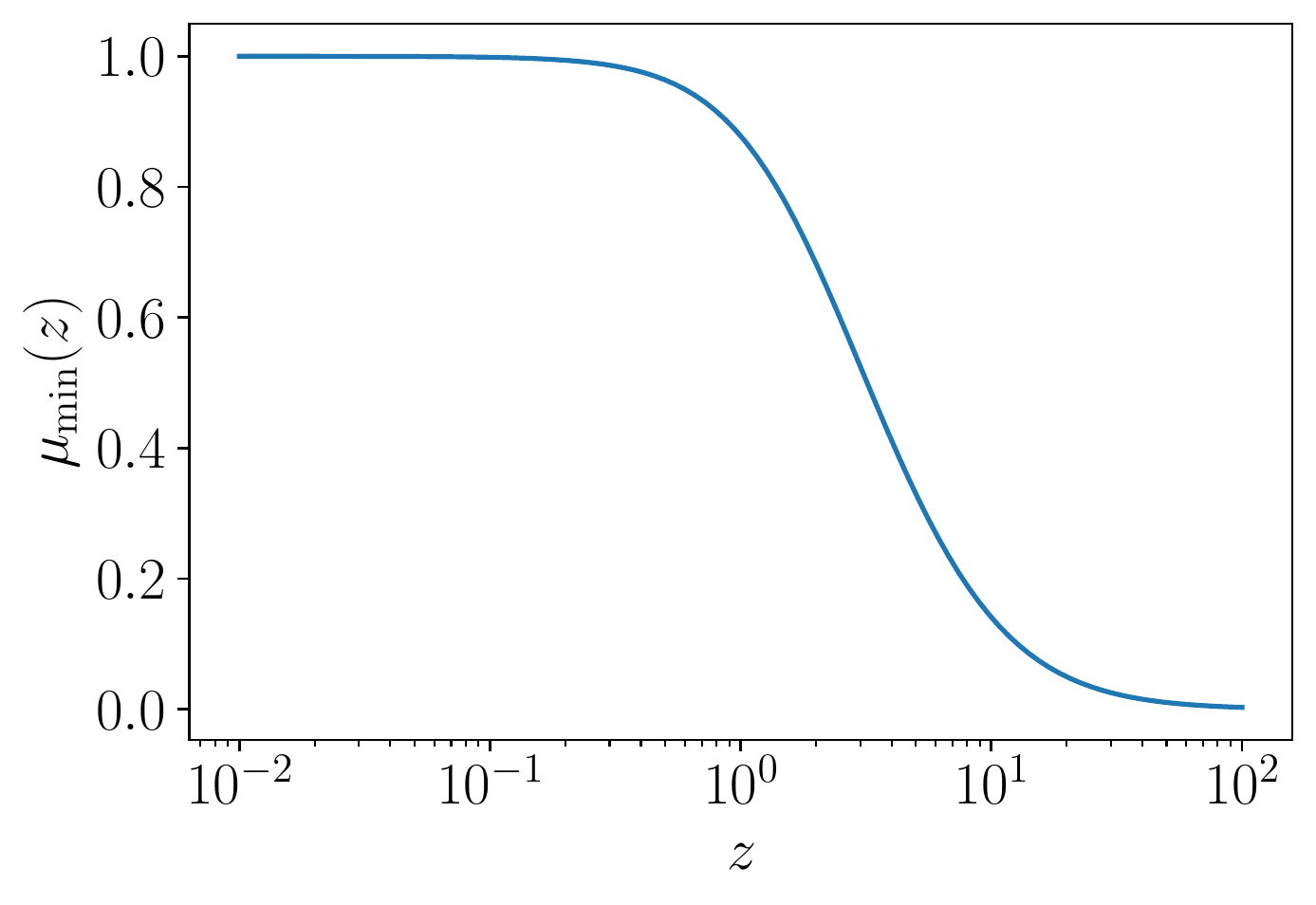}
	\hfill
	\includegraphics[width=0.49\columnwidth]{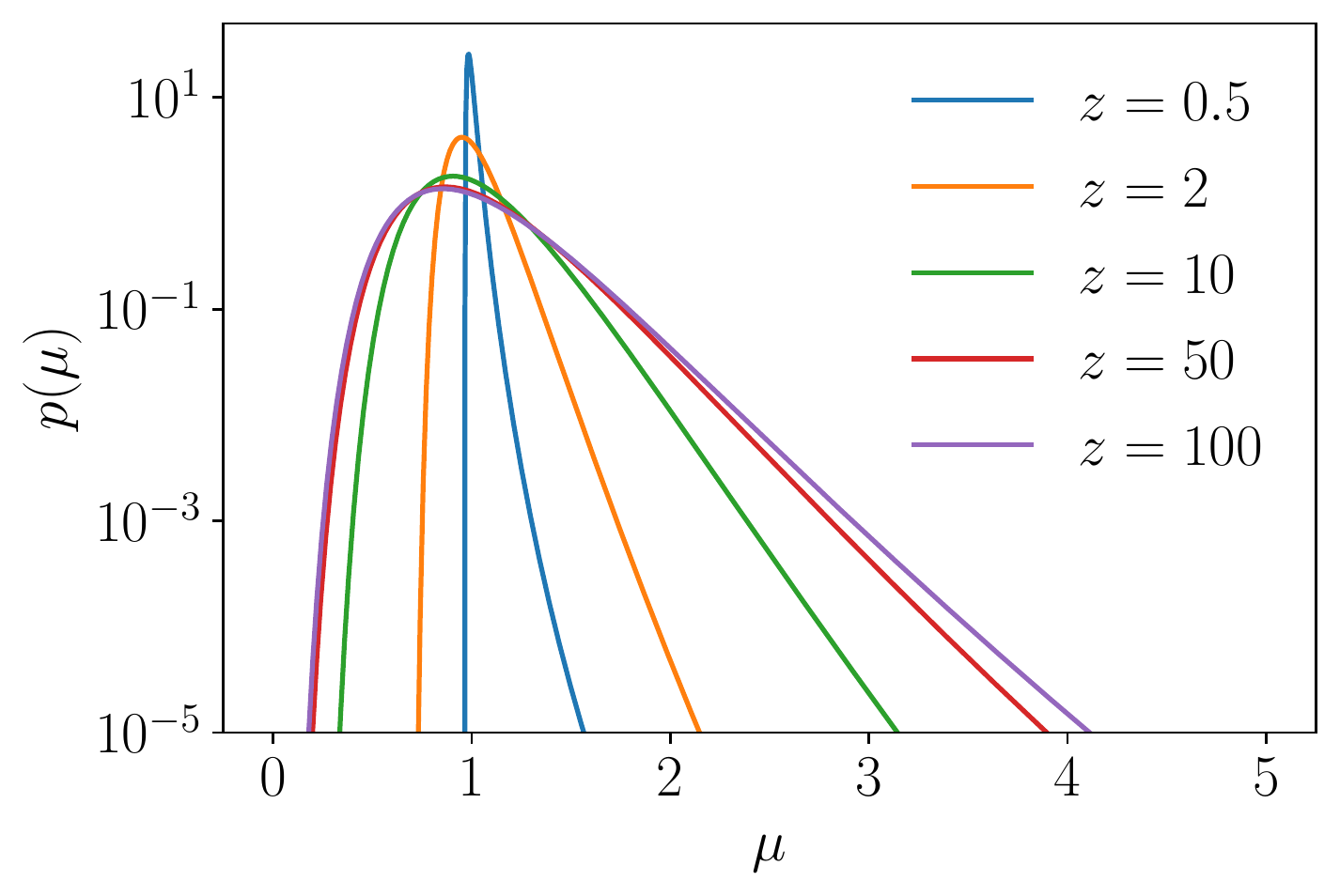}
	\caption{\textit{Left}: minimum magnification corresponding to Zel'dovich's empty-beam case, as defined in \cref{eq:empty_beam}. \textit{Right}: PDF of the magnification determined by the shifted lognormal model~\eqref{eq:model_PDF_mu} for different values of the source redshift~$z$.}
	\label{fig:mu_min_PDF_mu}
\end{figure}

We now turn to the full PDF~$p(\mu)$ of the magnification. This distribution must satisfy three physical requirements:
\begin{enumerate}
	\item It must vanish for $\mu<\mu_\mathrm{ min}$, where $\mu_\mathrm{ min}$ corresponds to the (de)magnification of Zel'dovich's empty beam~\cite{1964SvA.....8...13Z}. This is due to the fact that a bundle of null geodesics cannot be less focused than a beam propagating through pure vacuum. In that case, the angular diameter distance coincides with the affine parameter~$\lambda$ along the bundle; the minimum magnification thus reads
	\begin{equation}
		\label{eq:empty_beam}
		\mu_\mathrm{ min}(z) = \left[ \frac{\bar{D}(z)}{(1+z)^2\lambda(z)}\right]^2 < 1 \ ,
		\qquad
		\lambda(z) = \int_0^z \frac{\diff\zeta}{(1+\zeta)^2 H(\zeta)} \ .
	\end{equation}
	The evolution of $\mu_\mathrm{ min}$ with redshift $z$ is depicted in the left panel of \cref{fig:mu_min_PDF_mu}.
	\item The magnification averaged over sources must be unity~\cite{2008MNRAS.386..230W}
	\begin{equation}
		\label{eq:magnification_theorem}
		\langle \mu \rangle = \int_{\mu_\mathrm{ min}}^\infty \diff \mu \; \mu \, p(\mu) = 1 \ .
	\end{equation}
	This magnification theorem assumes that sources are homogeneously distributed in space. Note that the underlying averaging procedure -- over sources rather than over random directions in the sky -- is essential to this result. See \cite{2021A&A...655A..54B} and references therein for detailed discussions.
	\item The variance of the magnification should coincide with the one evaluated in \cref{sec:appendix:mockdata:lensing:variance_magnification},
	\begin{equation}
		\label{eq:magnification_variance_requirement}
		\langle \mu^2 \rangle - 1
		= \int_{\mu_\mathrm{ min}}^\infty \diff \mu \; \mu^2 \, p(\mu) - 1
		= 4\sigma_\kappa^2 \ .
	\end{equation}
\end{enumerate}

There are, of course, many models that would satisfy the above three requirements. We adopt, for simplicity, the following shifted lognormal model,
\begin{equation}
	\label{eq:model_PDF_mu}
	p(\mu) =
	\frac{1}{\sqrt{2\pi} \sigma (\mu - \mu_\mathrm{ min})} \,
	\exp\left\{- \frac{[\ln(\mu - \mu_\mathrm{ min}) - m]^2}{2\sigma^2} \right\} ,
\end{equation}
where $\mu_\mathrm{ min}$ is given by \cref{eq:empty_beam}, and the two free parameters $m, \sigma$ are fixed by the two conditions~\eqref{eq:magnification_theorem} and \eqref{eq:magnification_variance_requirement},
\begin{align}
	\sigma &= \sqrt{\ln\left[1 + \frac{4\sigma_\kappa^2}{(1-\mu_\mathrm{ min})^2}\right]} \ , \\
	m &= \ln(1 - \mu_\mathrm{ min}) - \frac{\sigma^2}{2} \ .
\end{align}

The resulting magnification PDF is depicted in the right panel of \cref{fig:mu_min_PDF_mu} for various values of the source redshift. Despite its simplicity, the shifted lognormal model mimics important properties of the expected magnification distribution in the inhomogeneous Universe. On the one hand, $p(\mu)$ peaks at $\mu<1$, thereby encoding the fact that most lines of sight have a lower column density than average, because voids occupy more volume in the Universe. On the other hand, $p(\mu)$ exhibits a longer tail towards high magnifications, which encodes that, albeit rare, overdensities can produce large magnifications. Note however that our model tends to underestimate the probability of those high magnifications compared to what is obtained with ray tracing in $N$-body simulations.

%

\bibliographystyle{JHEP}
\bibliography{biblio}

\providecommand{\href}[2]{#2}\begingroup\raggedright\begin{thebibliography}{100}

\bibitem{Scarcella:2020ssk}
F.~Scarcella, D.~Gaggero, R.~Connors, J.~Manshanden, M.~Ricotti and G.~Bertone,
  \emph{{Multi-wavelength detectability of isolated black holes in the Milky
  Way}},  \href{https://arxiv.org/abs/2012.10421}{{\ttfamily 2012.10421}}.

\bibitem{Scarcella:2021jzp}
F.~Scarcella, D.~Gaggero and J.~Garcia-Bellido, \emph{{Searching for isolated
  black holes in the Milky Way}},
  \href{https://doi.org/10.22323/1.395.0565}{\emph{PoS} {\bfseries ICRC2021}
  (2021) 565}.

\bibitem{Martinelli:2022elq}
M.~Martinelli, F.~Scarcella, N.B.~Hogg, B.J.~Kavanagh, D.~Gaggero and
  P.~Fleury, \emph{{Dancing in the dark: detecting a population of distant
  primordial black holes}},  \href{https://arxiv.org/abs/2205.02639}{{\ttfamily
  2205.02639}}.

\bibitem{Scarcella:2022pbh}
F.~Scarcella, D.~Gaggero and J.~Garcia-Bellido, \emph{{Searching for isolated
  black holes in the Milky Way: a Bayesian analysis}}, {\emph{In preparation}
  (2023) }.

\bibitem{Carter:1971zc}
B.~Carter, \emph{{Axisymmetric Black Hole Has Only Two Degrees of Freedom}},
  \href{https://doi.org/10.1103/PhysRevLett.26.331}{\emph{Phys. Rev. Lett.}
  {\bfseries 26} (1971) 331}.

\bibitem{Robinson:1975bv}
D.C.~Robinson, \emph{{Uniqueness of the Kerr black hole}},
  \href{https://doi.org/10.1103/PhysRevLett.34.905}{\emph{Phys. Rev. Lett.}
  {\bfseries 34} (1975) 905}.

\bibitem{Schwarzschild:1916uq}
K.~Schwarzschild, \emph{{On the gravitational field of a mass point according
  to Einstein's theory}}, {\emph{Sitzungsber. Preuss. Akad. Wiss. Berlin (Math.
  Phys. )} {\bfseries 1916} (1916) 189}
  [\href{https://arxiv.org/abs/physics/9905030}{{\ttfamily physics/9905030}}].

\bibitem{PhysRevLett.11.237}
R.P.~Kerr, \emph{Gravitational field of a spinning mass as an example of
  algebraically special metrics},
  \href{https://doi.org/10.1103/PhysRevLett.11.237}{\emph{Phys. Rev. Lett.}
  {\bfseries 11} (1963) 237}.

\bibitem{1964ApJ...140..796S}
E.E.~{Salpeter}, \emph{{Accretion of Interstellar Matter by Massive Objects.}},
  \href{https://doi.org/10.1086/147973}{\emph{\apj} {\bfseries 140} (1964)
  796}.

\bibitem{1972Natur.235...37W}
B.L.~{Webster} and P.~{Murdin}, \emph{{Cygnus X-1-a Spectroscopic Binary with a
  Heavy Companion ?}}, \href{https://doi.org/10.1038/235037a0}{\emph{\nat}
  {\bfseries 235} (1972) 37}.

\bibitem{1972Natur.235..271B}
C.T.~{Bolton}, \emph{{Identification of Cygnus X-1 with HDE 226868}},
  \href{https://doi.org/10.1038/235271b0}{\emph{\nat} {\bfseries 235} (1972)
  271}.

\bibitem{Abbott:2016blz}
{\scshape Virgo, LIGO Scientific} collaboration, \emph{{Observation of
  Gravitational Waves from a Binary Black Hole Merger}},
  \href{https://doi.org/10.1103/PhysRevLett.116.061102}{\emph{Phys. Rev. Lett.}
  {\bfseries 116} (2016) 061102}
  [\href{https://arxiv.org/abs/1602.03837}{{\ttfamily 1602.03837}}].

\bibitem{event2019first}
E.H.T.~Collaboration et~al., \emph{First m87 event horizon telescope results.
  i. the shadow of the supermassive black hole}, {\emph{arXiv preprint
  arXiv:1906.11238} (2019) }.

\bibitem{akiyama2022first}
K.~Akiyama, A.~Alberdi, W.~Alef, J.C.~Algaba, R.~Anantua, K.~Asada et~al.,
  \emph{First sagittarius a* event horizon telescope results. i. the shadow of
  the supermassive black hole in the center of the milky way}, {\emph{The
  Astrophysical Journal Letters} {\bfseries 930} (2022) L12}.

\bibitem{Bertone:2004pz}
G.~Bertone, D.~Hooper and J.~Silk, \emph{{Particle dark matter: Evidence,
  candidates and constraints}},
  \href{https://doi.org/10.1016/j.physrep.2004.08.031}{\emph{Phys. Rept.}
  {\bfseries 405} (2005) 279}
  [\href{https://arxiv.org/abs/hep-ph/0404175}{{\ttfamily hep-ph/0404175}}].

\bibitem{Profumo:2019ujg}
S.~Profumo, L.~Giani and O.F.~Piattella, \emph{{An Introduction to Particle
  Dark Matter}}, \href{https://doi.org/10.3390/universe5100213}{\emph{Universe}
  {\bfseries 5} (2019) 213} [\href{https://arxiv.org/abs/1910.05610}{{\ttfamily
  1910.05610}}].

\bibitem{Bertone:2016nfn}
G.~Bertone and D.~Hooper, \emph{{A History of Dark Matter}}, {\emph{Submitted
  to: Rev. Mod. Phys.} (2016) }
  [\href{https://arxiv.org/abs/1605.04909}{{\ttfamily 1605.04909}}].

\bibitem{2010dmp..book.....S}
R.H.~{Sanders}, \emph{{The Dark Matter Problem: A Historical Perspective}}
  (2010).

\bibitem{Begeman:1991iy}
K.G.~Begeman, A.H.~Broeils and R.H.~Sanders, \emph{{Extended rotation curves of
  spiral galaxies: Dark haloes and modified dynamics}},
  \href{https://doi.org/10.1093/mnras/249.3.523}{\emph{Mon. Not. Roy. Astron.
  Soc.} {\bfseries 249} (1991) 523}.

\bibitem{1970ApJ...159..379R}
V.C.~{Rubin} and J.~{Ford}, W.~Kent, \emph{{Rotation of the Andromeda Nebula
  from a Spectroscopic Survey of Emission Regions}},
  \href{https://doi.org/10.1086/150317}{\emph{\apj} {\bfseries 159} (1970)
  379}.

\bibitem{1975ApJ...201..327R}
M.S.~{Roberts} and R.N.~{Whitehurst}, \emph{{The rotation curve and geometry of
  M31 at large galactocentric distances.}},
  \href{https://doi.org/10.1086/153889}{\emph{\apj} {\bfseries 201} (1975)
  327}.

\bibitem{1978ApJ...225L.107R}
V.C.~{Rubin}, J.~{Ford}, W.~K. and N.~{Thonnard}, \emph{{Extended rotation
  curves of high-luminosity spiral galaxies. IV. Systematic dynamical
  properties, Sa -> Sc.}}, \href{https://doi.org/10.1086/182804}{\emph{\apjl}
  {\bfseries 225} (1978) L107}.

\bibitem{1970ApJ...160..811F}
K.C.~{Freeman}, \emph{{On the Disks of Spiral and S0 Galaxies}},
  \href{https://doi.org/10.1086/150474}{\emph{\apj} {\bfseries 160} (1970)
  811}.

\bibitem{1974Natur.250..309E}
J.~{Einasto}, A.~{Kaasik} and E.~{Saar}, \emph{{Dynamic evidence on massive
  coronas of galaxies}}, \href{https://doi.org/10.1038/250309a0}{\emph{\nat}
  {\bfseries 250} (1974) 309}.

\bibitem{1973A&A....26..483R}
M.S.~{Roberts} and A.H.~{Rots}, \emph{{Comparison of Rotation Curves of
  Different Galaxy Types}}, {\emph{\aap} {\bfseries 26} (1973) 483}.

\bibitem{1983ApJ...270..365M}
M.~{Milgrom}, \emph{{A modification of the Newtonian dynamics as a possible
  alternative to the hidden mass hypothesis.}},
  \href{https://doi.org/10.1086/161130}{\emph{\apj} {\bfseries 270} (1983)
  365}.

\bibitem{Bekenstein:2004ne}
J.D.~Bekenstein, \emph{{Relativistic gravitation theory for the MOND
  paradigm}}, \href{https://doi.org/10.1103/PhysRevD.70.083509}{\emph{Phys.
  Rev. D} {\bfseries 70} (2004) 083509}
  [\href{https://arxiv.org/abs/astro-ph/0403694}{{\ttfamily
  astro-ph/0403694}}].

\bibitem{Benito:2020lgu}
M.~Benito, F.~Iocco and A.~Cuoco, \emph{{Uncertainties in the Galactic Dark
  Matter distribution: An update}},
  \href{https://doi.org/10.1016/j.dark.2021.100826}{\emph{Phys. Dark Univ.}
  {\bfseries 32} (2021) 100826}
  [\href{https://arxiv.org/abs/2009.13523}{{\ttfamily 2009.13523}}].

\bibitem{1933AcHPh...6..110Z}
F.~{Zwicky}, \emph{{Die Rotverschiebung von extragalaktischen Nebeln}},
  {\emph{Helvetica Physica Acta} {\bfseries 6} (1933) 110}.

\bibitem{Markevitch:2001ri}
M.~Markevitch, A.H.~Gonzalez, L.~David, A.~Vikhlinin, S.~Murray, W.~Forman
  et~al., \emph{{A Textbook example of a bow shock in the merging galaxy
  cluster 1E0657-56}}, \href{https://doi.org/10.1086/339619}{\emph{Astrophys.
  J. Lett.} {\bfseries 567} (2002) L27}
  [\href{https://arxiv.org/abs/astro-ph/0110468}{{\ttfamily
  astro-ph/0110468}}].

\bibitem{Clowe:2003tk}
D.~Clowe, A.~Gonzalez and M.~Markevitch, \emph{{Weak lensing mass
  reconstruction of the interacting cluster 1E0657-558: Direct evidence for the
  existence of dark matter}},
  \href{https://doi.org/10.1086/381970}{\emph{Astrophys. J.} {\bfseries 604}
  (2004) 596} [\href{https://arxiv.org/abs/astro-ph/0312273}{{\ttfamily
  astro-ph/0312273}}].

\bibitem{Clowe:2006eq}
D.~Clowe, M.~Bradac, A.H.~Gonzalez, M.~Markevitch, S.W.~Randall, C.~Jones
  et~al., \emph{{A direct empirical proof of the existence of dark matter}},
  \href{https://doi.org/10.1086/508162}{\emph{Astrophys. J. Lett.} {\bfseries
  648} (2006) L109} [\href{https://arxiv.org/abs/astro-ph/0608407}{{\ttfamily
  astro-ph/0608407}}].

\bibitem{Bradac:2008eu}
M.~Bradac, S.W.~Allen, T.~Treu, H.~Ebeling, R.~Massey, R.G.~Morris et~al.,
  \emph{{Revealing the properties of dark matter in the merging cluster
  MACSJ0025.4-1222}}, \href{https://doi.org/10.1086/591246}{\emph{Astrophys.
  J.} {\bfseries 687} (2008) 959}
  [\href{https://arxiv.org/abs/0806.2320}{{\ttfamily 0806.2320}}].

\bibitem{Harvey:2015hha}
D.~Harvey, R.~Massey, T.~Kitching, A.~Taylor and E.~Tittley, \emph{{The
  non-gravitational interactions of dark matter in colliding galaxy clusters}},
  \href{https://doi.org/10.1126/science.1261381}{\emph{Science} {\bfseries 347}
  (2015) 1462} [\href{https://arxiv.org/abs/1503.07675}{{\ttfamily
  1503.07675}}].

\bibitem{2012MNRAS.426.2046A}
R.E.~{Angulo}, V.~{Springel}, S.D.M.~{White}, A.~{Jenkins}, C.M.~{Baugh} and
  C.S.~{Frenk}, \emph{{Scaling relations for galaxy clusters in the
  Millennium-XXL simulation}},
  \href{https://doi.org/10.1111/j.1365-2966.2012.21830.x}{\emph{\mnras}
  {\bfseries 426} (2012) 2046}
  [\href{https://arxiv.org/abs/1203.3216}{{\ttfamily 1203.3216}}].

\bibitem{Dietrich:2004kf}
J.P.~Dietrich, P.~Schneider, D.~Clowe, E.~Romano-Diaz and J.~Kerp, \emph{{Weak
  lensing evidence for a filament between the clusters A 222 and A 223 and its
  quantification}},
  \href{https://doi.org/10.1051/0004-6361:20041523}{\emph{Astron. Astrophys.}
  {\bfseries 440} (2005) 453}
  [\href{https://arxiv.org/abs/astro-ph/0406541}{{\ttfamily
  astro-ph/0406541}}].

\bibitem{Dietrich:2012mp}
J.P.~Dietrich, N.~Werner, D.~Clowe, A.~Finoguenov, T.~Kitching, L.~Miller
  et~al., \emph{{A filament of dark matter between two clusters of galaxies}},
  \href{https://doi.org/10.1038/nature11224}{\emph{Nature} {\bfseries 487}
  (2012) 202} [\href{https://arxiv.org/abs/1207.0809}{{\ttfamily 1207.0809}}].

\bibitem{Planck:2018vyg}
{\scshape Planck} collaboration, \emph{{Planck 2018 results. VI. Cosmological
  parameters}},
  \href{https://doi.org/10.1051/0004-6361/201833910}{\emph{Astron. Astrophys.}
  {\bfseries 641} (2020) A6}
  [\href{https://arxiv.org/abs/1807.06209}{{\ttfamily 1807.06209}}].

\bibitem{Taoso:2007qk}
M.~Taoso, G.~Bertone and A.~Masiero, \emph{{Dark Matter Candidates: A Ten-Point
  Test}}, \href{https://doi.org/10.1088/1475-7516/2008/03/022}{\emph{JCAP}
  {\bfseries 03} (2008) 022} [\href{https://arxiv.org/abs/0711.4996}{{\ttfamily
  0711.4996}}].

\bibitem{1983ApJ...274L...1W}
S.D.M.~{White}, C.S.~{Frenk} and M.~{Davis}, \emph{{Clustering in a
  neutrino-dominated universe}},
  \href{https://doi.org/10.1086/184139}{\emph{\apjl} {\bfseries 274} (1983)
  L1}.

\bibitem{1985ApJ...292..371D}
M.~{Davis}, G.~{Efstathiou}, C.S.~{Frenk} and S.D.M.~{White}, \emph{{The
  evolution of large-scale structure in a universe dominated by cold dark
  matter}}, \href{https://doi.org/10.1086/163168}{\emph{\apj} {\bfseries 292}
  (1985) 371}.

\bibitem{Frenk:2012ph}
C.S.~Frenk and S.D.M.~White, \emph{{Dark matter and cosmic structure}},
  \href{https://doi.org/10.1002/andp.201200212}{\emph{Annalen Phys.} {\bfseries
  524} (2012) 507} [\href{https://arxiv.org/abs/1210.0544}{{\ttfamily
  1210.0544}}].

\bibitem{2013MNRAS.428.1774B}
A.J.~{Benson}, A.~{Farahi}, S.~{Cole}, L.A.~{Moustakas}, A.~{Jenkins},
  M.~{Lovell} et~al., \emph{{Dark matter halo merger histories beyond cold dark
  matter - I. Methods and application to warm dark matter}},
  \href{https://doi.org/10.1093/mnras/sts159}{\emph{\mnras} {\bfseries 428}
  (2013) 1774} [\href{https://arxiv.org/abs/1209.3018}{{\ttfamily 1209.3018}}].

\bibitem{Kennedy:2013uta}
R.~Kennedy, C.~Frenk, S.~Cole and A.~Benson, \emph{{Constraining the warm dark
  matter particle mass with Milky Way satellites}},
  \href{https://doi.org/10.1093/mnras/stu719}{\emph{Mon. Not. Roy. Astron.
  Soc.} {\bfseries 442} (2014) 2487}
  [\href{https://arxiv.org/abs/1310.7739}{{\ttfamily 1310.7739}}].

\bibitem{Audren:2014bca}
B.~Audren, J.~Lesgourgues, G.~Mangano, P.D.~Serpico and T.~Tram,
  \emph{{Strongest model-independent bound on the lifetime of Dark Matter}},
  \href{https://doi.org/10.1088/1475-7516/2014/12/028}{\emph{JCAP} {\bfseries
  12} (2014) 028} [\href{https://arxiv.org/abs/1407.2418}{{\ttfamily
  1407.2418}}].

\bibitem{Tulin:2017ara}
S.~Tulin and H.-B.~Yu, \emph{{Dark Matter Self-interactions and Small Scale
  Structure}}, \href{https://doi.org/10.1016/j.physrep.2017.11.004}{\emph{Phys.
  Rept.} {\bfseries 730} (2018) 1}
  [\href{https://arxiv.org/abs/1705.02358}{{\ttfamily 1705.02358}}].

\bibitem{1985ApJ...299..633L}
C.G.~{Lacey} and J.P.~{Ostriker}, \emph{{Massive black holes in galactic halos
  ?}}, \href{https://doi.org/10.1086/163729}{\emph{\apj} {\bfseries 299} (1985)
  633}.

\bibitem{Boehm:2001hm}
C.~Boehm, A.~Riazuelo, S.H.~Hansen and R.~Schaeffer, \emph{{Interacting dark
  matter disguised as warm dark matter}},
  \href{https://doi.org/10.1103/PhysRevD.66.083505}{\emph{Phys. Rev. D}
  {\bfseries 66} (2002) 083505}
  [\href{https://arxiv.org/abs/astro-ph/0112522}{{\ttfamily
  astro-ph/0112522}}].

\bibitem{McDermott:2010pa}
S.D.~McDermott, H.-B.~Yu and K.M.~Zurek, \emph{{Turning off the Lights: How
  Dark is Dark Matter?}},
  \href{https://doi.org/10.1103/PhysRevD.83.063509}{\emph{Phys. Rev. D}
  {\bfseries 83} (2011) 063509}
  [\href{https://arxiv.org/abs/1011.2907}{{\ttfamily 1011.2907}}].

\bibitem{Sanchez-Salcedo:2010gfa}
F.J.~Sanchez-Salcedo, E.~Martinez-Gomez and J.~Magana, \emph{{On the fraction
  of dark matter in charged massive particles (CHAMPs)}},
  \href{https://doi.org/10.1088/1475-7516/2010/02/031}{\emph{JCAP} {\bfseries
  02} (2010) 031} [\href{https://arxiv.org/abs/1002.3145}{{\ttfamily
  1002.3145}}].

\bibitem{Arcadi:2017kky}
G.~Arcadi, M.~Dutra, P.~Ghosh, M.~Lindner, Y.~Mambrini, M.~Pierre et~al.,
  \emph{{The waning of the WIMP? A review of models, searches, and
  constraints}},
  \href{https://doi.org/10.1140/epjc/s10052-018-5662-y}{\emph{Eur. Phys. J. C}
  {\bfseries 78} (2018) 203}
  [\href{https://arxiv.org/abs/1703.07364}{{\ttfamily 1703.07364}}].

\bibitem{Abazajian:2017tcc}
K.N.~Abazajian, \emph{{Sterile neutrinos in cosmology}},
  \href{https://doi.org/10.1016/j.physrep.2017.10.003}{\emph{Phys. Rept.}
  {\bfseries 711-712} (2017) 1}
  [\href{https://arxiv.org/abs/1705.01837}{{\ttfamily 1705.01837}}].

\bibitem{Boyarsky:2018tvu}
A.~Boyarsky, M.~Drewes, T.~Lasserre, S.~Mertens and O.~Ruchayskiy,
  \emph{{Sterile neutrino Dark Matter}},
  \href{https://doi.org/10.1016/j.ppnp.2018.07.004}{\emph{Prog. Part. Nucl.
  Phys.} {\bfseries 104} (2019) 1}
  [\href{https://arxiv.org/abs/1807.07938}{{\ttfamily 1807.07938}}].

\bibitem{Lesgourgues:2006nd}
J.~Lesgourgues and S.~Pastor, \emph{{Massive neutrinos and cosmology}},
  \href{https://doi.org/10.1016/j.physrep.2006.04.001}{\emph{Phys. Rept.}
  {\bfseries 429} (2006) 307}
  [\href{https://arxiv.org/abs/astro-ph/0603494}{{\ttfamily
  astro-ph/0603494}}].

\bibitem{Irastorza:2018dyq}
I.G.~Irastorza and J.~Redondo, \emph{{New experimental approaches in the search
  for axion-like particles}},
  \href{https://doi.org/10.1016/j.ppnp.2018.05.003}{\emph{Prog. Part. Nucl.
  Phys.} {\bfseries 102} (2018) 89}
  [\href{https://arxiv.org/abs/1801.08127}{{\ttfamily 1801.08127}}].

\bibitem{Zeldovich:1967lct}
Y.B..N.~Zel'dovich, I.~D., \emph{{The Hypothesis of Cores Retarded during
  Expansion and the Hot Cosmological Model}}, {\emph{Soviet Astron. AJ (Engl.
  Transl. ),} {\bfseries 10} (1967) 602}.

\bibitem{Hawking:1971ei}
S.~Hawking, \emph{{Gravitationally collapsed objects of very low mass}},
  {\emph{Mon. Not. Roy. Astron. Soc.} {\bfseries 152} (1971) 75}.

\bibitem{Chapline:1975ojl}
G.F.~Chapline, \emph{{Cosmological effects of primordial black holes}},
  \href{https://doi.org/10.1038/253251a0}{\emph{Nature} {\bfseries 253} (1975)
  251}.

\bibitem{Carr:1974nx}
B.J.~Carr and S.W.~Hawking, \emph{{Black holes in the early Universe}},
  {\emph{Mon. Not. Roy. Astron. Soc.} {\bfseries 168} (1974) 399}.

\bibitem{1975ApJ...201....1C}
B.~{Carr}, \emph{{The primordial black hole mass spectrum}},
  \href{https://doi.org/10.1086/153853}{\emph{\apj} {\bfseries 201} (1975) 1}.

\bibitem{MACHO:1996qam}
{\scshape MACHO} collaboration, \emph{{The MACHO project LMC microlensing
  results from the first two years and the nature of the galactic dark halo}},
  \href{https://doi.org/10.1086/304535}{\emph{Astrophys. J.} {\bfseries 486}
  (1997) 697} [\href{https://arxiv.org/abs/astro-ph/9606165}{{\ttfamily
  astro-ph/9606165}}].

\bibitem{Fields:1999ar}
B.D.~Fields, K.~Freese and D.S.~Graff, \emph{{Chemical abundance constraints on
  white dwarfs as halo dark matter}},
  \href{https://doi.org/10.1086/308727}{\emph{Astrophys. J.} {\bfseries 534}
  (2000) 265} [\href{https://arxiv.org/abs/astro-ph/9904291}{{\ttfamily
  astro-ph/9904291}}].

\bibitem{MACHO:2000qbb}
{\scshape MACHO} collaboration, \emph{{The MACHO project: Microlensing results
  from 5.7 years of LMC observations}},
  \href{https://doi.org/10.1086/309512}{\emph{Astrophys. J.} {\bfseries 542}
  (2000) 281} [\href{https://arxiv.org/abs/astro-ph/0001272}{{\ttfamily
  astro-ph/0001272}}].

\bibitem{EROS-2:2006ryy}
{\scshape EROS-2} collaboration, \emph{{Limits on the Macho Content of the
  Galactic Halo from the EROS-2 Survey of the Magellanic Clouds}},
  \href{https://doi.org/10.1051/0004-6361:20066017}{\emph{Astron. Astrophys.}
  {\bfseries 469} (2007) 387}
  [\href{https://arxiv.org/abs/astro-ph/0607207}{{\ttfamily
  astro-ph/0607207}}].

\bibitem{LIGOScientific:2016aoc}
{\scshape LIGO Scientific, Virgo} collaboration, \emph{{Observation of
  Gravitational Waves from a Binary Black Hole Merger}},
  \href{https://doi.org/10.1103/PhysRevLett.116.061102}{\emph{Phys. Rev. Lett.}
  {\bfseries 116} (2016) 061102}
  [\href{https://arxiv.org/abs/1602.03837}{{\ttfamily 1602.03837}}].

\bibitem{Bird:2016dcv}
S.~Bird, I.~Cholis, J.B.~Muñoz, Y.~Ali-Haïmoud, M.~Kamionkowski, E.D.~Kovetz
  et~al., \emph{{Did LIGO detect dark matter?}},
  \href{https://doi.org/10.1103/PhysRevLett.116.201301}{\emph{Phys. Rev. Lett.}
  {\bfseries 116} (2016) 201301}
  [\href{https://arxiv.org/abs/1603.00464}{{\ttfamily 1603.00464}}].

\bibitem{Clesse:2016vqa}
S.~Clesse and J.~Garc\'ia-Bellido, \emph{{The clustering of massive Primordial
  Black Holes as Dark Matter: measuring their mass distribution with Advanced
  LIGO}}, \href{https://doi.org/10.1016/j.dark.2016.10.002}{\emph{Phys. Dark
  Univ.} {\bfseries 15} (2017) 142}
  [\href{https://arxiv.org/abs/1603.05234}{{\ttfamily 1603.05234}}].

\bibitem{PhysRevD.26.2681}
S.W.~Hawking, I.G.~Moss and J.M.~Stewart, \emph{Bubble collisions in the very
  early universe}, \href{https://doi.org/10.1103/PhysRevD.26.2681}{\emph{Phys.
  Rev. D} {\bfseries 26} (1982) 2681}.

\bibitem{1982PThPh..68.1979K}
H.~{Kodama}, M.~{Sasaki} and K.~{Sato}, \emph{{Abundance of Primordial Holes
  Produced by Cosmological First-Order Phase Transition}},
  \href{https://doi.org/10.1143/PTP.68.1979}{\emph{Progress of Theoretical
  Physics} {\bfseries 68} (1982) 1979}.

\bibitem{HAWKING1989237}
S.~Hawking, \emph{Black holes from cosmic strings},
  \href{https://doi.org/https://doi.org/10.1016/0370-2693(89)90206-2}{\emph{Physics
  Letters B} {\bfseries 231} (1989) 237}.

\bibitem{Jenkins:2020ctp}
A.C.~Jenkins and M.~Sakellariadou, \emph{{Primordial black holes from cusp
  collapse on cosmic strings}},
  \href{https://arxiv.org/abs/2006.16249}{{\ttfamily 2006.16249}}.

\bibitem{Carr:2020xqk}
B.~Carr and F.~Kuhnel, \emph{{Primordial Black Holes as Dark Matter: Recent
  Developments}},
  \href{https://doi.org/10.1146/annurev-nucl-050520-125911}{\emph{Ann. Rev.
  Nucl. Part. Sci.} {\bfseries 70} (2020) 355}
  [\href{https://arxiv.org/abs/2006.02838}{{\ttfamily 2006.02838}}].

\bibitem{Dolgov1993}
A.~Dolgov and J.~Silk, \emph{Baryon isocurvature fluctuations at small scales
  and baryonic dark matter},
  \href{https://doi.org/10.1103/PhysRevD.47.4244}{\emph{Phys. Rev. D}
  {\bfseries 47} (1993) 4244}.

\bibitem{Jedamzik:1996mr}
K.~Jedamzik, \emph{{Primordial black hole formation during the QCD epoch}},
  \href{https://doi.org/10.1103/PhysRevD.55.R5871}{\emph{Phys. Rev. D}
  {\bfseries 55} (1997) 5871}
  [\href{https://arxiv.org/abs/astro-ph/9605152}{{\ttfamily
  astro-ph/9605152}}].

\bibitem{Garcia-Bellido:1996mdl}
J.~Garcia-Bellido, A.D.~Linde and D.~Wands, \emph{{Density perturbations and
  black hole formation in hybrid inflation}},
  \href{https://doi.org/10.1103/PhysRevD.54.6040}{\emph{Phys. Rev. D}
  {\bfseries 54} (1996) 6040}
  [\href{https://arxiv.org/abs/astro-ph/9605094}{{\ttfamily
  astro-ph/9605094}}].

\bibitem{Garcia-Bellido:2017mdw}
J.~Garcia-Bellido and E.~Ruiz~Morales, \emph{{Primordial black holes from
  single field models of inflation}},
  \href{https://doi.org/10.1016/j.dark.2017.09.007}{\emph{Phys. Dark Univ.}
  {\bfseries 18} (2017) 47} [\href{https://arxiv.org/abs/1702.03901}{{\ttfamily
  1702.03901}}].

\bibitem{Ballesteros:2017fsr}
G.~Ballesteros and M.~Taoso, \emph{{Primordial black hole dark matter from
  single field inflation}},
  \href{https://doi.org/10.1103/PhysRevD.97.023501}{\emph{Phys. Rev.}
  {\bfseries D97} (2018) 023501}
  [\href{https://arxiv.org/abs/1709.05565}{{\ttfamily 1709.05565}}].

\bibitem{Pattison:2017mbe}
C.~Pattison, V.~Vennin, H.~Assadullahi and D.~Wands, \emph{{Quantum diffusion
  during inflation and primordial black holes}},
  \href{https://doi.org/10.1088/1475-7516/2017/10/046}{\emph{JCAP} {\bfseries
  10} (2017) 046} [\href{https://arxiv.org/abs/1707.00537}{{\ttfamily
  1707.00537}}].

\bibitem{Inman:2019wvr}
D.~Inman and Y.~Ali-Ha\"\i{}moud, \emph{{Early structure formation in
  primordial black hole cosmologies}},
  \href{https://doi.org/10.1103/PhysRevD.100.083528}{\emph{Phys. Rev. D}
  {\bfseries 100} (2019) 083528}
  [\href{https://arxiv.org/abs/1907.08129}{{\ttfamily 1907.08129}}].

\bibitem{Volonteri:2021sfo}
M.~Volonteri, M.~Habouzit and M.~Colpi, \emph{{The origins of massive black
  holes}}, \href{https://doi.org/10.1038/s42254-021-00364-9}{\emph{Nature Rev.
  Phys.} {\bfseries 3} (2021) 732}
  [\href{https://arxiv.org/abs/2110.10175}{{\ttfamily 2110.10175}}].

\bibitem{Lacki:2010zf}
B.C.~Lacki and J.F.~Beacom, \emph{{Primordial Black Holes as Dark Matter:
  Almost All or Almost Nothing}},
  \href{https://doi.org/10.1088/2041-8205/720/1/L67}{\emph{Astrophys. J.}
  {\bfseries 720} (2010) L67}
  [\href{https://arxiv.org/abs/1003.3466}{{\ttfamily 1003.3466}}].

\bibitem{Bertone:2019vsk}
G.~Bertone, A.M.~Coogan, D.~Gaggero, B.J.~Kavanagh and C.~Weniger,
  \emph{{Primordial Black Holes as Silver Bullets for New Physics at the Weak
  Scale}}, \href{https://doi.org/10.1103/PhysRevD.100.123013}{\emph{Phys. Rev.}
  {\bfseries D100} (2019) 123013}
  [\href{https://arxiv.org/abs/1905.01238}{{\ttfamily 1905.01238}}].

\bibitem{Adamek:2019gns}
J.~Adamek, C.T.~Byrnes, M.~Gosenca and S.~Hotchkiss, \emph{{WIMPs and
  stellar-mass primordial black holes are incompatible}},
  \href{https://arxiv.org/abs/1901.08528}{{\ttfamily 1901.08528}}.

\bibitem{Fujita:2014hha}
T.~Fujita, M.~Kawasaki, K.~Harigaya and R.~Matsuda, \emph{{Baryon asymmetry,
  dark matter, and density perturbation from primordial black holes}},
  \href{https://doi.org/10.1103/PhysRevD.89.103501}{\emph{Phys. Rev. D}
  {\bfseries 89} (2014) 103501}
  [\href{https://arxiv.org/abs/1401.1909}{{\ttfamily 1401.1909}}].

\bibitem{Hooper:2019gtx}
D.~Hooper, G.~Krnjaic and S.D.~McDermott, \emph{{Dark Radiation and Superheavy
  Dark Matter from Black Hole Domination}},
  \href{https://doi.org/10.1007/JHEP08(2019)001}{\emph{JHEP} {\bfseries 08}
  (2019) 001} [\href{https://arxiv.org/abs/1905.01301}{{\ttfamily
  1905.01301}}].

\bibitem{Gondolo:2020uqv}
P.~Gondolo, P.~Sandick and B.~Shams Es~Haghi, \emph{{Effects of primordial
  black holes on dark matter models}},
  \href{https://doi.org/10.1103/PhysRevD.102.095018}{\emph{Phys. Rev. D}
  {\bfseries 102} (2020) 095018}
  [\href{https://arxiv.org/abs/2009.02424}{{\ttfamily 2009.02424}}].

\bibitem{Cheek:2021cfe}
A.~Cheek, L.~Heurtier, Y.F.~Perez-Gonzalez and J.~Turner, \emph{{Primordial
  black hole evaporation and dark matter production. II. Interplay with the
  freeze-in or freeze-out mechanism}},
  \href{https://doi.org/10.1103/PhysRevD.105.015023}{\emph{Phys. Rev. D}
  {\bfseries 105} (2022) 015023}
  [\href{https://arxiv.org/abs/2107.00016}{{\ttfamily 2107.00016}}].

\bibitem{Green:2020jor}
A.M.~Green and B.J.~Kavanagh, \emph{{Primordial Black Holes as a dark matter
  candidate}}, \href{https://doi.org/10.1088/1361-6471/abc534}{\emph{J. Phys.
  G} {\bfseries 48} (2021) 043001}
  [\href{https://arxiv.org/abs/2007.10722}{{\ttfamily 2007.10722}}].

\bibitem{Byrnes:2021jka}
C.T.~Byrnes and P.S.~Cole, \emph{{Lecture notes on inflation and primordial
  black holes}},  \href{https://arxiv.org/abs/2112.05716}{{\ttfamily
  2112.05716}}.

\bibitem{Khlopov:1980mg}
M.Y.~Khlopov and A.G.~Polnarev, \emph{{PRIMORDIAL BLACK HOLES AS A COSMOLOGICAL
  TEST OF GRAND UNIFICATION}},
  \href{https://doi.org/10.1016/0370-2693(80)90624-3}{\emph{Phys. Lett. B}
  {\bfseries 97} (1980) 383}.

\bibitem{Harada:2016mhb}
T.~Harada, C.-M.~Yoo, K.~Kohri, K.-i.~Nakao and S.~Jhingan, \emph{{Primordial
  black hole formation in the matter-dominated phase of the Universe}},
  \href{https://doi.org/10.3847/1538-4357/833/1/61}{\emph{Astrophys. J.}
  {\bfseries 833} (2016) 61}
  [\href{https://arxiv.org/abs/1609.01588}{{\ttfamily 1609.01588}}].

\bibitem{PhysRevD.96.063507}
B.~Carr, T.~Tenkanen and V.~Vaskonen, \emph{Primordial black holes from
  inflaton and spectator field perturbations in a matter-dominated era},
  \href{https://doi.org/10.1103/PhysRevD.96.063507}{\emph{Phys. Rev. D}
  {\bfseries 96} (2017) 063507}.

\bibitem{Harada:2013epa}
T.~Harada, C.-M.~Yoo and K.~Kohri, \emph{{Threshold of primordial black hole
  formation}}, \href{https://doi.org/10.1103/PhysRevD.88.084051}{\emph{Phys.
  Rev. D} {\bfseries 88} (2013) 084051}
  [\href{https://arxiv.org/abs/1309.4201}{{\ttfamily 1309.4201}}].

\bibitem{Musco:2018rwt}
I.~Musco, \emph{{Threshold for primordial black holes: Dependence on the shape
  of the cosmological perturbations}},
  \href{https://doi.org/10.1103/PhysRevD.100.123524}{\emph{Phys. Rev. D}
  {\bfseries 100} (2019) 123524}
  [\href{https://arxiv.org/abs/1809.02127}{{\ttfamily 1809.02127}}].

\bibitem{Yoo:2020lmg}
C.-M.~Yoo, T.~Harada and H.~Okawa, \emph{{Threshold of Primordial Black Hole
  Formation in Nonspherical Collapse}},
  \href{https://doi.org/10.1103/PhysRevD.102.043526}{\emph{Phys. Rev. D}
  {\bfseries 102} (2020) 043526}
  [\href{https://arxiv.org/abs/2004.01042}{{\ttfamily 2004.01042}}].

\bibitem{Escriva:2021aeh}
A.~Escriv\`a, \emph{{PBH Formation from Spherically Symmetric Hydrodynamical
  Perturbations: A Review}},
  \href{https://doi.org/10.3390/universe8020066}{\emph{Universe} {\bfseries 8}
  (2022) 66} [\href{https://arxiv.org/abs/2111.12693}{{\ttfamily 2111.12693}}].

\bibitem{Byrnes:2018clq}
C.T.~Byrnes, M.~Hindmarsh, S.~Young and M.R.S.~Hawkins, \emph{{Primordial black
  holes with an accurate QCD equation of state}},
  \href{https://arxiv.org/abs/1801.06138}{{\ttfamily 1801.06138}}.

\bibitem{Carr:2019kxo}
B.~Carr, S.~Clesse, J.~Garc\'\i{}a-Bellido and F.~K\"uhnel, \emph{{Cosmic
  conundra explained by thermal history and primordial black holes}},
  \href{https://doi.org/10.1016/j.dark.2020.100755}{\emph{Phys. Dark Univ.}
  {\bfseries 31} (2021) 100755}
  [\href{https://arxiv.org/abs/1906.08217}{{\ttfamily 1906.08217}}].

\bibitem{Niemeyer:1997mt}
J.C.~Niemeyer and K.~Jedamzik, \emph{{Near-critical gravitational collapse and
  the initial mass function of primordial black holes}},
  \href{https://doi.org/10.1103/PhysRevLett.80.5481}{\emph{Phys. Rev. Lett.}
  {\bfseries 80} (1998) 5481}
  [\href{https://arxiv.org/abs/astro-ph/9709072}{{\ttfamily
  astro-ph/9709072}}].

\bibitem{Niemeyer:1999ak}
J.C.~Niemeyer and K.~Jedamzik, \emph{{Dynamics of primordial black hole
  formation}}, \href{https://doi.org/10.1103/PhysRevD.59.124013}{\emph{Phys.
  Rev. D} {\bfseries 59} (1999) 124013}
  [\href{https://arxiv.org/abs/astro-ph/9901292}{{\ttfamily
  astro-ph/9901292}}].

\bibitem{Musco:2004ak}
I.~Musco, J.C.~Miller and L.~Rezzolla, \emph{{Computations of primordial black
  hole formation}},
  \href{https://doi.org/10.1088/0264-9381/22/7/013}{\emph{Class. Quant. Grav.}
  {\bfseries 22} (2005) 1405}
  [\href{https://arxiv.org/abs/gr-qc/0412063}{{\ttfamily gr-qc/0412063}}].

\bibitem{Musco:2008hv}
I.~Musco, J.C.~Miller and A.G.~Polnarev, \emph{{Primordial black hole formation
  in the radiative era: Investigation of the critical nature of the collapse}},
  \href{https://doi.org/10.1088/0264-9381/26/23/235001}{\emph{Class. Quant.
  Grav.} {\bfseries 26} (2009) 235001}
  [\href{https://arxiv.org/abs/0811.1452}{{\ttfamily 0811.1452}}].

\bibitem{1974ApJ...187..425P}
W.H.~{Press} and P.~{Schechter}, \emph{{Formation of Galaxies and Clusters of
  Galaxies by Self-Similar Gravitational Condensation}},
  \href{https://doi.org/10.1086/152650}{\emph{\apj} {\bfseries 187} (1974)
  425}.

\bibitem{Planck:2018jri}
{\scshape Planck} collaboration, \emph{{Planck 2018 results. X. Constraints on
  inflation}}, \href{https://doi.org/10.1051/0004-6361/201833887}{\emph{Astron.
  Astrophys.} {\bfseries 641} (2020) A10}
  [\href{https://arxiv.org/abs/1807.06211}{{\ttfamily 1807.06211}}].

\bibitem{Ando:2018qdb}
K.~Ando, K.~Inomata and M.~Kawasaki, \emph{{Primordial black holes and
  uncertainties in the choice of the window function}},
  \href{https://doi.org/10.1103/PhysRevD.97.103528}{\emph{Phys. Rev. D}
  {\bfseries 97} (2018) 103528}
  [\href{https://arxiv.org/abs/1802.06393}{{\ttfamily 1802.06393}}].

\bibitem{Bullock:1996at}
J.S.~Bullock and J.R.~Primack, \emph{{NonGaussian fluctuations and primordial
  black holes from inflation}},
  \href{https://doi.org/10.1103/PhysRevD.55.7423}{\emph{Phys. Rev. D}
  {\bfseries 55} (1997) 7423}
  [\href{https://arxiv.org/abs/astro-ph/9611106}{{\ttfamily
  astro-ph/9611106}}].

\bibitem{Ivanov:1997ia}
P.~Ivanov, \emph{{Nonlinear metric perturbations and production of primordial
  black holes}}, \href{https://doi.org/10.1103/PhysRevD.57.7145}{\emph{Phys.
  Rev. D} {\bfseries 57} (1998) 7145}
  [\href{https://arxiv.org/abs/astro-ph/9708224}{{\ttfamily
  astro-ph/9708224}}].

\bibitem{Young:2015kda}
S.~Young and C.T.~Byrnes, \emph{{Signatures of non-gaussianity in the
  isocurvature modes of primordial black hole dark matter}},
  \href{https://doi.org/10.1088/1475-7516/2015/04/034}{\emph{JCAP} {\bfseries
  04} (2015) 034} [\href{https://arxiv.org/abs/1503.01505}{{\ttfamily
  1503.01505}}].

\bibitem{Franciolini:2018vbk}
G.~Franciolini, A.~Kehagias, S.~Matarrese and A.~Riotto, \emph{{Primordial
  Black Holes from Inflation and non-Gaussianity}},
  \href{https://doi.org/10.1088/1475-7516/2018/03/016}{\emph{JCAP} {\bfseries
  03} (2018) 016} [\href{https://arxiv.org/abs/1801.09415}{{\ttfamily
  1801.09415}}].

\bibitem{Atal:2018neu}
V.~Atal and C.~Germani, \emph{{The role of non-gaussianities in Primordial
  Black Hole formation}},
  \href{https://doi.org/10.1016/j.dark.2019.100275}{\emph{Phys. Dark Univ.}
  {\bfseries 24} (2019) 100275}
  [\href{https://arxiv.org/abs/1811.07857}{{\ttfamily 1811.07857}}].

\bibitem{Meerburg:2019qqi}
P.D.~Meerburg et~al., \emph{{Primordial Non-Gaussianity}},
  \href{https://arxiv.org/abs/1903.04409}{{\ttfamily 1903.04409}}.

\bibitem{Young:2019gfc}
S.~Young and C.T.~Byrnes, \emph{{Initial clustering and the primordial black
  hole merger rate}},
  \href{https://doi.org/10.1088/1475-7516/2020/03/004}{\emph{JCAP} {\bfseries
  03} (2020) 004} [\href{https://arxiv.org/abs/1910.06077}{{\ttfamily
  1910.06077}}].

\bibitem{Inomata:2021tpx}
K.~Inomata, E.~McDonough and W.~Hu, \emph{{Amplification of primordial
  perturbations from the rise or fall of the inflaton}},
  \href{https://doi.org/10.1088/1475-7516/2022/02/031}{\emph{JCAP} {\bfseries
  02} (2022) 031} [\href{https://arxiv.org/abs/2110.14641}{{\ttfamily
  2110.14641}}].

\bibitem{DeLuca:2019qsy}
V.~De~Luca, G.~Franciolini, A.~Kehagias, M.~Peloso, A.~Riotto and C.~\"Unal,
  \emph{{The Ineludible non-Gaussianity of the Primordial Black Hole
  Abundance}}, \href{https://doi.org/10.1088/1475-7516/2019/07/048}{\emph{JCAP}
  {\bfseries 07} (2019) 048}
  [\href{https://arxiv.org/abs/1904.00970}{{\ttfamily 1904.00970}}].

\bibitem{Young:2019yug}
S.~Young, I.~Musco and C.T.~Byrnes, \emph{{Primordial black hole formation and
  abundance: contribution from the non-linear relation between the density and
  curvature perturbation}},
  \href{https://doi.org/10.1088/1475-7516/2019/11/012}{\emph{JCAP} {\bfseries
  11} (2019) 012} [\href{https://arxiv.org/abs/1904.00984}{{\ttfamily
  1904.00984}}].

\bibitem{Kawasaki:2019mbl}
M.~Kawasaki and H.~Nakatsuka, \emph{{Effect of nonlinearity between density and
  curvature perturbations on the primordial black hole formation}},
  \href{https://doi.org/10.1103/PhysRevD.99.123501}{\emph{Phys. Rev. D}
  {\bfseries 99} (2019) 123501}
  [\href{https://arxiv.org/abs/1903.02994}{{\ttfamily 1903.02994}}].

\bibitem{Kinney:2005vj}
W.H.~Kinney, \emph{{Horizon crossing and inflation with large eta}},
  \href{https://doi.org/10.1103/PhysRevD.72.023515}{\emph{Phys. Rev. D}
  {\bfseries 72} (2005) 023515}
  [\href{https://arxiv.org/abs/gr-qc/0503017}{{\ttfamily gr-qc/0503017}}].

\bibitem{PhysRevD.50.7173}
P.~Ivanov, P.~Naselsky and I.~Novikov, \emph{Inflation and primordial black
  holes as dark matter},
  \href{https://doi.org/10.1103/PhysRevD.50.7173}{\emph{Phys. Rev. D}
  {\bfseries 50} (1994) 7173}.

\bibitem{Ezquiaga:2019ftu}
J.M.~Ezquiaga, J.~Garc\'\i{}a-Bellido and V.~Vennin, \emph{{The exponential
  tail of inflationary fluctuations: consequences for primordial black holes}},
  \href{https://doi.org/10.1088/1475-7516/2020/03/029}{\emph{JCAP} {\bfseries
  03} (2020) 029} [\href{https://arxiv.org/abs/1912.05399}{{\ttfamily
  1912.05399}}].

\bibitem{Braglia:2020eai}
M.~Braglia, D.K.~Hazra, F.~Finelli, G.F.~Smoot, L.~Sriramkumar and
  A.A.~Starobinsky, \emph{{Generating PBHs and small-scale GWs in two-field
  models of inflation}},
  \href{https://doi.org/10.1088/1475-7516/2020/08/001}{\emph{JCAP} {\bfseries
  08} (2020) 001} [\href{https://arxiv.org/abs/2005.02895}{{\ttfamily
  2005.02895}}].

\bibitem{Kannike:2017bxn}
K.~Kannike, L.~Marzola, M.~Raidal and H.~Veerm\"ae, \emph{{Single Field Double
  Inflation and Primordial Black Holes}},
  \href{https://doi.org/10.1088/1475-7516/2017/09/020}{\emph{JCAP} {\bfseries
  09} (2017) 020} [\href{https://arxiv.org/abs/1705.06225}{{\ttfamily
  1705.06225}}].

\bibitem{Green:2016xgy}
A.M.~Green, \emph{{Microlensing and dynamical constraints on primordial black
  hole dark matter with an extended mass function}},
  \href{https://doi.org/10.1103/PhysRevD.94.063530}{\emph{Phys. Rev.}
  {\bfseries D94} (2016) 063530}
  [\href{https://arxiv.org/abs/1609.01143}{{\ttfamily 1609.01143}}].

\bibitem{Gow:2020cou}
A.D.~Gow, C.T.~Byrnes and A.~Hall, \emph{{Accurate model for the primordial
  black hole mass distribution from a peak in the power spectrum}},
  \href{https://doi.org/10.1103/PhysRevD.105.023503}{\emph{Phys. Rev. D}
  {\bfseries 105} (2022) 023503}
  [\href{https://arxiv.org/abs/2009.03204}{{\ttfamily 2009.03204}}].

\bibitem{DeLuca:2020fpg}
V.~De~Luca, G.~Franciolini, P.~Pani and A.~Riotto, \emph{{Constraints on
  Primordial Black Holes: the Importance of Accretion}},
  \href{https://doi.org/10.1103/PhysRevD.102.043505}{\emph{Phys. Rev. D}
  {\bfseries 102} (2020) 043505}
  [\href{https://arxiv.org/abs/2003.12589}{{\ttfamily 2003.12589}}].

\bibitem{Mirbabayi:2019uph}
M.~Mirbabayi, A.~Gruzinov and J.~Nore\~na, \emph{{Spin of Primordial Black
  Holes}}, \href{https://doi.org/10.1088/1475-7516/2020/03/017}{\emph{JCAP}
  {\bfseries 03} (2020) 017}
  [\href{https://arxiv.org/abs/1901.05963}{{\ttfamily 1901.05963}}].

\bibitem{DeLuca:2019buf}
V.~De~Luca, V.~Desjacques, G.~Franciolini, A.~Malhotra and A.~Riotto,
  \emph{{The initial spin probability distribution of primordial black holes}},
  \href{https://doi.org/10.1088/1475-7516/2019/05/018}{\emph{JCAP} {\bfseries
  05} (2019) 018} [\href{https://arxiv.org/abs/1903.01179}{{\ttfamily
  1903.01179}}].

\bibitem{DeLuca:2020bjf}
V.~De~Luca, G.~Franciolini, P.~Pani and A.~Riotto, \emph{{The evolution of
  primordial black holes and their final observable spins}},
  \href{https://doi.org/10.1088/1475-7516/2020/04/052}{\emph{JCAP} {\bfseries
  04} (2020) 052} [\href{https://arxiv.org/abs/2003.02778}{{\ttfamily
  2003.02778}}].

\bibitem{Ali-Haimoud:2018dau}
Y.~Ali-Haïmoud, \emph{{Correlation function of high-threshold peaks and
  application to the initial (non)clustering of primordial black holes}},
  \href{https://arxiv.org/abs/1805.05912}{{\ttfamily 1805.05912}}.

\bibitem{Desjacques:2018wuu}
V.~Desjacques and A.~Riotto, \emph{{The Spatial Clustering of Primordial Black
  Holes}},  \href{https://arxiv.org/abs/1806.10414}{{\ttfamily 1806.10414}}.

\bibitem{Ballesteros:2018swv}
G.~Ballesteros, P.D.~Serpico and M.~Taoso, \emph{{On the merger rate of
  primordial black holes: effects of nearest neighbours distribution and
  clustering}},
  \href{https://doi.org/10.1088/1475-7516/2018/10/043}{\emph{JCAP} {\bfseries
  1810} (2018) 043} [\href{https://arxiv.org/abs/1807.02084}{{\ttfamily
  1807.02084}}].

\bibitem{Tada:2015noa}
Y.~Tada and S.~Yokoyama, \emph{{Primordial black holes as biased tracers}},
  \href{https://doi.org/10.1103/PhysRevD.91.123534}{\emph{Phys. Rev. D}
  {\bfseries 91} (2015) 123534}
  [\href{https://arxiv.org/abs/1502.01124}{{\ttfamily 1502.01124}}].

\bibitem{Carr:2017jsz}
B.~Carr, M.~Raidal, T.~Tenkanen, V.~Vaskonen and H.~Veermäe, \emph{{Primordial
  black hole constraints for extended mass functions}},
  \href{https://doi.org/10.1103/PhysRevD.96.023514}{\emph{Phys. Rev.}
  {\bfseries D96} (2017) 023514}
  [\href{https://arxiv.org/abs/1705.05567}{{\ttfamily 1705.05567}}].

\bibitem{Bellomo:2017zsr}
N.~Bellomo, J.L.~Bernal, A.~Raccanelli and L.~Verde, \emph{{Primordial Black
  Holes as Dark Matter: Converting Constraints from Monochromatic to Extended
  Mass Distributions}},
  \href{https://doi.org/10.1088/1475-7516/2018/01/004}{\emph{JCAP} {\bfseries
  1801} (2018) 004} [\href{https://arxiv.org/abs/1709.07467}{{\ttfamily
  1709.07467}}].

\bibitem{Carr:2020gox}
B.~Carr, K.~Kohri, Y.~Sendouda and J.~Yokoyama, \emph{{Constraints on
  primordial black holes}},
  \href{https://doi.org/10.1088/1361-6633/ac1e31}{\emph{Rept. Prog. Phys.}
  {\bfseries 84} (2021) 116902}
  [\href{https://arxiv.org/abs/2002.12778}{{\ttfamily 2002.12778}}].

\bibitem{Hawking:1975vcx}
S.W.~Hawking, \emph{{Particle Creation by Black Holes}},
  \href{https://doi.org/10.1007/BF02345020}{\emph{Commun. Math. Phys.}
  {\bfseries 43} (1975) 199}.

\bibitem{1974Natur.248...30H}
S.W.~{Hawking}, \emph{{Black hole explosions?}},
  \href{https://doi.org/10.1038/248030a0}{\emph{\nat} {\bfseries 248} (1974)
  30}.

\bibitem{1991ApJ...371..447M}
J.H.~{MacGibbon} and B.J.~{Carr}, \emph{{Cosmic Rays from Primordial Black
  Holes}}, \href{https://doi.org/10.1086/169909}{\emph{\apj} {\bfseries 371}
  (1991) 447}.

\bibitem{Carr:2009jm}
B.J.~Carr, K.~Kohri, Y.~Sendouda and J.~Yokoyama, \emph{{New cosmological
  constraints on primordial black holes}},
  \href{https://doi.org/10.1103/PhysRevD.81.104019}{\emph{Phys. Rev.}
  {\bfseries D81} (2010) 104019}
  [\href{https://arxiv.org/abs/0912.5297}{{\ttfamily 0912.5297}}].

\bibitem{Laha:2019ssq}
R.~Laha, \emph{{Primordial Black Holes as a Dark Matter Candidate Are Severely
  Constrained by the Galactic Center 511 keV $\gamma$ -Ray Line}},
  \href{https://doi.org/10.1103/PhysRevLett.123.251101}{\emph{Phys. Rev. Lett.}
  {\bfseries 123} (2019) 251101}
  [\href{https://arxiv.org/abs/1906.09994}{{\ttfamily 1906.09994}}].

\bibitem{Boudaud:2018hqb}
M.~Boudaud and M.~Cirelli, \emph{{Voyager 1 $e^\pm$ Further Constrain
  Primordial Black Holes as Dark Matter}},
  \href{https://doi.org/10.1103/PhysRevLett.122.041104}{\emph{Phys. Rev. Lett.}
  {\bfseries 122} (2019) 041104}
  [\href{https://arxiv.org/abs/1807.03075}{{\ttfamily 1807.03075}}].

\bibitem{Clark:2018ghm}
S.~Clark, B.~Dutta, Y.~Gao, Y.-Z.~Ma and L.E.~Strigari, \emph{{21 cm limits on
  decaying dark matter and primordial black holes}},
  \href{https://doi.org/10.1103/PhysRevD.98.043006}{\emph{Phys. Rev. D}
  {\bfseries 98} (2018) 043006}
  [\href{https://arxiv.org/abs/1803.09390}{{\ttfamily 1803.09390}}].

\bibitem{Poulin:2016anj}
V.~Poulin, J.~Lesgourgues and P.D.~Serpico, \emph{{Cosmological constraints on
  exotic injection of electromagnetic energy}},
  \href{https://doi.org/10.1088/1475-7516/2017/03/043}{\emph{JCAP} {\bfseries
  03} (2017) 043} [\href{https://arxiv.org/abs/1610.10051}{{\ttfamily
  1610.10051}}].

\bibitem{1987Natur.329..308M}
J.H.~{MacGibbon}, \emph{{Can Planck-mass relics of evaporating black holes
  close the Universe?}}, \href{https://doi.org/10.1038/329308a0}{\emph{\nat}
  {\bfseries 329} (1987) 308}.

\bibitem{Lehmann:2021ijf}
B.V.~Lehmann and S.~Profumo, \emph{{Black hole remnants are not too fast to be
  dark matter}},  \href{https://arxiv.org/abs/2105.01627}{{\ttfamily
  2105.01627}}.

\bibitem{1986ApJ...304....1P}
B.~{Paczynski}, \emph{{Gravitational Microlensing by the Galactic Halo}},
  \href{https://doi.org/10.1086/164140}{\emph{\apj} {\bfseries 304} (1986) 1}.

\bibitem{Macho:2000nvd}
{\scshape Macho} collaboration, \emph{{MACHO project limits on black hole dark
  matter in the 1-30 solar mass range}},
  \href{https://doi.org/10.1086/319636}{\emph{Astrophys. J. Lett.} {\bfseries
  550} (2001) L169} [\href{https://arxiv.org/abs/astro-ph/0011506}{{\ttfamily
  astro-ph/0011506}}].

\bibitem{2017Natur.548..183M}
P.~{Mr{\'o}z}, A.~{Udalski}, J.~{Skowron}, R.~{Poleski}, S.~{Koz{\l}owski},
  M.K.~{Szyma{\'n}ski} et~al., \emph{{No large population of unbound or
  wide-orbit Jupiter-mass planets}},
  \href{https://doi.org/10.1038/nature23276}{\emph{\nat} {\bfseries 548} (2017)
  183} [\href{https://arxiv.org/abs/1707.07634}{{\ttfamily 1707.07634}}].

\bibitem{Niikura:2019kqi}
H.~Niikura, M.~Takada, S.~Yokoyama, T.~Sumi and S.~Masaki, \emph{{Constraints
  on Earth-mass primordial black holes from OGLE 5-year microlensing events}},
  \href{https://doi.org/10.1103/PhysRevD.99.083503}{\emph{Phys. Rev. D}
  {\bfseries 99} (2019) 083503}
  [\href{https://arxiv.org/abs/1901.07120}{{\ttfamily 1901.07120}}].

\bibitem{Niikura:2017zjd}
H.~Niikura et~al., \emph{{Microlensing constraints on primordial black holes
  with Subaru/HSC Andromeda observations}},
  \href{https://doi.org/10.1038/s41550-019-0723-1}{\emph{Nature Astron.}
  {\bfseries 3} (2019) 524} [\href{https://arxiv.org/abs/1701.02151}{{\ttfamily
  1701.02151}}].

\bibitem{2015A&A...575A.107H}
M.R.S.~{Hawkins}, \emph{{A new look at microlensing limits on dark matter in
  the Galactic halo}},
  \href{https://doi.org/10.1051/0004-6361/201425400}{\emph{\aap} {\bfseries
  575} (2015) A107} [\href{https://arxiv.org/abs/1503.01935}{{\ttfamily
  1503.01935}}].

\bibitem{Green:2017qoa}
A.M.~Green, \emph{{Astrophysical uncertainties on stellar microlensing
  constraints on multi-Solar mass primordial black hole dark matter}},
  \href{https://doi.org/10.1103/PhysRevD.96.043020}{\emph{Phys. Rev.}
  {\bfseries D96} (2017) 043020}
  [\href{https://arxiv.org/abs/1705.10818}{{\ttfamily 1705.10818}}].

\bibitem{Calcino:2018mwh}
J.~Calcino, J.~Garcia-Bellido and T.M.~Davis, \emph{{Updating the MACHO
  fraction of the Milky Way dark halowith improved mass models}},
  \href{https://doi.org/10.1093/mnras/sty1368}{\emph{Mon. Not. Roy. Astron.
  Soc.} {\bfseries 479} (2018) 2889}
  [\href{https://arxiv.org/abs/1803.09205}{{\ttfamily 1803.09205}}].

\bibitem{Petac:2022rio}
M.~Peta\v{c}, J.~Lavalle and K.~Jedamzik, \emph{{Microlensing constraints on
  clustered primordial black holes}},
  \href{https://doi.org/10.1103/PhysRevD.105.083520}{\emph{Phys. Rev. D}
  {\bfseries 105} (2022) 083520}
  [\href{https://arxiv.org/abs/2201.02521}{{\ttfamily 2201.02521}}].

\bibitem{Gorton:2022fyb}
M.~Gorton and A.M.~Green, \emph{{Effect of clustering on primordial black hole
  microlensing constraints}},
  \href{https://arxiv.org/abs/2203.04209}{{\ttfamily 2203.04209}}.

\bibitem{Gaggero:2016dpq}
D.~Gaggero, G.~Bertone, F.~Calore, R.M.T.~Connors, M.~Lovell, S.~Markoff
  et~al., \emph{{Searching for Primordial Black Holes in the radio and X-ray
  sky}}, \href{https://doi.org/10.1103/PhysRevLett.118.241101}{\emph{Phys. Rev.
  Lett.} {\bfseries 118} (2017) 241101}
  [\href{https://arxiv.org/abs/1612.00457}{{\ttfamily 1612.00457}}].

\bibitem{Inoue:2017csr}
Y.~Inoue and A.~Kusenko, \emph{{New X-ray bound on density of primordial black
  holes}}, \href{https://doi.org/10.1088/1475-7516/2017/10/034}{\emph{JCAP}
  {\bfseries 10} (2017) 034}
  [\href{https://arxiv.org/abs/1705.00791}{{\ttfamily 1705.00791}}].

\bibitem{Manshanden:2018tze}
J.~Manshanden, D.~Gaggero, G.~Bertone, R.M.T.~Connors and M.~Ricotti,
  \emph{{Multi-wavelength astronomical searches for primordial black holes}},
  \href{https://doi.org/10.1088/1475-7516/2019/06/026}{\emph{JCAP} {\bfseries
  1906} (2019) 026} [\href{https://arxiv.org/abs/1812.07967}{{\ttfamily
  1812.07967}}].

\bibitem{1981MNRAS.194..639C}
B.J.~{Carr}, \emph{{Pregalactic black hole accretion and the thermal history of
  the universe}}, \href{https://doi.org/10.1093/mnras/194.3.639}{\emph{\mnras}
  {\bfseries 194} (1981) 639}.

\bibitem{Ricotti:2007au}
M.~Ricotti, J.P.~Ostriker and K.J.~Mack, \emph{{Effect of Primordial Black
  Holes on the Cosmic Microwave Background and Cosmological Parameter
  Estimates}}, \href{https://doi.org/10.1086/587831}{\emph{Astrophys. J.}
  {\bfseries 680} (2008) 829}
  [\href{https://arxiv.org/abs/0709.0524}{{\ttfamily 0709.0524}}].

\bibitem{Ali-Haimoud:2016mbv}
Y.~Ali-Haïmoud and M.~Kamionkowski, \emph{{Cosmic microwave background limits
  on accreting primordial black holes}},
  \href{https://doi.org/10.1103/PhysRevD.95.043534}{\emph{Phys. Rev.}
  {\bfseries D95} (2017) 043534}
  [\href{https://arxiv.org/abs/1612.05644}{{\ttfamily 1612.05644}}].

\bibitem{Poulin:2017bwe}
V.~Poulin, P.D.~Serpico, F.~Calore, S.~Clesse and K.~Kohri, \emph{{CMB bounds
  on disk-accreting massive primordial black holes}},
  \href{https://doi.org/10.1103/PhysRevD.96.083524}{\emph{Phys. Rev.}
  {\bfseries D96} (2017) 083524}
  [\href{https://arxiv.org/abs/1707.04206}{{\ttfamily 1707.04206}}].

\bibitem{Serpico:2020ehh}
P.D.~Serpico, V.~Poulin, D.~Inman and K.~Kohri, \emph{{Cosmic microwave
  background bounds on primordial black holes including dark matter halo
  accretion}},
  \href{https://doi.org/10.1103/PhysRevResearch.2.023204}{\emph{Phys. Rev.
  Res.} {\bfseries 2} (2020) 023204}
  [\href{https://arxiv.org/abs/2002.10771}{{\ttfamily 2002.10771}}].

\bibitem{Nakamura:1997sm}
T.~Nakamura, M.~Sasaki, T.~Tanaka and K.S.~Thorne, \emph{{Gravitational waves
  from coalescing black hole MACHO binaries}},
  \href{https://doi.org/10.1086/310886}{\emph{Astrophys. J. Lett.} {\bfseries
  487} (1997) L139} [\href{https://arxiv.org/abs/astro-ph/9708060}{{\ttfamily
  astro-ph/9708060}}].

\bibitem{Ioka:1998nz}
K.~Ioka, T.~Chiba, T.~Tanaka and T.~Nakamura, \emph{{Black hole binary
  formation in the expanding universe: Three body problem approximation}},
  \href{https://doi.org/10.1103/PhysRevD.58.063003}{\emph{Phys. Rev.}
  {\bfseries D58} (1998) 063003}
  [\href{https://arxiv.org/abs/astro-ph/9807018}{{\ttfamily
  astro-ph/9807018}}].

\bibitem{Sasaki:2016jop}
M.~Sasaki, T.~Suyama, T.~Tanaka and S.~Yokoyama, \emph{{Primordial Black Hole
  Scenario for the Gravitational-Wave Event GW150914}},
  \href{https://doi.org/10.1103/PhysRevLett.121.059901,
  10.1103/PhysRevLett.117.061101}{\emph{Phys. Rev. Lett.} {\bfseries 117}
  (2016) 061101} [\href{https://arxiv.org/abs/1603.08338}{{\ttfamily
  1603.08338}}].

\bibitem{Ali-Haimoud:2017rtz}
Y.~Ali-Ha\"\i{}moud, E.D.~Kovetz and M.~Kamionkowski, \emph{{Merger rate of
  primordial black-hole binaries}},
  \href{https://doi.org/10.1103/PhysRevD.96.123523}{\emph{Phys. Rev. D}
  {\bfseries 96} (2017) 123523}
  [\href{https://arxiv.org/abs/1709.06576}{{\ttfamily 1709.06576}}].

\bibitem{Kavanagh:2018ggo}
B.J.~Kavanagh, D.~Gaggero and G.~Bertone, \emph{{Merger rate of a subdominant
  population of primordial black holes}},
  \href{https://doi.org/10.1103/PhysRevD.98.023536}{\emph{Phys. Rev.}
  {\bfseries D98} (2018) 023536}
  [\href{https://arxiv.org/abs/1805.09034}{{\ttfamily 1805.09034}}].

\bibitem{Wong:2020yig}
K.W.K.~Wong, G.~Franciolini, V.~De~Luca, V.~Baibhav, E.~Berti, P.~Pani et~al.,
  \emph{{Constraining the primordial black hole scenario with Bayesian
  inference and machine learning: the GWTC-2 gravitational wave catalog}},
  \href{https://doi.org/10.1103/PhysRevD.103.023026}{\emph{Phys. Rev. D}
  {\bfseries 103} (2021) 023026}
  [\href{https://arxiv.org/abs/2011.01865}{{\ttfamily 2011.01865}}].

\bibitem{Raidal:2018bbj}
M.~Raidal, C.~Spethmann, V.~Vaskonen and H.~Veermäe, \emph{{Formation and
  Evolution of Primordial Black Hole Binaries in the Early Universe}},
  \href{https://doi.org/10.1088/1475-7516/2019/02/018}{\emph{JCAP} {\bfseries
  1902} (2019) 018} [\href{https://arxiv.org/abs/1812.01930}{{\ttfamily
  1812.01930}}].

\bibitem{Vaskonen:2019jpv}
V.~Vaskonen and H.~Veerm\"ae, \emph{{Lower bound on the primordial black hole
  merger rate}}, \href{https://doi.org/10.1103/PhysRevD.101.043015}{\emph{Phys.
  Rev. D} {\bfseries 101} (2020) 043015}
  [\href{https://arxiv.org/abs/1908.09752}{{\ttfamily 1908.09752}}].

\bibitem{Hall:2020daa}
A.~Hall, A.D.~Gow and C.T.~Byrnes, \emph{{Bayesian analysis of LIGO-Virgo
  mergers: Primordial vs. astrophysical black hole populations}},
  \href{https://doi.org/10.1103/PhysRevD.102.123524}{\emph{Phys. Rev. D}
  {\bfseries 102} (2020) 123524}
  [\href{https://arxiv.org/abs/2008.13704}{{\ttfamily 2008.13704}}].

\bibitem{Ananda:2006af}
K.N.~Ananda, C.~Clarkson and D.~Wands, \emph{{The Cosmological gravitational
  wave background from primordial density perturbations}},
  \href{https://doi.org/10.1103/PhysRevD.75.123518}{\emph{Phys. Rev. D}
  {\bfseries 75} (2007) 123518}
  [\href{https://arxiv.org/abs/gr-qc/0612013}{{\ttfamily gr-qc/0612013}}].

\bibitem{Baumann:2007zm}
D.~Baumann, P.J.~Steinhardt, K.~Takahashi and K.~Ichiki, \emph{{Gravitational
  Wave Spectrum Induced by Primordial Scalar Perturbations}},
  \href{https://doi.org/10.1103/PhysRevD.76.084019}{\emph{Phys. Rev. D}
  {\bfseries 76} (2007) 084019}
  [\href{https://arxiv.org/abs/hep-th/0703290}{{\ttfamily hep-th/0703290}}].

\bibitem{Domenech:2021ztg}
G.~Dom\`enech, \emph{{Scalar Induced Gravitational Waves Review}},
  \href{https://doi.org/10.3390/universe7110398}{\emph{Universe} {\bfseries 7}
  (2021) 398} [\href{https://arxiv.org/abs/2109.01398}{{\ttfamily
  2109.01398}}].

\bibitem{Gow:2020bzo}
A.D.~Gow, C.T.~Byrnes, P.S.~Cole and S.~Young, \emph{{The power spectrum on
  small scales: Robust constraints and comparing PBH methodologies}},
  \href{https://doi.org/10.1088/1475-7516/2021/02/002}{\emph{JCAP} {\bfseries
  02} (2021) 002} [\href{https://arxiv.org/abs/2008.03289}{{\ttfamily
  2008.03289}}].

\bibitem{Fixsen:1996nj}
D.J.~Fixsen, E.S.~Cheng, J.M.~Gales, J.C.~Mather, R.A.~Shafer and E.L.~Wright,
  \emph{{The Cosmic Microwave Background spectrum from the full COBE FIRAS data
  set}}, \href{https://doi.org/10.1086/178173}{\emph{Astrophys. J.} {\bfseries
  473} (1996) 576} [\href{https://arxiv.org/abs/astro-ph/9605054}{{\ttfamily
  astro-ph/9605054}}].

\bibitem{1994ApJ...420..439M}
J.C.~{Mather}, E.S.~{Cheng}, D.A.~{Cottingham}, J.~{Eplee}, R.~E.,
  D.J.~{Fixsen}, T.~{Hewagama} et~al., \emph{{Measurement of the Cosmic
  Microwave Background Spectrum by the COBE FIRAS Instrument}},
  \href{https://doi.org/10.1086/173574}{\emph{\apj} {\bfseries 420} (1994)
  439}.

\bibitem{Kohri:2014lza}
K.~Kohri, T.~Nakama and T.~Suyama, \emph{{Testing scenarios of primordial black
  holes being the seeds of supermassive black holes by ultracompact minihalos
  and CMB $\mu$-distortions}},
  \href{https://doi.org/10.1103/PhysRevD.90.083514}{\emph{Phys. Rev. D}
  {\bfseries 90} (2014) 083514}
  [\href{https://arxiv.org/abs/1405.5999}{{\ttfamily 1405.5999}}].

\bibitem{Nakama:2017xvq}
T.~Nakama, B.~Carr and J.~Silk, \emph{{Limits on primordial black holes from
  $\mu$ distortions in cosmic microwave background}},
  \href{https://doi.org/10.1103/PhysRevD.97.043525}{\emph{Phys. Rev. D}
  {\bfseries 97} (2018) 043525}
  [\href{https://arxiv.org/abs/1710.06945}{{\ttfamily 1710.06945}}].

\bibitem{Eda:2013gg}
K.~Eda, Y.~Itoh, S.~Kuroyanagi and J.~Silk, \emph{{New Probe of Dark-Matter
  Properties: Gravitational Waves from an Intermediate-Mass Black Hole Embedded
  in a Dark-Matter Minispike}},
  \href{https://doi.org/10.1103/PhysRevLett.110.221101}{\emph{Phys. Rev. Lett.}
  {\bfseries 110} (2013) 221101}
  [\href{https://arxiv.org/abs/1301.5971}{{\ttfamily 1301.5971}}].

\bibitem{Yue:2017iwc}
X.-J.~Yue and W.-B.~Han, \emph{{Gravitational waves with dark matter
  minispikes: the combined effect}},
  \href{https://arxiv.org/abs/1711.09706}{{\ttfamily 1711.09706}}.

\bibitem{Kavanagh:2020cfn}
B.J.~Kavanagh, D.A.~Nichols, G.~Bertone and D.~Gaggero, \emph{{Detecting dark
  matter around black holes with gravitational waves: Effects of dark-matter
  dynamics on the gravitational waveform}},
  \href{https://doi.org/10.1103/PhysRevD.102.083006}{\emph{Phys. Rev. D}
  {\bfseries 102} (2020) 083006}
  [\href{https://arxiv.org/abs/2002.12811}{{\ttfamily 2002.12811}}].

\bibitem{Babak:2017tow}
S.~Babak, J.~Gair, A.~Sesana, E.~Barausse, C.F.~Sopuerta, C.P.L.~Berry et~al.,
  \emph{{Science with the space-based interferometer LISA. V: Extreme
  mass-ratio inspirals}},
  \href{https://doi.org/10.1103/PhysRevD.95.103012}{\emph{Phys. Rev. D}
  {\bfseries 95} (2017) 103012}
  [\href{https://arxiv.org/abs/1703.09722}{{\ttfamily 1703.09722}}].

\bibitem{2021arXiv210804154C}
A.~{Coogan}, G.~{Bertone}, D.~{Gaggero}, B.J.~{Kavanagh} and D.A.~{Nichols},
  \emph{{Measuring the dark matter environments of black hole binaries with
  gravitational waves}}, {\emph{arXiv e-prints} (2021) arXiv:2108.04154}
  [\href{https://arxiv.org/abs/2108.04154}{{\ttfamily 2108.04154}}].

\bibitem{Hannuksela:2019vip}
O.A.~Hannuksela, K.C.Y.~Ng and T.G.F.~Li, \emph{{Extreme dark matter tests with
  extreme mass ratio inspirals}},
  \href{https://doi.org/10.1103/PhysRevD.102.103022}{\emph{Phys. Rev. D}
  {\bfseries 102} (2020) 103022}
  [\href{https://arxiv.org/abs/1906.11845}{{\ttfamily 1906.11845}}].

\bibitem{Gondolo:1999ef}
P.~Gondolo and J.~Silk, \emph{{Dark matter annihilation at the galactic
  center}}, \href{https://doi.org/10.1103/PhysRevLett.83.1719}{\emph{Phys. Rev.
  Lett.} {\bfseries 83} (1999) 1719}
  [\href{https://arxiv.org/abs/astro-ph/9906391}{{\ttfamily
  astro-ph/9906391}}].

\bibitem{Ullio:2001fb}
P.~Ullio, H.~Zhao and M.~Kamionkowski, \emph{{A Dark matter spike at the
  galactic center?}},
  \href{https://doi.org/10.1103/PhysRevD.64.043504}{\emph{Phys. Rev.}
  {\bfseries D64} (2001) 043504}
  [\href{https://arxiv.org/abs/astro-ph/0101481}{{\ttfamily
  astro-ph/0101481}}].

\bibitem{Zhao:2005zr}
H.-S.~Zhao and J.~Silk, \emph{{Mini-dark halos with intermediate mass black
  holes}}, \href{https://doi.org/10.1103/PhysRevLett.95.011301}{\emph{Phys.
  Rev. Lett.} {\bfseries 95} (2005) 011301}
  [\href{https://arxiv.org/abs/astro-ph/0501625}{{\ttfamily
  astro-ph/0501625}}].

\bibitem{volonteri2021origins}
M.~Volonteri, M.~Habouzit and M.~Colpi, \emph{The origins of massive black
  holes}, {\emph{Nature Reviews Physics} (2021) 1}.

\bibitem{Wanders:2014xia}
M.~Wanders, G.~Bertone, M.~Volonteri and C.~Weniger, \emph{{No WIMP Mini-Spikes
  in Dwarf Spheroidal Galaxies}},
  \href{https://doi.org/10.1088/1475-7516/2015/04/004}{\emph{JCAP} {\bfseries
  04} (2015) 004} [\href{https://arxiv.org/abs/1409.5797}{{\ttfamily
  1409.5797}}].

\bibitem{Merritt:2002vj}
D.~Merritt, M.~Milosavljevic, L.~Verde and R.~Jimenez, \emph{{Dark matter
  spikes and annihilation radiation from the galactic center}},
  \href{https://doi.org/10.1103/PhysRevLett.88.191301}{\emph{Phys. Rev. Lett.}
  {\bfseries 88} (2002) 191301}
  [\href{https://arxiv.org/abs/astro-ph/0201376}{{\ttfamily
  astro-ph/0201376}}].

\bibitem{1973A&A....24..337S}
N.I.~{Shakura} and R.A.~{Sunyaev}, \emph{{Black holes in binary systems.
  Observational appearance.}}, {\emph{\aap} {\bfseries 24} (1973) 337}.

\bibitem{1984ARA&A..22..471R}
M.J.~{Rees}, \emph{{Black Hole Models for Active Galactic Nuclei}},
  \href{https://doi.org/10.1146/annurev.aa.22.090184.002351}{\emph{\araa}
  {\bfseries 22} (1984) 471}.

\bibitem{Tauris:2003pf}
T.M.~Tauris and E.P.J.~van~den Heuvel, \emph{{Formation and evolution of
  compact stellar x-ray sources}},
  \href{https://arxiv.org/abs/astro-ph/0303456}{{\ttfamily astro-ph/0303456}}.

\bibitem{Corral-Santana:2015fud}
J.M.~Corral-Santana, J.~Casares, T.~Munoz-Darias, F.E.~Bauer,
  I.G.~Martinez-Pais and D.M.~Russell, \emph{{BlackCAT: A catalogue of
  stellar-mass black holes in X-ray transients}},
  \href{https://doi.org/10.1051/0004-6361/201527130}{\emph{Astron. Astrophys.}
  {\bfseries 587} (2016) A61}
  [\href{https://arxiv.org/abs/1510.08869}{{\ttfamily 1510.08869}}].

\bibitem{Wyrzykowski:2019jyg}
L.~Wyrzykowski and I.~Mandel, \emph{{Constraining the masses of microlensing
  black holes and the mass gap with Gaia DR2}},
  \href{https://doi.org/10.1051/0004-6361/201935842}{\emph{Astron. Astrophys.}
  {\bfseries 636} (2020) A20}
  [\href{https://arxiv.org/abs/1904.07789}{{\ttfamily 1904.07789}}].

\bibitem{Karolinski:2020jey}
N.~Karolinski and W.~Zhu, \emph{{Detecting isolated stellar-mass black holes in
  the absence of microlensing parallax effect}},
  \href{https://doi.org/10.1093/mnrasl/slaa121}{\emph{Mon. Not. Roy. Astron.
  Soc.} {\bfseries 498} (2020) L25}
  [\href{https://arxiv.org/abs/2006.02441}{{\ttfamily 2006.02441}}].

\bibitem{Calcino2018}
J.~Calcino, J.~Garc\'i­a-Bellido and T.M.~Davis, \emph{{Updating the MACHO
  fraction of the Milky Way dark halowith improved mass models}},
  \href{https://doi.org/10.1093/mnras/sty1368}{\emph{Monthly Notices of the
  Royal Astronomical Society} {\bfseries 479} (2018) 2889}
  [\href{https://arxiv.org/abs/https://academic.oup.com/mnras/article-pdf/479/3/2889/25149543/sty1368.pdf}{{\ttfamily
  https://academic.oup.com/mnras/article-pdf/479/3/2889/25149543/sty1368.pdf}}].

\bibitem{Park:2012cr}
K.~Park and M.~Ricotti, \emph{{Accretion onto Black Holes from Large Scales
  Regulated by Radiative Feedback. III. Enhanced Luminosity of Intermediate
  Mass Black Holes Moving at Supersonic Speeds}},
  \href{https://doi.org/10.1088/0004-637X/767/2/163}{\emph{Astrophys. J.}
  {\bfseries 767} (2013) 163}
  [\href{https://arxiv.org/abs/1211.0542}{{\ttfamily 1211.0542}}].

\bibitem{Sugimura:2020rdw}
K.~Sugimura and M.~Ricotti, \emph{{Structure and Instability of the Ionization
  Fronts around Moving Black Holes}},
  \href{https://doi.org/10.1093/mnras/staa1394}{\emph{Mon. Not. Roy. Astron.
  Soc.} {\bfseries 495} (2020) 2966}
  [\href{https://arxiv.org/abs/2003.05625}{{\ttfamily 2003.05625}}].

\bibitem{Hoyle1939}
F.~{Hoyle} and R.A.~{Lyttleton}, \emph{{The effect of interstellar matter on
  climatic variation}},
  \href{https://doi.org/10.1017/S0305004100021150}{\emph{Proceedings of the
  Cambridge Philosophical Society} {\bfseries 35} (1939) 405}.

\bibitem{Bondi1944}
H.~{Bondi} and F.~{Hoyle}, \emph{{On the mechanism of accretion by stars}},
  \href{https://doi.org/10.1093/mnras/104.5.273}{\emph{MNRAS} {\bfseries 104}
  (1944) 273}.

\bibitem{Fender:2013}
R.P.~{Fender}, T.J.~{Maccarone} and I.~{Heywood}, \emph{{The closest black
  holes}}, \href{https://doi.org/10.1093/mnras/sts688}{\emph{MNRAS} {\bfseries
  430} (2013) 1538} [\href{https://arxiv.org/abs/1301.1341}{{\ttfamily
  1301.1341}}].

\bibitem{Tsuna:2019kny}
D.~Tsuna and N.~Kawanaka, \emph{{Radio Emission from Accreting Isolated Black
  Holes in Our Galaxy}},
  \href{https://doi.org/10.1093/mnras/stz1809}{\emph{Mon. Not. Roy. Astron.
  Soc.} {\bfseries 488} (2019) 2099}
  [\href{https://arxiv.org/abs/1907.00792}{{\ttfamily 1907.00792}}].

\bibitem{Tsuna:2018oqt}
D.~Tsuna, N.~Kawanaka and T.~Totani, \emph{{X-ray Detectability of Accreting
  Isolated Black Holes in Our Galaxy}},
  \href{https://doi.org/10.1093/mnras/sty699}{\emph{Mon. Not. Roy. Astron.
  Soc.} {\bfseries 477} (2018) 791}
  [\href{https://arxiv.org/abs/1801.04667}{{\ttfamily 1801.04667}}].

\bibitem{Blandford:1999}
R.D.~{Blandford} and M.C.~{Begelman}, \emph{{On the fate of gas accreting at a
  low rate on to a black hole}},
  \href{https://doi.org/10.1046/j.1365-8711.1999.02358.x}{\emph{MNRAS}
  {\bfseries 303} (1999) L1}
  [\href{https://arxiv.org/abs/astro-ph/9809083}{{\ttfamily
  astro-ph/9809083}}].

\bibitem{Perna:2003}
R.~{Perna}, J.~{McDowell}, K.~{Menou}, J.~{Raymond} and M.V.~{Medvedev},
  \emph{{Chandra Observations of the Dwarf Nova WX Hydri in Quiescence}},
  \href{https://doi.org/10.1086/378855}{\emph{ApJ} {\bfseries 598} (2003) 545}
  [\href{https://arxiv.org/abs/astro-ph/0308081}{{\ttfamily
  astro-ph/0308081}}].

\bibitem{Pellegrini:2005}
S.~{Pellegrini}, \emph{{Nuclear Accretion in Galaxies of the Local Universe:
  Clues from Chandra Observations}},
  \href{https://doi.org/10.1086/429267}{\emph{ApJ} {\bfseries 624} (2005) 155}
  [\href{https://arxiv.org/abs/astro-ph/0502035}{{\ttfamily
  astro-ph/0502035}}].

\bibitem{2003ApJ...591..891B}
F.K.~{Baganoff}, Y.~{Maeda}, M.~{Morris}, M.W.~{Bautz}, W.N.~{Brandt}, W.~{Cui}
  et~al., \emph{{Chandra X-Ray Spectroscopic Imaging of Sagittarius A* and the
  Central Parsec of the Galaxy}},
  \href{https://doi.org/10.1086/375145}{\emph{\apj} {\bfseries 591} (2003) 891}
  [\href{https://arxiv.org/abs/astro-ph/0102151}{{\ttfamily
  astro-ph/0102151}}].

\bibitem{Narayan:1994}
R.~{Narayan} and I.~{Yi}, \emph{{Advection-dominated accretion: A self-similar
  solution}}, \href{https://doi.org/10.1086/187381}{\emph{ApJ Letters}
  {\bfseries 428} (1994) L13}
  [\href{https://arxiv.org/abs/astro-ph/9403052}{{\ttfamily
  astro-ph/9403052}}].

\bibitem{yqn03}
F.~{Yuan}, E.~{Quataert} and R.~{Narayan}, \emph{{Nonthermal Electrons in
  Radiatively Inefficient Accretion Flow Models of Sagittarius A*}},
  \href{https://doi.org/10.1086/378716}{\emph{\apj} {\bfseries 598} (2003) 301}
  [\href{https://arxiv.org/abs/astro-ph/0304125}{{\ttfamily
  astro-ph/0304125}}].

\bibitem{Esin1997}
A.A.~{Esin}, J.E.~{McClintock} and R.~{Narayan}, \emph{{Advection-Dominated
  Accretion and the Spectral States of Black Hole X-Ray Binaries: Application
  to Nova Muscae 1991}}, \href{https://doi.org/10.1086/304829}{\emph{\apj}
  {\bfseries 489} (1997) 865}
  [\href{https://arxiv.org/abs/astro-ph/9705237}{{\ttfamily
  astro-ph/9705237}}].

\bibitem{Fender:2001}
R.P.~{Fender}, \emph{{Powerful jets from black hole X-ray binaries in low/hard
  X-ray states}},
  \href{https://doi.org/10.1046/j.1365-8711.2001.04080.x}{\emph{MNRAS}
  {\bfseries 322} (2001) 31}
  [\href{https://arxiv.org/abs/astro-ph/0008447}{{\ttfamily
  astro-ph/0008447}}].

\bibitem{Plotkin2015}
R.M.~{Plotkin}, E.~{Gallo}, S.~{Markoff}, J.~{Homan}, P.G.~{Jonker},
  J.C.A.~{Miller-Jones} et~al., \emph{{Constraints on relativistic jets in
  quiescent black hole X-ray binaries from broad-band spectral modelling}},
  \href{https://doi.org/10.1093/mnras/stu2385}{\emph{\mnras} {\bfseries 446}
  (2015) 4098} [\href{https://arxiv.org/abs/1411.2973}{{\ttfamily 1411.2973}}].

\bibitem{Markoff2005}
S.~{Markoff}, M.A.~{Nowak} and J.~{Wilms}, \emph{{Going with the Flow: Can the
  Base of Jets Subsume the Role of Compact Accretion Disk Coronae?}},
  \href{https://doi.org/10.1086/497628}{\emph{\apj} {\bfseries 635} (2005)
  1203} [\href{https://arxiv.org/abs/astro-ph/0509028}{{\ttfamily
  astro-ph/0509028}}].

\bibitem{Corbel2000}
S.~{Corbel}, R.P.~{Fender}, A.K.~{Tzioumis}, M.~{Nowak}, V.~{McIntyre},
  P.~{Durouchoux} et~al., \emph{{Coupling of the X-ray and radio emission in
  the black hole candidate and compact jet source GX 339-4}}, {\emph{\aap}
  {\bfseries 359} (2000) 251}
  [\href{https://arxiv.org/abs/astro-ph/0003460}{{\ttfamily
  astro-ph/0003460}}].

\bibitem{Corbel2003}
S.~{Corbel}, M.A.~{Nowak}, R.P.~{Fender}, A.K.~{Tzioumis} and S.~{Markoff},
  \emph{{Radio/X-ray correlation in the low/hard state of GX 339-4}},
  \href{https://doi.org/10.1051/0004-6361:20030090}{\emph{\aap} {\bfseries 400}
  (2003) 1007} [\href{https://arxiv.org/abs/astro-ph/0301436}{{\ttfamily
  astro-ph/0301436}}].

\bibitem{Gallo2003}
E.~{Gallo}, R.P.~{Fender} and G.G.~{Pooley}, \emph{{A universal radio-X-ray
  correlation in low/hard state black hole binaries}},
  \href{https://doi.org/10.1046/j.1365-8711.2003.06791.x}{\emph{\mnras}
  {\bfseries 344} (2003) 60}
  [\href{https://arxiv.org/abs/astro-ph/0305231}{{\ttfamily
  astro-ph/0305231}}].

\bibitem{Corbel2008}
S.~{Corbel}, E.~{Koerding} and P.~{Kaaret}, \emph{{Revisiting the radio/X-ray
  flux correlation in the black hole V404 Cyg: from outburst to quiescence}},
  \href{https://doi.org/10.1111/j.1365-2966.2008.13542.x}{\emph{\mnras}
  {\bfseries 389} (2008) 1697}
  [\href{https://arxiv.org/abs/0806.3079}{{\ttfamily 0806.3079}}].

\bibitem{MillerJones2011}
J.C.A.~{Miller-Jones}, P.G.~{Jonker}, T.J.~{Maccarone}, G.~{Nelemans} and
  D.E.~{Calvelo}, \emph{{A Deep Radio Survey of Hard State and Quiescent Black
  Hole X-Ray Binaries}},
  \href{https://doi.org/10.1088/2041-8205/739/1/L18}{\emph{\apjl} {\bfseries
  739} (2011) L18} [\href{https://arxiv.org/abs/1106.0097}{{\ttfamily
  1106.0097}}].

\bibitem{Gallo2014}
E.~{Gallo}, J.C.A.~{Miller-Jones}, D.M.~{Russell}, P.G.~{Jonker}, J.~{Homan},
  R.M.~{Plotkin} et~al., \emph{{The radio/X-ray domain of black hole X-ray
  binaries at the lowest radio luminosities}},
  \href{https://doi.org/10.1093/mnras/stu1599}{\emph{\mnras} {\bfseries 445}
  (2014) 290} [\href{https://arxiv.org/abs/1408.3130}{{\ttfamily 1408.3130}}].

\bibitem{Merloni:2003}
A.~{Merloni}, S.~{Heinz} and T.~{di Matteo}, \emph{{A Fundamental Plane of
  black hole activity}},
  \href{https://doi.org/10.1046/j.1365-2966.2003.07017.x}{\emph{\mnras}
  {\bfseries 345} (2003) 1057}
  [\href{https://arxiv.org/abs/astro-ph/0305261}{{\ttfamily
  astro-ph/0305261}}].

\bibitem{F04}
H.~{Falcke}, E.~{K{\"o}rding} and S.~{Markoff}, \emph{{A scheme to unify
  low-power accreting black holes. Jet-dominated accretion flows and the
  radio/X-ray correlation}},
  \href{https://doi.org/10.1051/0004-6361:20031683}{\emph{\aap} {\bfseries 414}
  (2004) 895} [\href{https://arxiv.org/abs/astro-ph/0305335}{{\ttfamily
  astro-ph/0305335}}].

\bibitem{Plotkin:2012}
R.M.~Plotkin, S.~Markoff, B.C.~Kelly, E.~K\"{o}rding and S.F.~Anderson,
  \emph{{Using the Fundamental Plane of black hole activity to distinguish
  X-ray processes from weakly accreting black holes}},
  \href{https://doi.org/10.1111/j.1365-2966.2011.19689.x}{\emph{MNRAS}
  {\bfseries 419} (2012) 267}.

\bibitem{2019MNRAS.484.5734K}
J.M.D.~{Kruijssen}, J.E.~{Dale}, S.N.~{Longmore}, D.L.~{Walker},
  J.D.~{Henshaw}, S.M.R.~{Jeffreson} et~al., \emph{{The dynamical evolution of
  molecular clouds near the Galactic Centre - II. Spatial structure and
  kinematics of simulated clouds}},
  \href{https://doi.org/10.1093/mnras/stz381}{\emph{\mnras} {\bfseries 484}
  (2019) 5734} [\href{https://arxiv.org/abs/1902.01860}{{\ttfamily
  1902.01860}}].

\bibitem{Ferriere2007}
K.~{Ferri{\`e}re}, W.~{Gillard} and P.~{Jean}, \emph{{Spatial distribution of
  interstellar gas in the innermost 3 kpc of our galaxy}},
  \href{https://doi.org/10.1051/0004-6361:20066992}{\emph{AAP} {\bfseries 467}
  (2007) 611} [\href{https://arxiv.org/abs/astro-ph/0702532}{{\ttfamily
  astro-ph/0702532}}].

\bibitem{HeRG:2019}
C.-C.~{He}, M.~{Ricotti} and S.~{Geen}, \emph{{Simulating star clusters across
  cosmic time - I. Initial mass function, star formation rates, and
  efficiencies}}, \href{https://doi.org/10.1093/mnras/stz2239}{\emph{\mnras}
  {\bfseries 489} (2019) 1880}
  [\href{https://arxiv.org/abs/1904.07889}{{\ttfamily 1904.07889}}].

\bibitem{HeRG:2020}
C.-C.~{He}, M.~{Ricotti} and S.~{Geen}, \emph{{Simulating star clusters across
  cosmic time - II. Escape fraction of ionizing photons from molecular
  clouds}}, \href{https://doi.org/10.1093/mnras/staa165}{\emph{\mnras}
  {\bfseries 492} (2020) 4858}
  [\href{https://arxiv.org/abs/2001.06109}{{\ttfamily 2001.06109}}].

\bibitem{_zel_2010}
F.~Özel, D.~Psaltis, R.~Narayan and J.E.~McClintock, \emph{The black hole mass
  distribution in the galaxy},
  \href{https://doi.org/10.1088/0004-637x/725/2/1918}{\emph{The Astrophysical
  Journal} {\bfseries 725} (2010) 1918–1927}.

\bibitem{Farr_2011}
W.M.~Farr, N.~Sravan, A.~Cantrell, L.~Kreidberg, C.D.~Bailyn, I.~Mandel et~al.,
  \emph{The mass distribution of stellar-mass black holes},
  \href{https://doi.org/10.1088/0004-637x/741/2/103}{\emph{The Astrophysical
  Journal} {\bfseries 741} (2011) 103}.

\bibitem{2018}
A.G.A.~Brown, A.~Vallenari, T.~Prusti, J.H.J.~de~Bruijne, C.~Babusiaux,
  C.A.L.~Bailer-Jones et~al., \emph{Gaia data release 2},
  \href{https://doi.org/10.1051/0004-6361/201833051}{\emph{Astronomy \&
  Astrophysics} {\bfseries 616} (2018) A1}.

\bibitem{Hobbs_2005}
G.~Hobbs, D.R.~Lorimer, A.G.~Lyne and M.~Kramer, \emph{A statistical study of
  233 pulsar proper motions},
  \href{https://doi.org/10.1111/j.1365-2966.2005.09087.x}{\emph{Monthly Notices
  of the Royal Astronomical Society} {\bfseries 360} (2005) 974–992}.

\bibitem{2015ApJ...806...96L}
T.C.~{Licquia} and J.A.~{Newman}, \emph{{Improved Estimates of the Milky Way's
  Stellar Mass and Star Formation Rate from Hierarchical Bayesian
  Meta-Analysis}},
  \href{https://doi.org/10.1088/0004-637X/806/1/96}{\emph{\apj} {\bfseries 806}
  (2015) 96} [\href{https://arxiv.org/abs/1407.1078}{{\ttfamily 1407.1078}}].

\bibitem{McKee_2015}
C.F.~McKee, A.~Parravano and D.J.~Hollenbach, \emph{Stars, gas, and dark matter
  in the solar neighborhood},
  \href{https://doi.org/10.1088/0004-637x/814/1/13}{\emph{The Astrophysical
  Journal} {\bfseries 814} (2015) 13}.

\bibitem{Bird_2009}
A.J.~Bird, A.~Bazzano, L.~Bassani, F.~Capitanio, M.~Fiocchi, A.B.~Hill et~al.,
  \emph{The fourth ibis/isgri soft gamma-ray survey catalog},
  \href{https://doi.org/10.1088/0067-0049/186/1/1}{\emph{The Astrophysical
  Journal Supplement Series} {\bfseries 186} (2009) 1–9}.

\bibitem{Sofue_2013}
Y.~Sofue, \emph{Rotation curve and mass distribution in the galactic center
  —from black hole to entire galaxy—},
  \href{https://doi.org/10.1093/pasj/65.6.118}{\emph{Publications of the
  Astronomical Society of Japan} {\bfseries 65} (2013) 118}.

\bibitem{Irrgang_2013}
A.~Irrgang, B.~Wilcox, E.~Tucker and L.~Schiefelbein, \emph{Milky way mass
  models for orbit calculations},
  \href{https://doi.org/10.1051/0004-6361/201220540}{\emph{Astronomy \&
  Astrophysics} {\bfseries 549} (2013) A137}.

\bibitem{Sanders_2019}
J.L.~Sanders, L.~Smith, N.W.~Evans and P.~Lucas, \emph{Transverse kinematics of
  the galactic bar-bulge from vvv and gaia},
  \href{https://doi.org/10.1093/mnras/stz1630}{\emph{Monthly Notices of the
  Royal Astronomical Society} {\bfseries 487} (2019) 5188–5208}.

\bibitem{Hong:2016}
J.e.a.~{Hong}, \emph{{NuSTAR Hard X-Ray Survey of the Galactic Center Region.
  II. X-Ray Point Sources}},
  \href{https://doi.org/10.3847/0004-637X/825/2/132}{\emph{\apj} {\bfseries
  825} (2016) 132} [\href{https://arxiv.org/abs/1605.03882}{{\ttfamily
  1605.03882}}].

\bibitem{Bull:2018lat}
A.~Weltman et~al., \emph{{Fundamental physics with the Square Kilometre
  Array}}, \href{https://doi.org/10.1017/pasa.2019.42}{\emph{Publ. Astron. Soc.
  Austral.} {\bfseries 37} (2020) e002}
  [\href{https://arxiv.org/abs/1810.02680}{{\ttfamily 1810.02680}}].

\bibitem{2016ApJ...827..143C}
F.~{Calore}, M.~{Di Mauro}, F.~{Donato}, J.W.T.~{Hessels} and C.~{Weniger},
  \emph{{Radio Detection Prospects for a Bulge Population of Millisecond
  Pulsars as Suggested by Fermi-LAT Observations of the Inner Galaxy}},
  \href{https://doi.org/10.3847/0004-637X/827/2/143}{\emph{\apj} {\bfseries
  827} (2016) 143} [\href{https://arxiv.org/abs/1512.06825}{{\ttfamily
  1512.06825}}].

\bibitem{2008ApJS..174..481L}
T.J.W.~{Lazio} and J.M.~{Cordes}, \emph{{A VLA Survey for Compact Radio Sources
  in the Galactic Center}}, \href{https://doi.org/10.1086/521676}{\emph{\apjs}
  {\bfseries 174} (2008) 481}.

\bibitem{Russell2013}
D.M.~{Russell}, S.~{Markoff}, P.~{Casella}, A.G.~{Cantrell}, R.~{Chatterjee},
  R.P.~{Fender} et~al., \emph{{Jet spectral breaks in black hole X-ray
  binaries}}, \href{https://doi.org/10.1093/mnras/sts377}{\emph{\mnras}
  {\bfseries 429} (2013) 815}
  [\href{https://arxiv.org/abs/1211.1655}{{\ttfamily 1211.1655}}].

\bibitem{Kashlinsky:2019kac}
A.~Kashlinsky et~al., \emph{{Electromagnetic probes of primordial black holes
  as dark matter}},  \href{https://arxiv.org/abs/1903.04424}{{\ttfamily
  1903.04424}}.

\bibitem{Salucci:2018hqu}
P.~Salucci, \emph{{The distribution of dark matter in galaxies}},
  \href{https://doi.org/10.1007/s00159-018-0113-1}{\emph{Astron. Astrophys.
  Rev.} {\bfseries 27} (2019) 2}
  [\href{https://arxiv.org/abs/1811.08843}{{\ttfamily 1811.08843}}].

\bibitem{Navarro:1996gj}
J.F.~Navarro, C.S.~Frenk and S.D.M.~White, \emph{{A Universal density profile
  from hierarchical clustering}},
  \href{https://doi.org/10.1086/304888}{\emph{Astrophys. J.} {\bfseries 490}
  (1997) 493} [\href{https://arxiv.org/abs/astro-ph/9611107}{{\ttfamily
  astro-ph/9611107}}].

\bibitem{Governato:2009bg}
F.~Governato et~al., \emph{{At the heart of the matter: the origin of bulgeless
  dwarf galaxies and Dark Matter cores}},
  \href{https://doi.org/10.1038/nature08640}{\emph{Nature} {\bfseries 463}
  (2010) 203} [\href{https://arxiv.org/abs/0911.2237}{{\ttfamily 0911.2237}}].

\bibitem{Binney:2001wu}
J.J.~Binney and N.W.~Evans, \emph{{Cuspy dark-matter haloes and the Galaxy}},
  \href{https://doi.org/10.1046/j.1365-8711.2001.04968.x}{\emph{Mon. Not. Roy.
  Astron. Soc.} {\bfseries 327} (2001) L27}
  [\href{https://arxiv.org/abs/astro-ph/0108505}{{\ttfamily
  astro-ph/0108505}}].

\bibitem{Hong:2016qjq}
J.~Hong et~al., \emph{{NuSTAR Hard X-ray Survey of the Galactic Center Region
  II: X-ray Point Sources}},
  \href{https://doi.org/10.3847/0004-637X/825/2/132}{\emph{Astrophys. J.}
  {\bfseries 825} (2016) 132}
  [\href{https://arxiv.org/abs/1605.03882}{{\ttfamily 1605.03882}}].

\bibitem{1916SPAW.......688E}
A.~{Einstein}, \emph{{N{\"a}herungsweise Integration der Feldgleichungen der
  Gravitation}}, {\emph{Sitzungsberichte der K{\"o}niglich Preu{\ss}ischen
  Akademie der Wissenschaften (Berlin} (1916) 688}.

\bibitem{Hulse:1974eb}
R.A.~Hulse and J.H.~Taylor, \emph{{Discovery of a pulsar in a binary system}},
  \href{https://doi.org/10.1086/181708}{\emph{Astrophys. J. Lett.} {\bfseries
  195} (1975) L51}.

\bibitem{1982ApJ...253..908T}
J.H.~{Taylor} and J.M.~{Weisberg}, \emph{{A new test of general relativity -
  Gravitational radiation and the binary pulsar PSR 1913+16}},
  \href{https://doi.org/10.1086/159690}{\emph{\apj} {\bfseries 253} (1982)
  908}.

\bibitem{Franciolini:2021tla}
G.~Franciolini, V.~Baibhav, V.~De~Luca, K.K.Y.~Ng, K.W.K.~Wong, E.~Berti
  et~al., \emph{{Quantifying the evidence for primordial black holes in
  LIGO/Virgo gravitational-wave data}},
  \href{https://arxiv.org/abs/2105.03349}{{\ttfamily 2105.03349}}.

\bibitem{Franciolini:2021xbq}
G.~Franciolini, R.~Cotesta, N.~Loutrel, E.~Berti, P.~Pani and A.~Riotto,
  \emph{{How to assess the primordial origin of single gravitational-wave
  events with mass, spin, eccentricity, and deformability measurements}},
  \href{https://arxiv.org/abs/2112.10660}{{\ttfamily 2112.10660}}.

\bibitem{Hutsi:2020sol}
G.~H\"utsi, M.~Raidal, V.~Vaskonen and H.~Veerm\"ae, \emph{{Two populations of
  LIGO-Virgo black holes}},
  \href{https://doi.org/10.1088/1475-7516/2021/03/068}{\emph{JCAP} {\bfseries
  03} (2021) 068} [\href{https://arxiv.org/abs/2012.02786}{{\ttfamily
  2012.02786}}].

\bibitem{1931ApJ....74...81C}
S.~{Chandrasekhar}, \emph{{The Maximum Mass of Ideal White Dwarfs}},
  \href{https://doi.org/10.1086/143324}{\emph{{Astrophys.~J.}} {\bfseries 74}
  (1931) 81}.

\bibitem{LIGOScientific:2019kan}
{\scshape LIGO Scientific, Virgo} collaboration, \emph{{Search for Subsolar
  Mass Ultracompact Binaries in Advanced LIGO\textquoteright{}s Second
  Observing Run}},
  \href{https://doi.org/10.1103/PhysRevLett.123.161102}{\emph{Phys. Rev. Lett.}
  {\bfseries 123} (2019) 161102}
  [\href{https://arxiv.org/abs/1904.08976}{{\ttfamily 1904.08976}}].

\bibitem{LIGOScientific:2021job}
{\scshape LIGO Scientific, VIRGO, KAGRA} collaboration, \emph{{Search for
  subsolar-mass binaries in the first half of Advanced LIGO and Virgo's third
  observing run}},  \href{https://arxiv.org/abs/2109.12197}{{\ttfamily
  2109.12197}}.

\bibitem{Lasky:2015lej}
P.D.~Lasky et~al., \emph{{Gravitational-wave cosmology across 29 decades in
  frequency}}, \href{https://doi.org/10.1103/PhysRevX.6.011035}{\emph{Phys.
  Rev. X} {\bfseries 6} (2016) 011035}
  [\href{https://arxiv.org/abs/1511.05994}{{\ttfamily 1511.05994}}].

\bibitem{PhysRevD.30.272}
E.~Witten, \emph{Cosmic separation of phases},
  \href{https://doi.org/10.1103/PhysRevD.30.272}{\emph{Phys. Rev. D} {\bfseries
  30} (1984) 272}.

\bibitem{Kamionkowski:1993fg}
M.~Kamionkowski, A.~Kosowsky and M.S.~Turner, \emph{{Gravitational radiation
  from first order phase transitions}},
  \href{https://doi.org/10.1103/PhysRevD.49.2837}{\emph{Phys. Rev. D}
  {\bfseries 49} (1994) 2837}
  [\href{https://arxiv.org/abs/astro-ph/9310044}{{\ttfamily
  astro-ph/9310044}}].

\bibitem{LIGOScientific:2021nrg}
{\scshape LIGO Scientific, Virgo, KAGRA} collaboration, \emph{{Constraints on
  Cosmic Strings Using Data from the Third Advanced LIGO\textendash{}Virgo
  Observing Run}},
  \href{https://doi.org/10.1103/PhysRevLett.126.241102}{\emph{Phys. Rev. Lett.}
  {\bfseries 126} (2021) 241102}
  [\href{https://arxiv.org/abs/2101.12248}{{\ttfamily 2101.12248}}].

\bibitem{Morras:2021atg}
G.~Morr\'as, J.~Garc\'\i{}a-Bellido and S.~Nesseris, \emph{{Search for black
  hole hyperbolic encounters with gravitational wave detectors}},
  \href{https://doi.org/10.1016/j.dark.2021.100932}{\emph{Phys. Dark Univ.}
  {\bfseries 35} (2022) 100932}
  [\href{https://arxiv.org/abs/2110.08000}{{\ttfamily 2110.08000}}].

\bibitem{Garcia-Bellido:2021jlq}
J.~Garc\'\i{}a-Bellido, S.~Jaraba and S.~Kuroyanagi, \emph{{The stochastic
  gravitational wave background from close hyperbolic encounters of primordial
  black holes in dense clusters}},
  \href{https://doi.org/10.1016/j.dark.2022.101009}{\emph{Phys. Dark Univ.}
  {\bfseries 36} (2022) 101009}
  [\href{https://arxiv.org/abs/2109.11376}{{\ttfamily 2109.11376}}].

\bibitem{Caprini:2018mtu}
C.~Caprini and D.G.~Figueroa, \emph{{Cosmological Backgrounds of Gravitational
  Waves}}, \href{https://doi.org/10.1088/1361-6382/aac608}{\emph{Class. Quant.
  Grav.} {\bfseries 35} (2018) 163001}
  [\href{https://arxiv.org/abs/1801.04268}{{\ttfamily 1801.04268}}].

\bibitem{Maggiore:1999vm}
M.~Maggiore, \emph{{Gravitational wave experiments and early universe
  cosmology}}, \href{https://doi.org/10.1016/S0370-1573(99)00102-7}{\emph{Phys.
  Rept.} {\bfseries 331} (2000) 283}
  [\href{https://arxiv.org/abs/gr-qc/9909001}{{\ttfamily gr-qc/9909001}}].

\bibitem{NANOGrav:2020bcs}
{\scshape NANOGrav} collaboration, \emph{{The NANOGrav 12.5 yr Data Set: Search
  for an Isotropic Stochastic Gravitational-wave Background}},
  \href{https://doi.org/10.3847/2041-8213/abd401}{\emph{Astrophys. J. Lett.}
  {\bfseries 905} (2020) L34}
  [\href{https://arxiv.org/abs/2009.04496}{{\ttfamily 2009.04496}}].

\bibitem{enwiki:1084886995}
{Wikipedia contributors}, ``Apsis --- {Wikipedia}{,} the free encyclopedia.''
  \url{https://en.wikipedia.org/w/index.php?title=Apsis&oldid=1084886995},
  2022.

\bibitem{BNSmerger}
{LIGO Scientific Collaboration}, {Virgo Collaboration}, B.P.~Abbott et~al.,
  \emph{{GW170817: Observation of Gravitational Waves from a Binary Neutron
  Star Inspiral}},
  \href{https://doi.org/10.1103/PhysRevLett.119.161101}{\emph{Phys. Rev. Lett.}
  {\bfseries 119} (2017) 161101}.

\bibitem{Belgacem:2019tbw}
E.~Belgacem, Y.~Dirian, S.~Foffa, E.J.~Howell, M.~Maggiore and T.~Regimbau,
  \emph{{Cosmology and dark energy from joint gravitational wave-GRB
  observations}},
  \href{https://doi.org/10.1088/1475-7516/2019/08/015}{\emph{JCAP} {\bfseries
  08} (2019) 015} [\href{https://arxiv.org/abs/1907.01487}{{\ttfamily
  1907.01487}}].

\bibitem{Maggiore2007}
M.~Maggiore, \emph{Gravitational Waves: Volume 1: Theory and Experiments},
  Oxford University Press (2007).

\bibitem{2021NatRP...3..344B}
M.~{Bailes}, B.K.~{Berger}, P.R.~{Brady}, M.~{Branchesi}, K.~{Danzmann},
  M.~{Evans} et~al., \emph{{Gravitational-wave physics and astronomy in the
  2020s and 2030s}},
  \href{https://doi.org/10.1038/s42254-021-00303-8}{\emph{Nature Reviews
  Physics} {\bfseries 3} (2021) 344}.

\bibitem{2015CQGra..32g4001L}
{LIGO Scientific Collaboration}, \emph{{Advanced LIGO}},
  \href{https://doi.org/10.1088/0264-9381/32/7/074001}{\emph{Classical and
  Quantum Gravity} {\bfseries 32} (2015) 074001}
  [\href{https://arxiv.org/abs/1411.4547}{{\ttfamily 1411.4547}}].

\bibitem{VIRGO:2014yos}
{\scshape VIRGO} collaboration, \emph{{Advanced Virgo: a second-generation
  interferometric gravitational wave detector}},
  \href{https://doi.org/10.1088/0264-9381/32/2/024001}{\emph{Class. Quant.
  Grav.} {\bfseries 32} (2015) 024001}
  [\href{https://arxiv.org/abs/1408.3978}{{\ttfamily 1408.3978}}].

\bibitem{KAGRA:2018plz}
{\scshape KAGRA} collaboration, \emph{{KAGRA: 2.5 Generation Interferometric
  Gravitational Wave Detector}},
  \href{https://doi.org/10.1038/s41550-018-0658-y}{\emph{Nature Astron.}
  {\bfseries 3} (2019) 35} [\href{https://arxiv.org/abs/1811.08079}{{\ttfamily
  1811.08079}}].

\bibitem{Dooley:2014nga}
{\scshape LIGO Scientific} collaboration, \emph{{Status of GEO 600}},
  \href{https://doi.org/10.1088/1742-6596/610/1/012015}{\emph{J. Phys. Conf.
  Ser.} {\bfseries 610} (2015) 012015}
  [\href{https://arxiv.org/abs/1411.6588}{{\ttfamily 1411.6588}}].

\bibitem{Punturo:2010zz}
M.~Punturo et~al., \emph{{The Einstein Telescope: A third-generation
  gravitational wave observatory}},
  \href{https://doi.org/10.1088/0264-9381/27/19/194002}{\emph{Class. Quant.
  Grav.} {\bfseries 27} (2010) 194002}.

\bibitem{LIGOScientific:2016wof}
{\scshape LIGO Scientific} collaboration, \emph{{Exploring the Sensitivity of
  Next Generation Gravitational Wave Detectors}},
  \href{https://doi.org/10.1088/1361-6382/aa51f4}{\emph{Class. Quant. Grav.}
  {\bfseries 34} (2017) 044001}
  [\href{https://arxiv.org/abs/1607.08697}{{\ttfamily 1607.08697}}].

\bibitem{Klein:2015hvg}
A.~Klein et~al., \emph{{Science with the space-based interferometer eLISA:
  Supermassive black hole binaries}},
  \href{https://doi.org/10.1103/PhysRevD.93.024003}{\emph{Phys. Rev. D}
  {\bfseries 93} (2016) 024003}
  [\href{https://arxiv.org/abs/1511.05581}{{\ttfamily 1511.05581}}].

\bibitem{Amaro-Seoane2017}
P.e.a.~{Amaro-Seoane}, \emph{{Laser Interferometer Space Antenna}},
  {\emph{arXiv e-prints} (2017) arXiv:1702.00786}
  [\href{https://arxiv.org/abs/1702.00786}{{\ttfamily 1702.00786}}].

\bibitem{Kawamura:2020pcg}
S.~Kawamura et~al., \emph{{Current status of space gravitational wave antenna
  DECIGO and B-DECIGO}},
  \href{https://doi.org/10.1093/ptep/ptab019}{\emph{PTEP} {\bfseries 2021}
  (2021) 05A105} [\href{https://arxiv.org/abs/2006.13545}{{\ttfamily
  2006.13545}}].

\bibitem{Gong:2021gvw}
Y.~Gong, J.~Luo and B.~Wang, \emph{{Concepts and status of Chinese space
  gravitational wave detection projects}},
  \href{https://doi.org/10.1038/s41550-021-01480-3}{\emph{Nature Astron.}
  {\bfseries 5} (2021) 881} [\href{https://arxiv.org/abs/2109.07442}{{\ttfamily
  2109.07442}}].

\bibitem{1978SvA....22...36S}
M.V.~{Sazhin}, \emph{{Opportunities for detecting ultralong gravitational
  waves}}, {\emph{\sovast} {\bfseries 22} (1978) 36}.

\bibitem{1979ApJ...234.1100D}
S.~{Detweiler}, \emph{{Pulsar timing measurements and the search for
  gravitational waves}}, \href{https://doi.org/10.1086/157593}{\emph{\apj}
  {\bfseries 234} (1979) 1100}.

\bibitem{Hobbs:2013aka}
G.~Hobbs, \emph{{The Parkes Pulsar Timing Array}},
  \href{https://doi.org/10.1088/0264-9381/30/22/224007}{\emph{Class. Quant.
  Grav.} {\bfseries 30} (2013) 224007}
  [\href{https://arxiv.org/abs/1307.2629}{{\ttfamily 1307.2629}}].

\bibitem{Kramer:2013kea}
M.~Kramer and D.J.~Champion, \emph{{The European Pulsar Timing Array and the
  Large European Array for Pulsars}},
  \href{https://doi.org/10.1088/0264-9381/30/22/224009}{\emph{Class. Quant.
  Grav.} {\bfseries 30} (2013) 224009}.

\bibitem{Brazier:2019mmu}
A.~Brazier et~al., \emph{{The NANOGrav Program for Gravitational Waves and
  Fundamental Physics}},  \href{https://arxiv.org/abs/1908.05356}{{\ttfamily
  1908.05356}}.

\bibitem{Manchester:2013ndt}
R.N.~Manchester, \emph{{The International Pulsar Timing Array}},
  \href{https://doi.org/10.1088/0264-9381/30/22/224010}{\emph{Class. Quant.
  Grav.} {\bfseries 30} (2013) 224010}
  [\href{https://arxiv.org/abs/1309.7392}{{\ttfamily 1309.7392}}].

\bibitem{LIGOScientific:2022myk}
{\scshape LIGO Scientific, VIRGO, KAGRA} collaboration, \emph{{First joint
  observation by the underground gravitational-wave detector, KAGRA, with
  GEO600}},  \href{https://arxiv.org/abs/2203.01270}{{\ttfamily 2203.01270}}.

\bibitem{LIGOScientific:2021psn}
{\scshape LIGO Scientific, VIRGO, KAGRA} collaboration, \emph{{The population
  of merging compact binaries inferred using gravitational waves through
  GWTC-3}},  \href{https://arxiv.org/abs/2111.03634}{{\ttfamily 2111.03634}}.

\bibitem{LIGOScientific:2020kqk}
{\scshape LIGO Scientific, Virgo} collaboration, \emph{{Population Properties
  of Compact Objects from the Second LIGO-Virgo Gravitational-Wave Transient
  Catalog}}, \href{https://doi.org/10.3847/2041-8213/abe949}{\emph{Astrophys.
  J. Lett.} {\bfseries 913} (2021) L7}
  [\href{https://arxiv.org/abs/2010.14533}{{\ttfamily 2010.14533}}].

\bibitem{LIGOScientific:2020iuh}
{\scshape LIGO Scientific, Virgo} collaboration, \emph{{GW190521: A Binary
  Black Hole Merger with a Total Mass of $150 M_{\odot}$}},
  \href{https://doi.org/10.1103/PhysRevLett.125.101102}{\emph{Phys. Rev. Lett.}
  {\bfseries 125} (2020) 101102}
  [\href{https://arxiv.org/abs/2009.01075}{{\ttfamily 2009.01075}}].

\bibitem{LIGOScientific:2020ufj}
{\scshape LIGO Scientific, Virgo} collaboration, \emph{{Properties and
  Astrophysical Implications of the 150 M$_\odot$ Binary Black Hole Merger
  GW190521}}, \href{https://doi.org/10.3847/2041-8213/aba493}{\emph{Astrophys.
  J. Lett.} {\bfseries 900} (2020) L13}
  [\href{https://arxiv.org/abs/2009.01190}{{\ttfamily 2009.01190}}].

\bibitem{LIGOScientific:2020zkf}
{\scshape LIGO Scientific, Virgo} collaboration, \emph{{GW190814: Gravitational
  Waves from the Coalescence of a 23 Solar Mass Black Hole with a 2.6 Solar
  Mass Compact Object}},
  \href{https://doi.org/10.3847/2041-8213/ab960f}{\emph{Astrophys. J. Lett.}
  {\bfseries 896} (2020) L44}
  [\href{https://arxiv.org/abs/2006.12611}{{\ttfamily 2006.12611}}].

\bibitem{Zevin:2020gbd}
M.~Zevin, S.S.~Bavera, C.P.L.~Berry, V.~Kalogera, T.~Fragos, P.~Marchant
  et~al., \emph{{One Channel to Rule Them All? Constraining the Origins of
  Binary Black Holes Using Multiple Formation Pathways}},
  \href{https://doi.org/10.3847/1538-4357/abe40e}{\emph{Astrophys. J.}
  {\bfseries 910} (2021) 152}
  [\href{https://arxiv.org/abs/2011.10057}{{\ttfamily 2011.10057}}].

\bibitem{Mapelli:2021gyv}
M.~Mapelli, Y.~Bouffanais, F.~Santoliquido, M.A.~Sedda and M.C.~Artale,
  \emph{{The cosmic evolution of binary black holes in young, globular, and
  nuclear star clusters: rates, masses, spins, and mixing fractions}},
  \href{https://doi.org/10.1093/mnras/stac422}{\emph{Mon. Not. Roy. Astron.
  Soc.} {\bfseries 511} (2022) 5797}
  [\href{https://arxiv.org/abs/2109.06222}{{\ttfamily 2109.06222}}].

\bibitem{Suyama:2019cst}
T.~Suyama and S.~Yokoyama, \emph{{Clustering of primordial black holes with
  non-Gaussian initial fluctuations}},
  \href{https://doi.org/10.1093/ptep/ptz105}{\emph{PTEP} {\bfseries 2019}
  (2019) 103E02} [\href{https://arxiv.org/abs/1906.04958}{{\ttfamily
  1906.04958}}].

\bibitem{PhysRevD.99.063532}
T.~Bringmann, P.F.~Depta, V.~Domcke and K.~Schmidt-Hoberg, \emph{Towards
  closing the window of primordial black holes as dark matter: The case of
  large clustering},
  \href{https://doi.org/10.1103/PhysRevD.99.063532}{\emph{Phys. Rev. D}
  {\bfseries 99} (2019) 063532}.

\bibitem{Caputo:2020irr}
A.~Caputo, L.~Sberna, A.~Toubiana, S.~Babak, E.~Barausse, S.~Marsat et~al.,
  \emph{{Gravitational-wave detection and parameter estimation for accreting
  black-hole binaries and their electromagnetic counterpart}},
  \href{https://doi.org/10.3847/1538-4357/ab7b66}{\emph{Astrophys. J.}
  {\bfseries 892} (2020) 90}
  [\href{https://arxiv.org/abs/2001.03620}{{\ttfamily 2001.03620}}].

\bibitem{DeLuca:2020qqa}
V.~De~Luca, G.~Franciolini, P.~Pani and A.~Riotto, \emph{{Primordial Black
  Holes Confront LIGO/Virgo data: Current situation}},
  \href{https://doi.org/10.1088/1475-7516/2020/06/044}{\emph{JCAP} {\bfseries
  06} (2020) 044} [\href{https://arxiv.org/abs/2005.05641}{{\ttfamily
  2005.05641}}].

\bibitem{DeLuca:2020jug}
V.~De~Luca, V.~Desjacques, G.~Franciolini and A.~Riotto, \emph{{The clustering
  evolution of primordial black holes}},
  \href{https://doi.org/10.1088/1475-7516/2020/11/028}{\emph{JCAP} {\bfseries
  11} (2020) 028} [\href{https://arxiv.org/abs/2009.04731}{{\ttfamily
  2009.04731}}].

\bibitem{Jedamzik:2020ypm}
K.~Jedamzik, \emph{{Primordial Black Hole Dark Matter and the LIGO/Virgo
  observations}},
  \href{https://doi.org/10.1088/1475-7516/2020/09/022}{\emph{JCAP} {\bfseries
  09} (2020) 022} [\href{https://arxiv.org/abs/2006.11172}{{\ttfamily
  2006.11172}}].

\bibitem{Jedamzik:2020omx}
K.~Jedamzik, \emph{{Consistency of Primordial Black Hole Dark Matter with
  LIGO/Virgo Merger Rates}},
  \href{https://doi.org/10.1103/PhysRevLett.126.051302}{\emph{Phys. Rev. Lett.}
  {\bfseries 126} (2021) 051302}
  [\href{https://arxiv.org/abs/2007.03565}{{\ttfamily 2007.03565}}].

\bibitem{Tkachev:2020uin}
M.~Tkachev, S.~Pilipenko and G.~Yepes, \emph{{Dark Matter Simulations with
  Primordial Black Holes in the Early Universe}},
  \href{https://doi.org/10.1093/mnras/staa3103}{\emph{Mon. Not. Roy. Astron.
  Soc.} {\bfseries 499} (2020) 4854}
  [\href{https://arxiv.org/abs/2009.07813}{{\ttfamily 2009.07813}}].

\bibitem{Mouri:2002mc}
H.~Mouri and Y.~Taniguchi, \emph{{Runaway merging of black holes: analytical
  constraint on the timescale}},
  \href{https://doi.org/10.1086/339472}{\emph{Astrophys. J. Lett.} {\bfseries
  566} (2002) L17} [\href{https://arxiv.org/abs/astro-ph/0201102}{{\ttfamily
  astro-ph/0201102}}].

\bibitem{1989ApJ...343..725Q}
G.D.~{Quinlan} and S.L.~{Shapiro}, \emph{{Dynamical Evolution of Dense Clusters
  of Compact Stars}}, \href{https://doi.org/10.1086/167745}{\emph{\apj}
  {\bfseries 343} (1989) 725}.

\bibitem{Korol:2019jud}
V.~Korol, I.~Mandel, M.C.~Miller, R.P.~Church and M.B.~Davies, \emph{{Merger
  rates in primordial black hole clusters without initial binaries}},
  \href{https://doi.org/10.1093/mnras/staa1644}{\emph{Mon. Not. Roy. Astron.
  Soc.} {\bfseries 496} (2020) 994}
  [\href{https://arxiv.org/abs/1911.03483}{{\ttfamily 1911.03483}}].

\bibitem{Franciolini:2022ewd}
G.~Franciolini, K.~Kritos, E.~Berti and J.~Silk, \emph{{Primordial black hole
  mergers from three-body interactions}},
  \href{https://arxiv.org/abs/2205.15340}{{\ttfamily 2205.15340}}.

\bibitem{Chisholm:2005vm}
J.R.~Chisholm, \emph{{Clustering of primordial black holes: basic results}},
  \href{https://doi.org/10.1103/PhysRevD.73.083504}{\emph{Phys. Rev.}
  {\bfseries D73} (2006) 083504}
  [\href{https://arxiv.org/abs/astro-ph/0509141}{{\ttfamily
  astro-ph/0509141}}].

\bibitem{Chisholm:2011kn}
J.R.~Chisholm, \emph{{Clustering of Primordial Black Holes. II. Evolution of
  Bound Systems}},
  \href{https://doi.org/10.1103/PhysRevD.84.124031}{\emph{Phys. Rev. D}
  {\bfseries 84} (2011) 124031}
  [\href{https://arxiv.org/abs/1110.4402}{{\ttfamily 1110.4402}}].

\bibitem{gw-horizon-plot}
E.~Hall, ``{gw-horizon-plot [Git Repository]}.''
  \url{https://git.ligo.org/evan.hall/gw-horizon-plot}, 2019.

\bibitem{Oguri:2017ock}
M.~Oguri, J.M.~Diego, N.~Kaiser, P.L.~Kelly and T.~Broadhurst,
  \emph{{Understanding caustic crossings in giant arcs: characteristic scales,
  event rates, and constraints on compact dark matter}},
  \href{https://doi.org/10.1103/PhysRevD.97.023518}{\emph{Phys. Rev. D}
  {\bfseries 97} (2018) 023518}
  [\href{https://arxiv.org/abs/1710.00148}{{\ttfamily 1710.00148}}].

\bibitem{PBHbounds}
B.J.~Kavanagh, ``{PBHbounds [Code v1.0, accessed 03/03/2022]}.''
  \url{https://github.com/bradkav/PBHbounds},
  \href{https://doi.org/10.5281/zenodo.3538998}{DOI:10.5281/zenodo.3538998},
  2022.
\newblock 10.5281/zenodo.3538998.

\bibitem{2012Sci...337..444S}
H.~{Sana}, S.E.~{de Mink}, A.~{de Koter}, N.~{Langer}, C.J.~{Evans},
  M.~{Gieles} et~al., \emph{{Binary Interaction Dominates the Evolution of
  Massive Stars}},
  \href{https://doi.org/10.1126/science.1223344}{\emph{Science} {\bfseries 337}
  (2012) 444} [\href{https://arxiv.org/abs/1207.6397}{{\ttfamily 1207.6397}}].

\bibitem{2020FrASS...7...38M}
M.~{Mapelli}, \emph{{Binary black hole mergers: formation and populations}},
  \href{https://doi.org/10.3389/fspas.2020.00038}{\emph{Frontiers in Astronomy
  and Space Sciences} {\bfseries 7} (2020) 38}
  [\href{https://arxiv.org/abs/2105.12455}{{\ttfamily 2105.12455}}].

\bibitem{Celoria:2018mzr}
M.~Celoria, R.~Oliveri, A.~Sesana and M.~Mapelli, \emph{{Lecture notes on black
  hole binary astrophysics}},  7, 2018
  [\href{https://arxiv.org/abs/1807.11489}{{\ttfamily 1807.11489}}].

\bibitem{Barack:2018yly}
L.~Barack et~al., \emph{{Black holes, gravitational waves and fundamental
  physics: a roadmap}},
  \href{https://doi.org/10.1088/1361-6382/ab0587}{\emph{Class. Quant. Grav.}
  {\bfseries 36} (2019) 143001}
  [\href{https://arxiv.org/abs/1806.05195}{{\ttfamily 1806.05195}}].

\bibitem{2013A&ARv..21...59I}
N.~{Ivanova}, S.~{Justham}, X.~{Chen}, O.~{De Marco}, C.L.~{Fryer},
  E.~{Gaburov} et~al., \emph{{Common envelope evolution: where we stand and how
  we can move forward}},
  \href{https://doi.org/10.1007/s00159-013-0059-2}{\emph{\aapr} {\bfseries 21}
  (2013) 59} [\href{https://arxiv.org/abs/1209.4302}{{\ttfamily 1209.4302}}].

\bibitem{Fragos:2019box}
T.~Fragos, J.J.~Andrews, E.~Ramirez-Ruiz, G.~Meynet, V.~Kalogera, R.E.~Taam
  et~al., \emph{{The Complete Evolution of a Neutron-Star Binary through a
  Common Envelope Phase Using 1D Hydrodynamic Simulations}},
  \href{https://doi.org/10.3847/2041-8213/ab40d1}{\emph{Astrophys. J. Lett.}
  {\bfseries 883} (2019) L45}
  [\href{https://arxiv.org/abs/1907.12573}{{\ttfamily 1907.12573}}].

\bibitem{Samsing:2013kua}
J.~Samsing, M.~MacLeod and E.~Ramirez-Ruiz, \emph{{The Formation of Eccentric
  Compact Binary Inspirals and the Role of Gravitational Wave Emission in
  Binary-Single Stellar Encounters}},
  \href{https://doi.org/10.1088/0004-637X/784/1/71}{\emph{Astrophys. J.}
  {\bfseries 784} (2014) 71} [\href{https://arxiv.org/abs/1308.2964}{{\ttfamily
  1308.2964}}].

\bibitem{Rodriguez:2016kxx}
C.L.~Rodriguez, S.~Chatterjee and F.A.~Rasio, \emph{{Binary Black Hole Mergers
  from Globular Clusters: Masses, Merger Rates, and the Impact of Stellar
  Evolution}}, \href{https://doi.org/10.1103/PhysRevD.93.084029}{\emph{Phys.
  Rev. D} {\bfseries 93} (2016) 084029}
  [\href{https://arxiv.org/abs/1602.02444}{{\ttfamily 1602.02444}}].

\bibitem{Antonini:2016gqe}
F.~Antonini and F.A.~Rasio, \emph{{Merging black hole binaries in galactic
  nuclei: implications for advanced-LIGO detections}},
  \href{https://doi.org/10.3847/0004-637X/831/2/187}{\emph{Astrophys. J.}
  {\bfseries 831} (2016) 187}
  [\href{https://arxiv.org/abs/1606.04889}{{\ttfamily 1606.04889}}].

\bibitem{1975MNRAS.173..729H}
D.C.~{Heggie}, \emph{{Binary evolution in stellar dynamics.}},
  \href{https://doi.org/10.1093/mnras/173.3.729}{\emph{\mnras} {\bfseries 173}
  (1975) 729}.

\bibitem{1975AJ.....80..809H}
J.G.~{Hills}, \emph{{Encounters between binary and single stars and their
  effect on the dynamical evolution of stellar systems.}},
  \href{https://doi.org/10.1086/111815}{\emph{\aj} {\bfseries 80} (1975) 809}.

\bibitem{DiCarlo:2019pmf}
U.N.~Di~Carlo, N.~Giacobbo, M.~Mapelli, M.~Pasquato, M.~Spera, L.~Wang et~al.,
  \emph{{Merging black holes in young star clusters}},
  \href{https://doi.org/10.1093/mnras/stz1453}{\emph{Mon. Not. Roy. Astron.
  Soc.} {\bfseries 487} (2019) 2947}
  [\href{https://arxiv.org/abs/1901.00863}{{\ttfamily 1901.00863}}].

\bibitem{SFR_MadauReview}
P.~Madau and M.~Dickinson, \emph{{Cosmic Star Formation History}},
  \href{https://doi.org/10.1146/annurev-astro-081811-125615}{\emph{Ann. Rev.
  Astron. Astrophys.} {\bfseries 52} (2014) 415}
  [\href{https://arxiv.org/abs/1403.0007}{{\ttfamily 1403.0007}}].

\bibitem{Vangioni:2014axa}
E.~Vangioni, K.A.~Olive, T.~Prestegard, J.~Silk, P.~Petitjean and V.~Mandic,
  \emph{{The Impact of Star Formation and Gamma-Ray Burst Rates at High
  Redshift on Cosmic Chemical Evolution and Reionization}},
  \href{https://doi.org/10.1093/mnras/stu2600}{\emph{Mon. Not. Roy. Astron.
  Soc.} {\bfseries 447} (2015) 2575}
  [\href{https://arxiv.org/abs/1409.2462}{{\ttfamily 1409.2462}}].

\bibitem{Nagamine:2003bd}
K.~Nagamine, V.~Springel and L.~Hernquist, \emph{{Abundance of damped
  Lyman-alpha absorbers in cosmological SPH simulations}},
  \href{https://doi.org/10.1111/j.1365-2966.2004.07393.x}{\emph{Mon. Not. Roy.
  Astron. Soc.} {\bfseries 348} (2004) 421}
  [\href{https://arxiv.org/abs/astro-ph/0302187}{{\ttfamily
  astro-ph/0302187}}].

\bibitem{Mapelli:2018wys}
M.~Mapelli and N.~Giacobbo, \emph{{The cosmic merger rate of neutron stars and
  black holes}}, \href{https://doi.org/10.1093/mnras/sty1613}{\emph{Mon. Not.
  Roy. Astron. Soc.} {\bfseries 479} (2018) 4391}
  [\href{https://arxiv.org/abs/1806.04866}{{\ttfamily 1806.04866}}].

\bibitem{Belczynski:2017gds}
K.~Belczynski et~al., \emph{{Evolutionary roads leading to low effective spins,
  high black hole masses, and O1/O2 rates for LIGO/Virgo binary black holes}},
  \href{https://doi.org/10.1051/0004-6361/201936528}{\emph{Astron. Astrophys.}
  {\bfseries 636} (2020) A104}
  [\href{https://arxiv.org/abs/1706.07053}{{\ttfamily 1706.07053}}].

\bibitem{Dvorkin:2016wac}
I.~Dvorkin, E.~Vangioni, J.~Silk, J.-P.~Uzan and K.A.~Olive,
  \emph{{Metallicity-constrained merger rates of binary black holes and the
  stochastic gravitational wave background}},
  \href{https://doi.org/10.1093/mnras/stw1477}{\emph{Mon. Not. Roy. Astron.
  Soc.} {\bfseries 461} (2016) 3877}
  [\href{https://arxiv.org/abs/1604.04288}{{\ttfamily 1604.04288}}].

\bibitem{Mukherjee:2021ags}
S.~Mukherjee and J.~Silk, \emph{{Can we distinguish astrophysical from
  primordial black holes via the stochastic gravitational wave background?}},
  \href{https://doi.org/10.1093/mnras/stab1932}{\emph{Mon. Not. Roy. Astron.
  Soc.} {\bfseries 506} (2021) 3977}
  [\href{https://arxiv.org/abs/2105.11139}{{\ttfamily 2105.11139}}].

\bibitem{Belczynski:2016obo}
K.~Belczynski, D.E.~Holz, T.~Bulik and R.~O'Shaughnessy, \emph{{The first
  gravitational-wave source from the isolated evolution of two 40-100 Msun
  stars}}, \href{https://doi.org/10.1038/nature18322}{\emph{Nature} {\bfseries
  534} (2016) 512} [\href{https://arxiv.org/abs/1602.04531}{{\ttfamily
  1602.04531}}].

\bibitem{Chen:2019irf}
Z.-C.~Chen and Q.-G.~Huang, \emph{{Distinguishing Primordial Black Holes from
  Astrophysical Black Holes by Einstein Telescope and Cosmic Explorer}},
  \href{https://arxiv.org/abs/1904.02396}{{\ttfamily 1904.02396}}.

\bibitem{Hild:2011np}
S.~Hild, \emph{{Beyond the Second Generation of Laser-Interferometric
  Gravitational Wave Observatories}},
  \href{https://doi.org/10.1088/0264-9381/29/12/124006}{\emph{Class. Quant.
  Grav.} {\bfseries 29} (2012) 124006}
  [\href{https://arxiv.org/abs/1111.6277}{{\ttfamily 1111.6277}}].

\bibitem{Vitale:2016icu}
S.~Vitale and M.~Evans, \emph{{Parameter estimation for binary black holes with
  networks of third generation gravitational-wave detectors}},
  \href{https://doi.org/10.1103/PhysRevD.95.064052}{\emph{Phys. Rev. D}
  {\bfseries 95} (2017) 064052}
  [\href{https://arxiv.org/abs/1610.06917}{{\ttfamily 1610.06917}}].

\bibitem{2011GReGr..43..623B}
M.G.~{Beker}, G.~{Cella}, R.~{Desalvo}, M.~{Doets}, H.~{Grote}, J.~{Harms}
  et~al., \emph{{Improving the sensitivity of future GW observatories in the
  1-10 Hz band: Newtonian and seismic noise}},
  \href{https://doi.org/10.1007/s10714-010-1011-7}{\emph{General Relativity and
  Gravitation} {\bfseries 43} (2011) 623}.

\bibitem{Koushiappas:2017kqm}
S.M.~Koushiappas and A.~Loeb, \emph{{Maximum redshift of gravitational wave
  merger events}},
  \href{https://doi.org/10.1103/PhysRevLett.119.221104}{\emph{Phys. Rev. Lett.}
  {\bfseries 119} (2017) 221104}
  [\href{https://arxiv.org/abs/1708.07380}{{\ttfamily 1708.07380}}].

\bibitem{DeLuca:2021wjr}
V.~De~Luca, G.~Franciolini, P.~Pani and A.~Riotto, \emph{{Bayesian Evidence for
  Both Astrophysical and Primordial Black Holes: Mapping the GWTC-2 Catalog to
  Third-Generation Detectors}},
  \href{https://doi.org/10.1088/1475-7516/2021/05/003}{\emph{JCAP} {\bfseries
  05} (2021) 003} [\href{https://arxiv.org/abs/2102.03809}{{\ttfamily
  2102.03809}}].

\bibitem{DeLuca:2021hde}
V.~De~Luca, G.~Franciolini, P.~Pani and A.~Riotto, \emph{{The Minimum Testable
  Abundance of Primordial Black Holes at Future Gravitational-Wave Detectors}},
   \href{https://arxiv.org/abs/2106.13769}{{\ttfamily 2106.13769}}.

\bibitem{Ng:2021sqn}
K.K.Y.~Ng, S.~Chen, B.~Goncharov, U.~Dupletsa, S.~Borhanian, M.~Branchesi
  et~al., \emph{{On the single-event-based identification of primordial black
  hole mergers at cosmological distances}},
  \href{https://arxiv.org/abs/2108.07276}{{\ttfamily 2108.07276}}.

\bibitem{Atal:2022zux}
V.~Atal, J.J.~Blanco-Pillado, A.~Sanglas and N.~Triantafyllou,
  \emph{{Constraining changes in the merger history of (P)BH binaries with the
  stochastic gravitational wave background}},
  \href{https://arxiv.org/abs/2201.12218}{{\ttfamily 2201.12218}}.

\bibitem{Braglia:2022icu}
M.~Braglia, J.~Garcia-Bellido and S.~Kuroyanagi, \emph{{Tracking the origin of
  black holes with the stochastic gravitational wave background popcorn
  signal}},  \href{https://arxiv.org/abs/2201.13414}{{\ttfamily 2201.13414}}.

\bibitem{Aghanim:2018eyx}
{\scshape Planck} collaboration, \emph{{Planck 2018 results. VI. Cosmological
  parameters}},  \href{https://arxiv.org/abs/1807.06209}{{\ttfamily
  1807.06209}}.

\bibitem{Abbott:2016xvh}
B.P.~Abbott et~al., \emph{{Sensitivity of the Advanced LIGO detectors at the
  beginning of gravitational wave astronomy}},
  \href{https://doi.org/10.1103/PhysRevD.93.112004}{\emph{Phys. Rev. D}
  {\bfseries 93} (2016) 112004}
  [\href{https://arxiv.org/abs/1604.00439}{{\ttfamily 1604.00439}}].

\bibitem{Li:2013lza}
T.G.F.~Li, \emph{{Extracting Physics from Gravitational Waves: Testing the
  Strong-field Dynamics of General Relativity and Inferring the Large-scale
  Structure of the Universe}}, Ph.D. thesis, Vrije U., Amsterdam, 2013.

\bibitem{Bayes:1764vd}
R.~Bayes, \emph{{An essay toward solving a problem in the doctrine of
  chances}}, \href{https://doi.org/10.1098/rstl.1763.0053}{\emph{Phil. Trans.
  Roy. Soc. Lond.} {\bfseries 53} (1764) 370}.

\bibitem{Torrado:2020dgo}
J.~Torrado and A.~Lewis, \emph{{Cobaya: Code for Bayesian Analysis of
  hierarchical physical models}},
  \href{https://doi.org/10.1088/1475-7516/2021/05/057}{\emph{JCAP} {\bfseries
  05} (2021) 057} [\href{https://arxiv.org/abs/2005.05290}{{\ttfamily
  2005.05290}}].

\bibitem{Lewis:2019xzd}
A.~Lewis, \emph{{GetDist: a Python package for analysing Monte Carlo samples}},
   \href{https://arxiv.org/abs/1910.13970}{{\ttfamily 1910.13970}}.

\bibitem{Trashorras:2020mwn}
M.~Trashorras, J.~Garc\'\i{}a-Bellido and S.~Nesseris, \emph{{The clustering
  dynamics of primordial black boles in $N$-body simulations}},
  \href{https://doi.org/10.3390/universe7010018}{\emph{Universe} {\bfseries 7}
  (2021) 18} [\href{https://arxiv.org/abs/2006.15018}{{\ttfamily 2006.15018}}].

\bibitem{Canas-Herrera:2021qxs}
G.~Ca\~nas Herrera, O.~Contigiani and V.~Vardanyan, \emph{{Learning How to
  Surf: Reconstructing the Propagation and Origin of Gravitational Waves with
  Gaussian Processes}},
  \href{https://doi.org/10.3847/1538-4357/ac09e3}{\emph{Astrophys. J.}
  {\bfseries 918} (2021) 20}
  [\href{https://arxiv.org/abs/2105.04262}{{\ttfamily 2105.04262}}].

\bibitem{deSouza2011Aaanda}
R.S.~{de Souza}, N.~{Yoshida} and K.~{Ioka}, \emph{{Populations III.1 and III.2
  gamma-ray bursts: constraints on the event rate for future radio and X-ray
  surveys}}, \href{https://doi.org/10.1051/0004-6361/201117242}{\emph{\aap}
  {\bfseries 533} (2011) A32}
  [\href{https://arxiv.org/abs/1105.2395}{{\ttfamily 1105.2395}}].

\bibitem{Ng:2020qpk}
K.K.Y.~Ng, S.~Vitale, W.M.~Farr and C.L.~Rodriguez, \emph{{Probing multiple
  populations of compact binaries with third-generation gravitational-wave
  detectors}}, \href{https://doi.org/10.3847/2041-8213/abf8be}{\emph{Astrophys.
  J. Lett.} {\bfseries 913} (2021) L5}
  [\href{https://arxiv.org/abs/2012.09876}{{\ttfamily 2012.09876}}].

\bibitem{Ng:2022agi}
K.K.Y.~Ng, G.~Franciolini, E.~Berti, P.~Pani, A.~Riotto and S.~Vitale,
  \emph{{Constraining high-redshift stellar-mass primordial black holes with
  next-generation ground-based gravitational-wave detectors}},
  \href{https://arxiv.org/abs/2204.11864}{{\ttfamily 2204.11864}}.

\bibitem{Gunn:1972sv}
J.E.~Gunn and J.R.~Gott, III, \emph{{On the Infall of Matter into Clusters of
  Galaxies and Some Effects on Their Evolution}},
  \href{https://doi.org/10.1086/151605}{\emph{Astrophys. J.} {\bfseries 176}
  (1972) 1}.

\bibitem{1984ApJ...281....1F}
J.A.~{Fillmore} and P.~{Goldreich}, \emph{{Self-similar gravitational collapse
  in an expanding universe}}, \href{https://doi.org/10.1086/162070}{\emph{\apj}
  {\bfseries 281} (1984) 1}.

\bibitem{1985ApJS...58...39B}
E.~{Bertschinger}, \emph{{Self-similar secondary infall and accretion in an
  Einstein-de Sitter universe}},
  \href{https://doi.org/10.1086/191028}{\emph{\apjs} {\bfseries 58} (1985) 39}.

\bibitem{Quinlan:1996bw}
G.D.~Quinlan, \emph{{The Time scale for core collapse in spherical star
  clusters}}, \href{https://doi.org/10.1016/S1384-1076(96)00018-8}{\emph{New
  Astron.} {\bfseries 1} (1996) 255}
  [\href{https://arxiv.org/abs/astro-ph/9606182}{{\ttfamily
  astro-ph/9606182}}].

\bibitem{1980lssu.book.....P}
P.J.E.~{Peebles}, \emph{{The large-scale structure of the universe}}, Princeton
  University Press (1980).

\bibitem{Zyla:2020zbs}
{\scshape Particle Data Group} collaboration, \emph{{Review of Particle
  Physics}}, \href{https://doi.org/10.1093/ptep/ptaa104}{\emph{PTEP} {\bfseries
  2020} (2020) 083C01}.

\bibitem{Canas-Herrera:2019npr}
G.~Ca\~nas Herrera, O.~Contigiani and V.~Vardanyan, \emph{{Cross-correlation of
  the astrophysical gravitational-wave background with galaxy clustering}},
  \href{https://doi.org/10.1103/PhysRevD.102.043513}{\emph{Phys. Rev. D}
  {\bfseries 102} (2020) 043513}
  [\href{https://arxiv.org/abs/1910.08353}{{\ttfamily 1910.08353}}].

\bibitem{Mukherjee2021}
S.~Mukherjee, B.D.~Wandelt, S.M.~Nissanke and A.~Silvestri, \emph{Accurate
  precision cosmology with redshift unknown gravitational wave sources},
  \href{https://doi.org/10.1103/PhysRevD.103.043520}{\emph{Phys. Rev. D}
  {\bfseries 103} (2021) 043520}
  [\href{https://arxiv.org/abs/2007.02943}{{\ttfamily 2007.02943}}].

\bibitem{Nuttall2018}
L.K.~{Nuttall}, \emph{{Characterizing transient noise in the LIGO detectors}},
  \href{https://doi.org/10.1098/rsta.2017.0286}{\emph{Philosophical
  Transactions of the Royal Society of London Series A} {\bfseries 376} (2018)
  20170286} [\href{https://arxiv.org/abs/1804.07592}{{\ttfamily 1804.07592}}].

\bibitem{LIGO:2021ppb}
{\scshape LIGO} collaboration, \emph{{LIGO detector characterization in the
  second and third observing runs}},
  \href{https://doi.org/10.1088/1361-6382/abfd85}{\emph{Class. Quant. Grav.}
  {\bfseries 38} (2021) 135014}
  [\href{https://arxiv.org/abs/2101.11673}{{\ttfamily 2101.11673}}].

\bibitem{Straumann:2013spu}
N.~Straumann, \emph{{General Relativity}}, Graduate Texts in Physics, Springer,
  Dordrecht (2013),
  \href{https://doi.org/10.1007/978-94-007-5410-2}{10.1007/978-94-007-5410-2}.

\bibitem{Dalang:2019rke}
C.~Dalang, P.~Fleury and L.~Lombriser, \emph{{Horndeski gravity and standard
  sirens}}, \href{https://doi.org/10.1103/PhysRevD.102.044036}{\emph{Phys. Rev.
  D} {\bfseries 102} (2020) 044036}
  [\href{https://arxiv.org/abs/1912.06117}{{\ttfamily 1912.06117}}].

\bibitem{2014A&A...568A..22B}
M.e.a.~{Betoule}, \emph{{Improved cosmological constraints from a joint
  analysis of the SDSS-II and SNLS supernova samples}},
  \href{https://doi.org/10.1051/0004-6361/201423413}{\emph{Astronomy and
  Astrophysics} {\bfseries 568} (2014) A22}
  [\href{https://arxiv.org/abs/1401.4064}{{\ttfamily 1401.4064}}].

\bibitem{Sathyaprakash:2009xt}
B.S.~Sathyaprakash, B.F.~Schutz and C.~Van Den~Broeck, \emph{{Cosmography with
  the Einstein Telescope}},
  \href{https://doi.org/10.1088/0264-9381/27/21/215006}{\emph{Class. Quant.
  Grav.} {\bfseries 27} (2010) 215006}
  [\href{https://arxiv.org/abs/0906.4151}{{\ttfamily 0906.4151}}].

\bibitem{Zhao:2010sz}
W.~Zhao, C.~Van Den~Broeck, D.~Baskaran and T.G.F.~Li, \emph{{Determination of
  Dark Energy by the Einstein Telescope: Comparing with CMB, BAO and SNIa
  Observations}}, \href{https://doi.org/10.1103/PhysRevD.83.023005}{\emph{Phys.
  Rev.} {\bfseries D83} (2011) 023005}
  [\href{https://arxiv.org/abs/1009.0206}{{\ttfamily 1009.0206}}].

\bibitem{Cai2016}
R.-G.~Cai and T.~Yang, \emph{Estimating cosmological parameters by the
  simulated data of gravitational waves from the einstein telescope},
  {\emph{Physical Review D} {\bfseries 95} (2016) }
  [\href{https://arxiv.org/abs/arxiv:1608.08008v2}{{\ttfamily
  arxiv:1608.08008v2}}].

\bibitem{Du:2018tia}
M.~Du, W.~Yang, L.~Xu, S.~Pan and D.F.~Mota, \emph{{Future constraints on
  dynamical dark-energy using gravitational-wave standard sirens}},
  \href{https://doi.org/10.1103/PhysRevD.100.043535}{\emph{Phys. Rev.}
  {\bfseries D100} (2019) 043535}
  [\href{https://arxiv.org/abs/1812.01440}{{\ttfamily 1812.01440}}].

\bibitem{Jin:2020hmc}
S.-J.~Jin, D.-Z.~He, Y.~Xu, J.-F.~Zhang and X.~Zhang, \emph{{Forecast for
  cosmological parameter estimation with gravitational-wave standard siren
  observation from the Cosmic Explorer}},
  \href{https://arxiv.org/abs/2001.05393}{{\ttfamily 2001.05393}}.

\bibitem{Hogg:2020ktc}
N.B.~Hogg, M.~Martinelli and S.~Nesseris, \emph{{Constraints on the distance
  duality relation with standard sirens}},
  \href{https://doi.org/10.1088/1475-7516/2020/12/019}{\emph{JCAP} {\bfseries
  12} (2020) 019} [\href{https://arxiv.org/abs/2007.14335}{{\ttfamily
  2007.14335}}].

\bibitem{2010MNRAS.405..535J}
J.~{J{\"o}nsson}, M.~{Sullivan}, I.~{Hook}, S.~{Basa}, R.~{Carlberg},
  A.~{Conley} et~al., \emph{{Constraining dark matter halo properties using
  lensed Supernova Legacy Survey supernovae}},
  \href{https://doi.org/10.1111/j.1365-2966.2010.16467.x}{\emph{\mnras}
  {\bfseries 405} (2010) 535}
  [\href{https://arxiv.org/abs/1002.1374}{{\ttfamily 1002.1374}}].

\bibitem{1964SvA.....8...13Z}
Y.B.~{Zel'dovich}, \emph{{Observations in a Universe Homogeneous in the Mean}},
  {\emph{Sov. Astron. Lett.} {\bfseries 8} (1964) 13}.

\bibitem{2008MNRAS.386..230W}
O.~{Wucknitz}, \emph{{From planes to spheres: about gravitational lens
  magnifications}},
  \href{https://doi.org/10.1111/j.1365-2966.2008.13017.x}{\emph{\mnras}
  {\bfseries 386} (2008) 230}
  [\href{https://arxiv.org/abs/0801.3758}{{\ttfamily 0801.3758}}].

\bibitem{2021A&A...655A..54B}
M.-A.~{Breton} and P.~{Fleury}, \emph{{Theoretical and numerical perspectives
  on cosmic distance averages}},
  \href{https://doi.org/10.1051/0004-6361/202040140}{\emph{\aap} {\bfseries
  655} (2021) A54} [\href{https://arxiv.org/abs/2012.07802}{{\ttfamily
  2012.07802}}].

\end{thebibliography}\endgroup
\cleardoublepage




\newpage
\mbox{}

\end{document}